\documentclass[a4paper, twoside=true,  fontsize=12pt,  bibliography=totoc,DIV=15, BCOR = 11 mm]{scrbook}

\usepackage[USenglish]{babel}
\DeclareOldFontCommand{\rm}{\normalfont\rmfamily}{\mathrm}
\DeclareOldFontCommand{\sf}{\normalfont\sffamily}{\mathsf}
\DeclareOldFontCommand{\tt}{\normalfont\ttfamily}{\mathtt}
\DeclareOldFontCommand{\bf}{\normalfont\bfseries}{\mathbf}
\DeclareOldFontCommand{\it}{\normalfont\itshape}{\mathit}
\DeclareOldFontCommand{\sl}{\normalfont\slshape}{\@nomath\sl}
\DeclareOldFontCommand{\sc}{\normalfont\scshape}{\@nomath\sc}
\makeatother

\usepackage[utf8]{inputenc}
\usepackage[table]{xcolor}

\usepackage{sidecap}
\usepackage{graphicx,float,cite,amssymb,subcaption,amsmath,slashed,enumitem}
\usepackage{mathtools}
\usepackage{hyperref}
\usepackage{caption}
\usepackage{xcolor}
\usepackage{dsfont}
\newcommand{\comments}[1]{}

\newcommand\beq{\begin{equation}}
\newcommand\eeq{\end{equation}}

\newcommand{\dd}{\mathrm{d}}

\newcommand\eps{\epsilon}

\DeclareMathOperator{\arcsinh}{arcsinh}

\DeclareMathOperator{\arccot}{arccot}

\DeclareMathOperator{\arcsinhh}{arcsin(h)}
\DeclareMathOperator{\sinhh}{sin(h)}
\renewcommand{\dd}{\text{d}}

\def\MO{{M_5}}

 \usepackage{cancel}
\usepackage{bm}
\title{Non-equilibrium dynamics in Holography}
\author{Sebastian Grieninger }
\date{December 2019}
\usepackage{setspace}
\usepackage{mdframed}
\usepackage{lmodern}

\newlength\longest
\def\coeff#1#2{{\textstyle {\frac {#1}{#2}}}}

\begin{document}
\pagenumbering{roman}
\begin{titlepage}

\begin{center}
\vspace*{3 cm}
\textsf{\textbf{\huge \mbox{Non-equilibrium dynamics in Holography}}}

\vspace{3 cm}
{\huge Dissertation} \\ {\Large zur Erlangung des akademischen Grades \\ {\textit{ doctor rerum naturalium}} (Dr. rer. nat.)}

\end{center}

\begin{figure}[h!]
    \centering
	\includegraphics[width=16cm]{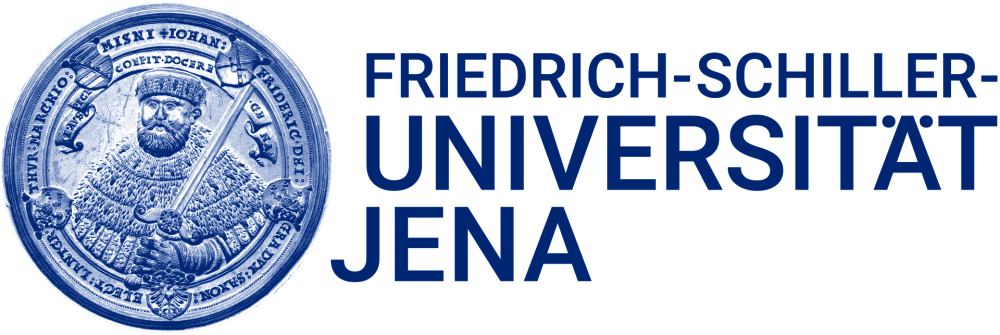}
\end{figure}

\vspace*{\fill}

\noindent
\textbf{vorgelegt dem Rat der Physikalisch-Astronomischen Fakult\"at\\
der Friedrich-Schiller Universit\"at Jena}

\vspace{1cm}
\noindent
\textbf{von M.Sc. Sebastian Leonard Grieninger\\
geboren am 14.08.1991 in Ha\ss furt}

\newpage
\thispagestyle{empty}

\vspace*{\fill}

\minisec{Gutachter}
\begin{enumerate}
  \item Prof. Dr. Martin Ammon  \\ 
  Theoretisch-Physikalisches Institut\\
  Friedrich-Schiller Universität Jena

  \item Prof. Dr. Johanna Erdmenger  \\ 
  Theoretische Physik III\\
  Julius-Maximilian Universität W\"urzburg
  \item  Dr. Daniel Ar\'ean  \\ 
  Instituto de Física Teórica UAM-CSIC \\
  Universidad Autónoma de Madrid
\end{enumerate}

Tag der Disputation: 09.07.2020 

\newpage
\end{titlepage}

\clearpage

\thispagestyle{empty}
\null\vfill

\settowidth\longest{\itshape\hspace{3cm} A witty saying proves nothing.}
%\centering
\begin{center}
\parbox{\longest}{%
  \raggedright{\itshape%
  \hspace{3cm} A witty saying proves nothing. \par\bigskip
  }   
  \raggedleft\MakeUppercase{\hspace{3.3cm} Voltaire}\par%
}
\end{center}

\vfill\vfill

\clearpage
\thispagestyle{empty}
\newpage\null\thispagestyle{empty}\newpage 
\section*{\huge Abstract}\thispagestyle{empty}
In this thesis, we investigate aspects of non-equilibrium dynamics of strongly coupled field theories within holography. While quantum field theories are systematically accessible in the weakly coupled regime, we still lack a standard approach to strongly coupled quantum field theories. One approach which is particularly well suited to study the non-equilibrium dynamics in strongly coupled field theories is holography. Within holography, we relate strongly coupled quantum field theories to weakly coupled gravity in anti de-Sitter space. In the following, we give a brief summary of the main results. 

So far, hydrodynamics -- an effective long range description of systems at finite temperature -- was only considered in the limit of weak magnetic fields. We establish a hydrodynamic description for anomalous quantum field theories subject to strong external field for the first time in the literature. By means of Einstein-Maxwell-Chern-Simons theory in AdS$_5$, we explicitly demonstrate which transport coefficients are non-vanishing due to the chiral anomaly and thus important for the transport behavior.

Spontaneously broken continuous symmetries lead to exciting phenomena such as the appearance of Goldstone bosons in the low energy spectrum. The hydrodynamics for spontaneously broken translational invariance is considered in textbooks; in this thesis, we show that this is more subtle and the textbook treatment has to be revised since the description is missing a novel thermodynamic coefficient. Within the holographic dual to systems with broken translations -- holographic massive gravity -- we lay out a road map for extensions of hydrodynamics to momentum dissipation. Furthermore, we study the imprint of spontaneously broken translations beyond linear response theory in terms of periodically driven strongly coupled quantum field theories.

Another important non-equilibrium scenario specially important for the understanding of our universe is quantum gravity in de-Sitter. Recently, the bold claim of the so-called swampland conjectures has attracted great interest since it banishes all stable theories of quantum gravity on de-Sitter with matter into swampland. Within the well-defined framework of the DS/dS correspondence, we set out to derive consistency conditions on the matter content in de-Sitter. Surprisingly, our proposed bound is violated by any reasonable form of matter. In our discussion, we find a novel one-parameter family of entangling surfaces which interpolates between the two solutions known so far.

The last chapter of this chapter is dedicated to solvable irrelevant deformations in quantum field theory -- the $T\bar T$ deformation. Within holography, we derive the entanglement entropies for generic subintervals in a $T\bar T$ deformed quantum field theory on a sphere. We also resolve the confusion in the literature about a seeming mismatch between the holographic and field theory results for the entanglement entropy in general dimensions.

The thesis is based on my research partly published in~\cite{Ammon:2019apj,Baggioli:2019abx,Baggioli:2019mck,Ammon:2020xyv,Ammon:2020rvg,Geng:2019bnn,Grieninger:2019zts,Baggioli:2020edn}.
\newpage\null\thispagestyle{empty}\newpage
\newpage\section*{\huge Zusammenfassung}
\thispagestyle{empty}
In dieser Dissertation werden Aspekte der Nichtgleichgewichtsdynamik in stark gekoppelten Quantenfeldtheorien im Rahmen von Holographie untersucht. Bisher wurde, im Gegensatz zur perturbativen Entwicklung in schwach gekoppelten Quantenfeldtheorien, noch kein systematischer Zugang zu stark gekoppelten Quantenfeldtheorien gefunden. Ein Zugang, der besonders effektiv für die Nichtgleichgewichtsdynamik funktioniert, ist Holographie. Im Rahmen von Holographie werden stark gekoppelte Quantenfeldtheorien mit schwach gekoppelten Gravitationstheorien in einer Anti-de Sitter Raumzeit in Verbindung setzt. Im Folgenden wird eine kurz Zusammenfassung der Ergebnisse gegeben.

Die Hydrodynamik, welche eine für große Distanzen gültige effektive Beschreibung bei endlicher Temperatur ist, wurde bisher nur für schwache Magnetfelder ausgearbeitet. In dieser Dissertation wird die bisherige Literatur um eine hydrodynamische Beschreibung für anomale Quantenfeldtheorien in starken externen Magnetfeldern erweitert. Des Weiteren werden Transportgrößen in Einstein-Maxwell-Chern-Simons Theorie in AdS$_5$ berechnet und explizit gezeigt, welche neuen Transportgrößen aufgrund der chiralen Anomalie auftreten.

Die spontane Brechung kontinuierlicher Symmetrien führt zu interessanten Effekten wie das Auftreten von Goldstonebosonen im Niederenergiespektrum. Dabei wird die hydrodynamische Theorie für spontan gebrochene Translationsinvarianz als Lehrbuchmaterial angesehen. In der vorliegenden Dissertation wird jedoch gezeigt, dass die Beschreibung nicht vollständig ist und ein neuartiger Transportkoeffizient mit in die Betrachtung einbezogen werden muss. Des Weiteren wird die lineare Antwort für Theorien mit explizit und pseudo-spontan gebrochener Translationsinvarianz  innerhalb von holographisch massiver Gravitation, welche die duale Beschreibung für Theorien mit gebrochener Translationsinvarianz darstellt, ausgearbeitet. Dies stellt einen wichtigen Fortschritt für die Entwicklung einer auf hydrodynamischen Methoden basierenden Beschreibung von Theorien mit schwacher Impulsdissipation dar. Anschließend werden die Effekte von spontan gebrochener Translationsinvarianz auf stark gekoppelte, periodisch getriebene Quantenfeldtheorien ohne die Beschränkung auf lineare Antworttheorie untersucht.

Ein Nichtgleichgewichtsszenarium, welches insbesondere für die Beschreibung unseres Universums wichtig ist, stellt Quantengravitation in de Sitter dar. Diese Diskussion wurde insbesondere durch die kontroversen Sumpfland Vermutungen angeheizt. Die Sumpfland Vermutungen verbannen alle stabilen Quantentgravitationstheorien für de Sitter die Materie beinhalten ins Sumpfland. Diese Vermutung wird in dem wohldefinierten Konzept der DS/dS-Korrespondenz untersucht und eine Konsistenzbedingung an den Materieinhalt postuliert, die sich auf informationstheoretische Argumente stützt. Überaschenderweise wird diese Konsistenzbedingung von jeglicher Form von Materie, die die Null-Energiebedingung der allgemeinen Relativitätstheorie erfüllt, verletzt. Während der Diskussion wurde eine bisher unbekannte einparametrige Schaar von Minimalflächen in de Sitter gefunden, die alle dieselbe Verschränkungsentropie liefern und zwischen den beiden bisher bekannten Lösungen interpolieren. 

Im letzten Kapitel dieser Dissertation wird eine exakt lösbare irrelevante Deformation von Quantenfeldtheorien, die so genannte $T\bar T$-Deformation diskutiert. Unter Zuhilfenahme der neu gefundenen einparametrigen Schaar von Minimalflächen werden die Verschränkunsentropie für beliebige Teilintervalle einer mit einer $T\bar T$ deformierten Quantenfeldtheorie auf einer Sphäre in einem holographischen Modell brechnet. Dabei wird die Unstimmigkeit in der Literatur bezüglich den Resultaten für die Verschränkungentropie gegenüberliegender Punkte gelöst. Bisher gab es eine scheinbare Diskrepanz zwischen den Resultaten von Holographie und Feldtheorie in höheren Dimensionen.  

Diese Dissertation basiert auf meinen Forschungsarbeiten, die bereits teilweise in den folgenden Referenzen veröffentlicht wurden~\cite{Ammon:2019apj,Baggioli:2019abx,Baggioli:2019mck,Ammon:2020xyv,Ammon:2020rvg,Geng:2019bnn,Grieninger:2019zts,Baggioli:2020edn}.
\thispagestyle{empty}\newpage
\setcounter{page}{1}
\pagenumbering{arabic}
\tableofcontents

\chapter{Introduction}
\vspace{-0.25cm}
Modern theoretical physics is based on the universal language of quantum fields, with applications ranging from particle physics, cosmology, and the early universe to condensed matter and statistical physics. For example, the Standard Model of particle physics~\cite{PhysRevLett.19.1264,Glashow:1961tr,Salam:1968rm} (see also references in~\cite{Weinberg_2004}) is based on a relativistic quantum field theory (QFT) which successfully describes all elementary particles and their interactions by bringing together three of the four known fundamental forces (electromagnetic, strong- and weak interactions but not the gravitational force). The predictions of the Standard Model have been experimentally verified up to high precision -- most prominently the discovery of the Higgs boson at the Large Hadron Collider (LHC) in 2012~\cite{Aad:2012tfa,Chatrchyan:2012xdj,Higgs:1964ia,PhysRevLett.13.508,PhysRev.145.1156,PhysRevLett.13.321}.

Moreover, QFT also plays an important role in statistical and condensed matter physics, for example by explaining the universality associated with critical exponents of second order phase transitions. Universality is a prediction of the renormalization group theory of phase transitions, which states that the properties of a system near a phase transition depend only on a small number of features, such as dimensionality and symmetry, and are insensitive to the underlying microscopic properties of the system. 

Within QFT, the coupling constant determines the strength of the associated force and thus the physical interactions. For sufficiently small coupling constants, we may expand the QFT  perturbatively order by order in the coupling constant and determine the scattering amplitudes or other observables. Even though the perturbative approach is very powerful, it is not sufficient for deriving very basic quantities which are only accessible by non-perturbative approaches. For example, using perturbative methods we cannot compute the masses of protons, neutrons, and other bound states, such as mesons and hadrons, from first principles. Another open question in particle physics is how to obtain the phase diagram of quantum chromodynamics at finite temperature and density. In condensed matter, the renowned BCS theory of superconductivity~\cite{PhysRev.106.162,PhysRev.108.1175,PhysRev.97.1724,PhysRev.104.1189} describes the phenomenon of superconductivity in conventional (weakly coupled) superconductors in terms of Cooper pairs. For a certain class of superconductors which exhibit superconductivity up to high temperatures BCS theory breaks down and the corresponding systems are likely strongly coupled and standard methods are no longer applicable.

While we have standard methods to solve problems in weakly coupled QFTs, a standard approach to the strong coupling regime of QFT is still an open research question. Unfortunately, there is no one-size-fits-all approach for calculations in the non-perturbative regime of QFTs, but rather a custom tailoring of a lot of seemingly different methods to specific problems. The standard approaches are lattice QFT techniques, the (functional) renormalization group, and the so-called anti de-Sitter/conformal field theory (AdS/CFT) correspondence.

One particular challenging task within theoretical physics is to describe the out-of-equilibrium dynamics of strongly coupled quantum fields;  this includes the far-from equilibrium dynamics and the near-equilibrium dynamics in the context of linear response theory. In recent years, the central research question in the understanding of the mentioned strongly coupled condensed matter systems was how they react to a time dependent coupling, i.e. a quantum quench~\cite{Ammon:2016fru,Grieninger:2017jxz,Buchel:2013gba,Das:2011nk,Buchel:2014gta,Astaneh:2014fga,Hung:2013dka,Nozaki:2013wia,Buchel:2013lla,Mandal:2013id,Marolf:2013ioa,Buchel:2012gw,Calabrese:2007rg,Sotiriadis:2010si,Albash:2010mv,Basu:2013vva,Basu:2011ft,Das:2019cgl,Ugajin:2013xxa,Das:2010yw,Das:2019qaj,Goykhman:2018ihr,Dymarsky:2017awt,Myers:2017sxr,Caputa:2017ixa,Das:2016eao,Calabrese:2016xau,Basu:2012gg,Das:2016lla,Myers:2016wsu,Coser:2014gsa,Hashimoto:2014yza}. In particular, does the system equilibrate and if so, does it thermalize to a new ground state? The non-equilibrium dynamics is not only interesting for questions within condensed matter physics but also in particle physics fundamental questions require us to study the real-time dynamics of strongly coupled QFTs. The Quark-Gluon Plasma (QGP) has been extensively studied at the RHIC and the LHC with Heavy-Ion Collisions~\cite{Adams:2005dq,Karch:2011ds}. A fundamental problem in the theoretical understanding of the QGP is why it reaches its equilibrium state so fast. 

Unfortunately, the powerful conventional approaches to strongly coupled field theories, for instance lattice field theories, are limited regarding real-time calculations and hence also within non-equilibrium dynamics. For example, the application of statistical Monte-Carlo methods to perform the path integral for lattice gauge theories only works for imaginary times and real actions; this restriction makes the computation of the phase diagram of strong interactions at finite density, and studying out-of-equilibrium dynamics very challenging.\footnote{Note that there has been made made progress in studying real time dynamics on the lattice in so-called quantum simulations~\cite{Labuhn_2016,Bernien_2017,PhysRevX.4.041037,PhysRevX.8.021070}.} 

There is an approach to strongly coupled quantum field theories which is is particularly well suited for studying out-of-equilibrium phenomena from first principles: the AdS/CFT correspondence~\cite{Maldacena:1997re,Gubser:1998bc,Witten:1998qj}. The AdS/CFT correspondence maps certain strongly coupled supersymmetric quantum field theories to classical \mbox{(super-)}gravity theories in weakly curved asymptotically Anti-de Sitter (AdS) spacetimes. Even though there is no proof for the AdS/CFT correspondence in a strict mathematical sense, but a lot of evidence in favour of it, it is an ideal theoretical laboratory for deepening our understanding of strongly coupled quantum field theory and the diverse phenomena it describes. 
The AdS/CFT correspondence emerged as an appropriate playground for strongly coupled condensed matter systems~\cite{Hartnoll:2009sz,Herzog:2009xv,Hartnoll:2016apf} starting with the holographic superconductors and superfluids~\cite{Hartnoll:2008kx,Hartnoll:2008vx,Ammon:2008fc,Ammon:2009fe,Ammon:2009xh,Ammon:2010pg,Albash:2008eh,Maeda:2008ir,Ammon:2009wc,Gubser:2008wv,Amado:2013lia,Arias:2016nww,Horowitz:2008bn,Arean:2015sqa,Arean:2014oaa,Landea:2014naa,Erdmenger:2013zaa,Erdmenger:2012zu,Hoyos:2014nua,Chapman:2013qpa,Sonner:2014tca,Hoyos:2013eha,Erdmenger:2008rm,Bhaseen:2012gg,Arias:2012py,Arean:2013mta,Erdmenger:2011tj,Amado:2013aea,Gauntlett:2009bh,Gauntlett:2009dn,Sonner:2009fk,Amado:2013xya,Nishioka:2009zj,Horowitz:2010gk,Jensen:2018hse}, the Kondo model\cite{Erdmenger:2013dpa,OBannon:2015cqy,Erdmenger:2015spo,Erdmenger:2015xpq,Erdmenger:2016vud,Erdmenger:2016jjg,Erdmenger:2016msd,Erdmenger:2018xqz,Erdmenger:2020hug}, topological fractional insulators \cite{Karch:2010mn,Ammon:2012dd,HoyosBadajoz:2010ac,Maciejko:2010tx}, and external electromagnetic sources~\cite{DHoker:2009mmn,DHoker:2009ixq,DHoker:2010onp,DHoker:2010zpp,Jensen:2010vd,Ammon:2011je,Fuini:2015hba,Grieninger:2017jxz,Ammon:2016fru,Endrodi:2018ikq,Bu:2012mq,Cartwright:2019opv}. 

The equilibrium behavior of strongly coupled field theories at finite temperature is -- in terms of the AdS/CFT correspondence -- captured by the thermodynamics of black holes in the gravitational theory. If we want to study the out-of-equilibrium response of a physical system, we have to drive it out of its equilibrium state. The linear response regime, where we consider small perturbations about the thermal equilibrium, is well established in holography and mapped to computing the so-called quasi-normal modes of a black hole \cite{Kovtun:2005ev}. Not only can we study the near equilibrium behavior within the AdS/CFT correspondence, but we also have a powerful tool to investigate the full nonlinear far-from-equilibrium dynamics. In particular, we might get insights into whether systems equilibrate or what turbulence in a normal fluid is. The non-equilibrium dynamics of strongly coupled QFTs translates in the language of AdS/CFT into time-dependent problems in classical general relativity which is tractable by methods developed within numerical relativity. As we will see, the price we have to pay for re-formulating the problem in classical gravity is the so-called large $N$ limit in the field theory.

Throughout this thesis, we will employ the AdS/CFT correspondence to shed light into different non-equilibrium phenomena:
    \begin{mdframed}[backgroundcolor=gray!20] 
    \begin{itemize}
        \item \textbf{Broken translational invariance and transport:} (ch. \ref{chapter:brokenspacetime}) \newline
        What is the imprint of spontaneously and explicitly broken spacetime symmetries on the transport properties of strongly coupled QFTs? Can we describe the physics in terms of a consistent hydrodynamic theory?
        \item \textbf{Anomalous hydrodynamics in strong external magnetic fields:} (ch. \ref{sec:anotranspo})\newline
        How do strong external magnetic fields influence the various transport coefficients such as the viscosity tensor of strongly coupled (anomalous) QFTs? Are novel transport coefficients present as an effect of the chiral anomaly? 
        \item \textbf{Entanglement entropy and non-equilibrium dynamics in dS:} (ch. \ref{section:EEandtheswampland})\newline
        Phenomenological swampland conjectures banish quantum theories with matter in de-Sitter into swampland. Can we provide a swampland bound within a well defined microscopic framework?
        \item\textbf{$T\bar T$ deformed QFTs:} (ch. \ref{sec:TTbarandcutoff})\newline
        How does the entanglement entropy for generic subintervals in a QFT on a sphere change along the $T\bar T$ trajectory? Can we derive the trajectory for field theories dual to dS in general dimensions and match the field theory calculation and the computation within holography?
    \end{itemize}
  \end{mdframed}
\vspace{-0.2cm}
\subsection*{Broken translational invariance and transport}
 In the first part of thesis, we will study the effects of broken symmetries on the out-of-equilibrium behavior of strongly coupled field theories. First, we restrict ourselves to the long range dynamics for small perturbations about the equilibrium state -- the hydrodynamic regime. 
Hydrodynamics is an effective theory and based on the symmetries of the system under consideration. In particular, the hydrodynamic equations are based on the conservation equations -- for example energy and momentum -- and are supplemented by so-called constitutive relations. The conservation of energy and momentum is via Noether's theorem intimately related with its associated symmetry -- the invariance under spacetime translations. 

Translational invariance is one of the fundamental symmetries in nature; it is however broken in many condensed matter systems. The spontaneous breakdown of a continuous symmetry leads to exciting phenomena such as the appearance of so-called Goldstone phonons. However, breaking the translational symmetry explicitly also implies momentum non-conservation and we can no longer apply the standard textbook hydrodynamics. In order to extend hydrodynamics beyond the standard regime, we need a playground to test and verify our results. The AdS/CFT correspondence provides us with a suitable framework to deepen and verify our knowledge about effective field theories such as hydrodynamics. The successful implementation of momentum relaxation into the AdS/CFT correspondence via massive gravity theories~\cite{Vegh:2013sk,Donos:2013eha,Andrade:2013gsa,Blake:2013owa} and successively the identification of holographic phonons~\cite{Alberte:2017cch,Alberte:2017oqx} fuelled extensive work on (pseudo)-spontaneously and explicitly broken translations~\cite{Vegh:2013sk,Donos:2013eha,Andrade:2013gsa,Blake:2013owa,Blake:2013bqa,Donos:2013gda,Davison:2013jba,Ling:2014saa,Donos:2014oha,Baggioli:2014roa,Davison:2014lua,Andrade:2014xca,Donos:2014cya,Gouteraux:2014hca,Andrade:2016tbr,Andrade:2015iyf,Baggioli:2015zoa,Blake:2015ina,Donos:2015bxe,Blake:2015epa,Amoretti:2016bxs,Baggioli:2015gsa,Baggioli:2015dwa,Alberte:2015isw,Alberte:2016xja,Gouteraux:2012yr,Blake:2017qgd,Andrade:2017cnc,Andrade:2018gqk,Baggioli:2016rdj,Baggioli:2019sio,Andrade:2017jmt,Alberte:2017cch,Alberte:2017oqx,Davison:2018ofp,Delacretaz:2017zxd,Krikun:2018agd,Andrade:2015hpa,Gouteraux:2018wfe,Amoretti:2019cef,Amoretti:2018tzw,Baggioli:2016pia,Gouteraux:2016wxj,Amoretti:2017axe,Delacretaz:2016ivq,Baggioli:2018bfa,Donos:2017ihe,Donos:2018kkm,Amoretti:2017frz,Alberte:2018doe,Delacretaz:2019wzh,Andrade:2017leb,Andrade:2019bky,Balm:2019dxk,Romero-Bermudez:2019lzz,Esposito:2017qpj,Watanabe:2011dk,Donos:2019tmo,Donos:2019txg,Donos:2019hpp,Ammon:2020xyv,Ammon:2019apj,Baggioli:2020nay,Alberte:2020eil,Nicolis:2017eqo,Baggioli:2019aqf,Baggioli:2019elg,Baggioli:2019abx,Baggioli:2019mck,Ammon:2019wci,Andrade:2019zey,Amoretti:2019kuf}.
In terms of effective field theories and hydrodynamics, we have at least some field theoretic understanding of the near-equilibrium physics with broken translational symmetry. In the nonlinear regime, however, the approaches are still based on phenomenological models (such as \cite{doi:10.1098/rstl.1867.0004,christensen2013theory}). Within holography, we investigate the influence of broken symmetries on the far-from-equilibrium behavior.
\vspace{-0.2cm}
\subsection*{Anomalous hydrodynamics in strong external magnetic fields}
Another interesting aspect of the influence of broken symmetries on the non-equilibrium behavior are anomalies in field theories. Within this thesis, we will develop a consistent hydrodynamic framework for quantifying the effects of the chiral anomaly on the transport behavior in strongly coupled anomalous QFTs in the presence of strong external magnetic fields. In order to show which transport coefficients are non-zero in the presence of the anomaly (and if the novel transport coefficients generically contribute), we probe the system within holography. Studying anomalies within the AdS/CFT correspondence is especially interesting since the underlying physics is universal i.e. phenomena such as the chiral-magnetic effect for example are totally determined in terms of the anomaly coefficient. Chiral anomalies, as they appear in so-called top-down constructions for four dimensional $\mathcal N=4$ super-Yang Mill theory, have a crucial impact on the non-equilibrium behavior~\cite{Ammon:2017ded,Ammon:2016mwa,Ammon:2016fru,Kharzeev:2009pj,Fukushima:2008xe,Kharzeev:2004ey,Grozdanov:2016ala,Bai:2012ci,Kalaydzhyan:2011vx,Landsteiner:2019kxb,Ammon:2016szz,Liu:2018djq,Liu:2018bye,Landsteiner:2016stv,Sun:2016gpy,Landsteiner:2015pdh,Jimenez-Alba:2015awa,Landsteiner:2014vua,Fernandez-Pendas:2019rkh,Jensen:2010em,Copetti:2019rfp,Landsteiner:2017lwm,Bu:2016vum,Copetti:2017ywz,Landsteiner:2017hye,Copetti:2016ewq,Cortijo:2016wnf,Erdmenger:2015qqa,Fukushima:2010vw,Rubakov:2010qi,Hoyos:2011us,Grieninger:2017jxz,Hoyos:2013qwa,Bu:2019mow,Buividovich:2016ulp,Landsteiner:2015lsa,Rebhan:2009vc,Jensen:2013rga,Jensen:2013kka,Jensen:2013vta,Jensen:2012kj,Newman:2005hd,Jensen:2012jy,Zhang:2018wzm,Jensen:2011xb,Jimenez-Alba:2014iia,Buividovich:2009wi,Chang:2014jna,Aharony:1999rz,Bilal:1999ph,Bu:2018psl,Landsteiner:2013aba,Landsteiner:2013sja,Bu:2018drd,Landsteiner:2012kd,Bu:2018trt,Bu:2016oba,Landsteiner:2012dm,Landsteiner:2011tg,Landsteiner:2011tf,Haack:2018ztx,Landsteiner:2011iq,Bu:2018vcp,Landsteiner:2011cp,Amado:2011zx,Gynther:2010ed} (see \cite{Landsteiner:2016led} for a review) and we cannot neglect them as it is overwhelmingly done in the present literature.

\vspace{-0.2cm}
\subsection*{Entanglement entropy and non-equilibrium dynamics in de-Sitter}
The Planck data from 2018 confirmed that our universe exhibits a slightly positive cosmological constant ($\Lambda=(2.846\pm0.076)\,10^{-122}\,m_\text{Pl}^2$~\cite{Aghanim:2018eyx}). Together with a positive vacuum, the positive cosmological constant implies a negative pressure and thus an accelerated expansion of the universe. So far, we focused on non-equilibrium properties of strongly coupled field theories in flat spacetimes. In de-Sitter (dS) space, the absence of a globally timelike Killing vector makes the tasks at hand inherently a non-equilibrium scenario. In the following, we want to study non-local observables in the context of dS. 

While quantum gravity in AdS is accessible in terms of the AdS/CFT correspondence, its definition in dS is still unknown albeit its crucial importance for the understanding of our universe. The question about dS gravity was fueled by the bold swampland conjectures~\cite{Obied:2018sgi} claiming that quantum gravity in dS requires a scalar potential with large negative mass squared, rendering dS unstable, or a non-vanishing derivative destroying dS completely~\cite{Ooguri:2018wrx,Garg:2018reu}. In short, the swampland conjectures ban all stable solutions for dS quantum gravity into swampland. Since this is a big assertion, we want to investigate the conjecture within a well defined framework for the microscopic theory. 

\begin{figure}
    \centering
   \includegraphics[width=5cm]{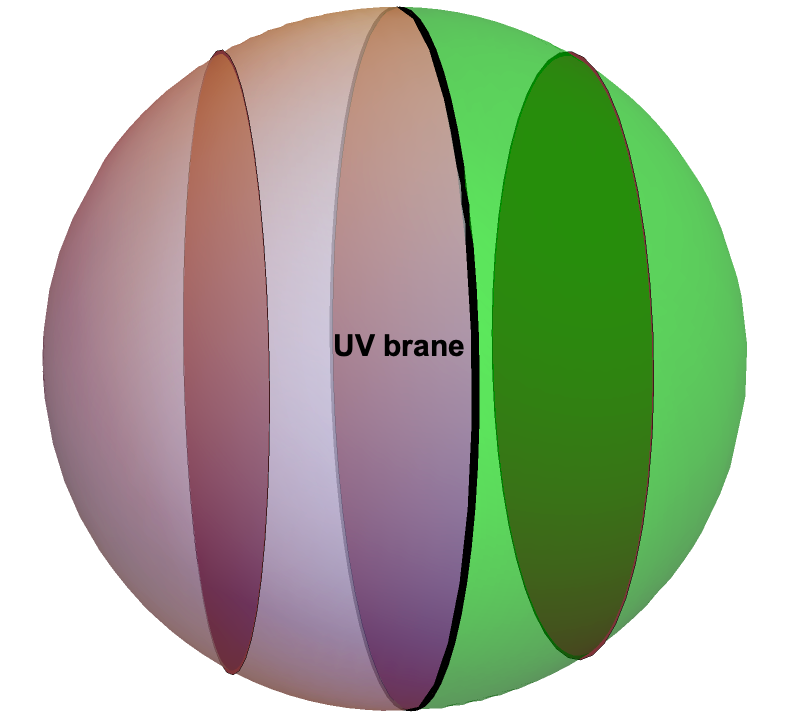}
        \caption{Euclidean dS$_3$ at a fixed time. The two asymptotic AdS$_3$ regions are depicted in purple and green, respectively. The UV brane is the middle slice and corresponds to the maximum dS$_2$ static patch (and we exemplarily show two additional static patches). 
        }
        \label{pic:dsds}
\end{figure}

One way to obtain such a framework is to try to embed dS into the framework holography~\cite{Strominger:2001pn,Anninos:2011ui,Alishahiha:2005dj,Alishahiha:2004md,Dong:2011uf,Freivogel:2006xu,Maldacena:2002vr,Harlow:2011ke,Silverstein:2016ggb,Anninos:2012qw}; this may be done in terms of the DS/dS correspondence~\cite{Alishahiha:2004md,Alishahiha:2005dj} (see \cite{Dong:2010pm,Dong:2018cuv,Gorbenko:2018oov,Grieninger:2019zts,Lewkowycz:2019xse,Geng:2019bnn,Geng:2019yxo,Geng:2019ruz} for recent developments) or the dS/CFT correspondence~\cite{Strominger:2001pn}. Throughout this thesis, we will employ the former. In contrast to the AdS/CFT correspondence, the DS/dS correspondence consists of two asymptotic infrared AdS regions glued together in the middle by means of a ultraviolet (UV) brane (see figure \ref{pic:dsds}). The two CFTs -- dual to the two dS$_{d+1}$ regions -- are located on the UV brane and are coupled to dynamical gravity since the graviton is localized on the UV slice. 

In the second part of this thesis, we want to probe the swampland conjectures for dS within a well defined microscopic framework by means of an information theoretic measure -- the entanglement entropy.
The entanglement of quantum state is one remarkable aspect of quantum theories which gained a lot of interest over the recent years.
Entanglement entropy may be defined in the universal language of quantum fields, even though explicit calculations are notoriously difficult. Ryu and Takayanagi \cite{Ryu:2006bv,Ryu:2006ef} (see \cite{Rangamani:2016dms} for a review) showed that the entanglement entropy for a subsystem of a strongly coupled field theory is the holographic dual to computing the area of the minimal surface enclosing the subsystem in the spacetime of the gravitational theory. The notion of holographic entanglement entropy was extensively studied in AdS but we may also extend the discussion to dS~\cite{Maldacena:1998ih,Balasubramanian:2001nb,Geng:2019bnn,Grieninger:2019zts,Dong:2018cuv,Sato:2015tta,Tetradis:2019vjn,Narayan:2017xca,Narayan:2015oka,Geng:2019ruz,Giataganas:2019wkd,Narayan:2019pjl,Arias:2019pzy,Fernandes:2019ige}. The central question we want to investigate is: can we formulate constraints on the matter content in dS in terms of consistency conditions from entanglement entropy?
\vspace{-0.2cm}
\subsection*{$T\bar T$ deformed QFTs}
While within the DS/dS correspondence there are many open questions about how to relate the behavior in the gravitational theory to the field theory on dS$_d$, we have much better control about this process within the AdS/CFT correspondence. In the deep infrared (IR), AdS and dS are indistinguishable and the AdS/CFT correspondence provides us with a field theory dual to (A)dS. 

In \cite{Gorbenko:2018oov} the authors showed that a particular irrelevant deformation of QFTs might help us to understand the embedding from the DS/dS correspondence in the AdS/CFT correspondence better -- the $T\bar T$ deformation. The $T\bar T$ deformation, where $T$ refers to the energy-momentum tensor, is an irrelevant deformation which is exactly solvable~\cite{Smirnov:2016lqw,Cavaglia:2016oda,Zamolodchikov:2004ce}
and thus attracted great interest in field theory and holography~\cite{ McGough:2016lol,  Dubovsky:2017cnj,Jeong:2019ylz, Cardy:2018sdv, Apolo:2019yfj,Cottrell:2018skz,Guica:2019nzm, Kraus:2018xrn, Donnelly:2018bef, Hartman:2018tkw, Taylor:2018xcy, Bonelli:2018kik,LeFloch:2019rut, Aharony:2018vux,Chakraborty:2019mdf, Gross:2019ach,Aharony:2018bad, Datta:2018thy, Giveon:2017nie,Shyam:2017znq,Murdia:2019fax,Shyam:2018sro,Jiang:2019tcq,Park:2018snf, Guica:2017lia, Baggio:2018rpv, Chang:2018dge,Chen:2019mis,Ota:2019yfe,Cardy:2019qao,Banerjee:2019ewu,Caputa:2019pam,Gorbenko:2018oov,2019arXiv190809299S,Chakraborty:2018kpr,He:2019ahx,He:2019glx,Aguilera-Damia:2019tpe,Geng:2019yxo,Donnelly:2019pie,Lewkowycz:2019xse,Ireland:2019vvj,Tolley:2019nmm,Asrat:2019end,Jiang:2019hxb}.
The holographic dual to the $T\bar T$ deformation is to simply chop off part of the spacetime and to ``move the CFT into the bulk''~\cite{McGough:2016lol}. 
To embed the DS/dS correspondence into AdS/CFT, we move the boundary with $T\bar T$ deformations inwards to the IR, where we know the field theory dual in terms of the AdS/CFT correspondence. From the IR, we may trigger the flow towards the UV with a $T\bar T$ deformation of opposite sign and ``grow back'' the UV region. The last section of this thesis is dedicated to deepen our understanding of this intriguing irrelevant deformation and its realization within holography.

\chapter{The AdS/CFT correspondence}
In the seminal paper~\cite{Maldacena:1997re}, Maldacena conjectured a duality between two theories which might look very surprising at first sight: $\mathcal N=4\ SU(N)$ super-Yang Mills theory in 3+1 dimensional Minkowski space is \textit{equivalent} to type IIB superstring theory on AdS$_5\times$S$^5$.
We may not only write down AdS/CFT dualities for different spacetime dimensions in the form AdS$_{d+1}$/CFT$_d$ but also for less supersymmetric and non-relativistic field theories~\cite{Aharony:2008ug,Son:2008ye,Balasubramanian:2008dm,Goldberger:2008vg,Barbon:2008bg}. The AdS/CFT correspondence is an example of a deeper underlying physical principle: holography. Rooted in the pioneering work of Bekenstein and Hawking on the area law of the black hole entropy \cite{PhysRevD.7.2333,Hawking:1974rv}, the holographic principle states that the complete information of a spacetime volume element may be stored on its lower dimensional spatial boundary. The AdS/CFT correspondence realizes the holographic principle in terms of equating the dynamics of an (effectively) $d+1$ dimensional theory of gravity with a $d$ dimensional quantum theory on the spatial boundary of the spacetime.\footnote{When we speak of spacetime, we usually refer to the spacetime of the gravitational theory.} 

Another intriguing feature of the AdS/CFT correspondence is that it is a strong/weak duality; we may obtain insights into a strongly coupled theory by studying its conceptually much easier accessible weakly coupled dual by means of the AdS/CFT correspondence.
In order to apply the AdS/CFT correspondence to study a specific system in a strongly coupled field theory there are two very different approaches on how to obtain the corresponding gravitational theory. The first approach is the so-called \textit{top-down} approach; we start with a model in string theory and truncate the field content of the gravitational theory to AdS$_{d+1}$. This approach is preferable since we know the exact field theory dual but it is extremely difficult to find consistent truncations from string theory and its dual field theory. There is another approach, however, where we motivate the gravitational action phenomenologically -- the so-called \textit{bottom-up} approach. In \textit{bottom-up} holography, we write down a gravitational toy model which models the core features of the dual field theory. The recipe of how to obtain the gravitational toy model is formulated in the AdS/CFT dictionary which is the topic of the next section.

In this chapter, we present the foundations of the AdS/CFT correspondence. In particular, we formulate the so-called holographic dictionary which translates the quantities in the gravitational theory into the quantities in the dual field theory and vice verse. In order to get an intuition about how the dictionary works in practice, we discuss two examples: the massive scalar field and correlation functions in AdS, and the boundary energy-momentum tensor. We then proceed to generalize the AdS/CFT correspondence to finite density and finite temperature. At finite temperature, we discuss applications of the AdS/CFT correspondence for example within linear response theory. We close the chapter by discussing non-local observables such as entanglement entropy and a generalization of the holographic concept to de-Sitter space.

\section{The AdS/CFT dictionary}\label{sec:dict}
In this section, we formulate the AdS/CFT correspondence in general terms on the level of the partition functions of both theories. We also write down the field operator map at the conformal boundary where both theories are defined. 

For a generic CFT, we consider the generating functional $W_\text{CFT}[\gamma_{\mu\nu},A_\mu,\phi_{(\bm{s})}]$ for the connected Green's functions of the conserved energy-momentum tensor $T_{\mu\nu}$, global symmetry currents $J_\mu$ and any composite operator $\mathcal O$ in the dual field theory. For any operator $T_{\mu\nu}, J^\mu$, and $\mathcal O$, we have the corresponding source terms, the induced metric $\gamma_{\mu\nu}$ at the conformal boundary, a gauge field $A_\mu$, and the scalar source $\phi_{(\bm{s})}$.
 In CFT language, the partition function $Z_\text{CFT}=\text{exp}(-W_\text{CFT})$ thus reads
\begin{equation}
       Z_\text{CFT}[\gamma_{\mu\nu},A_\mu,\phi_{(\bm{s})}]\!=\!\left\langle\!\text{exp}\!\left(\int\dd^dx\sqrt{-\gamma}\left(\frac12\,T^{\mu\nu}(x)\,\gamma_{\mu\nu}(x)+J^\mu(x)\, A_\mu(x) +\mathcal O(x)\, \phi_{(\bm{s})}(x)\!\right)\!\right)\!\right\rangle\!.\label{eq:dictionaryads}
\end{equation}
The energy-momentum tensor, conserved current- and the composite operator may be extracted from eq. \eqref{eq:dictionaryads} by varying the generating functional with respect to the corresponding source, i.e.
\begin{align}\label{eq:extracvev}
  & \langle \mathcal O(x)\rangle=\left.\frac{\delta W_\text{CFT}}{\delta\phi_{(\bm{s})}(x)} \right|_{\substack{\gamma_{\mu\nu}(x)=\eta_{\mu\nu}\\A_\mu(x)=0\\\phi_{(\bm{s})}(x)=0}},  \quad\quad\langle T^{\mu\nu}(x)\rangle=\left.\frac{2}{\sqrt{-\gamma}}\frac{\delta W_\text{CFT}}{\delta\gamma_{\mu\nu}(x)}\right|_{\substack{\gamma_{\mu\nu}(x)=\eta_{\mu\nu}\\A_\mu(x)=0\\\phi_{(\bm{s})}(x)=0}}, \\& \langle J^\mu(x)\rangle=\left.\frac{\delta W_\text{CFT}}{\delta A_\mu(x)} \right|_{\substack{\gamma_{\mu\nu}(x)=\eta_{\mu\nu}\\A_\mu(x)=0\\\phi_{(\bm{s})}(x)=0}}.
\end{align}
It is important to emphasize that the field theory content in eq. \eqref{eq:dictionaryads} is not restricted to one current- or composite operator only. We can include $N$ current operators $J_\mu^N$ and $M$ composite operators $\mathcal O^M$ into our prescription by including them with their source terms $ A_\mu^N$ and $\phi^M_{(\bm{s})}$, respectively.  

To simplify the following discussion, we restrict the field content in eq. \eqref{eq:dictionaryads} to one composite operator $\mathcal O$. The generalization to multiple operators and to including the energy-momentum tensor and current operators is straightforward.
In this case, eq. \eqref{eq:dictionaryads} reads
\begin{equation}\label{eq:CFTpart}
    Z_\text{CFT}[\phi_{(\bm{s})}]=\text{exp}(-W_\text{CFT}[\phi_{(\bm{s})}])=\left\langle\text{exp}\!\left(\int\dd^dx\,\phi_{(\bm{s})}(x)\,\mathcal O(x)\right)\right\rangle.
\end{equation}
To understand the equivalent of the CFT partition function in the gravitational theory, we consider a general propagating field $\phi$.\footnote{$\phi$ does not have to be a scalar field though we will we suppress all indices for simplicity.}  We parametrize the AdS spacetime in so-called Poincar\'e coordinates
\begin{equation}
    \dd s^2=\frac{L^2}{u^2}\,\left(\dd u^2+\eta_{\mu\nu}\dd x^\mu\,\dd x^\nu\right),\label{eq:poincareM}
\end{equation}
where $u$ denotes the extra dimension and the conformal boundary is located at $u=0$.
The greek indices sum over field theory dimensions $\mu\in\{0,1,\dots,d-1\}$, while latin indices are the indices including the radial coordinate $u$ and sum $i\in\{0,1,\ldots,d\}$. In order to formulate the map between the gravity fields and the dual field theory operators, we expand the gravitational fields near the conformal boundary -- where both theories are defined. Generally, the source term for the dual field theory operator is identified with the leading term of the asymptotic near-boundary expansion of the corresponding gravitational field (in standard quantization). By comparing the symmetries, we may find the dual pairs to the gravitational fields (see \cite{Ammon:2015wua} for a detailed and comprehensive explanation) at the conformal boundary. 

The claim of the AdS/CFT correspondence is the equality of the partition functions of both theories at the conformal boundary and is known as the Gubser-Klebanov-Polyakov-Witten (GKPW) formula \cite{Gubser:1998bc,Witten:1998qj}
\begin{equation}
    Z_\text{CFT}[\phi_{(\bm{s})}]=\left. Z_\text{string}\right|_{\lim_{u\to0}\left(\phi(u,x)\,u^{\Delta-d}=\phi_{(\bm{s})}(x)\right)}\label{eq:adscorr}.
\end{equation}
In eq. \eqref{eq:adscorr}, we identified the leading behavior $\phi_{(\bm{s})}$ of the gravitational field $\phi$ with the source of the composite operator $\mathcal O$ in the dual CFT. Additionally, we identify the coefficient $\Delta$ which appears in the asymptotic expansion of the gravitational field with the dimension of the dual field theory operator $\mathcal O$ (this will be more clear in example 1). The subleading mode of the supergravity field determines the dynamics of the expectation value of the dual operator $\mathcal O$.

The equivalence of the partition functions \eqref{eq:adscorr} is referred to as the strongest form of the AdS/CFT correspondence. However, it is very hard to perform calculations for some generic parameter in the strongest form; in particular, the partition function of type IIb string theory $Z_\text{string}$ is not known explicitly. It is thus practical to weaken the form of the duality in order to make it better tractable. We may approximate $Z_\text{string}$ in terms of supergravity by performing a saddle point approximation. On the field theory side this is equivalent to taking the large $N$ limit, where $N$ is the rank of the gauge group. The weak form of the correspondence \eqref{eq:adscorr} equates the partition function of a strongly coupled conformal field theory in the large $N$ limit with a supergravity theory
\begin{equation}
    \left.Z_\text{CFT}[\phi_{(\bm{s})}]\right|_\text{$N$ large}\approx \left.\text{exp}\left(-S_\text{grav}\right)\right|_{\lim_{u\to0}\left(\phi(u,x)\,u^{\Delta-d}=\phi_{(\bm{s})}(x)\right)}.\label{eq:AdSweak}
\end{equation}
Let us take stake at where we are at: eq. \eqref{eq:AdSweak} is highly non-trivial statement; it equates the partition function of a $d+1$ dimensional classical theory of gravity to a $d$ dimensional quantum field theory (in flat space).

\subsection*{Example 1: Scalar fields and correlation functions in AdS}
The prescription in eq. \eqref{eq:adscorr} looks very abstract at first. In order to illustrate the underlying consequences of this statement, we consider the simplest example --  a massive scalar toy model in AdS. For simplicity, we also restrict our discussion to CFTs in the large $N$ limit dual to supergravity on AdS$_{d+1}$~\cite{Ammon:2010zz,Ramallo:2013bua,Ammon:2010ely,Fraga:2008va}.\footnote{By choosing spherical harmonics on the sphere, we may decompose the supergravity fields into Kaluza-Klein towers on the S$^5$. Since we do not source the Kaluza-Klein modes, they are gapped and we neglect them.} 

The action of a real scalar field $\phi$ with mass $m$ in AdS$_{d+1}$ is given by
\begin{equation}\label{eq:sgravscalar}
    S_\text{grav}=-\frac C2\int\dd u\,\dd^dx\,\sqrt{-g}\,\left(g^{mn}\partial_m\phi\,\partial_n\phi+m^2\phi^2\right),
\end{equation}
with the AdS metric in the Poincare patch given by eq. \eqref{eq:poincareM}.

The equation of motion for the scalar field on the curved AdS background follows from the Euler-Lagrange equations
\begin{equation}
    (\Box_g-m^2)\,\phi=0, \quad \left.\Box_g\right|_\text{AdS}=\frac{1}{L^2}\,\left(u^2\,\partial_u^2-(d-1)\,u\,\partial_u+u^2\,\eta_{\mu\nu}\,\partial^\mu\partial^\nu\right).
\label{eq:scalareom}\end{equation}
By solving eq. \eqref{eq:scalareom} for the second radial derivative (and transforming to Fourier space $\phi(u,\bm{x})=1/(2\pi)^{d}\int\dd^{d}k\,\text{exp}(i\,k^\mu x_\mu)\,\phi_k(u)$), we see that $u=0$ is a singular point of the differential equation
\begin{equation}
\partial^2_u\phi_k(u)-p(u)\,\partial_u\phi_k(u)-q(u)\,\phi_k(u)=0, \quad p(u)=(d-1)/u,\quad q(u)=m^2L^2/u^2+k^2.\label{eq:dgl1}
\end{equation}
However, $p(u)$ and $q(u)$ are analytical after multiplying with appropriate powers of $u$ and $u=0$ is only a regular singular point. For regular singular points, Frobenius' theorem guarantees the existence of a solution at this point in terms of a power series with two undetermined coefficients $\phi_{(\bm{s})}(x)$ and $\phi_{(\bm{v})}(x)$
\begin{equation}
    \phi(u,x)\sim \phi_{(\bm{s})}(x)\,u^{\Delta_-}+c_1 u^{\Delta_-+1}\ldots+\phi_{(\bm{v})}(x)\,u^{\Delta_+}+\ldots.\label{eq:Frobeniusseries}
\end{equation}
We will refer to the mode $\phi_{(\bm{s})}(x)$ as leading (or non-normalizable) mode and to $\phi_{(\bm{v})}(x)$ as the sub-leading (or normalizable) mode.\footnote{If $(\Delta_+-\Delta_-)\in\mathbb{Z}$, the power series \eqref{eq:Frobeniusseries} may contain logarithmic terms $\sim u^{\Delta_+}\,\log(u)$ For simplicity, we omit this technical detail in eq. \eqref{eq:Frobeniusseries}.} The exponents of the undetermined coefficients follow immediately by plugging the ansatz $\phi(u,x)\sim u^\Delta$ in eq. \eqref{eq:dgl1} and solving the quadratic equation to leading order in $u$, which gives
\begin{equation}
    \Delta_\pm=\frac d2\pm\sqrt{\frac{d^2}{4}+m^2L^2}.
\end{equation}
The remaining unknown coefficients appearing in the power series \eqref{eq:Frobeniusseries} are totally determined in terms of the two undetermined coefficients by solving the differential equation order by order in $u$. The two undetermined coefficients $\phi_{(\bm{s})}(x)$ and $\phi_{(\bm{v})}(x)$ are the input parameters of the underlying theory and may be related to field theory quantities by dimensional analysis \cite{Ammon:2015wua};  recall that $\phi_{(\bm{v})}(x)$ corresponds to the expectation value of the dual scalar field operator $\mathcal O$ of dimension $\Delta\equiv\Delta_+$. Similarly, the leading mode $\phi_{(\bm{s})}(x)=\lim_{u\to 0}\phi(u,x)u^{-\Delta_-}$ acts as source term for the dual scalar operator in eq. \eqref{eq:CFTpart}.
In particular, by comparing the representations of the supergravity field $\phi$ and the CFT operators we find the important relation
\begin{equation}
m^2L^2  =  \Delta(\Delta-d),
\end{equation}
which relates the mass $m$ of the supergravity scalar field to the conformal dimension $\Delta\equiv\Delta_+$ of the dual field theory operator $\mathcal O$. For positive masses $m$, the conformal dimension of the dual operator is bigger than $d$ and the operator corresponds to an irrelevant deformation. In the case of massless scalars, the scaling dimension is $\Delta=d$ corresponding to a marginal operator. In flat space, negative masses lead to tachyonic instabilities; interestingly and in contrast to that, we have to give the scalar fields in AdS a negative (but not too negative) mass in order to source a relevant operator in the dual field theory\footnote{This may be seen by rescaling the scalar $\phi=u^{d/2}\tilde\phi$ and introducing a new variable $y=\log(u)$ in eq.\eqref{eq:poincareM} followed by an integration by parts. In the new variables, we may find the action of a scalar field in flat space by introducing the effective mass $m_\text{eff}^2L^2=m^2L^2+d^2/4$.}
\begin{equation}
    m^2L^2\ge-d^2/4
\end{equation}
which is the famous Breitenlohner-Freedman bound \cite{Breitenlohner:1982bm}. For masses in the range \begin{equation}
   -d^2/4<m^2L^2\le -d^2/4+1,
\end{equation}
we may interchange the definition of source and expectation value since both modes are normalizable. In that case, the former expectation value $\phi_{(\bm{v})}(x)$ acts as source term and vice versa. We thus may use either of them to quantize the theory which implies that there exist two possible conformal field theories dual to the same classical AdS action.

So far, we discussed the one-point function of the composite operator $\mathcal O$. From eq. \eqref{eq:extracvev} and \eqref{eq:adscorr}, it is also clear how to calculate $n$-point functions in the dual quantum field theory. The variation of the CFT partition function with respect to the CFT sources translates within the AdS/CFT correspondence to varying the gravitational partition function with respect to the leading mode in the asymptotic expansion
\begin{equation}
    \langle\mathcal O_1(x_1)\,\mathcal O_2(x_2)\,\ldots\,\mathcal O(x_n)\rangle=-\left.\frac{\delta^{n}S^\text{ren}_\text{grav}[\phi]}{\delta\phi^1_{(\bm{s})}(x_1)\,\delta\phi^2_{(\bm{s})}(x_2)\ldots\delta\phi^n_{(\bm{s})}(x_n)}\right|_{\delta\phi^i_{(\bm{s})}=0},
\end{equation}
where $S^\text{ren}_\text{grav}$ refers to the renormalized action~\cite{deHaro:2000vlm,Skenderis:2002wp}.
\subsection*{Example 2: The holographic stress tensor}\label{section:stresstensor}
In this subsection, we show how to extract the renormalized stress tensor on a cutoff slice with radial position $r=r_c$. For the holographic renormalization procedure, we usually set $1/r_c=\varepsilon\ll1$. However, with regards to our discussion about $T\bar T$ deformations in section \ref{sec:TTbarandcutoff}, we will keep the cutoff surface at an arbitrary radial distance for now. 

The bare quantum field theory quantities will be divergent \textit{per se}. On the gravity side, the divergences are manifest in terms of the infinite volume of AdS~\cite{deHaro:2000vlm,Skenderis:2002wp}.
 Similar to the renormalization procedures developed for quantum field theories, we have a recipe available on how to cure the divergences in AdS: the holographic renormalization~\cite{deHaro:2000vlm,Bianchi:2001kw,Skenderis:2002wp,Papadimitriou:2004ap,Papadimitriou:2005tia}. 
 Within the AdS/CFT correspondence, low energies correspond to the region deep in the bulk while high energies correspond to the asymptotic region. In this sense, the radial coordinate may be viewed as the energy scale of the dual field theory and the UV divergences of the dual field theory are mapped to the infinite extension of AdS in the radial direction. In order to regularize quantities such as correlation functions, holographic renormalization tells us to move the boundary slightly into the bulk and consider the cutoff surface $1/r_c=\epsilon\ll 1$ to read off the field theory quantities instead of the conformal boundary located at $u=0$.
 To ensure that the quantities are finite in the limit $\varepsilon\to 0$, we have to supplement the Einstein-Hilbert action of the gravitational theory with counter terms~\cite{deHaro:2000vlm,Skenderis:2002wp}. With the extrinsic curvature $K$, the $d+1$ dimensional Ricci scalar $R$, the AdS curvature radius $L$ and the cosmological constant $\Lambda$, the renormalized action reads
\begin{equation}
S_\text{tot}=S_\text{EH}+S_\text{surf}+S_\text{ct},\label{actionS}
\end{equation}
with \begin{align}
    & S_\text{EH}=-\frac{1}{16\pi\,G_N}\int\dd^{d+1}x\,\sqrt{g}\,\left(R-2\Lambda\right),\quad S_\text{surf}=-\frac{1}{8\pi\,G_N}\int\dd^dx\,\sqrt{\gamma}\,K,\\
    &S_\text{ct}=\frac{1}{16\pi\,G_N}\!\int\!\dd^dx\,\sqrt{\gamma}\left(\!2c_1\frac{d-1}{L}+\frac{c_2\,L}{d-2}\,\tilde R+\frac{c_3\,L^3}{(d-4)\,(d-2)^2}\!\left(\!\tilde R_{\mu\nu}\tilde R^{\mu\nu}\!-\frac{d}{4\,(d-1)}\tilde R^2\right)\!\right)\!,\label{counter}
\end{align}
In eq. \eqref{counter}, we denote the $d$ dimensional quantities with tildes, i.e. $\tilde R,\,\tilde R_{\mu\nu}$ are the Ricci scalar- and tensor, respectively, on the cutoff slice $u=\varepsilon$ with the induced metric $\gamma$. Furthermore, $c_1=1$ for $d\ge 2$, $c_2=1$ for $d\ge 3$ and $c_3=1$ for $d\ge 5$ and zero otherwise.

On the cutoff slice $r=r_c$, the stress tensor of the boundary field theory $T_{\mu\nu}^\text{bdy}$ is related to the bulk stress tensor computed by varying eq. \eqref{actionS} with respect to the metric in terms of the rescaling $T^\text{BY}_{\mu\nu}=r_c^{d-2}\,T^\text{bdy}_{\mu\nu}$. Similarly, the induced metric on the cutoff slice is related to the metric of the CFT by $g_{\mu\nu}(r=r_c,x)=\gamma_{\mu\nu}(x)=r_c^2\,\gamma_{\mu\nu}^\text{bdy}(x)$. In particular, for $r_c\to\infty$ the CFT metric is flat. The complete dictionary to translate the quantities at the cutoff surface into field theory quantities is described in \cite{Hartman:2018tkw}. For simplicity, we set $r_c=1$ from now on.

For clarity and comprehensibility, we split the renormalized stress tensor in two parts $T^\text{ren}_{\mu\nu}[\gamma]=T_{\mu\nu}[\gamma]+C_{\mu\nu}[\gamma]$, the standard holographic stress tensor on the cutoff surface $r=r_c$, $T_{\mu\nu}$, and the corresponding curvature contributions of the counterterms eq. \eqref{counter}, denoted by $C_{\mu\nu}$~\cite{deHaro:2000vlm,Balasubramanian:1999re,Hartman:2018tkw,Banerjee:2019ewu}
\begin{align}
    T_{\mu\nu}=&\frac{1}{8 \pi G_N}\left(K_{\mu\nu}-K\,\gamma_{\mu\nu}-c_1\,\frac{d-1}{L}\,\gamma_{\mu\nu}+\frac{c_2\,L}{d-2}\,\tilde G_{\mu\nu}\right.\nonumber\\
    &\left.\ +\frac{c_3\,L^3}{(d-4)(d-2)^2}\,\left(2\,\left(\tilde R_{\mu\nu\rho\sigma}-\frac 14\,\gamma_{\mu\nu}\,\tilde R_{\rho\sigma}\right)\,\tilde R^{\rho\sigma}-\frac{d}{2\,(d-1)}\,\left(\tilde R_{\mu\nu}-\frac 14\,\tilde R\,\gamma_{\mu\nu}\right )\tilde R\right.\right.\nonumber\\
    &\ \left. \left. -\frac{1}{2\,(d-1)}\left(\gamma_{\mu\nu}\,\Box \tilde R+(d-2)\,\nabla_\mu\nabla_\nu\tilde R\right)+\Box\tilde R_{\mu\nu}\right)\right).\label{EMtensor}
\end{align}
With eq. \eqref{counter} the curvature dependent counterterms give thus rise to the contribution (in $d\ge 3$)~\cite{deHaro:2000vlm,Skenderis:2002wp}
\begin{align}
C_{\mu\nu}=&-\frac{1}{8 \pi G_N}\left(\!c_2\,\tilde G_{\mu\nu}+c_3\,b_d\left[2\left(\tilde R_{\mu\nu\rho\sigma}-\frac14\,\gamma_{\mu\nu}\,\tilde R_{\rho\sigma}\!\right)\!\tilde R^{\rho\sigma}-\frac{d}{2(d-1)}\left(\tilde R_{\mu\nu}-\frac14\,\tilde R\,\gamma_{\mu\nu}\!\right)\!\tilde R\right.\right.\nonumber\\
&\left.\left. -\frac{1}{2\,(d-1)}\left(\gamma_{\mu\nu}\,\Box \tilde R+(d-2)\,\nabla_\mu\nabla_\nu\tilde R\right)+\Box\tilde R_{\mu\nu}\right]\right).\label{EMcounter}
\end{align}
Note that for a finite radial cutoff, the stress tensor is inherently regularized. The divergences become apparent in the limit $r_c\to \infty$.

\section{Generalizations of the AdS/CFT correspondence}
So far, we have discussed the AdS/CFT correspondence only at vanishing temperature. However, by including the option to study strongly coupled QFTs at finite temperature and finite density, we are opening the possibility to strengthen our knowledge about strongly coupled field theories in the areas where lattice QFT is limited, for example the phase diagram at finite temperature and finite density.

\subsection{Finite temperature}
The gravity duals of QFTs at finite temperature are black branes in AdS. In classical general relativity, black branes are thermodynamic objects which may be associated with a temperature: the Hawking-temperature $T_H$; in terms of the AdS/CFT correspondence the Hawking temperature of the black object in the gravitational theory corresponds to the temperature of the field theory.

At finite temperature $T$, we do not only observe quantum fluctuations of the field theory operators but also thermodynamic fluctuations. The macroscopic quantity, to which we refer to as temperature within thermodynamics, are statistical averages of the thermodynamic fluctuations in the microscopic theory. In this section, we restrict ourselves to the so-called canonical ensemble where the particle number is fixed. It is straightforward to adapt the results for the grand canonical ensemble, where the temperature $T$, the volume $V$, and the particle number $N$ are kept fixed \cite{Ammon:2015wua}. In field theory, the probability operator for finding a system in a thermal state at temperature $T$ is formally equivalent to the time evolution operator if we complexify the time according to $t=i\beta$, where $\beta=1/T$~\cite{Ammon:2015wua,ZinnJustin:2000dr}.

On the gravity side, we will see that the dual to the temperature $T$ in the field theory is the Hawking temperature of black holes or black branes by considering the metric of D$3$ branes
\begin{equation}\label{eq:blackD3braneAdS5}
   \dd s^2=H^{-1/2}\left(-f\,\dd t^2+\dd \bm{x}^2\right)+H^{1/2}\left(\frac{\dd r^2}{f}+r^2\,\dd\Omega_5^2\right)\!, \ \ \ \ \ \ \ f(r)=1-\left(\frac{r_h}{r}\right)^4,
\end{equation}
where $H(r)=1+L^4/r^4$. After a change of coordinates $u\equiv L/r$ and Wick rotating $\tau=i\,t$, we find for the AdS part of the metric in the near horizon limit $u\to u_h$
\begin{equation}
    \dd s^2\approx\frac{4\rho^2}{u_h^2}\,\dd \tau^2+\frac{L^2}{u_h^2}\,\dd\bm{x}^2+\dd\rho^2,
\end{equation}
where we also introduced the coordinate $u=u_h(1-\rho^2/L^2)$ which measures the distance from the horizon $u_h$. In order to avoid conical singularities, the time coordinate has to be periodic $2\tau/u_h\sim 2\tau/u_h+2\pi$ or in other words $\Delta \tau=u_h \pi$. However, we already know this periodicity from the temperature of the dual field theory. For a discussion about how to implement a finite density in terms of the AdS/CFT correspondence, we refer the reader to the literature on that topic (see e.g.~\cite{Ammon:2015wua}).

\subsection{Linear response in field theory and holography}\label{sec:linrespoandhol}
By including a finite temperature $T$ to the AdS/CFT machinery, we are able to study the thermodynamical properties of a QFT at equilibrium. Small fluctuations about this equilibrium state lie within the range of linear response theory. By slightly perturbing the system with a small external source it is possible study the equilibration process. In QFT, the response of a system subject to the presence of external influences $\varphi_I$ coupled to a set of operators $\mathcal O^I(x)$ is given in terms of~\cite{Ammon:2015wua}
\begin{equation}
    \delta\hat H=-\int \dd^dx\,\varphi_I(t,\bm{x})\,\mathcal O^I(t,\bm{x}).
\end{equation}
The external field $\varphi^I(t,\bm{x})$ induces a shift in the expectation values of the corresponding operators 
\begin{equation}\label{eq:vevshift}
    \delta\langle\mathcal O^I(x)\rangle=\int \dd^dy\,G_R^{IJ}(x,y)\,\varphi_J(y)+\mathcal O(\varphi^I),
\end{equation}
where the retarded Green's function is defined as
\begin{equation}
    G^R_{IJ}(x,x')=i\,\theta(t-t')\,\langle\{ \mathcal{\hat O}^I(x),  \mathcal{\hat O}^J(x')\}_\pm\rangle.
\end{equation}
The cases $\pm$ denote the commutator and anticommutator for bosonic and fermionic operators, respectively. Using the retarded Green's function means that only sources in the past can influence the physics. This prescription implements causality in a natural way. At time $t$, the shift of the vacuum expectation value $\delta \langle \mathcal O^I(t,x)\rangle$ is only caused by sources $\varphi^I(t',x')$ with $t'<t$. In Fourier space, eq. \eqref{eq:vevshift} is given by~\cite{Ammon:2015wua,Baggioli:2019rrs}
\begin{equation}
\delta\langle\mathcal O^I(k)\rangle=G^{IJ}(k)\,\varphi_J(k)+\mathcal O(\varphi^2)\ \ \ \ \ \ \Rightarrow\ \ \  \ \ \ G^{IJ}(k)=\frac{\delta\langle\mathcal O^I(k)\rangle}{\varphi_J(k)}+\mathcal O(\varphi^2).    
\end{equation}
In the following, we want to work out how to implement the prescription of retarded Green's function in the AdS/CFT correspondence and how to compute the corresponding object in the gravitational theory. 
The black brane in eq. \eqref{eq:blackD3braneAdS5} in AdS$_5$ reads in coordinates with $u=r_h^2/r^2, H(r)=L^4/r^4$ and $r_h=\pi T$
\begin{equation}
    \dd s^2=\frac{(\pi\,T\,L)^2}{u}\,\left(-f(u)\,\dd t^2+\dd\bm{x}^2\right)+\frac{L^2}{4\,u^2\,f(u)}\,\dd u^2,\ \ \ \ \ \ \ \ \ f(u)=1-u^2.
\end{equation}
We now consider a massive scalar field on top of this background with equation of motion $(\Box-m^2)\,\phi=0$. In Fourier space $\phi(k,u)=\int\dd^4k/(2\pi)^4\,e^{i\,k_\mu x^\mu}$, the equation of motion for the scalar field reads
\begin{equation}
    4u^3\,\partial_u\left(\frac{f(u)}{u}\,\partial_u\phi(u,k)\right)+\frac{u}{(\pi T)^2\,f(u)}\,(\omega^2-|\bm{k}|^2\,f(u))\,\phi(u,k)-m^2L^2\,\phi(u,k)=0,\label{eq:scalaringreens}
\end{equation}
with the near-boundary behavior as outlined in eq. \eqref{eq:Frobeniusseries}
\begin{equation}
    \phi(u,k)\sim\phi_{(\bm{s})}(k)\,u^{(d-\Delta)/2}(1+\mathcal O(u))+\phi_{(\bm{v})}(k)\,u^{\Delta/2}(1+\mathcal O(u)).\label{eq:leadingsubleading}
\end{equation}
The horizon, which is a zero of $f(u)$, is the second regular singular point of the differential equation and similarly to eq. \eqref{eq:Frobeniusseries}, we make a power series ansatz
\begin{equation}
    \phi_k(u)\sim(1-u)^\kappa \quad\quad\quad \Rightarrow\quad\quad\quad \kappa=\pm i\,\omega/(4\,\pi\, T).\label{eq:infallinggreens}
\end{equation}
The boundary solution with ``+'' corresponds to outgoing waves at the horizon while the ``-'' solution correspond to the infalling solution. Note that the infalling solution can only be influenced by boundary sources in the past. If $\phi_k(u)$ is a solution to the equation of motion \eqref{eq:scalaringreens} which satisfies the the infalling condition at the horizon \eqref{eq:infallinggreens} and with leading and sub-leading mode at the conformal boundary denoted as in eq. \eqref{eq:leadingsubleading}, the retarded Green's function is (up to contact terms) defined by~\cite{Blake:2019otz}
\begin{equation}
    G^R_{\phi\phi}(k)=L^{d-1}\,(2\Delta-d)\,\frac{\phi_{(\bm{v})}(k)}{\phi_{(\bm{s})}(k)}.\label{eq:Greensfunctiongrav}
\end{equation}
In this section, we identified the thermal equilibrium of a QFT in terms of the AdS/CFT correspondence with a black object in the gravitational theory. 
Small perturbation of the black holes in asymptotically AdS spacetimes are dual to small perturbations of the QFT about its equilibrium state. If we perturb black holes or black branes, the surrounding geometry will ring and settle back down to equilibrium. The frequencies of the ``ringing'' and the relaxation time back to equilibrium are independent of the perturbation and totally determined by the properties of the black object. In context of the AdS/CFT correspondence, we are only interested in perturbations subject to ingoing boundary conditions at the horizon since these solutions satisfy the causality requirement in the dual CFT. By restricting the solution to ingoing waves, we neglect the solution for outgoing waves at the horizon which renders the boundary value problem non-hermitian and the corresponding frequencies are complex. At the conformal boundary, we subject the fluctuations to Dirichlet boundary conditions, since we do not explicitly source them. Note that black hole horizons and the ingoing boundary conditions lead to matter falling into the black hole and thus to dissipation (see \cite{Parikh:1997ma,PhysRevD.33.915,1986bhmp.book.....T,Donos:2015gia,Iqbal:2008by,Kovtun:2003wp,Vishveshwara:1970zz,Horowitz:1999jd} for the so-called membrane paradigm and its implementation into holography). This is exactly what we want to describe in the dual field theory! Small fluctuations in the field theory are damped by dissipation and the system settles back to its equilibrium state. Dissipation introduces singularities in the retarded Green's functions throughout the complex frequency plane. In terms of the AdS/CFT correspondence, this corresponds to the gravitational fluctuations which satisfy Dirichlet boundary conditions at the conformal boundary since the denominator of eq. \eqref{eq:Greensfunctiongrav} vanishes.
\subsection{Hydrodynamics}\label{sec:hydro}
One interesting regime within linear response is the hydrodynamic regime where the length scales of interested are much larger than the characteristic length scale of the system, i.e. we study long-wavelength fluctuations close to thermal equilibrium. In this section, we want to get an intuition about hydrodynamics by considering the simplest scenario: the relativistic fluid, following the references~\cite{Kovtun:2012rj,Rangamani:2009xk} closely. We will explore the hydrodynamics of more complex systems in the context of broken symmetries, i.e. spontaneous symmetry breaking and anomalies in later chapters. The hydrodynamic equations are based on the conservation of the energy-momentum tensor and the present global charges. In particular, for continuous symmetries (of the fundamental microscopic theory) Noether's theorem implies the existence of conserved currents. We consider a theory with translational invariance and conservation of the particle number with the associated conserved currents
\begin{align}
  \partial_\mu \langle T^{\mu\nu}\rangle=0, && \partial_\mu \langle J^\mu\rangle=0.
\end{align}
In the following, we neglect the indication of $\langle\ldots\rangle$ throughout this section as convention in hydrodynamics, even though quantities such as $T^{\mu\nu}$ and $J^\mu$ are operator valued.
The decomposition of the energy-momentum tensor $T^{\mu\nu}$ and the current $J^\mu$ in terms of the hydrodynamic variables, i.e. the local temperature $T(x)$, the local fluid velocity $v(x)$ and the chemical potential $\mu(x)$ is referred to as the constitutive relations. For an arbitrary timelike vector $u^\mu$, the decomposition reads
\begin{align}\label{eq:hydrodecompositionequilibrium}
    T^{\mu\nu}={\cal E}\,u^\mu u^\nu +{\cal P}\,\Delta^{\mu\nu}+({\cal Q}^\mu \, u^\nu+{\cal Q}^\nu\,u^\mu)+{\cal T}^{\mu\nu},&& J^\mu={\cal N}\,u^\mu+{\cal J}^\mu,
\end{align}
where $\Delta^{\mu\nu}=g^{\mu\nu}+u^{\mu\nu}$ is the projector on the spatial dimensions, ${\cal E},\,{\cal P},\,{\cal N}$ are scalars, ${\cal Q}^\mu,\,{\cal J}^\mu$ transverse vectors and ${\cal T}^{\mu\nu}$ a transverse, symmetric, and traceless tensor.
Within hydrodynamics, we write the coefficients appearing in the decomposition in terms of the hydrodynamic variables $u_\mu,\,T,\,$ and $\mu$ (constitutive relations).

In the hydrodynamic expansion, we expand the constitutive relations in a derivative expansion in the hydrodynamical variables. The zeroth order of the hydrodynamic expansion corresponds to ideal hydrodynamics (thermodynamics) and no dissipative terms are present. The first order computes the dissipative corrections to ideal hydrodynamics.

\subsubsection{Zeroth order}
The transverse vectors and tensor ${\cal Q}^\mu,\,{\cal J}^\mu,{\cal T}^{\mu\nu}$ can only be constructed in terms of derivatives. The scalars ${\cal E},\,{\cal P},\,{\cal N}$, however, are functions of the hydrodynamic variables. The zeroth order coefficients of the expansion corresponds to thermal equilibrium and are thus determined by means of the energy-momentum tensor $T^{\mu\nu}=\text{diag}(\bar\varepsilon,p,\ldots,p)$ and current $J^\mu=(n,0,\ldots,0)$ in static equilibrium. In these expressions, $\bar\varepsilon$ corresponds to the energy density, $p$ is the thermodynamic pressure and $n$ the equilibrium charge density, respectively. In case of a fluid flowing with constant velocity $u^\mu(x)$ the local thermodynamic quantities correspond to the scalars, i.e. ${\cal E}(x)=\varepsilon(x),\,{\cal P}(x)=p(x),\,{\cal N}(x)=n(x)$. We may extract the charge density $n$, the entropy density $s$, and the energy density $\bar\varepsilon$ form the equilibrium equations of state $p(T,\mu)$ by $n=\partial p/\partial\mu,$ $s=\partial p/\partial T$, and $\bar\varepsilon=-p+s\, T+n\,\mu$. Note that the entropy does not increase in ideal (zeroth order) hydrodynamics.

\subsubsection{First order}
The first order takes dissipative effects of the equilibration process of the fluid into account and corresponds to a non-equilibrium process. Out-of-equilibrium, the local thermodynamic quantities are ambiguous; specifically, their definition depends on the frame we choose. The transport coefficients, which are physical observables of the system, have to be frame invariant. One method to fix the redefinition of the variables is the so-called Landau frame (no energy flow in the local rest frame).
In this frame, the constitutive relations read to first order 
\begin{align}
    &T^{\mu\nu}=\varepsilon\,u^\mu u^\nu+p\Delta^{\mu\nu}\!-\eta\,\Delta^{\mu\alpha}\Delta^{\nu\beta}\!\left(\partial_\alpha u_\beta+\partial_\beta u_\alpha-\frac 2d\,\eta_{\alpha\beta}\,\partial_\mu  u^\mu\!\right)-\zeta\Delta^{\mu\nu}\partial_\lambda u^\lambda+\mathcal O(\partial^2)\nonumber\\
    &J^\mu =n\,u^\mu-\sigma\,T\,\Delta^{\mu\nu}\,\partial_\nu(\mu/T)+\chi_T\,\Delta^{\mu\nu}\partial_\nu T+\mathcal O(\partial^2),
\end{align}
where we introduced the coefficients $\sigma$ (charge conductivity),$\,\chi_T$, $\zeta$ (bulk viscosity) and $\eta$ (shear viscosity) which have to be determined from the underlying microscopic theory.

\subsubsection{Correlation functions from hydrodynamics}
In this section, we connect the hydrodynamic formalism to the retarded Green's function which we considered within linear response theory. In general, the source $\lambda_a$ causes a shift in the expectation values of the hydrodynamic variables $\varphi_a$
\begin{equation}
    \delta\langle \varphi_a(\omega,\bm{k})\rangle=G_{ab}^R(\omega,\bm{k})\,\lambda_b(\omega,\bm{k}).
\end{equation}
At $t=0$, the hydrodynamic variables $\varphi_a$ and the sources $\lambda_a$ are related (in Fourier space) by the thermodynamic susceptibilities
\begin{equation}
    \varphi_a(\omega=0,\bm{k}\to0)=\chi_{ab}\,\lambda_b(\omega=0,\bm{k}\to 0), \ \ \chi_{ab}=\left(\frac{\partial\varphi_a}{\partial\lambda_b}\right).
\end{equation}
This concept is easy to generalize in order to obtain the hydrodynamic dispersion relations. For simplicity, we restrict ourselves to $\mu=0$ and 2+1 dimensions. In this case, the hydrodynamic equations of the preceding section read
\begin{align}
    \dot\varepsilon+i\,k\,\pi_\|=0,&& \dot\pi_\|+i\,k\left(\frac{\partial p}{\partial\bar\varepsilon}\right)\varepsilon+\frac{\eta+\zeta}{\bar\varepsilon+p}\,k^2 \pi_\|=0,&& \dot\pi_\perp+\frac{\eta}{\bar\varepsilon+p}\,k^2\,\pi_\perp=0,
\end{align}
where we denote the equilibrium energy density by $\bar \varepsilon$ in order to distinguish it from the hydrodynamic variable $\varepsilon$.
In the transverse sector, it is straightforward to write down the Green's function since the hydrodynamic equation of motion is formally a diffusion equation
\begin{equation}\label{eq:transverse_fluid}
    G^R_{\pi_\perp\pi_\perp}(\omega,k)=-\frac{\eta\,k^2}{i\,\omega-\eta\,k^2/(\bar\epsilon+p)}.
\end{equation}
Taking the imaginary part of eq. \eqref{eq:transverse_fluid}, we find the Kubo formula for the shear viscosity
\begin{equation}
    \eta=-\lim\limits_{\omega\to 0}\lim\limits_{k\to 0}\frac{\omega}{k^2}\,\text{Im}\,G^R_{\pi_\perp\pi_\perp}(\omega,k).
\end{equation}
Similarly to the transverse sector, we are able to derive the Green's function in the longitudinal sector which consists of the two coupled equations for energy- and longitudinal momentum conservation in the order $\varphi_a=(\varepsilon,\pi_\|),\,\lambda_a=(\delta T/T,v_\|)$
\begin{equation}
    G_{ab}^R(\omega,k)=-\frac{\eta/(\bar\varepsilon+p)}{\omega^2-k^2\,\partial p/\partial\bar\varepsilon+i\,\omega\,(\eta+\zeta)/(\bar\varepsilon+p)\,k^2}\begin{pmatrix}k^2&\omega\,k\\\omega\,k& k^2\,\frac{\eta}{\bar\varepsilon+p}-i\,\omega\,\frac{\eta+\zeta}{\bar\varepsilon+p}\,k^2\end{pmatrix}.
\end{equation}

\section{Entanglement Entropy from QFT and Holography}
After discussing local observables within the AdS/CFT correspondence in terms of local operators, we will now switch to non-local observables.
In quantum field theories, the notion of entanglement is very natural.
If we divide a quantum system into two parts, the natural question arises: how entangled are they? As we will see throughout this chapter, entanglement is directly related to the degrees of freedom of the theory. In a two-dimensional CFT, for example, the entanglement entropy is proportional to the central charge $c$. For dimensions greater than $d=2$, computing the entanglement entropy is a daunting task even in free theories. Using the AdS/CFT correspondence, however, we can translate the complicated task of doing the calculation in quantum field theory to a problem of differential geometry in a classical theory of gravity.  

\subsection{Entanglement Entropy in Field Theory}
In this subsection, we review the basic definition and how to compute entanglement entropy in quantum theories. At zero temperature, the system is in the pure ground state $|\Psi\rangle$, which we assume to be non-degenerated. In order to compute the von Neumann entropy of the quantum system, we have to compute the density matrix $\rho_\text{tot}=|\Psi\rangle\langle\Psi|$.

In the ground state, the von Neumann entropy is zero by $S_\text{tot}=-\text{tr }\rho_\text{tot}\,\log\rho_\text{tot}=0.$  If we divide the system in two parts $A$ and $B$, where $\mathcal H_\text{tot}=\mathcal H_A\otimes\mathcal H_B$, an observer restricted to subsystem $A$ will only see part of the density matrix describing the full system
\begin{equation}\label{eq:reddens}
\rho_A=\text{tr}_B\rho_\text{tot}.
\end{equation}
Using the reduced density matrix eq. \eqref{eq:reddens}, the entanglement entropy of subsystem $A$ is defined as the von Neumann entropy in the reduced Hilbert space
\begin{equation}
S_A=-\text{tr}_A\rho_A\,\log\rho_A. \label{eq:vonNeumann}
\end{equation}
Another important quantity is the so-called R\'enyi entropy $(n\neq1)$
\begin{equation}\label{eq:Renyi}
    S^{(n)}=\frac{1}{1-n}\log\text{tr}\,\rho_A^n,
\end{equation}
which reduces to the von Neumann entropy \eqref{eq:vonNeumann} in the limit $n\to 1$.
With the notion of entanglement entropy at hand, we can work out the procedure of how to compute it. In general, it is not possible to compute the right side in \eqref{eq:vonNeumann} for a generic subsystem $A$ explicitly which we need in order to apply \eqref{eq:vonNeumann}. Calabrese and Cardy realized that we can adapt a trick from statistical physics -- the replica trick. Let $\lambda_i$ be the $i$-th eigenvalue of the reduced density matrix $\rho_A$ and we furthermore assume that all eigenvalues lie in the range $\lambda_i\in[0,1]$ and the sum is normalized to $\sum_i\lambda_i=1$. Then we find that the sum $\text{tr}\rho^n_A=\sum_i\lambda_i^n$ is analytic for Re$(n)>1$, even for non-integer $n$\cite{Calabrese:2009qy}
\begin{equation}
-\frac{\partial}{\partial n}\left.\text{tr}_A\,\rho^n_A\right|_{n=1}=-\frac{\partial}{\partial n}\left.\log\text{tr}_A\,\rho^n_A\right|_{n=1}=\left.\frac{\text{tr}_A\,\rho^n_A-1}{1-n}\right|_{n=1}=S_A.\label{eq:EEQFT}
\end{equation}
The last equality holds because the R\'enyi entropy \eqref{eq:Renyi} converges in the limit $n\to 1^+$  to the value of the entanglement entropy \eqref{eq:vonNeumann} if \eqref{eq:vonNeumann} is finite.
\begin{figure}[h] 
	\centering
	\includegraphics[width=8.5cm]{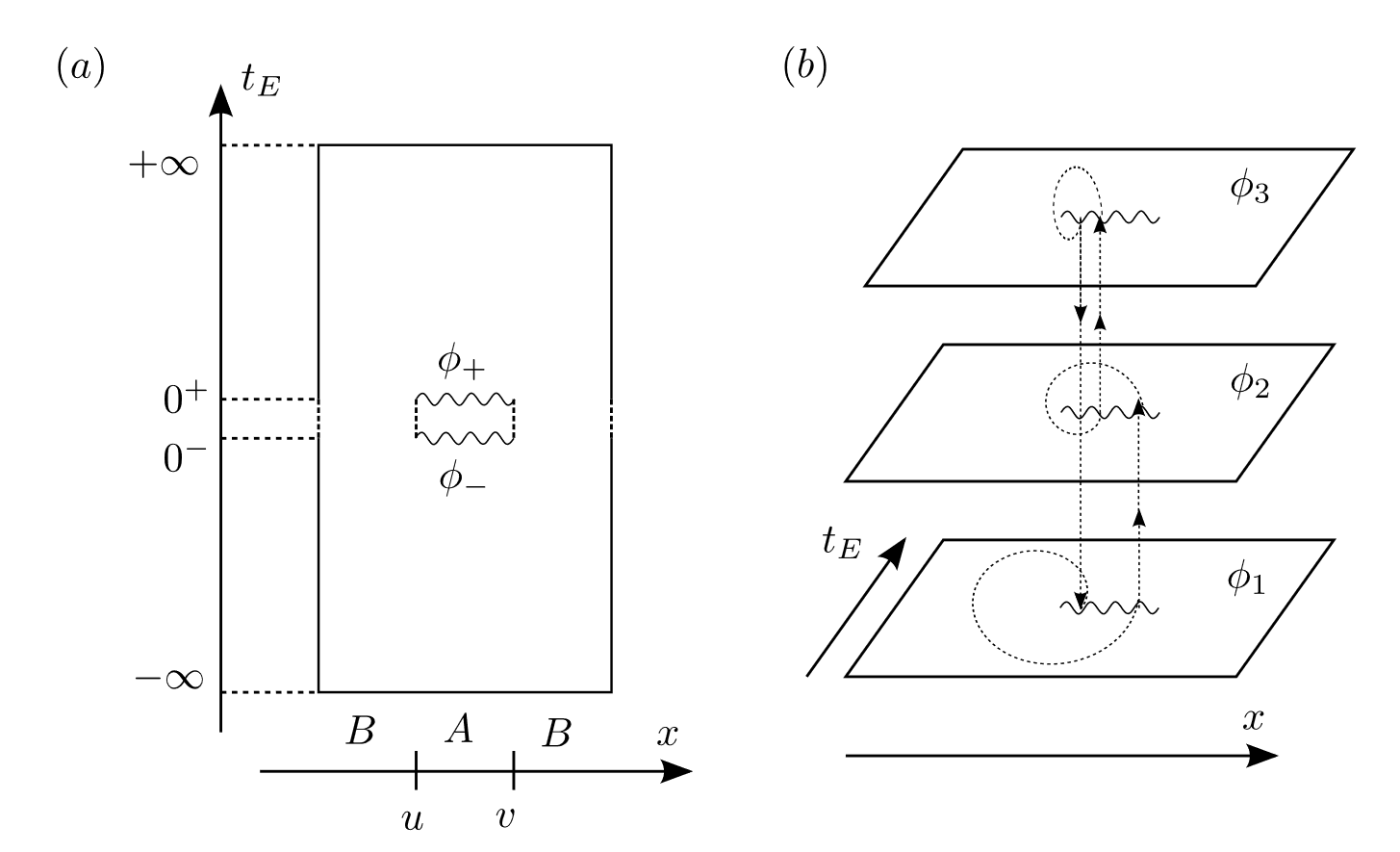}
	\caption{Cartoon of the $n$ copies sewn together to form a $n$-sheeted Riemann surface. Graphic taken from~\cite{Nishioka:2009un}.}\label{fig:EEreplicaglue}
\end{figure}
According to eq. \eqref{eq:EEQFT}, we have to compute the trace of the $n$ reduced density matrices $\rho_A^n$ for a given subsystem $A$ within the QFT. In the following, we consider a two-dimensional QFT, where $A$ is the single interval $x\in[u,v]$ at $t_E=0$ in Euclidean signature (see the left side of figure \ref{fig:EEreplicaglue} for a graphical representation). The ground state follows from by integrating the partition function over all possible field configurations $\phi$ from $t_E=-\infty$ to $t_E=0$
\begin{equation}
\Psi(\phi_0(x))=\int_{t_E=-\infty}^{\phi(t_E=0,x)=\phi_0(x)} \mathcal D\phi\,e^{-S(\phi)}.\label{eq:gsRR}\end{equation}
The density matrix is given in terms of the ground state eq. \eqref{eq:gsRR} by $\rho(\Psi(\phi_0)\bar\Psi(\phi_0'))$. We have to integrate out the degrees of freedom in region $B$ since we are interested in obtaining the reduced density matrix $\rho_A$. 
At $t_E=0$, we integrate out $\phi(t_E=0,x)$ on $B$ by imposing the boundary condition $\phi_0(x)=\phi'_0(x)$ for all $x\in B$ and $0^\pm$ as defined in figure \ref{fig:EEreplicaglue}
\begin{equation}
    \rho_A(\phi_+\phi_-)=(Z_1)^{-1}\int_{t_E=-\infty}^{t_E=\infty}\mathcal D\phi\,e^{-S[\phi]}\prod\limits_{x\in A}\delta(\phi(0^+,x)-\phi_+(x))\,\delta(\phi(0^-,x)-\phi_-(x)),
\end{equation}
where $Z_1$ is the vacuum partition function we use to normalize the reduced density matrix $\text{tr}_A\,\rho_A=1$. In order to apply the replica trick as explained in eq. \eqref{eq:EEQFT}, we have to make $n$ copies of the reduced density matrix and trace over them successively
\begin{equation}
    \text{tr}_A\,\rho_A^n=\text{tr}_A\left(\rho_A(\phi_{1+}\phi_{1-})\dots\rho_A(\phi_{n+}\phi_{n-})\right)=Z_1^{-n}\int_{(t_E,x)\in\mathcal R_n}\mathcal D\phi\,e^{-S[\phi]}=\frac{Z_n}{(Z_1)^n},
\end{equation}
where we are gluing the $n-$copies $\{\phi_{i\pm}(x)\}$ together as $\phi_{i-}(x)=\phi_{(i+1)+}(x)$ and integrate over $\phi_{i+}(x)$ (see the right side of figure \ref{fig:EEreplicaglue} for a cartoon).

\subsection{Holographic Entanglement Entropy}
In holography, the notion of entanglement entropy is an example of how the emergent extra radial direction organizes the degrees of freedom in the dual CFT. Motivated by the Bekenstein-Hawking formula for the entropy of black hole horizons, Ryu and Takayanagi suggested that the generalization of the Bekenstein-Hawking formula to the area $A$ of minimal surfaces $\gamma_A$ is the holographic dual of the entanglement entropy in quantum field theory
\begin{equation}\label{eq:RTformula}
    S_\text{EE}=\frac{A(\gamma_A)}{4\,G_N},
\end{equation}
where the $d-2$ dimensional manifold $\partial\gamma_A=\partial A$ corresponds to the boundary of subsystem $A$. In order to illustrate the Ryu-Takayanagi formula, we calculate the entanglement entropy of a spherical region of radius $R$~\cite{Casini:2011kv,Rangamani:2016dms,Headrick:2019eth,Hartnoll:2016apf}. For spherical symmetry, the boundary metric in a constant time slice reads $\dd s^2_\text{bdy}=\dd \rho^2+\rho^2\,\dd\Omega_{d-2}^2$, where $\rho$ is the radial coordinate of the boundary. In these coordinates, the entangling surfaces are parametrized by $z=z(\rho)$ and we can write the Lagragian for the minimal area surfaces in terms of the induced metric
\begin{equation}
    \mathcal L(z(\rho),\rho)
=\frac{L^{d}}{z^{d}}\,\rho^{d-1}\,\sqrt{1+\left(z'(\rho)\right)^2}.\label{eq:minimalsurfacespherical}
\end{equation}
After imposing the boundary conditions $z(R)=0$ and $z'(0)=0$, we find the solution
    $z(\rho)=\sqrt{R^2-\rho^2}.$ 
The regularized area follows from integrating the induced metric evaluated on the solution \eqref{eq:minimalsurfacespherical}
\begin{align}\label{eq:areagamma}
     A(\gamma_A)&=L^{d-1}\Omega_{d-2}\,\int_{0}^{R-\epsilon}\dd\rho\,\frac{R\,\rho^{d-2}}{(R^2-\rho^2)^{d/2}}\\&=\frac{L^{d-1}\Omega_{d-2}}{d-1}\left(1-\frac{\varepsilon}{R}\right)^{d-1}\,{}_2F_1\!\left(\frac{d-1}{2},\frac d2,\frac{d+1}{2},\left(1-\frac{\varepsilon}{R}\right)^2\right),
\end{align}
where $\text{}_2F_1$ is the hypergeometric function $\text{}_2F_1(a,b;c;z)$ and $\varepsilon\ll1$ is the cutoff. In the $\varepsilon\to 0$ limit of \eqref{eq:areagamma}, we see that the leading term of the entanglement entropy \eqref{eq:RTformula} depends on the dimension of the spacetime and is given by~\cite{Casini:2011kv}
\begin{align}
    S_\text{EE}\sim\begin{cases}\frac{L^{d-1}}{G_N}\left(\frac R\varepsilon\right)^{d-2}+\text{subleading}& d>2,\\
    \frac{L}{2G_N}\,\log\frac R\varepsilon+\text{subleading}&d=2\,.\end{cases}
\end{align}
In $d=2$, we find with the central charge $c=3L/(2\,G_N)$ the universal result
\begin{equation}\label{eq:universald2}
    S_{\text{EE}}=\frac{c}{3}\log\frac{R}{\varepsilon}+\ldots\,.
\end{equation}

\section{The DS/dS-correspondence}
One attempt to resolve the puzzle of quantum gravity in de Sitter (dS) -- using the powerful tools of holography -- is the DS/dS-correspondence \cite{Alishahiha:2004md} which is based on uplifting the AdS/CFT correspondence \cite{Alishahiha:2005dj,Dong:2010pm,Freivogel:2006xu,Dong:2011uf}. In contrast to AdS, dS may be viewed as two asymptotic AdS IR regions glued together on their UV slice~\cite{Cotler:2019dcj,Cotler:2019nbi}. This scenario corresponds to a Randall-Sundrum type setup \cite{Randall:1999vf} with a localized graviton on the UV slice \cite{Karch:2003em}. The DS/dS-correspondence conjectures that warped dS$_{d+1}$ spacetime is dual to two $d$-dimensional CFTs which are coupled to one another and are living on the central UV slice. Due to the localized graviton on the central slice, the CFTs are also coupled to gravity.

The basic idea of DS/dS becomes apparent by writing the $(d+1)$-dimensional (A)dS as a warped spacetime in $d$-dimensional dS$_{d}$
\begin{equation}
\dd s^2= \dd r^2 + e^{2A(r)}\, \dd s^2_{dS_{d}},\ \ \ e^{A(r)}=\begin{cases}L\,\sin(r/L)&\text{for dS}_{d+1}\\L\,\sinh(r/L)&\text{for AdS}_{d+1}\end{cases}\label{Adsds}
\end{equation}
where we denote the radial direction by $r$. While the warpfactor of AdS grows without bound for $r\to\infty$, the warpfactor of dS reaches a maximum on the central (UV) slice located at $r/L=\pi/2$. This inherently implies that the dS spacetime has a built-in UV cutoff and all quantities are automatically regularized. Both warp factors vanish on the horizon $r=0$. Note that the warp factor of dS has a second zero at $r=\pi L$, indicating a second horizon. In the highly redshifted region $r/L\ll 1$, the warp factors both vanish linearly and the spacetimes are indistinguishable. For dS$_d$ sliced AdS$_{d+1}$, the AdS/CFT-correspondence provides us with a dual CFT. Since the spacetimes are identical in the infrared region, the authors of \cite{Alishahiha:2004md} conjectured, that this is also the dual CFT to dS$_{d+1}$. We will see later that this dual may be constructed for the whole spacetime by deforming the dual field theory with a so-called $T\bar T$ deformation \cite{Gorbenko:2018oov}. Concretely, the proposal of \cite{Alishahiha:2004md} suggests that the holographic dual of dS$_{d+1}$ consists of two $d$-dimensional CFTs which live on the central UV slice. In particular:
    \begin{mdframed}[backgroundcolor=gray!20]\begin{center} 
Quantum Gravity on dS$_{d+1}$ (+matter)\\ is \textit{equivalent} to\\
Two $d$ dimensional CFTs with UV cutoff + dynamical gravity on dS$_d$.
\end{center}    \end{mdframed}
The two CFTs act only as a pair at fairly low energies and are coupled in terms of irrelevant interactions~\cite{Dong:2018cuv}
\begin{equation}
    S_\text{mix}\sim\int\dd^dx\,\sqrt{-\gamma}\,L^{2\Delta-d}_\text{dS}\mathcal O_1\mathcal O_2+\ldots,
\end{equation}
where $\gamma$ is the metric of the $d$-dimensional gravity theory on the central slice.

\subsection{Density matrix in 2d CFT with cutoff}
We consider a CFT in $d=2$ with large central charge dual to classical gravity. In the energy spectrum, there is a transition $E_{\text{CFT}\star}=\Delta-\frac{c}{12}=\frac{c}{12}$, where the entanglement entropy starts following a Cardy formula~\cite{Dong:2018cuv}. The number of states in the CFT Hilbert space -- up to energy $ E_{\text{CFT}\star}$ -- is to leading order in the central charge $c$ given by
$\text{dim}(H_{\Delta\le c/6})=e^{\pi c/3}.$ The strong interactions between the two CFTs lead to an approximately maximally entangled state with density matrix given by
\begin{equation}
    \rho_{1, \text{max}}\simeq\mathbb{I}\frac{1}{ \text{dim}(H_{\Delta\le c/6})}.
\end{equation}
The entanglement entropy of this maximally entangled state is given by tracing out one of the CFTs~\cite{Dong:2018cuv}
\begin{equation}\label{eq:cftcutoff}
  S=\log\text{dim}(H_{\Delta\le c/6})=\frac{\pi c}{3}.
\end{equation}

\subsection{Holographic entanglement entropies in DS/dS}
For the dS$_{d+1}$ metric in the coordinates of \eqref{Adsds}, the minimal areas minimize
\begin{equation}\label{eq:dsdsdsdsds}
\mathcal L=L^{d-1}\,\cos^{d-2}(\beta)\,\sin^{d-2}\left(\frac {r(\beta)}{L}\right)\,\sqrt{(r'(\beta)^2+L^2\,\sin^2\left(\frac {r(\beta)}{L}\right)}.
\end{equation}
where we used the dS$_d$ metric \mbox{$\dd s^2_{\text{dS}_d}=- \sin^2 \beta \dd \tau^2 + \dd \beta^2 + \cos^2 \beta \, \dd\Omega^2_{d-2}$} in the static patch $\tau=0$. One consistent solution minimizing eq. \eqref{eq:dsdsdsdsds} is given in terms of a great circle with $r/L\equiv\frac\pi2$.
The constant solution for the entangling surface leads to a volume law for the entanglement entropy~\eqref{eq:RTformula}.
For a volume of the size of de-Sitter $A=L_\text{dS}$, the volume is given by $A_{d-1}(\gamma_A)=2\pi^{d/2}\,L_\text{dS}^{d-1}/\Gamma(d/2)$
\begin{equation}
    S_{d=2,\,\text{EE}}=\frac{2\pi\,L_\text{dS}}{4\,G_N}=\frac{\pi c}{3},
\end{equation}
where we used the expression $c=3\,L_\text{AdS}/(2\,G_N)$ and identified $L_\text{dS}=L_\text{AdS}$ in the infrared. Note that this result is in perfect agreement with the universal CFT result for CFTs with cutoff \eqref{eq:cftcutoff}. The result is finite within the field theory and holography because the Hilbert space of the former and the radial distance of the latter are finite.

\chapter{Broken spacetime symmetries}\label{chapter:brokenspacetime}
The spontaneous breakdown of a global symmetry, such as translational invariance, leads to the appearance of gapless degrees of freedom -- the Goldstone bosons. Goldstone bosons were first introduced in the context of the BCS theory of superconductivity \cite{PhysRev.117.648,Goldstone:1961eq,PhysRev.127.965}. Imagine, the global symmetry is not an exact symmetry but lightly broken by a small explicit source term. Breaking this not-exact symmetry spontaneously leads still to \mbox{(pseudo-)}Goldstone bosons in the spectrum which acquire a small mass. In this chapter, we study the effects of broken translational invariance on the hydrodynamic spectrum. First, we discuss the set-up in terms of spontaneously broken translations and establish the hydrodynamic and holographic description and compare the predictions within our holographic model. After discussing the spectrum of the hydrodynamic modes, we proceed to break translations first explicitly and then discuss the so-called pseudo-spontaneous regime. To conclude the discussion about broken translational invariance, we go beyond the linear response regime and discuss the full time-dependent response. This chapter is based on my work published in~\cite{Ammon:2019apj,Baggioli:2019abx,Baggioli:2019mck,Ammon:2020xyv} in collaboration with Martin Ammon, Matteo Baggioli, Se\'an Gray, Akash Jain, and Hesam Soltanpanahi.

\section{Fluids and solids -- a symmetry consideration}
In this section, we consider a medium extended throughout the whole spatial volume of the flat spacetime. In $d$ spacetime dimensions, we may define an invertible mapping $\phi$ between the physical coordinates and the internal space by introducing $(d-1)$-scalar fields $\phi^I(t,\bm{x})$, $I
 \in[1,\dots,d]$~\cite{Dubovsky:2011sj,Dubovsky:2011sk,Bhattacharya:2012zx,Son:2002zn,Dubovsky:2005xd,Esposito:2017qpj,Endlich:2012vt,Nicolis:2015sra,Nicolis:2013lma}. The $\phi^I$ are the comoving coordinates which characterize the position of local fluid elements at a given time in terms  of the spacetime coordinates $\bm{x}$ and $t$. 
 The comoving coordinates $\phi^I$ are related to the normalized hydrodynamic velocity field $u_\mu(x)$ in terms of an orthogonality constraint
 \begin{equation}
     \frac{\dd }{\dd\tau}\phi^I(x(\tau))\equiv u^\mu\,\partial_\mu \phi^I=0,
 \end{equation}
 where $u_\mu$ satisfies $u_\mu u^\mu=-1$. Note that the comoving coordinates $\phi^I$ stay constant along the lines of $u_\mu$.
 
 The symmetry group of fluids is very large and includes for example shifts and rotations of the comoving coordinates. The comoving coordinates of fluids respect internal volume preserving diffeomorphisms (VPDiffs) given by
 \begin{equation}
     \phi^I\to\xi^I(\phi^J), \ \quad \det(\partial\xi^I/\partial\phi^J)=1.\label{eq:VDiff}
 \end{equation}
In equilibrium, we may choose the comoving coordinates to agree with the physical ones by setting (where $i,j,I,J$ sum over the spatial dimensions in this section)
\begin{equation}
    \langle \phi^I\rangle= x^i\,\delta_i^I. \label{eq:equil}
\end{equation}
By making this particular frame choice, the field configuration breaks the internal symmetries down to a linear combination of internal- and spacetime translations and rotations. At sufficiently low energies and momenta, fluctuations about the equilibrium state are totally dominated by the Goldstone bosons associated with the breakdown of the continuous symmetries by the ground state. To summarize the symmetries of the system, let us denote the generators of spacetime translations, rotations, and Lorentz boosts by $P_\mu, J_{ij}$ and $K_i$, respectively. We also refer to internal rotations as $L_{IJ}$. The following symmetries are (un)broken\footnote{Note that we do not consider charge in our description. This may be done by introducing and additional scalar field $\phi^0$ which transforms under the U(1) symmetry. In that case, the symmetry group is bigger and we have to include shifts in the chemical potential in the category of broken symmetries.}~\cite{Nicolis:2013lma,Nicolis:2013sga}:
\begin{align*}
    \text{unbroken}&=\begin{cases}
    \bar P_\mu\equiv P_\mu+a_\mu&\ \text{translations}\\
    \bar J_{ij}\equiv J_{ij}+L_{ij}&\ \text{rotations}
    \end{cases}\quad\quad
    \text{broken}&=\begin{cases}
    K_i&\ \text{boosts}\\
   a_\mu&\ \text{constant shifts}\\
    M_{ij}&\ \text{special linear}
    \end{cases} 
\end{align*}
What changes if we want to describe a solid instead of a fluid? The solid is only invariant under a small subset of the symmetry group defining the fluid eq. \eqref{eq:VDiff}. In particular, the solid is invariant under constant internal translations $T_\phi$ and rotations $R^I_{\ J}\in\,O(d-1)$
\begin{equation}
    T_\phi(a^I):\ \phi^I(t,x)\to\phi^I(t,x)+a^I, \quad \phi^I\to R^I_{\ J}\,\phi^J. 
\end{equation}
Even though the ground state breaks translations spontaneously, there is still a combination of internal and ordinary translations left invariant. For the translation operator $T_x(c):\ \phi^I(x,t)\to\phi^I(x+c,t)$, the combination $T_x(c)-T_\phi(c^I)$ is conserved and represents the momentum operator. The broken generators $T_x+T_\phi$ give rise to the $(d-1)$ Goldstone bosons of the spontaneously broken translational symmetry. The Goldstone bosons have to satisfy the commutation relations
\begin{equation}
    [\phi^I(x),\pi^J(y)]=i\,\delta^{(d-1)}(x-y)\,\left(\delta^{IJ}+\delta^{Jj}\, \partial_j\phi^I\right).
\end{equation}
The smaller internal symmetry group of the solid compared to the fluid requires us to replace $M_{ij}$ with $L_{ij}$ in the chart of symmetry breaking. In order to implement the translational and rotational invariance into the hydrodynamic description, we construct it from the derivatives of $\phi^I$ and demand the theory to be rotationally invariant.

\section{Elasticity theory and Phonons}
In elasticity theory, deforming a medium with a strain, is given in terms of the strain tensor~\cite{chaikin_lubensky_1995}
\begin{equation}
    u_{ij}=\frac 12\,(\partial_iu_j+\partial_j u_i),\label{eq:defo}
\end{equation}
where we introduced the displacement variable $\bm{u}.$
For simplicity, we want to restrict ourselves to homogeneous and isotropic mediums. The impact of the elastic deformations on the medium is captured by Hooke's law, which relates the stress tensor to the strain tensor
\begin{equation}
    \sigma_{ij}=\frac{\partial f}{\partial u_{ij}}=\kappa\,u^k_{\ k}\,\delta_{ij}+2\,G\left(u_{ij}-\frac 1d\,u^k_{\ k}\,\delta_{ij}\right),\label{eq:Hooke}
\end{equation}
where $f$ denotes the free energy. The parameter $G$ is the shear elastic modulus and $\kappa$ is the bulk modulus, respectively. By integrating with respect to $u_{ij}$, we find the free energy
\begin{equation}
    f=\frac 12\,\kappa \,(u^j_{\ j})^2+2\,G\,\left(u_{ij}u^{ij}-\frac1d \,(u^j_{\ j})^2\right)+c,\label{eq:freeenergy}
\end{equation}
where $c$ is a constant.
In the following, we recast the deformations in field theory language for two spatial dimensions. The effective action of the low energy theory reads
\begin{equation}
    S=\int \dd^{2+1}x\, F(X,Z),
\end{equation}
where $X$ and $Z$ are the building blocks
\begin{equation}
    X=\text{tr}\!\left(I^{IJ}\right), \quad Z=\det\!\left( I^{IJ}\right), \ I^{IJ}=\partial_\mu\phi^I\partial^\mu\phi^J.
\end{equation}
The indices $I,J$ run over the two spatial dimensions $I,J\in\{x,y\}$ and are raised and lowered with the Kronecker delta $\delta^{IJ}$. The homogeneous and isotropic equilibrium configuration is given by eq. \eqref{eq:equil}.
Under small deformations out of the equilibrium configuration, the $\phi^I$ transform as
\begin{equation}
    \phi^I=\langle\phi^I\rangle+ \Phi^I,\ I^{IJ}
\to \langle I^{IJ}\rangle+\partial_\mu\langle\phi^I\rangle\,\partial^\mu\Phi^J+\partial_\mu\Phi^I\,\partial^\mu\langle\phi^J\rangle+\mathcal O(\Phi^2),\end{equation}
with the equilibrium values $\langle I^{IJ}\rangle=\delta^{IJ}, \ \partial_\mu\langle\phi^I\rangle=\delta_\mu^{\ I}$. From this transformation property, we identify the elastic deformation defined in eq. \eqref{eq:defo} with
\begin{equation}
    u^{IJ}\equiv \frac 12\,\left(I^{IJ}-\langle I^{IJ}\rangle\right),
\end{equation}
where the displacements from the equilibrium configuration are encoded in $\Phi$. In this language, the free energy \eqref{eq:freeenergy} reads
\begin{equation}
    f=\frac12\,(\kappa+G)\,\lambda_{\|}^2+\frac12\,G\,\lambda_\perp^2+b\,\delta\epsilon\,\lambda_\|,\label{eq:freeenergy2}
\end{equation}
where we introduced variables for the transverse and longitudinal part of the Goldstone operator $\Phi^I$ by $\lambda_\|=\nabla\cdot\phi$ and $\lambda_\perp=\nabla\times\phi$.\footnote{The Kubo formulas in our holographic model indicate that $b=0$ and we thus neglect $b$ for simplicity.} For fluids, the elastic shear modulus $G$ is zero and we immediately see from \eqref{eq:freeenergy2} that the coupling of the Goldstones to the transverse part vanishes.

In the following, we want to embed the phonon physics into an hydrodynamic description.
The hydrodynamic description contains five hydrodynamic variables $\varphi^A$ and their corresponding sources $s_A$~\cite{Delacretaz:2017zxd,Ammon:2019apj,PhysRevA.6.2401,PhysRevB.22.2514}
\begin{equation}
    \varphi^A=\frac{\partial f}{\partial  s_A}=\{\delta \varepsilon,\,\pi_\|,\,\pi_\perp,\,\lambda_\|,\,\lambda_\perp\},\ \quad\quad\quad s_A=\{\delta T,\,v_\|,\,v_\perp,\,s_\|,\,s_\perp\},
    \label{eq:hydrovar}
\end{equation}
which are the energy density, the longitudinal and transverse momentum density, and the longitudinal and transverse part of the phonon. The hydrodynamic variables are accompanied by their sources, the temperature, velocity, and displacement. As we discussed in section \ref{sec:hydro}, the relation between $\varphi^A$ and $s^A$ is given by the thermodynamic susceptibilities
\begin{equation}\label{eq:suscept}
    \varphi_A=\chi_{AB}\,s^B,\ \ \chi_{AB}=\frac{\partial\varphi_A}{\partial s^B}, \ \ \quad\quad \chi_{AB}=\text{diag}(c_V,\chi_{\pi\pi},\chi_{\pi\pi},(\kappa+G)^{-1},G^{-1}),
\end{equation}
with the specific heat $c_V=\partial\varepsilon/\partial T$.
From eq. \eqref{eq:suscept} it is clear that the $G\to 0$ limit is subtle and must already be taken on the level of the free energy \eqref{eq:freeenergy2}.
The conservation equations of our model are energy conservation and momentum conservation
\begin{equation}\label{eq:conservation_spontaneous}
    \dot{\epsilon}+\partial^j\tau^0_{\ j}=0, \quad \dot\pi_i+\partial^j\tau_{ij}=0,
\end{equation}
respectively. The conservation equations are accompanied by the so-called Josephson relations. The Josephson relations, known from the hydrodynamic theory of superfluidity, are the equations of motions for the Goldstones. In equilibrium, the system evolves as $\dot\phi_i=\dot u_i= v_i$ which provides a non-dissipative coupling between the elastic displacement variable $u\sim s$ and the momentum density $\pi$.\footnote{The coupling is non-dissipative since it describes a steady state.} This implies that we treat $\lambda\sim s\sim \partial\phi$ as zeroth order in the derivative expansion.\footnote{Note that this situation is similar to the superfluid where the equation of motion for the Goldstones are related to changes of the free energy with respect to the particle number $\dot\phi=-\mu$.} Out-of-equilibrium, the Josephson relation reads to first order in the derivatives
\begin{equation}\label{eq:josephson_full}
    \dot\phi_i=v_i+\gamma_2\,\partial_iT+\xi_\perp\,\partial^j(\partial_j\phi_i-\partial_i\phi_j)+ \xi_\|\,\partial_i\partial^j\phi_j,
\end{equation}
where $\gamma_2,\xi_\perp,\xi_\|$ are the dissipative coefficients.
By taking the divergence and the rotation of eq. \eqref{eq:josephson_full}, we find the Josephson relation for the longitudinal and transverse Goldstones. To complete the hydrodynamic description, we have to write down the constitutive relations which relate the currents to their respective thermodynamic variables.

The linear displacements $u$ act as sources for the energy momentum conservation. By splitting them in their transverse $s_\perp$ and longitudinal $s_\|$ part, the (symmetrized) energy-momentum tensor takes the form
 \begin{align}\label{eq:const_em_pho}
 \tau^0_i=&\chi_{\pi \pi}\,v_i-\bar\kappa_0\,\partial_iT-T\,\gamma_2\,\partial_i s_\|+\mathcal O(\partial^2),\\
 \tau_{ij}=&\delta_{ij}\,\left[ p-(\kappa+G)\,\partial\cdot\Phi\right]- 2G\,\left[\partial_{(i}\Phi_{j)}-\delta_{ij}\,\partial\cdot\Phi\right]-\sigma_{ij}+\mathcal O(\partial^2),
 \end{align}
 where $\bar\kappa_0$ and $\gamma_2$ are transport coefficients, $p$ the thermodynamic pressure, and $\sigma_{ij}$ are the one-derivative corrections. The  $\sigma_{ij}$ have to respect the same symmetries as the elastic deformations in eq. \eqref{eq:Hooke}
\begin{equation}
    \sigma_{ij}=\eta\,\left(\partial_iv_j+\partial_jv_i-\delta_{ij}\,(\nabla\cdot v)\right),
\end{equation}
where we introduced the shear viscosity $\eta$ and already set the bulk viscosity to zero as it vanishes in CFTs \cite{Son:2007vk}.
\section{The holographic model}
After laying out the field theoretical framework, we want to construct the holographic gravity dual incorporating broken translations.
In order to describe the dual of strongly coupled crystal and fluid dynamics, we employ the simple bottom up model consisting of the Einstein-Hilbert action with a mass term for the graviton $\mathcal L_\phi$ \cite{Andrade:2013gsa,Vegh:2013sk,Baggioli:2014roa}
\begin{equation}
    S=M_p^2\,\int\dd^4x\sqrt{-g}\,\left(\frac R2+\frac{3}{L^2}+\mathcal L_\phi\right)+M_p^2\,\int\dd^3x\,\sqrt{-\gamma}\,K.
\end{equation}
The mass term for the graviton $\mathcal L_\phi$ breaks the translational invariance of the theory leaving us with a Lorentz violating gravity theory. However, adding generic mass terms for the graviton is a highly non-trivial task since the theory is plagued by various instabilities -- some such as the Boulware-Deser ghost only appearing at the nonlinear level~\cite{PhysRevD.6.3368}. In 2010, de Rahm, Gabadaze, and Tolley (dRGT) successfully constructed a stable massive gravity theory for asymptotically flat space~\cite{deRham:2010kj} (see \cite{RevModPhys.84.671} for a review). In general \cite{Alberte:2015isw,Vegh:2013sk,Rubakov:2004eb}, the mass terms couple via the metric perturbations $h^{\mu\nu}$ to the action as $\mathcal L_\phi\sim1/2(m_0(r)\,(h^{tt})^2+2m_1^2(r)\,(h^{ti})^2-m_2(r)\,(h^{ij})^2+\ldots),$ where the ellipsis denotes the remaining combinations. The mass term breaks the diffeomorphism invariance of the spacetime explicitly since the metric fluctuations are not invariant under coordinate transformations $x^\mu\to\tilde x^\mu(x)$. There is an elegant way to restore diffeomorphism invariance in the physical coordinates -- the St\"uckelberg trick. We introduce the St\"uckelberg scalars $\phi^I$, which are the physical coordinates in order to write the mass term as a gauge invariant combination of the spacetime metric
\begin{equation}
     I^{IJ}\equiv \eta^{\mu\nu}\,\partial_\mu\phi^I\,\partial_\nu\phi^J
\end{equation}
together with the reference metric in the configuration space of the St\"uckelbergs $f_{AB}(\phi^C)$. The breakdown of a subset of the diffeomorphism invariance in our gravity theory becomes apparent in the unitary gauge where the St\"uckelberg scalars read
\begin{equation}
    \langle\phi^I\rangle=x^i\delta_i^I,\label{eq:scalarprof}
\end{equation}
which spontaneously breaks the diffeomorphism invariance and internal symmetries up to a diagonal subgroup. The ground state is left invariant by the diagonal subgroup of transformations
\begin{equation}
    \phi^I\to\phi^I-\tilde\phi^I,\ \quad x^\mu\to x^\mu+\tilde x^\mu
\end{equation}
for the ground state \eqref{eq:scalarprof} since $\langle \phi^I\rangle\to\langle\phi^I\rangle+\tilde x^\mu-\tilde \phi^I$.
The St\"uckelberg scalars which correspond to the dual Goldstone operators at the conformal boundary render the graviton massive and in equilibrium their spatially linear profile \eqref{eq:scalarprof} breaks the translational invariance~\cite{Andrade:2013gsa,Vegh:2013sk}. In the following, we will focus our attention on the two fundamental phases of matter: solids and fluids. As argued in the field theory section, we use a set of bulk scalars $\phi^I(\bm{x},t,u)$ as building blocks
\begin{equation}
    I^{IJ}=\partial_\mu\phi^I\partial^\mu\phi^J,\quad X=\text{tr}(I^{IJ}),\quad Z=\text{det}(I^{IJ}),
\end{equation}
where potentials of the form $V(X,Z)$ describe solids and $V(Z)$ fluids. It is important to notice that the bulk St\"uckelberg scalars are generally different than the Goldstone operators in the dual field theory. All in all, the holographic model is given by the action~\cite{Baggioli:2014roa}
 \begin{equation}\label{eq:model}
     S=M_p^2\int\dd^4x\,\sqrt{-g}\,\left(\frac{R}{2}+\frac{3}{L^2}-m^2\,V(X,Z)\right),
 \end{equation}
 where $m$ is the coupling of the scalars to gravity. In order to ensure that our gravity theory is not plagued by the severe pathologies which might arise in massive gravity theories, we have to impose some restrictions on the form of the potential $V(X)$.
 As shown in \cite{Baggioli:2014roa}, the absence of ghosts in the theory imposes the constraint $V'(\bar X)>0$ on the potential, where $\bar X$ is evaluated on the equilibrium configuration. To make sure that the theory does not include gradient instabilities, we also demand that $1+\bar X\,V''(\bar X)/V'(\bar X)>0$.
 
 Finally, to complete the holographic model, we have to make an ansatz for the metric; we want to consider a strongly coupled CFTs in 2+1 dimensions at finite temperature; on the gravity side this is realized by a black brane in asymptotically AdS$_4$
 \begin{equation}
     \dd s^2=\frac{1}{u^2}\,\left(-f(u)\,\dd t^2-2\,\dd t\,\dd u+\dd x^2+\dd y^2\right),
 \end{equation}
 where $u$ denotes the radial direction. Note, that we chose Eddington-Finkelstein coordinates for numerical convenience and set the AdS curvature radius $L=1$. The horizon $u_h$ is the zero of the blackening factor $f(u)$ and the conformal boundary is located at $u=0$. 
 The energy momentum tensor for our setup follows by varying the action with respect to the metric and reads with $V_Y=\partial V/\partial Y$
 \begin{equation}
     T_{\mu\nu}^\phi=-\frac{2}{\sqrt{-g}}\frac{\delta S_\phi}{\delta g^{\mu\nu}}=-g_{\mu\nu}V+\partial_\mu\phi^I\partial_\nu\phi_I\,V_X+2\,(\partial_\mu\phi^I\,\partial_\nu\phi_I\, I^{J}_J-\partial_\mu\phi^I\,\partial_\nu\phi^J\, I_{IJ})\,V_Z\label{EMtensor:HMG}.
 \end{equation}
 The big advantage of those simple holographic models compared to similar approaches is that we may solve the background Einstein's equations analytically. This allows us to compute the thermodynamic properties of our system exactly. By integrating the background Einstein's equations with respect to the radial coordinate, we find the emblackening factor in terms of the scalar potential $V(X,Z)$
 \begin{equation}
     f(u)=u^3\int_{u}^{u_h}\frac{\dd \zeta}{\zeta^4}\left(3-m^2\, V(X(\zeta^2),Z(\zeta^4))\right).
 \end{equation}
 With the emblackening factor at hand it is straightforward to compute the temperature of the dual CFT given by
 \begin{equation}
     T=-\frac{f'(u_h)}{4\pi}=\frac{3-m^2\,V(X(u_h^2),Z(u_h^2))}{4\pi\,u_h}.
 \end{equation}
 The Einstein's equations are supplemented by the equations of motion for the St\"uckelberg scalars $\phi^I$
 \begin{equation}
     \partial_\mu\left(\sqrt{-g}\,g^{\mu\nu}\,\partial_\nu\phi^I\,\frac{\partial V}{\partial I^{IJ}}\right)=0.
 \end{equation}
In section \ref{sec:linrespoandhol}, we discussed that transport properties are encoded in the response of the system to small fluctuations. In holography, we find the linear response of the dual quantum field theory by solving the linearized equations for the fluctuations. Concretely, we consider perturbations of the metric and the scalar fields about the equilibrium state ($\phi^I=x^I,g_{mn}$)
 \begin{equation}
     \phi^I(t,\bm{x},u)=x^I+\epsilon\,\delta\phi^{I}(u)\,e^{i\, (ky-\omega t)}, \quad g_{mn}(t,\bm{x},u)=\hat g_{mn}+\epsilon\,h_{mn}(u)\,e^{i\,(ky-\omega t)},
 \end{equation}
 where we chose the momentum to point in the $y$-direction.
 The Goldstone operators on the dual field theory side are encoded in the asymptotic behavior of the St\"uckelberg scalars $\phi^I$ at the conformal boundary
\begin{equation}
\phi^I\sim\phi^I_{(0)}(\bm{x},u)+\phi^I_{(1)}(\bm{x},u)\,u^\alpha+\phi^I_\text{log}(\phi^I_{(0)}(\bm{x},u))\,u^\alpha\,\log(u)+\ldots,\label{asym}
\end{equation}
where the exponent depends on the exact form of the potential $V(X,Z)$. For polynomial potentials of the form $V(X)=X^N$ and $V(Z)=Z^M$, we find \mbox{$\alpha=5-2N$} and \mbox{$\alpha=5-4 M$}, respectively. The term $\phi^I_{(0)}(\bm{x},u)$ is always non-zero evaluated on the equilibrium configuration  since $\phi^I_{(0)}(\bm{x},u)=x^I$. 

According to the AdS/CFT-dictionary, the leading term in the asymptotic expansion \eqref{asym} corresponds to a source for the dual field theory operator $\mathcal O^I$, while the subleading free coefficient in \eqref{asym} corresponds to the vacuum expectation value of $\mathcal O^I$.\footnote{In the following, we only consider the standard quantization; in the case of the alternative quantization or mixed boundary conditions the discussion has to be altered appropriately.} However, which of the two coefficients in \eqref{asym} is the leading term is not a priori clear but rather depends on the value of $\alpha$ and thus on the form of the potential $V(X,Z)$. If $
\alpha>0$, $\phi^I_{(0)}$ acts as source term for the dual operator $\mathcal O^I$ and thus the translational symmetry is broken explicitly. In the case of $\alpha<0$, $\phi^I_{(0)}$ is associated with the vacuum expectation value for the dual operator $\mathcal O$ in the field theory and $\phi_{(1)}$ is the source term. In that case, the translational symmetry breaking is spontaneous and totally dynamical for the equilibrium solution $\phi^I=x^I$. To summarize, for potentials of the form 
\begin{equation}V(X,Z)=\begin{cases}X^N,\quad N>5/2\\ Z^N,\quad N>5/4
\end{cases}\label{eq:potentials}
\end{equation} the field theory operator $\mathcal O^I$ is not explicitly sourced and translations are broken spontaneously. In the following, we will focus on the transport properties in systems where the translational invariance is broken explicitly (EXB) or spontaneously (SSB) and the interplay of EXB and SSB \cite{Alberte:2017cch,Alberte:2017oqx,Ammon:2019apj,Ammon:2019wci,Baggioli:2019abx}. 

\section{Spontanous symmetry breaking}
In the last section, we laid out the fundamentals about the physics of spontaneously broken translations. The spontaneous breakdown of the continuous translational symmetry leads to Goldstone bosons in the spectrum~\cite{Goldstone1961,PhysRev.127.965}.  In this section, we want to investigate the hydrodynamic modes from a holographic perspective and explicitly compute the dispersion relations and transport coefficients. For simplicity, we consider a momentum pointing in $y$-direction and classify the direction with respect to their orientation to the momentum. 

\subsection*{Transverse sector}
In the transverse sector, the hydrodynamic variables (introduced in eq. \eqref{eq:hydrovar}) are the transverse phonon $\lambda_\perp$ and momentum density $\pi_\perp$. The corresponding hydrodynamic equations are given in terms of momentum conservation eq.~\eqref{eq:conservation_spontaneous} and the transverse part of the Josephson relation eq. \eqref{eq:josephson_full}. By evaluating the hydrodynamic equations for the constitutive relations of the energy-momentum tensor eq. \eqref{eq:const_em_pho} in Fourier space, it is straightforward to obtain the dispersion relations in the transverse sector; in particular, since the sector is composed of two hydrodynamic variables, we are expecting two modes consisting of the two propagating diffusive sound-like modes -- the transverse phonons ~\cite{chaikin_lubensky_1995,Ammon:2019apj}
\begin{equation}
    \omega=\pm c_\text{T}\,k-i\, D_\text{T}k^2,
\end{equation}
with the transverse speed of sound $c_\text{T}$ and the transverse diffusion constant $D_\text{T}$ given by
\begin{equation}
    c_\text{T}^2=\frac{G}{\chi_{\pi\pi}}, \quad\quad\quad\quad D_\text{T}=\frac{1}{2}\,\left(G\,\xi_\perp+\frac{\eta}{\chi_{\pi\pi}}\right).\label{eq:transdisp}
\end{equation}
Recall that $G$ is the shear elastic modulus, $\chi_{\pi\pi}$ the momentum susceptibility, $\eta$ the shear viscosity and $\xi_\perp$.
The quantities on the right hand side of eq. \eqref{eq:transdisp} are encoded in Kubo formulas which we may compute in terms of linear response theory within holography. The left hand side, i.e. the speed of sound and attenuation constant are encoded in the quasi-normal modes. This means, we can verify the hydrodynamical predictions by using the hydrodynamic formulas with the transport coefficients and thermodynamical quantities extracted from our holographic model and compare to the dispersion relations obtained by means of QNM. The Kubo formulas relevant for the sound modes are given by
\begin{align}\label{defd}
    &G\,=\,\lim_{\omega \rightarrow 0}\,\lim_{k \rightarrow 0} \, \textrm{Re}  \left[G^R_{T_{xy}T_{xy}}(\omega,k)\right],
    \\ \label{defd2}
    &\chi_{\pi\pi} = \lim_{\omega \to 0} \omega \lim_{k \to 0} \frac{1}{k} \mathrm{Re}[G^R_{\varepsilon\pi_\|}(\omega,k)],\\ \label{defd3}
    & \xi_\perp\,=\,\lim_{\omega\, \rightarrow 0}\,\omega\,\lim_{k \to 0}\,\mathrm{Im}\left[G^R_{\Phi^{(1)}\Phi^{(1)}}\left(\omega,k\right)\,\right],\\ \label{defd4}
    & \eta\,=-\,\lim_{\omega\, \rightarrow 0}\,\frac{1}{\omega}\,\lim_{k \to 0}\,\mathrm{Im}\left[G^R_{T_{xy}T_{xy}}\left(\omega,k\right)\,\right],
\end{align}
where $G^R$ refers to the the retarded Green's function.

By computing the Kubo formulas, we notice one important difference between the models for solids $V(X,Z)$ and the fluid model $V(Z)$ -- the shear elastic modulus vanishes in the latter. According to \eqref{eq:transdisp}, this has a drastic effect on the transverse phonons; without shear modulus they are non-propagating and thus purely diffusive. As we already noted in the field theory section, this is clear on the level of the free energy \eqref{eq:freeenergy}; the transverse part of the strain decouples and the time derivative of the transverse momentum density vanishes to zeroth order. In this case, the physics we observe in the transverse sector is the same as for the relativistic fluid -- a purely diffusive mode which we discussed in section \ref{sec:hydro} (see e.g. eq. \ref{eq:transverse_fluid}).
The dispersion relation for the solid model was first considered qualitatively in \cite{Alberte:2017oqx}.

\begin{figure}[h]
    \centering
 \includegraphics[width=5.1cm]{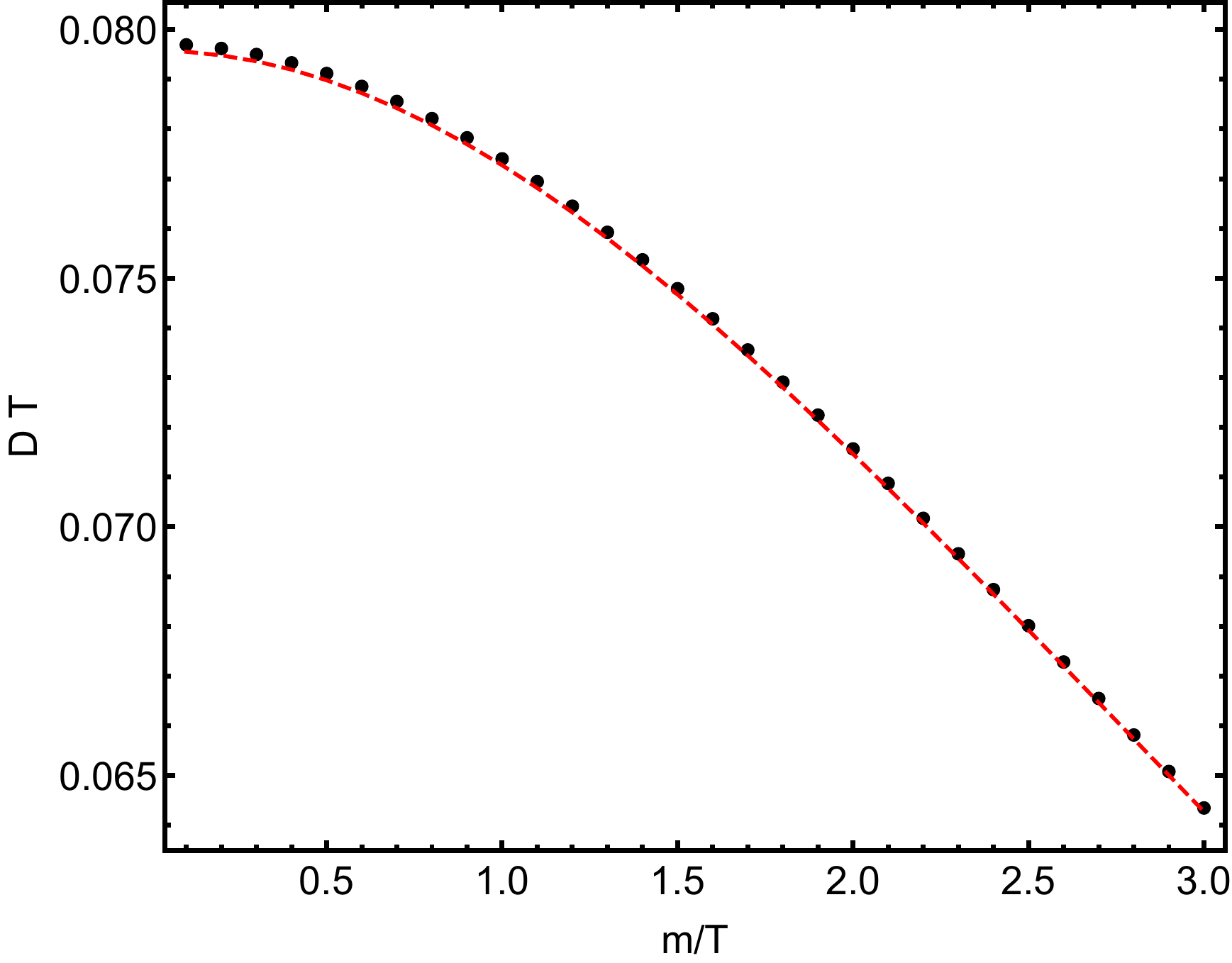}\quad\includegraphics[width=5.1cm]{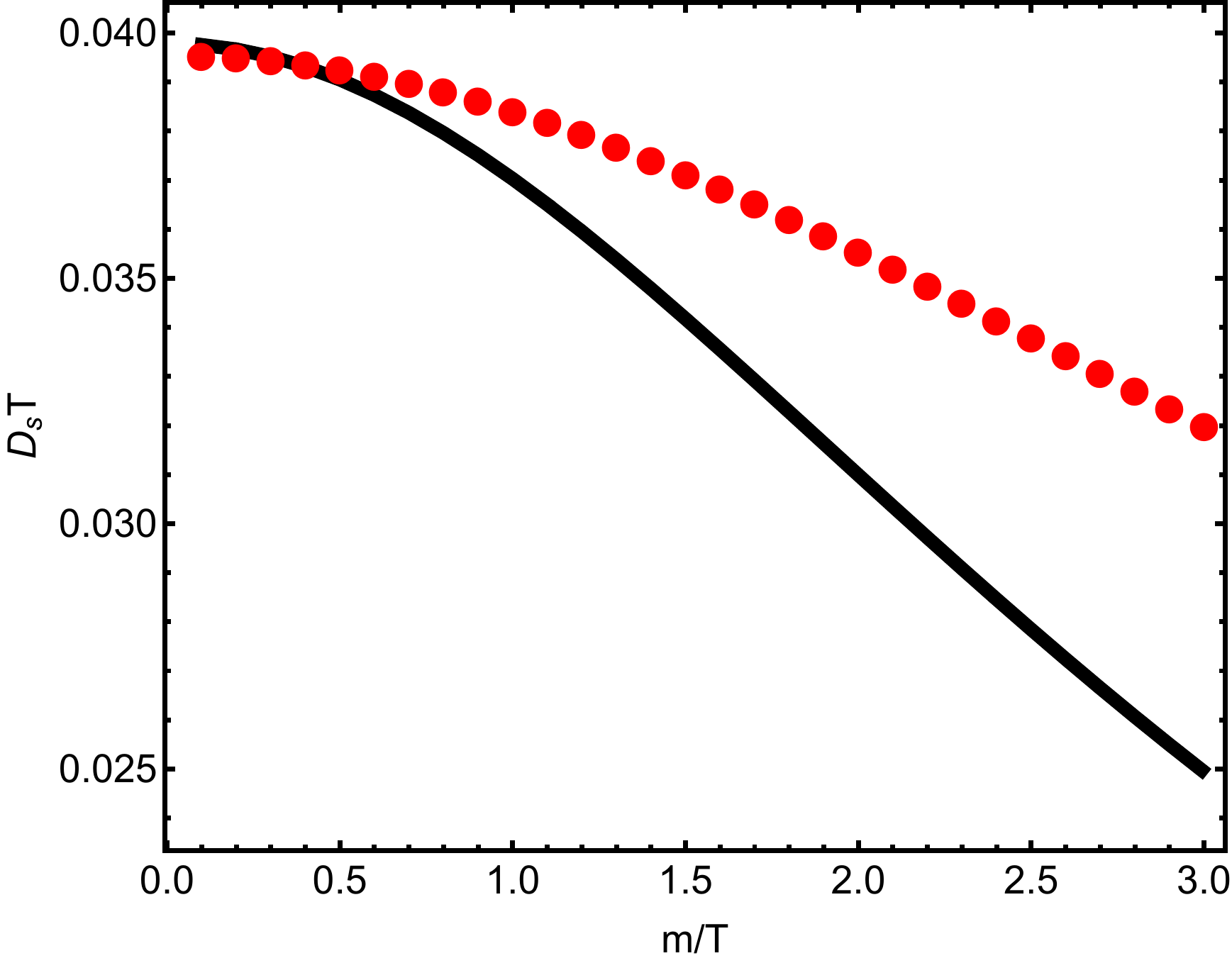}
        \caption{\textbf{Left: } The diffusion constant of the transverse sound modes in terms of the dimensionless $m/T$ for a potential of the form $V(Z)=Z^2$. The dashed line corresponds to the hydrodynamic formula \eqref{eq:transdisp} evaluated on the quantities obtained by means of Kubo formulas. \textbf{Right: }The sound attenuation constant $D_p$ in the longitudinal sector as function of $m/T$ for a potential of the form $V(Z)=Z^2$. The black line computed in terms of the hydrodynamic formula \eqref{eq:soundattenu}. Within the precision of the numerical data, we find good agreement for $m/T \ll 1$.}\label{fig:trdiff}
\end{figure}

In the left side of figure \ref{fig:trdiff}, we match the diffusion constant obtained in the fluid model $V(Z)$ \cite{Baggioli:2019abx} quantitatively to the expression derived from hydrodynamics eq. \eqref{eq:transdisp}. To do this comparison, we have to compute two sets of data: on the one hand, we compute the dispersion relations in terms of QNMs. For given $m/T$, we compute the QNMs for several values of $k$ and extract the coefficient scaling like $k^2$ by doing a polynomial fit. The red dashed line, on the other hand, we obtain by computing the transport coefficients appearing in \eqref{eq:transdisp} directly in terms of the Kubo formulas \eqref{defd}-\eqref{defd4}. To extract Kubo formulas, such as the elastic shear modulus $G$ \eqref{defd}, we have to solve the linearized equations for the fluctions in the presence of sources. For example, to compute $\chi_{\pi\pi}$ according to eq. \eqref{defd3}, we have to set the source of the metric fluctuation $h_{ty}(t,y)$ (encoding the hydrodynamic variable $\pi_\|$) to 1, solve the equations in dependence of the momentum $k$ and the frequency $\omega$ and extract the expectation value of the metric fluctuation $h_{tt}$ (encoding the hydrodynamic variable $\varepsilon$). All other sources are zero. At fixed $m/T$, we perform a scan in $k$ for each value of $\omega$ to determine the coefficient linear in $k$ and subsequently scan $\omega$ to extract the $1/\omega$ behavior for those values. Note that $k$ and $\omega$ are restricted to small values and the order of the limits is important. The comparison of both predictions for the diffusion constant is shown in figure \ref{fig:trdiff} and both results match for our holographic model. We solve all equations of motion in this chapter by means of so-called pseudo-spectral methods (see appendix \ref{ref:appnumrel} for an introduction to the numerical methods and appendix \ref{app:sectionconv} and \ref{app1} for convergence and details on equations of motion).
\subsection*{Longitudinal sector}
The longitudinal sector consists of three hydrodynamic variables, $\lambda_\|,\pi_\|,\varepsilon$, the longitudinal phonon, the longitudinal momentum density, and all scalar quantities (in our case the energy density $\varepsilon$). In Fourier space, we may solve the three hydrodynamic equations, the two conservation equations for energy and momentum conservation \eqref{eq:conservation_spontaneous} and the longitudinal part of the Josephson equation~ \eqref{eq:josephson_full}. This yields three modes in the spectrum; two propagating, diffusive sound modes -- the longitudinal phonons -- and a purely diffusive mode. The dispersion relation of the longitudinal phonons are to diffusive propagating modes given in terms of
\begin{equation}
    \omega = \pm c_\text{L}\, k - i\, D_\text{p}\,k^2
\end{equation}
with the corresponding longitudinal speed of sound $c_\text{L}$ is given by
\begin{equation}\label{eq:speedlongi}
    c_L^2 = \frac{\partial p}{ \partial \varepsilon} + \frac{\kappa +G}{\chi_{\pi\pi}}=\frac{1}{2}+c_T^2=\frac{1}{2}+\frac{G}{\chi_{\pi\pi}},
\end{equation}
where we explicitly took the derivative of the thermodynamic pressure with respect to the energy density.
The attenuation constant $D_\text{p}$ is given by
\begin{equation}
 D_p = \frac{1}{2}\,\frac{\eta}{\chi_{\pi\pi}} + \frac{1}{2}\,\frac{c_V (\kappa+G)^2\xi_\| - (\kappa+G) c_V (\partial p/\partial \varepsilon) \bar{T} \gamma_2 + (\partial p/\partial \varepsilon)  \bar{\kappa}_0 \chi_{\pi\pi} -\gamma_2 (\kappa+G) \chi_{\pi\pi}}{c_V (\kappa + G + (\partial p/\partial \varepsilon)  \chi_{\pi\pi})},\label{eq:soundattenu}
\end{equation}
where $c_V\equiv\partial\epsilon/\partial T$ is the specific heat.
Similarly, we find for the purely diffusive mode the dispersion relation
\begin{equation}
    \omega = -i\, D_\Phi\, k^2,
\end{equation}
with diffusion constant given by
\begin{equation}\label{eq:crystaldiffusionEnergy}
    D_\Phi = (\kappa + G) \,\frac{\bar{\kappa}_0 + \gamma_2 \chi_{\pi\pi} + c_V (\partial p/\partial \varepsilon)  (\bar{T} \gamma_2 + \xi_\| \chi_{\pi\pi}) }{c_V (\kappa + G + (\partial p/\partial \varepsilon)  \chi_{\pi\pi})}.
    \end{equation}

The relevant Kubo formulas for the hydrodynamic expressions in the longitudinal sector may be extracted from the following correlators
\begin{align}
 \xi_\|\,&=\,\lim_{\omega\, \rightarrow 0}\,\omega\,\lim_{k \to 0}\,\mathrm{Im}\left[G^R_{\Phi^{(y)}\Phi^{(y)}}\left(\omega,k\right)\,\right],\\
    \bar{\kappa}_0 &= -\lim_{\omega \to 0} \omega \lim_{k \to 0} \frac{1}{k^2} \mathrm{Im}\left[G^R_{\varepsilon\varepsilon}(\omega,k)\right], \label{eq:kuboKappaMain}\\
    T \gamma_2 &= -\lim_{\omega \to 0} \omega \lim_{k \to 0} \frac{1}{k} \mathrm{Re}\left[G^R_{\varepsilon\Phi^{(y)}}(\omega,k)\right]. \label{eq:kuboGammaMain}
\end{align}

\begin{figure}[h]
\centering
\includegraphics[width=5.5cm]{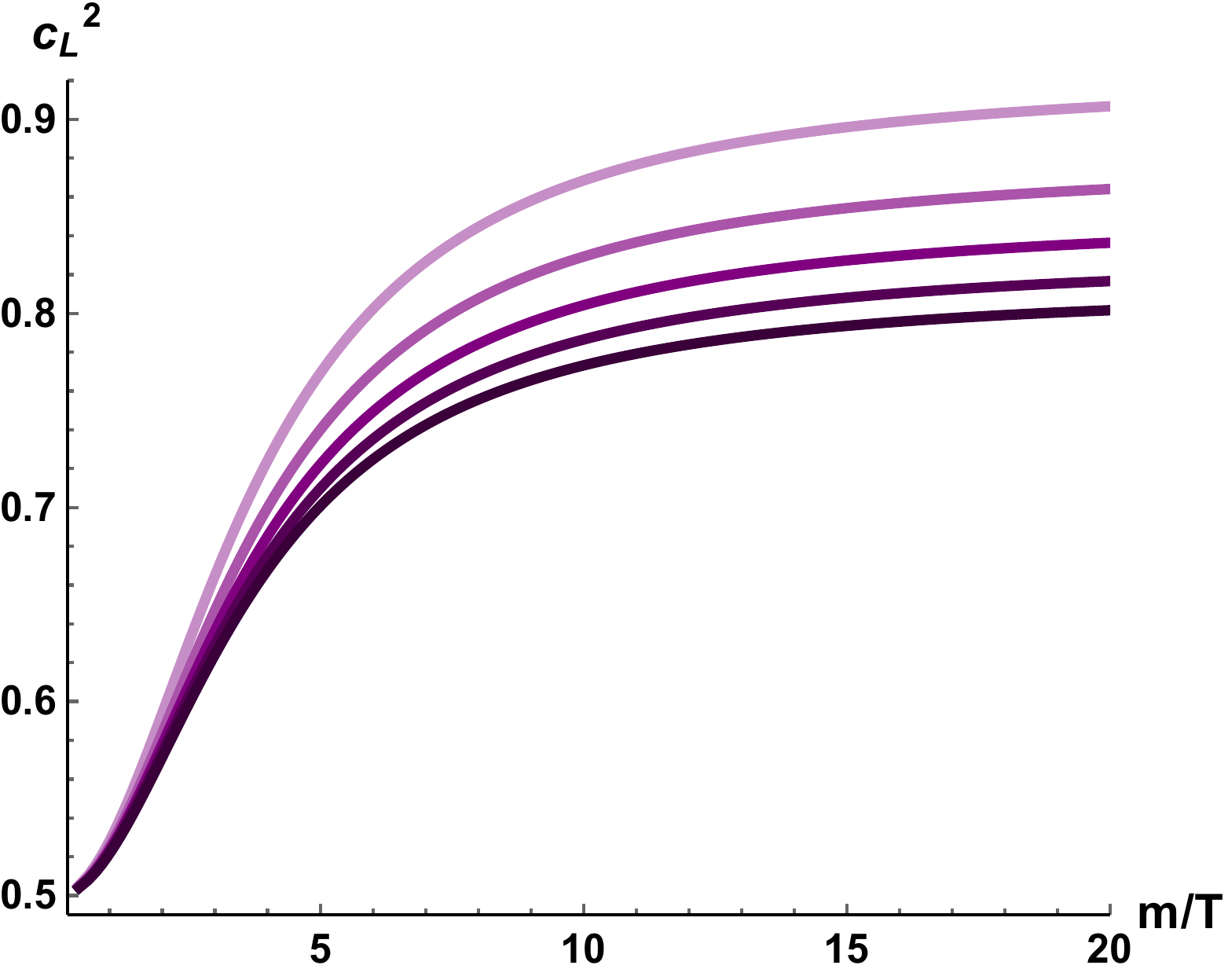}
\quad
\includegraphics[width=5cm]{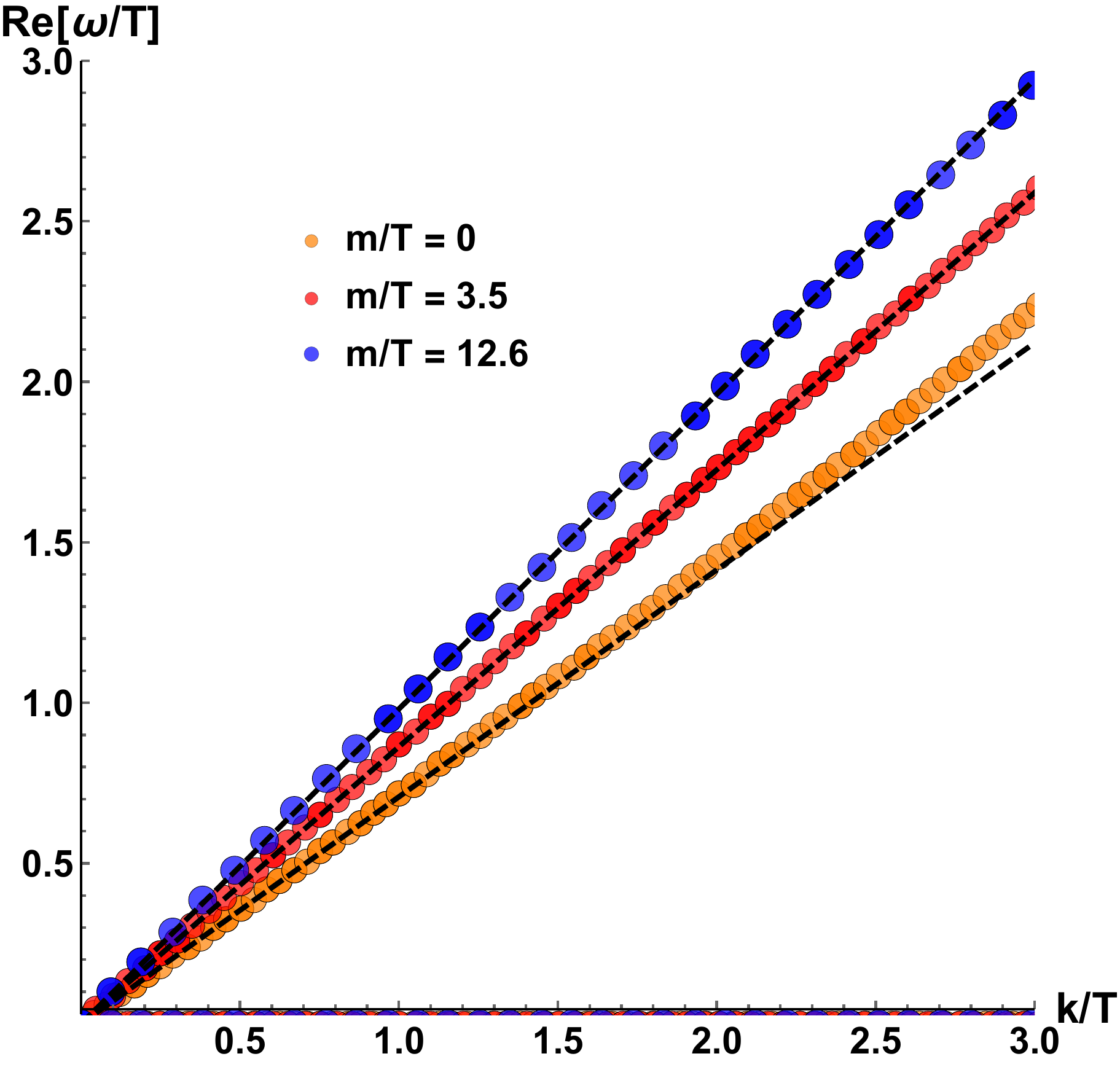}
\caption{\textbf{Left: }The speed of the longitudinal phonons for various potentials \mbox{$N=3,4,5,6,7,8$} (from lighter to darker color) as function of the dimensionless spontaneous symmetry breaking scale $m/T$. \textbf{Right: }Real part of the dispersion relation of the longitudinal phonons determining the speed of sound for the potential $N=3$ and various $m/T\in \{0, 3.5, 12.6\}$. The dashed black lines are the comparisons with the theoretical formula \eqref{eq:speedlongi}.}
\label{fig:speedfig}
\end{figure}

 With these expressions at hand, we are able to check the hydrodynamic expressions by means of our holographic model. The longitudinal sound mode in the solid model \cite{Ammon:2019apj} is depicted in left side figure \ref{fig:speedfig} for various exponents $N$ of the solid model potential $V(X)=X^N$. We note that the behavior is for all exponents qualitatively the same, where the exact values depend on the underlying theory determined by $N$. In the right side of figure \ref{fig:speedfig}, we demonstrate that the hydrodynamic formula eq. \eqref{eq:speedlongi} matches the real part of the sound QNM. We see that the deviation of the speed of sound from the result for a CFT is governed by the magnitude of the shear elastic modulus $G$. In the case of the fluid, this inherently implies that the longitudinal speed of sound always matches 1/2, independent of the symmetry breaking strength $m/T$.
This is exactly what we observe when we compute the longitudinal speed of sound in the holographic fluid $V(Z)$~\cite{Baggioli:2019abx}. Note that in contrast to the transverse sector, the longitudinal phonons are coupled to the system even at $G=0$ in terms of  the bulk shear modulus $\kappa$.

\begin{figure}[h]
    \centering
    \includegraphics[width=6.2cm]{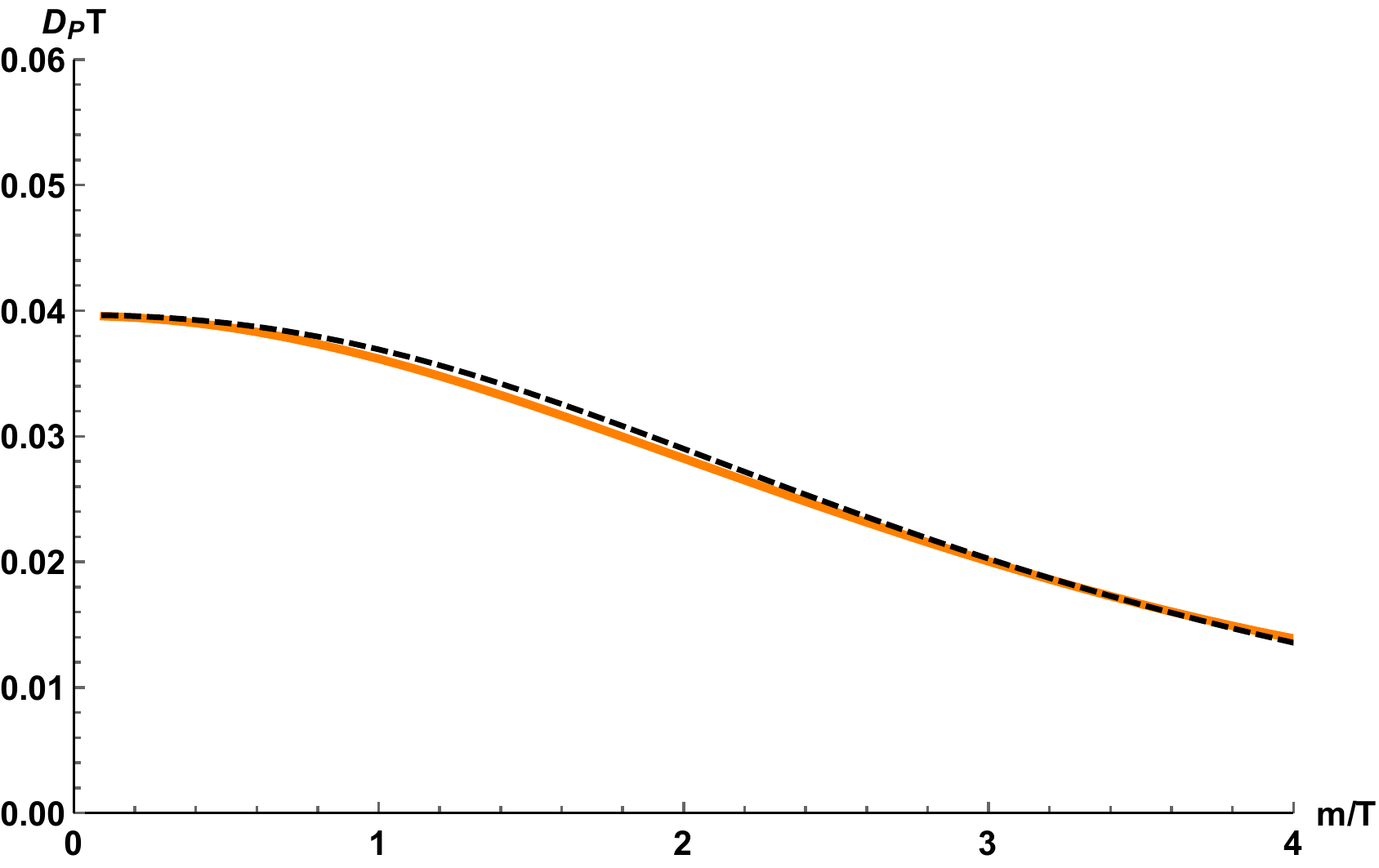}\quad\includegraphics[width=6.2cm]{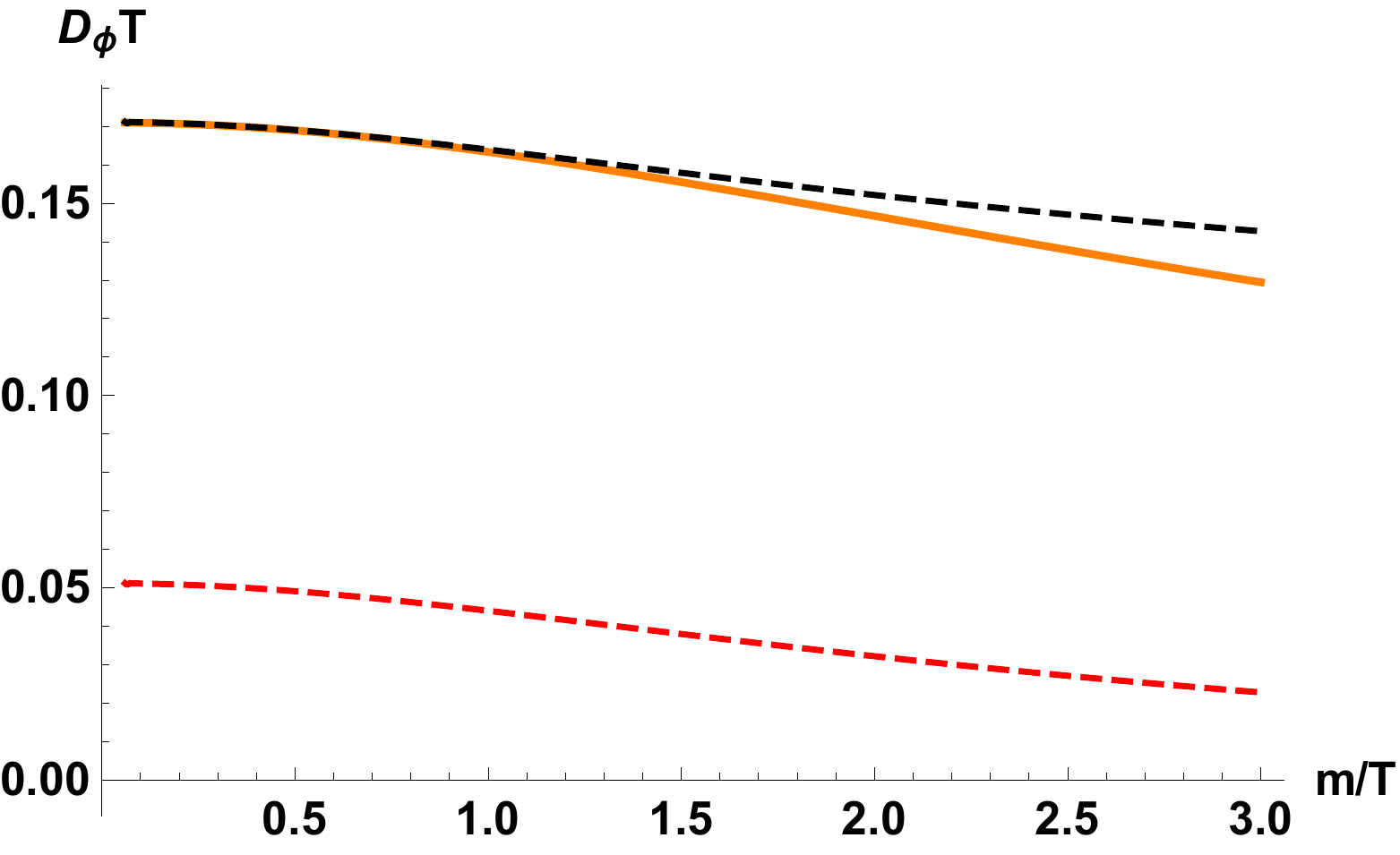}
    \caption{\textbf{Left: }The comparison between the damping coefficient $D_p$ extracted from the numerical data (solid line) and the hydrodynamic formula \eqref{eq:soundattenu}, for the specific $N=5$. We found similar results for several $N>5/2$. \textbf{Right:} The comparison between the numerical value of the diffusion constant $D_\Phi$ for $N=5$ (solid line) and the predictions given by the hydrodynamic formula \eqref{eq:crystaldiffusionEnergy} (lower, dashed line). The upper dashed black line is the formula \eqref{eq:crystaldiffusionEnergy} with an additional constant shift. The agreement between the shifted hydrodynamic formula and the data is evident and valid until quite large values of $m/T$.}
    \label{figcheck1}
\end{figure}

The attenuation constant of the sound modes is depicted in the left side figure \ref{figcheck1} for a potential of the form $V(X,Z)=X^5$. We find excellent agreement between the hydrodynamics prediction (black dashed line) and our results obtained from holography (orange line). We also verified that the sound attenuation constant fits the hydrodynamic formula eq. \eqref{eq:soundattenu} for the fluid model $V(Z)=Z^2$~\cite{Baggioli:2019abx} as shown in the right side of figure \ref{fig:trdiff}.

The last aspect to check for spontaneous symmetry breaking is the diffusion constant of the purely diffusive mode. The results are shown in the right side of figure \ref{figcheck1}. The QNM results are depicted in orange while the hydrodynamic prediction is given in terms of the red dashed line. Surprisingly, the results do not agree. If we shift the line by a constant, however, we find very good agreement between the hydrodynamic prediction and the holographic results. We verified that this shift is independent of the precise form of the potential, i.e. independent of the exponent $N$. Even in the limit of small $m/T$, the results do not agree. In this limit, we can take the probe approximation and study the holographic side analytically. In this limit, the scalar equation of motions reads 
\begin{equation}
    \delta \phi_{\parallel}' \left(\frac{f'}{f}+\frac{2 i \omega }{f}+\frac{2 N}{u}-\frac{4}{u}\right)+\delta \phi_{\parallel} \left(-\frac{k^2 N}{f}+\frac{2 i N \omega }{u
   f}-\frac{4 i \omega }{u f}\right)+\delta \phi_{\parallel}''\,=\,0,\label{uno1}
\end{equation}
with the background given in terms of
\begin{equation}
    f(u)\,=\,1\,-\,\left(\frac{u}{u_h}\right)^3.
\end{equation}
By matching the horizon and boundary asymptotics, we are able to obtain the dispersion relations perturbatively in $k$ which yields
\begin{equation}\label{eq:diffusionAnalytic}
    \tilde{D}_\phi = \frac{N}{2N-3}\,u_h\,,
\end{equation}
where $N$ is the exponent of the potential $V(X)$ which defines the model. 
The hydrodynamic prediction reads for small $m/T$
\begin{equation}\label{eq:diffusionLeading}
    D_\Phi \approx (\kappa+G) \,\xi_\| +\mathcal O(m^2)=\frac{3}{2}\frac{1}{2N-3}u_h+\mathcal O(m^2).
\end{equation}
By comparing \eqref{eq:diffusionAnalytic} and \eqref{eq:diffusionLeading}, we find
\begin{equation}
    \tilde D_\phi\,T=D_\Phi T+\frac{3}{8\pi}
\end{equation}
which is indeed independent of the exponent of the potential. We will comment on the resolution of this disagreement in the conclusions. 

\begin{figure}
\centering
\includegraphics[width=5.5cm]{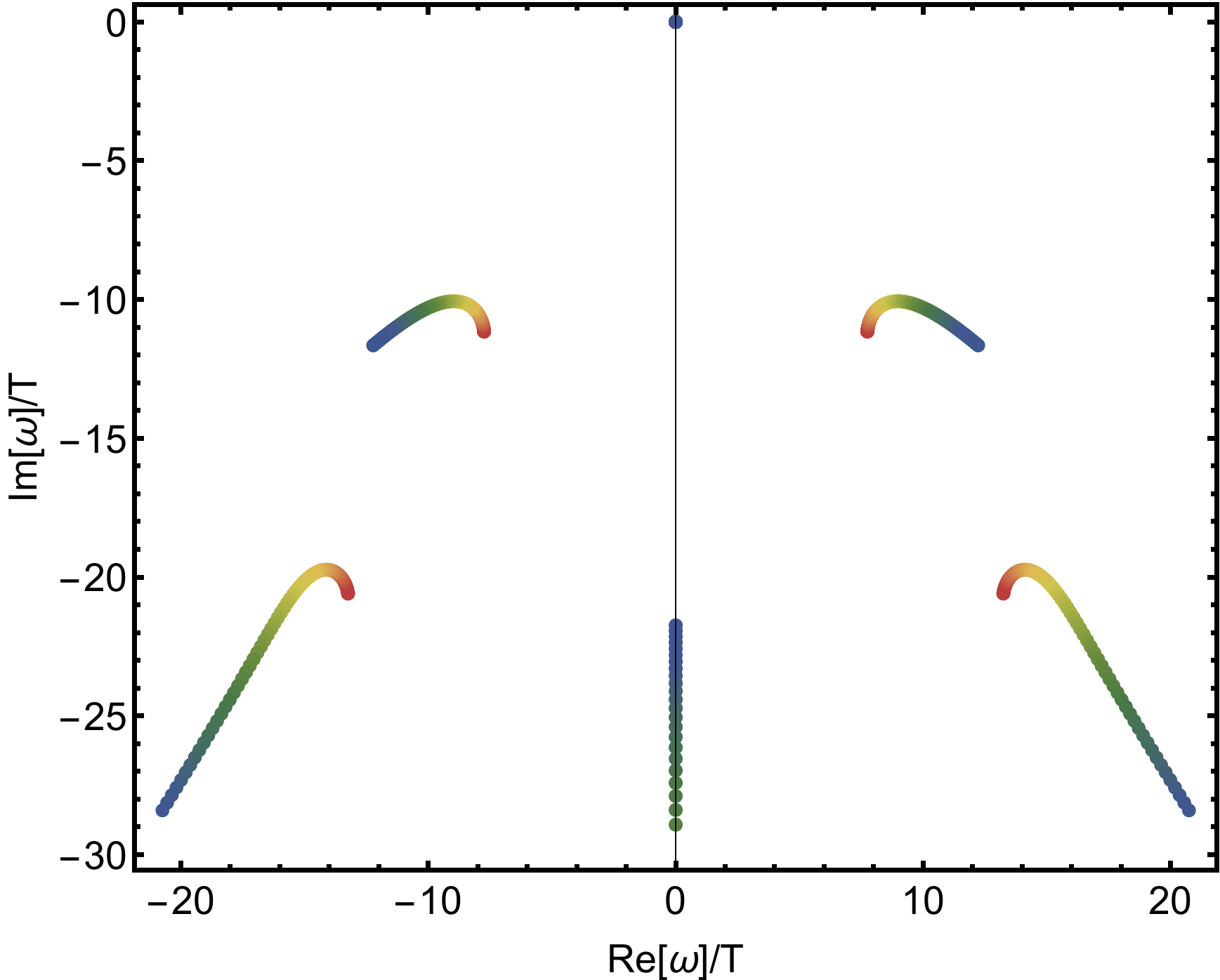}
\quad
\includegraphics[width=5.5cm]{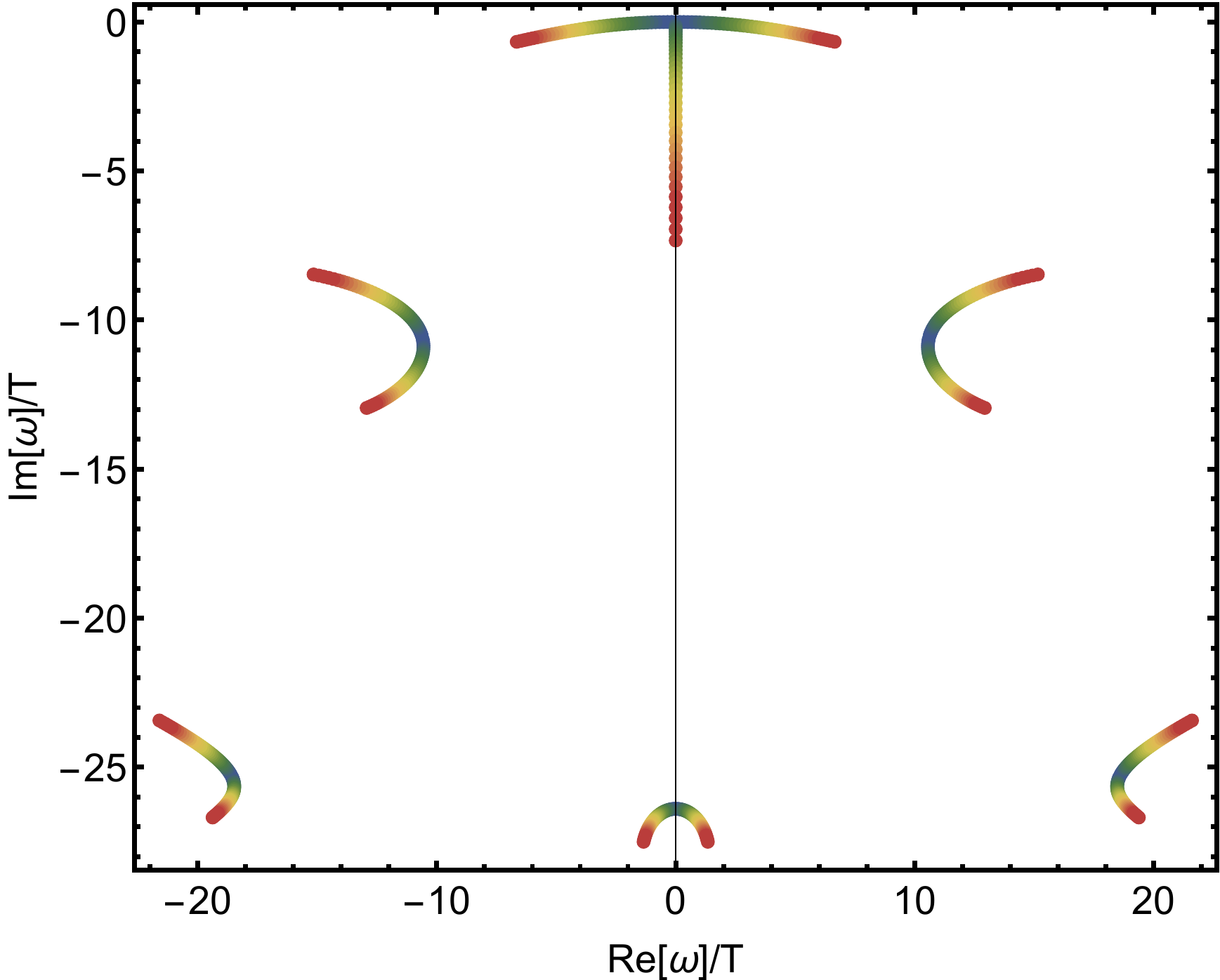}
\caption{Spectrum of higher QNMs for the potential $V(X)=X^3$. \textbf{Left:} The QNM spectrum for vanishing momentum as a function of the dimensionless (inverse) temperature $m/T \in [0, 6.5].$ Blue dots denote low temperatures while red dots refer to high temperatures. \textbf{Right:} The QNM spectrum for fixed temperature $m/T=0.179$ as a function of the dimensionless momentum $k/T \in [0.186,7.45].$ \label{pic:moving}}
\end{figure}

Hydrodynamics is only a subset of the parameter space of linear response theory. In holography, we can go beyond the hydrodynamic regime and compute the full regime of linear response in terms of the QNM spectrum. In the left side of figure \ref{pic:moving}, we draw a cartoon of the QNM spectrum at vanishing momentum in dependence of the spontaneous symmetry breaking scale. In the right side of figure \ref{pic:moving}, we present the first higher QNMs in the spectrum as well as the QNMs for larger values of the momentum $k$ in the longitudinal sector of the solid model $V(X,Z)=X^3$.

\section{Explicit symmetry breaking}
In this section, we focus on breaking the translational symmetry explicitly. Unlike for sponanteous translational symmetry breaking, the momentum is no longer conserved when translations are broken explicitly. We may implement this into our hydrodynamic formalism by ``breaking'' textbook hydrodynamics and relaxing the momentum conservation equation \cite{Davison:2013jba,Davison:2014lua}
    \begin{align}
    \dot{\pi}_i(t,{y}) + \partial^j \tau_{ij}(t,{y}) &= -\Gamma \pi_i(t,y), \label{eq:momentumconservation}
\end{align}
where we introduced the momentum dissipation rate $\Gamma$. We assume that $\Gamma$ is sufficiently small so that the Josephson equations discussed in the spontaneous case still hold. It is important to note that eq. \eqref{eq:momentumconservation} is beyond hydrodynamics and motivated by phenomenological arguments. In this section, we will use holography, to demonstrate that these phenomenological arguments are indeed consistent in the regime $\Gamma\ll 1.$ In eq. \eqref{eq:momentumconservation}, we have two very different regimes:
\begin{itemize}
    \item $\Gamma\ll k\ll 1$: Since $\Gamma$ is neglectable compared to $k$, momentum is approximately conserved and the physics is analog to the spontaneous case.
    \item $k\ll\Gamma\ll  1$: In this regime, even for small momentum dissipation rates momentum will have dissipated significantly at large distances. The physics is totally dominated by the remaining conserved quantity -- energy -- and thus purely diffusive. 
\end{itemize}
In the intermediate regime between the two regimes, the system transitions from the purely diffusive physics to the physics of spontaneously broken translations with two sound-like modes after undergoing a collision at $k^\star$. 

From the equation for the momentum non-conservation \eqref{eq:momentumconservation} and the energy conservation \eqref{eq:conservation_spontaneous}, it is straightforward to derive the two-point functions using the approach of Kadanoff and Martin as explained in section \ref{sec:hydro}. In particular, we find \cite{Davison:2014lua,Davison:2013jba,Andrade:2013gsa}
\begin{align}
    &&G^R_{\pi_\perp\pi_\perp}=\chi_{\pi\pi}\,\frac{\Gamma+k^2\,\eta/\chi_{\pi\pi}}{-i\omega+\Gamma+k^2\,\eta/\chi_{\pi\pi}},
    &&G^R_{\pi_\|\pi_\|}=\chi_{\pi\pi}\,\frac{k^2\frac{\partial p}{\partial\varepsilon}-i\,\omega\,\left(\Gamma+k^2\,\eta/\chi_{\pi\pi}\right)}{i\,\omega\left(-i\,\omega+\Gamma+k^2\,\eta/\chi_{\pi\pi}\right)\,-k^2\,\partial p/\partial\varepsilon}.\label{eq:KMformula}
\end{align}
For example, the heat conductivity $\kappa$ and the so-called transverse momentum conductivity $\kappa_\perp$ are given by taking first the $k\to0$ and then the $\omega\to 0$ of~\cite{Hartnoll:2007ih,Davison:2014lua}
\begin{align}
   \hspace{-0.25cm} \kappa(\omega,k)\!=\!\frac{i}{\omega\,T}\,(G^R_{\pi_\|\pi_\|}\!(\omega,k)-G^R_{\pi_\|\pi_\|}\!(0,k)),\ \, \kappa_\perp(\omega,k)\!=\!\frac{i}{\omega\,T}\,(G^R_{\pi_\perp\pi_\perp}\!(\omega,k)-G^R_{\pi_\perp\pi_\perp}\!(0,k))\label{eq:heatconduct}.
\end{align}
The dispersion relations in the longitudinal and transverse sector are encoded in the poles of the Green's functions and we obtain them with the method of Kadanoff and Martin by solving the denominator of eq. \eqref{eq:heatconduct} for $\omega.$

\subsection*{Transverse sector}
\begin{figure}[h]
    \centering \includegraphics[width=4.9cm]{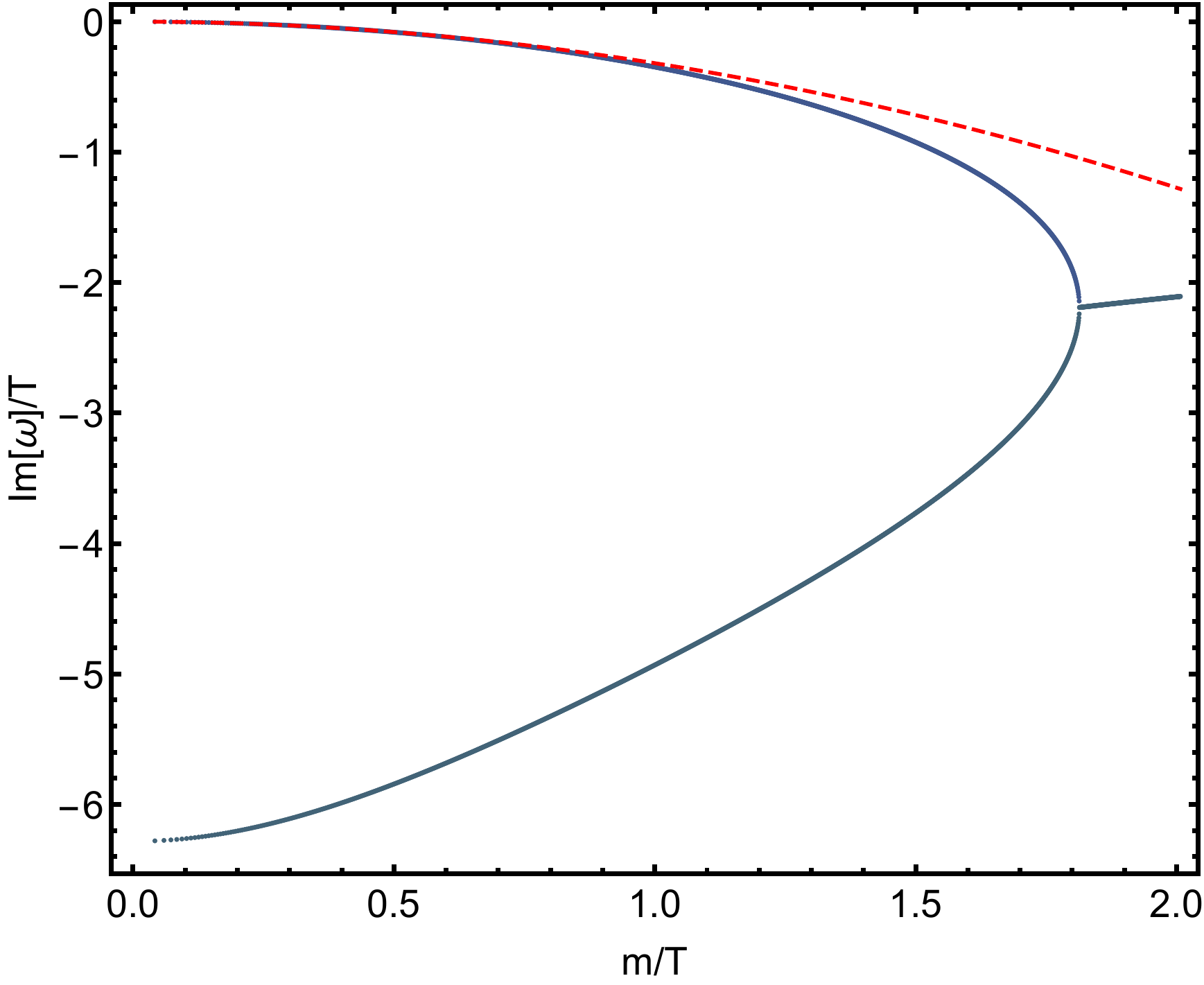}
    \includegraphics[width=5.0cm]{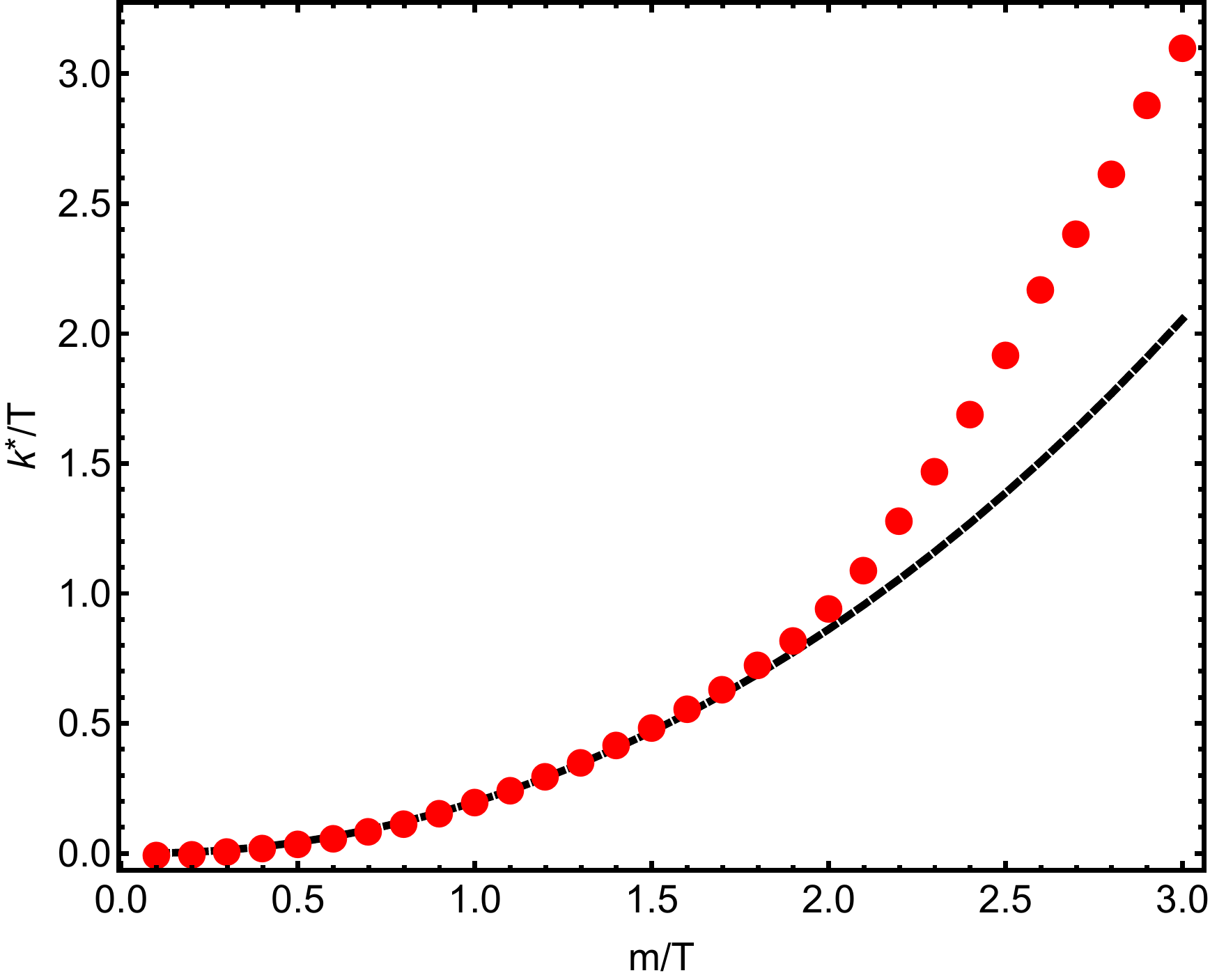}
        \caption{\textbf{Left: } The imaginary part of the lowest mode of the transverse spectrum for $V(Z)=Z$ as function of $m/T$. The red dashed line is eq. \eqref{d1}. Around $m/T \approx 1.8$ the mode collides with another pole. \textbf{Right: }The momentum of the poles collision $k^\star$ extracted from the numerical data (red bullets). The dashed line is the hydrodynamic approximation of eq. \eqref{eq:prediciton_collission} which show good agreement for $m/T \ll 1$.}
        \label{figuno}
\end{figure}

In this section, we want to explicitly compute the dispersion relations within the holographic fluid model $V(Z)=Z$ in the presence of explicit symmetry breaking and match the QNM data to expressions obtained from hydrodynamics. For small explicit breaking (which corresponds to small masses $m$), we may implement the momentum non-conservation as described in eq. \eqref{eq:momentumconservation}. The modes in the transverse sector are given by (using eq. \eqref{eq:heatconduct})~\cite{Davison:2014lua,Davison:2013jba,Amoretti:2018tzw}
\begin{equation}
    \omega=-i\,\Gamma-i\,D_p\,k^2+\mathcal{O}(k^4), \quad \Gamma=\frac{m^2\,(V_X+2\,V_Z)}{2\pi\,T}+\mathcal{O}(m^4),\quad D_p=\frac{\eta}{\chi_{\pi\pi}}+\dots \label{d1}.
\end{equation}
To leading order in $m$, the diffusion constant is unchanged but we observe that the mode is gapped with $\omega=-i\Gamma$. Increasing $m$ and hence the strength of the explicit symmetry breaking, the mode eventually collides with another mode at $\Gamma\sim T$ ($m/T\approx \sqrt{\pi}$). After this collision, the response is no longer dominated by a single long-lived pole close to the origin of the complex frequency plane. In the left side of figure \ref{figuno}, we probe this regime and the transition by computing the lowest QNM at zero momentum for increasing strength of the explicit translational symmetry breaking. The QNM data is depicted in blue while the gap of the lowest mode $\Gamma$ -- obtained from equation \eqref{d1} -- is represented by the red dashed line. We find excellent agreement for values of $m/T\lesssim1$. We also note that the transition is located at approximately $m/T\approx 1.8$ in agreement with $\Gamma\sim T$. The physics observed in the transverse sector of the fluid model $V(Z)=Z$ with explicitly broken translations is very similar to previous studies in the linear axion model representing solids~\cite{Davison:2014lua,Kim:2014bza}.

\begin{figure}[h]
    \centering
   \includegraphics[width=5cm]{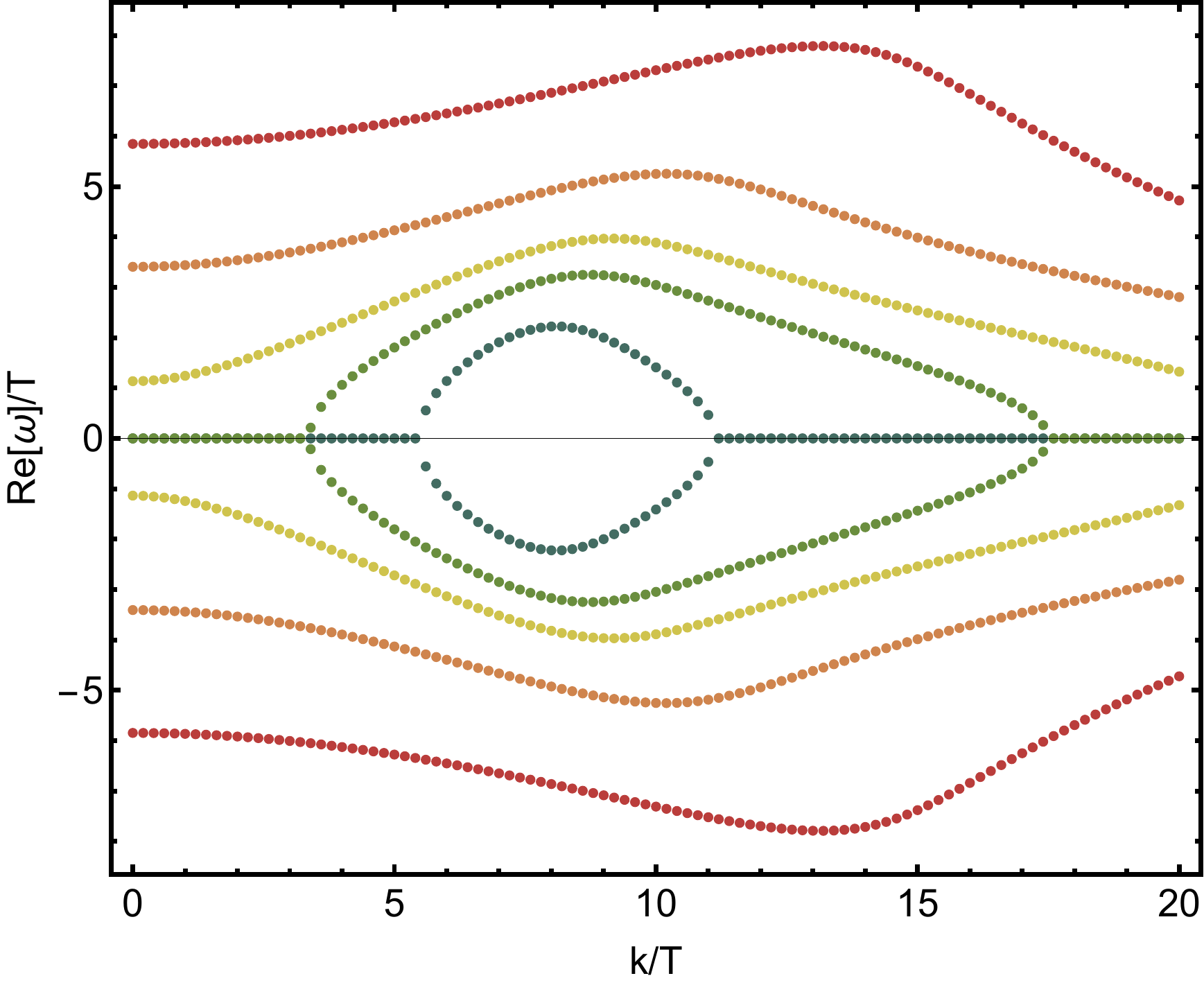}\quad  \includegraphics[width=5.1cm]{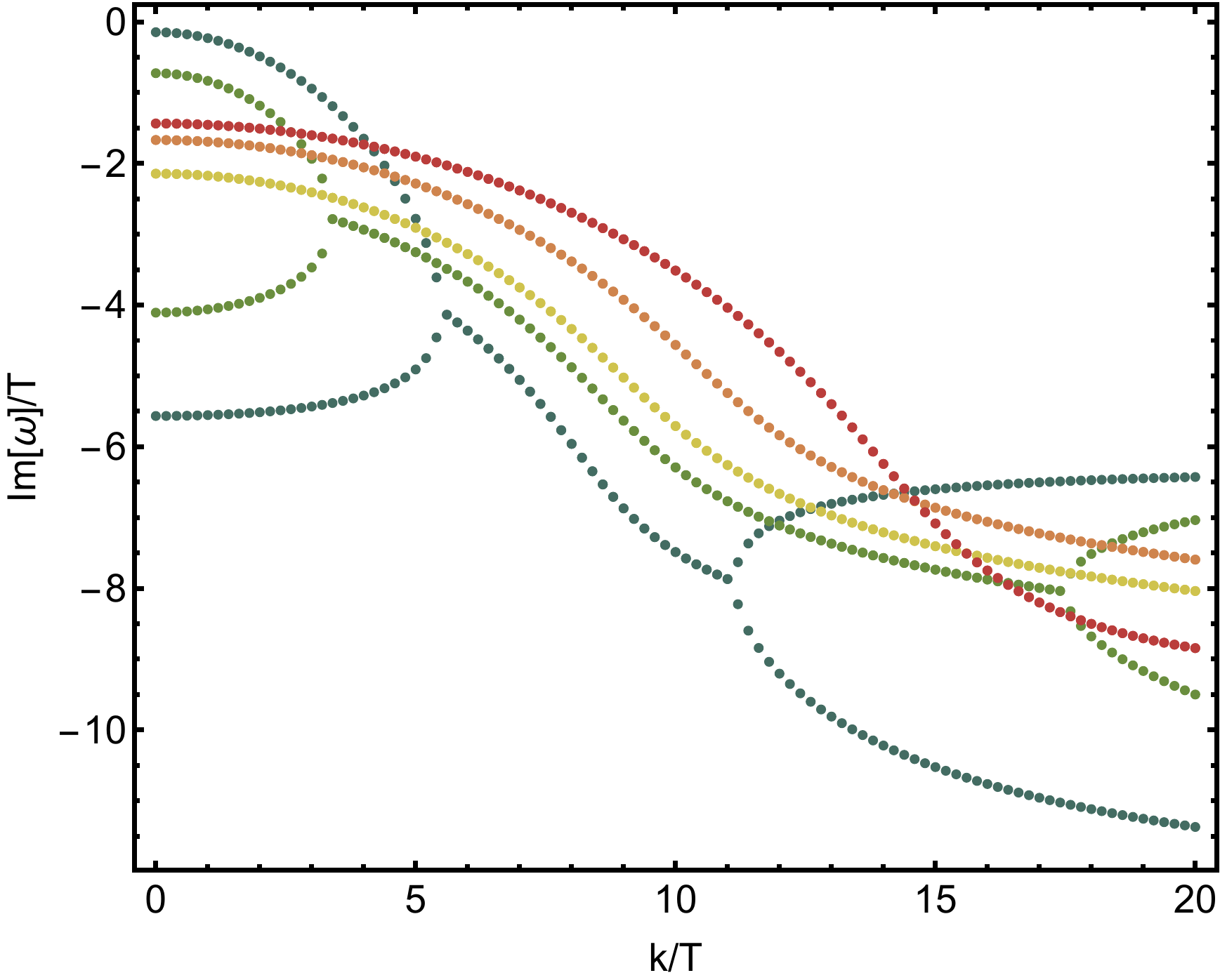}
        \caption{The two lowest modes in the transverse spectrum of the fluid model $V(Z)=Z$ for  $m/T \in [0.67,6.28]$ (from blue to red).}
        \label{figdueb}
\end{figure}

In figure \ref{figdueb}, we present the real and imaginary part of the two lowest modes in the QNM spectrum for several values of $m/T$ beyond the limit $k/T\ll 1$ and the transition at $\Gamma\sim T$. In this limit, the dispersion eq. \eqref{d1} is no longer applicable. However, we observe a very interesting interplay between the first two modes. For the smallest value of $m/T$ (blue in figure \ref{figdueb}), the modes collide at a certain momentum $k_1$ and exhibit a finite real part after the collision. This is referred to as a $k$-gap like behavior and was already observed in the similar solid model~\cite{Alberte:2016xja}. Following the the modes to even large $k/T$, the modes collide again and the real part ceases to exist. Increasing the explicit breaking scale, the first collision moves to smaller momenta until it reaches the origin. When it reaches the origin, we note that the imaginary part shows a large gap at $k=0$, as we observed in figure \ref{figuno} after the transition.
\subsection*{Longitudinal sector}
Similar to the transverse sector, the physics in the longitudinal sector depends on the value of the momentum compared to the explicit symmetry breaking. In the longitudinal sector, the dispersion relation follows in terms of the Kadanoff and Martin procedure in terms of the heat conductivity. The generic dispersion relation of the lowest mode reads using eq. \eqref{eq:heatconduct}~\cite{Davison:2014lua}
\begin{equation}
        \omega\,=\,\pm\,\sqrt{\frac{\partial p}{\partial \epsilon}\,k^2-\,\frac{1}{4}\,\left(\Gamma+\frac{\eta}{\chi_{\pi\pi}}\,k^2\right)^2}\,-\,\frac{i}{2}\,\left(\Gamma\,+\,\frac{\eta}{\chi_{\pi\pi}}\,k^2\right)\,+\,\dots.\label{disp2}
    \end{equation}
The sound mode of relativistic hydrodynamics is recovered in absence of momentum dissipation $\Gamma=0$~\cite{Policastro:2002tn}.
At small momenta, $k/\Gamma \ll 1$, sound is destroyed by momentum dissipation and the relevant hydrodynamic poles are\footnote{The ellipsis in all the following expressions represent higher order corrections in $\omega/T,k/T$.}
\begin{equation}
    \omega\,=\,-\,i\,\frac{\partial p}{\partial \epsilon}\,\Gamma^{-1}\,k^2\,+\,\dots\,,\quad \quad \omega\,=\,-\,i\,\Gamma\,+\,i\,k^2\,\left(\frac{\partial p}{\partial \epsilon}\,\Gamma^{-1}\,-\,\frac{\eta}{\epsilon\,+\,p}\right)\,+\,\dots\label{eq:nosound}
\end{equation}
which is obtained by expanding eq. \eqref{disp2} for small $k/\Gamma$. 
Increasing the momentum, these two modes collide with each other at $k=k^\star$ when the square root in eq. \eqref{disp2} vanishes \begin{equation}
    \Gamma\, +\frac{\eta}{\chi_{\pi\pi}}\,(k^\star)^2-2\,k^\star\,\sqrt{\frac{\partial p}{\partial \epsilon}}=0.\label{eq:prediciton_collission}
\end{equation}
After the collision the modes form a propagating sound-like mode
\begin{equation}
    \omega=\pm\frac{\partial p}{\partial \epsilon}\,k\,-\,i\,\left(\Gamma\,+\,\frac{\eta}{\chi_{\pi\pi}}\right)\,k^2+\,\dots.\label{eq:aftercoll}
\end{equation}
In the right side of figure \ref{figuno}, we depict the collision at $k^\star$ where the crossover predicted by hydrodynamic methods in eq. \eqref{eq:prediciton_collission} happens. The data extracted from the QNM data is shown as red dots while the hydrodynamic prediction is the black dashed line. Interestingly, at $m/T\approx 1.8$  $k^\star/T$ starts scaling linearly in $m/T$.
\begin{figure}[h]
    \centering
   \includegraphics[width=5cm]{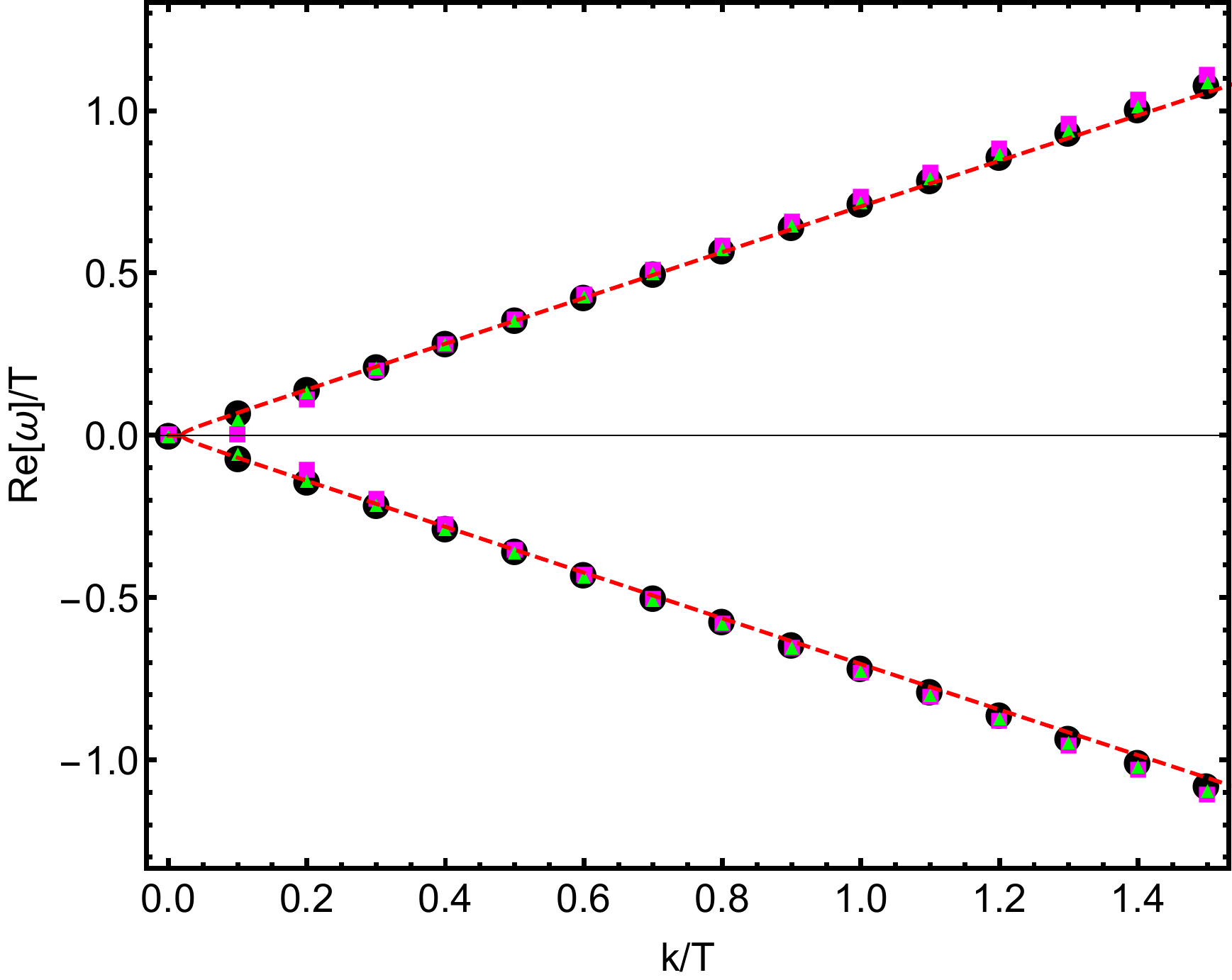}\quad  \includegraphics[width=5.1cm]{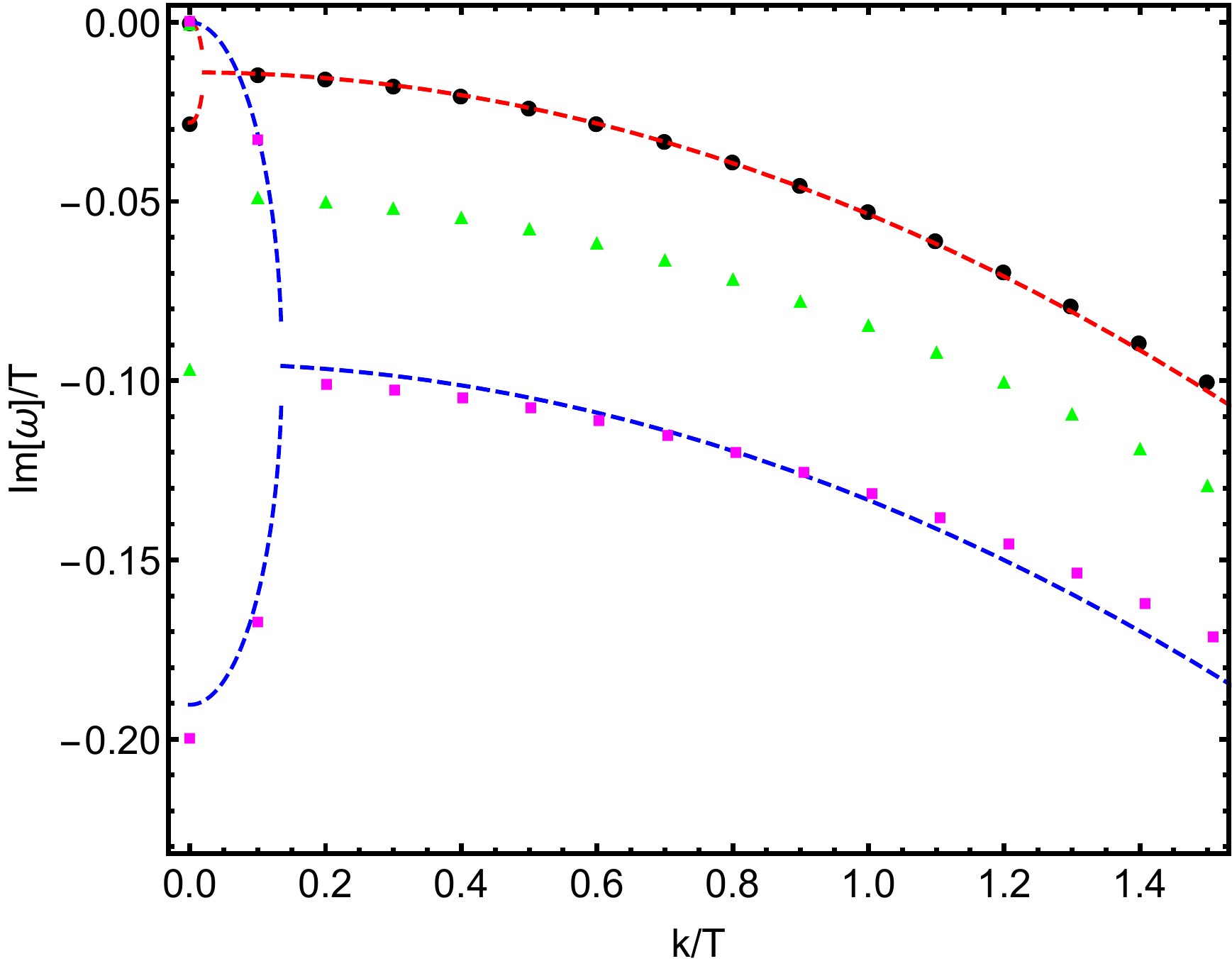}
        \caption{The dispersion relation of the lowest modes for $m/T=\{0.297, 0.773, 0.544\}$ (from black to magenta). The dashed lines are the hydrodynamic formula \eqref{disp2} which works well for $m/T \ll 1$.}
        \label{figsmall}
\end{figure}
The collision of the modes is shown explicitly in figure \ref{figsmall} for three different values of $m/T$. The behavior before, at, and after the collision is well described by equation eq. \eqref{disp2}. For small $k$, the physics is completely determined by the energy conservation. We observe purely diffusive modes with dispersion relation eq. \eqref{eq:nosound}. After the collision sound-like behavior is restored and we observe a pair of modes with dispersion relation given by \eqref{eq:aftercoll}.

\section{Interplay between EXB and SSB}
The interplay of the explicit and spontaneous symmetry breaking has a crucial impact on the phonons; they acquire a mass term i.e. a finite real part at $k=0$. Since the phonons are no longer hydrodynamic modes in a strict sense, we will refer to them as pseudo-phonons and denote their mass by the pinning-frequency $\omega_0^2$. The mass of the pions in chiral theory is determined by the same mechanism~\cite{Burgess:1998ku}. Within the framework of chiral symmetry breaking Gell-Mann-Oakes-Renner showed in 1968 \cite{PhysRev.175.2195} that the mass depends on both symmetry breaking scales, the explicit $\langle$EXB$\rangle$ and the spontaneous $\langle$SSB$\rangle$ symmetry breaking scale $ \omega_o^2=\langle\text{EXB}\rangle\langle\text{SSB}\rangle.$
The validity of this relation in the context of holographic phonons for broken translations was verified in \cite{Baggioli:2019abx,Amoretti:2016bxs,Alberte:2017cch,Andrade:2017cnc,Li:2018vrz,Donos:2019tmo,Musso:2018wbv,Amoretti:2018tzw,Ammon:2019wci}.

In this section, we will combine the concepts of the last two sections and focus on the so-called pseudo-spontaneous limit where the ratio of the EXB to the SSB is much smaller than 1. In our holographic model, we realize this by considering
\begin{align}
    V(X)&=\alpha X+\beta X^5 \ (\text{solid}),\quad\ V(Z)=\alpha Z+\beta Z^2\ (\text{fluid}),\label{eq:potentials2}
\end{align}
with $\alpha/\beta\ll 1$. The physics we described in the last two sections are the limiting cases $\alpha=0$ and $\beta=0$. For $\alpha=0$, the translational symmetry breaking is spontaneous while for $\beta=0$ the associated translational symmetry breaking is explicit. 

Let us briefly summarize what we have observed so far. In the transverse sector of the fluid model with spontaneously broken translations, we found a  non-propagating diffusive mode while the explicit symmetry breaking yields a purely diffusive mode in the limit of small momentum dissipation. In order to model this with hydrodynamic methods, we relaxed the momentum conservation equation to account for momentum dissipation; by switching on spontaneous symmetry breaking in addition to the explicit breaking, we have to take yet another mechanism into account -- the phase relaxation of the phonons $\bar\Omega$ \cite{Baggioli:2019abx,Amoretti:2018tzw,Amoretti:2019kuf,Donos:2019txg,Ammon:2019wci,Delacretaz:2017zxd}. 

On the one hand, the explicit symmetry breaking related to momentum non-conservation required to relax the momentum conservation equation
    \begin{align}
    \dot{\pi}_i(t,{y}) + \partial^j \tau_{ij}(t,{y}) &= -\Gamma \pi_i(t,y). 
\end{align}
On the other hand, taking also the spontaneous symmetry breaking into account, we have to relax the Josephson-equations since the spontaneous breakdown of the translational symmetry yields a shift in the Goldstone fields
\begin{align}
    (\partial_t  +\ \bar\Omega_{\|\,})\,\lambda_{\|\,}&=\nabla\cdot v+\ldots\ \quad\ \text{and}\ \quad
    (\partial_t\ +\ \bar\Omega_\perp)\,\lambda_\perp=\nabla\times v+\ldots\,.\label{eq:phaserelax}
\end{align}
On the gravity side, the momentum dissipation rate is encoded in the momentum operator which is holographically realized by metric fluctuations in the gravitational sector. In the case of breaking the translational invariance explicitly, the system dissipates momentum with finite momentum dissipation rate, i.e. $\Gamma\not =0$. By switching on spontaneous symmetry breaking, we observe a second relaxation scale $\bar{\Omega}$, captured by the scalar sector, since the internal shift symmetry of the St\"uckelberg fields is broken resulting in a finite phase relaxation, i.e. $\bar{\Omega}\neq0$.
In the pseudo-spontaneous limit, both relaxation mechanisms are present since the sectors are coupled in terms of $m$ (see eq. \eqref{eq:model}). This novel relaxation scale is only small in the pseudo-spontaneous limit and it depends on the relaxation scales as
\begin{equation}
    \bar\Omega\sim\frac{\langle\text{EXB}\rangle}{\langle\text{SSB}\rangle}.
\end{equation}
In this section, we test the validity of previous results for the transverse sector in our fluid model. We also provide insights into the more complex longitudinal sector of the fluid and the solid model. 
\begin{figure}
    \centering\includegraphics[width=5.4cm]{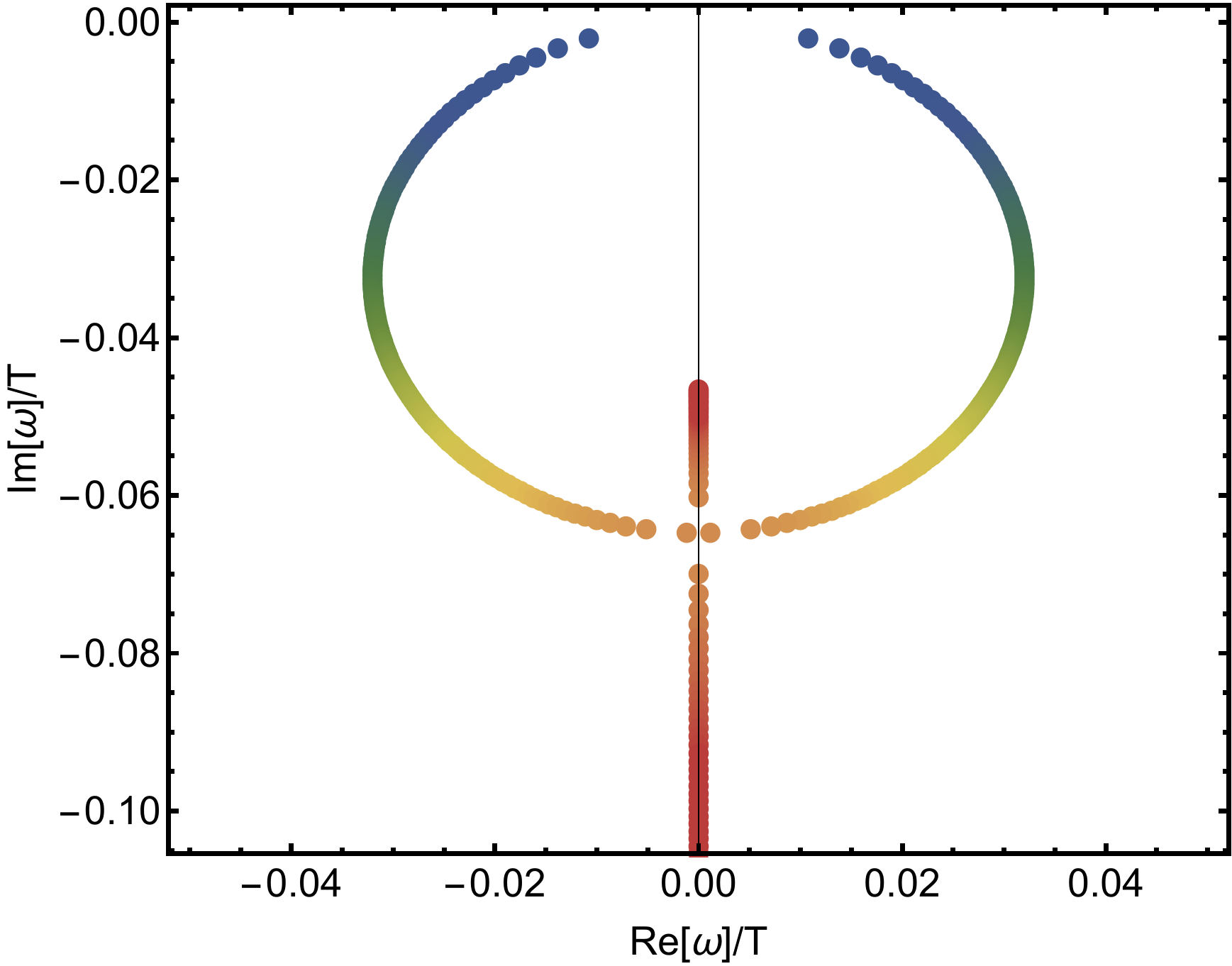}
 \quad   \includegraphics[width=5.4cm]{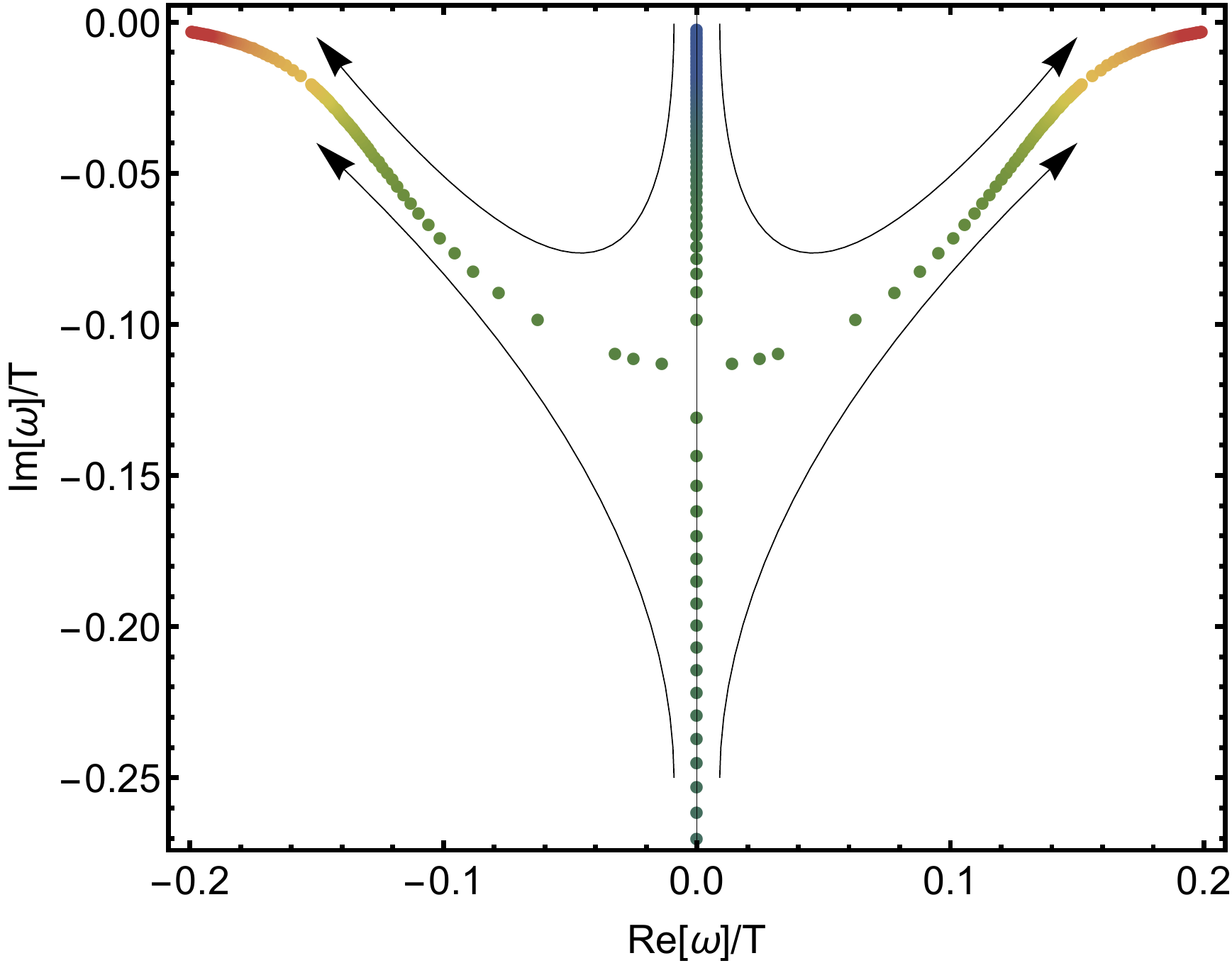}\quad
        \caption{\textbf{Left: } Lowest modes in the transverse sector of the fluid model at $k=0$ at fixed $m/T = 0.1$, $\beta = 5$ and increasing $\alpha$ (from blue to red). \textbf{Right:} The dynamics of the lowest two modes in the transverse sector for $m/T=0.3,\ \alpha=0.05$ and increasing $\beta$ from zero to large values. The collision between the two poles happens at $\beta \sim 1$ and it produces the two modes with finite real part.}
        \label{pic:kzeropseudo}
\end{figure}

\subsection*{Transverse sector}
\begin{figure}[H]
    \centering
   \includegraphics[width=4.8cm]{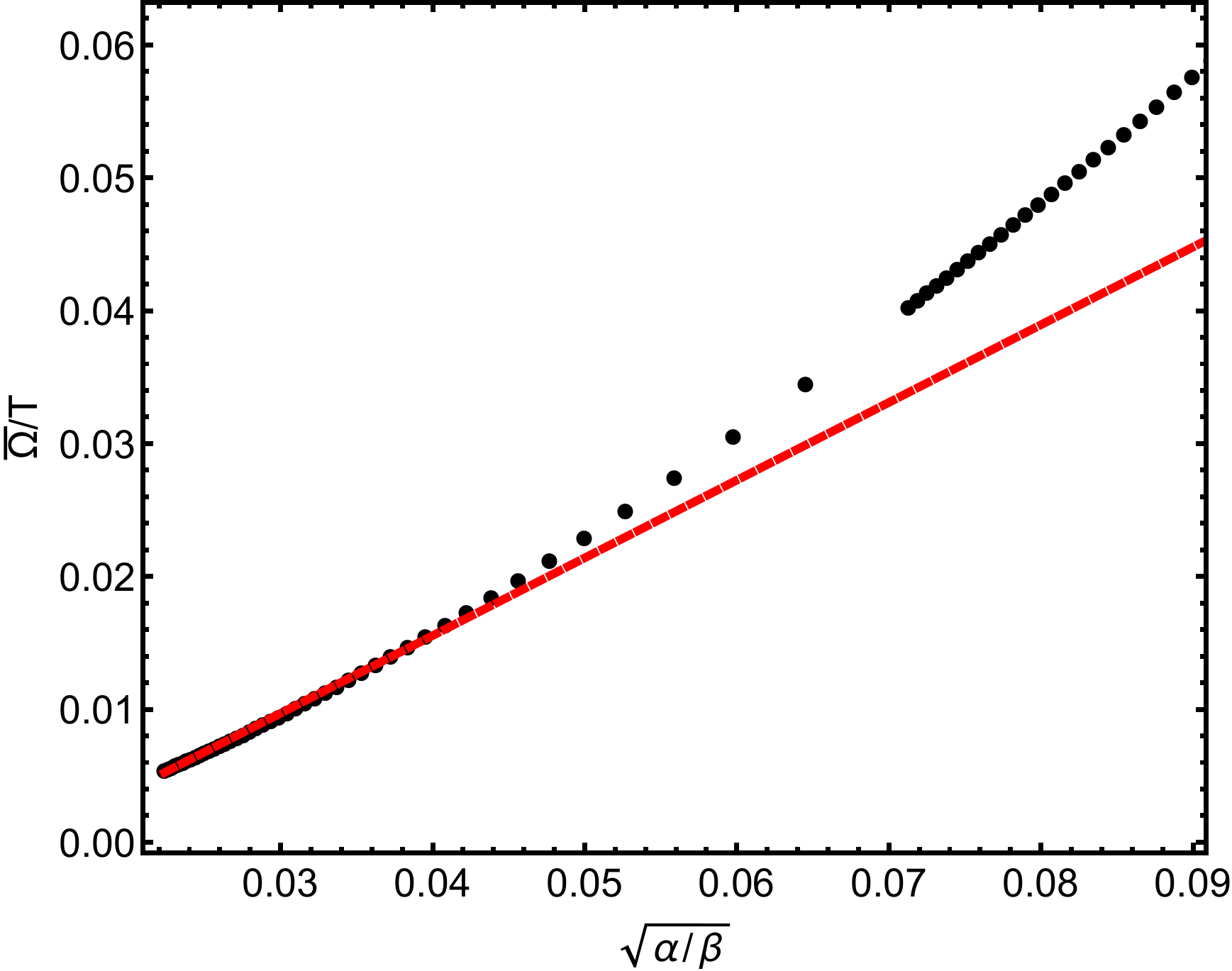}\quad  \includegraphics[width=4.8cm]{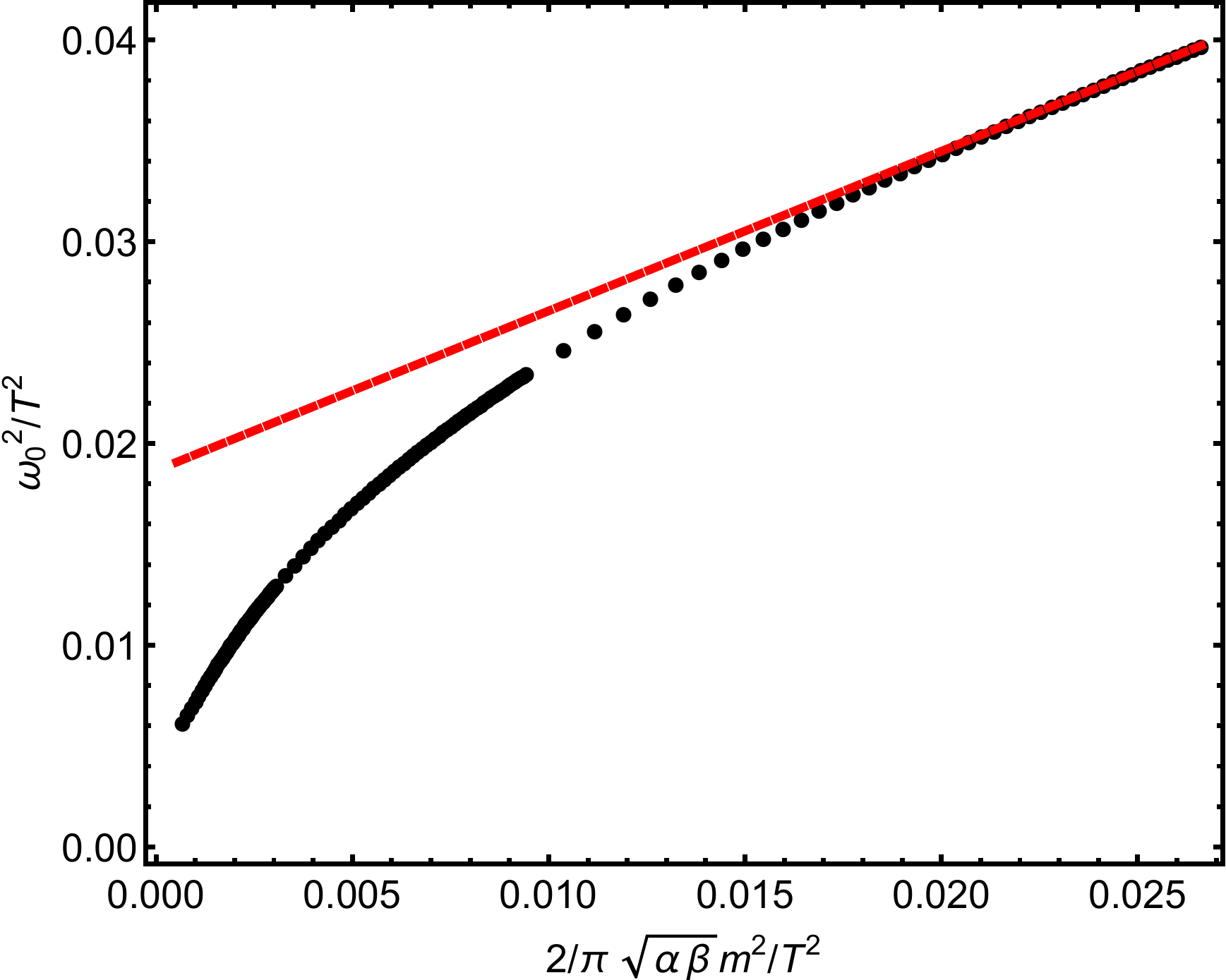}\quad   \includegraphics[width=5.0cm]{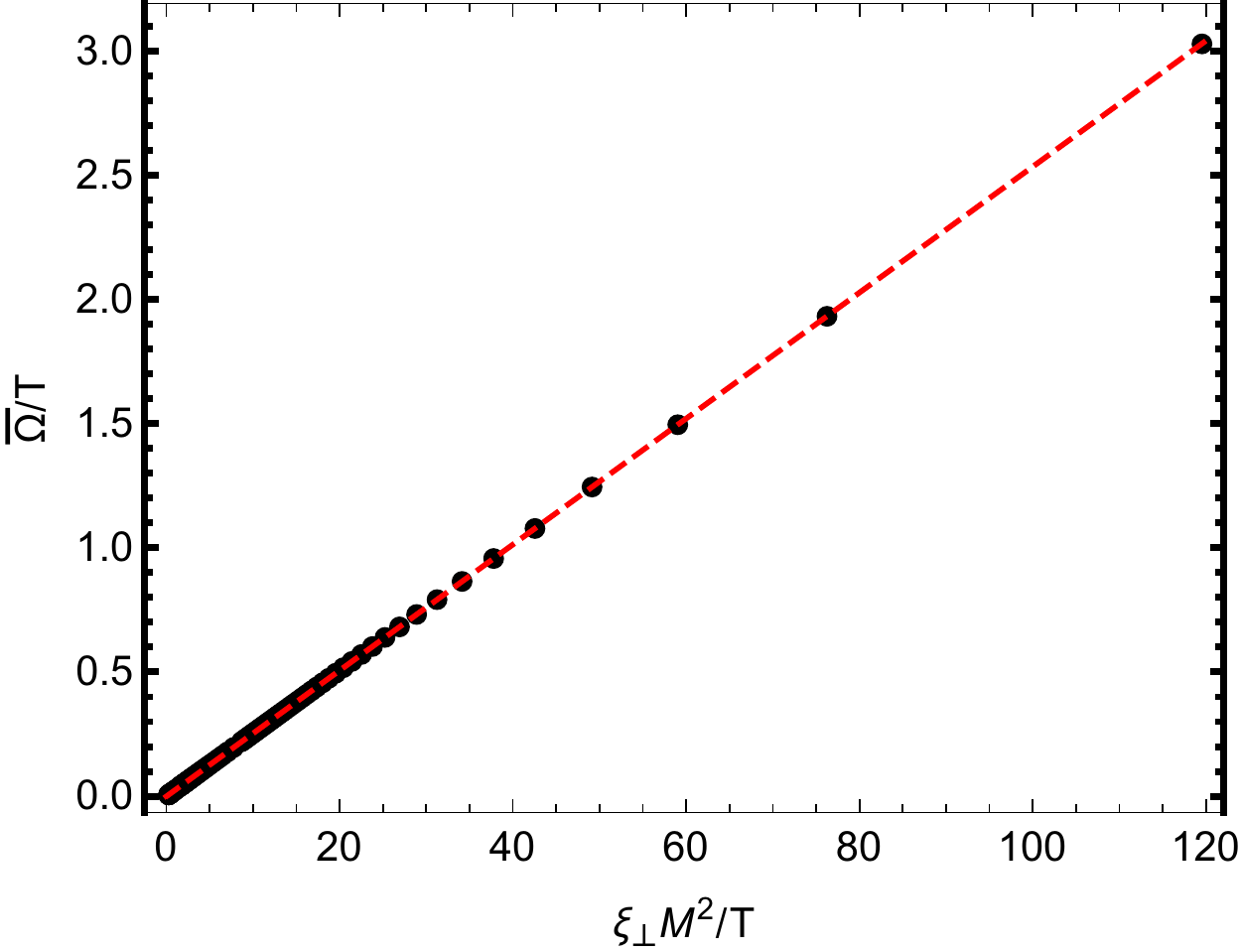}
        \caption{\textbf{Left and Middle:} Dependence of the novel relaxation scale $\bar{\Omega}$ and the pinning frequency $\omega_0$ in function of the SSB parameter $\langle SSB \rangle \sim \sqrt{\beta}$ for fixed $m/T=0.3,\ \alpha=0.05$. The red lines indicate linear scalings. \textbf{Right:} The phase relaxation $\bar\Omega$ vs \eqref{universe} (dashed, red).}
        \label{pic:GMOR}
\end{figure}

Using the modified Josephson relations \eqref{eq:phaserelax} together with the non-conservation of the momentum \eqref{eq:momentumconservation} and the energy conservation equation, we  may write down the momentum-momentum correlators in analogy to eq. \eqref{eq:heatconduct}. From the poles, we find the retarded Green's function which yields at vanishing momentum $k=0$ in the transverse sector~\cite{Delacretaz:2017zxd,Amoretti:2019kuf}
\begin{equation}
    \omega=\frac12\,\left(-i\,(\Gamma+\bar\Omega)\pm\sqrt{4\,\omega_o^2-(\Gamma-\bar\Omega)^2}\right),\label{eq:k0}
\end{equation}
where we introduced the so-called pinning frequency $\omega_0$ which corresponds to the mass gap of the pseudo-phonons.
Depending on the values of $\omega_o^2,\,\Gamma$ and $\bar\Omega$, we have three different scenarios. 
\begin{enumerate}[label=(\roman*)]
    \item $4 \omega_o^2>(\Gamma-\bar\Omega)^2$: Deep in the pseudo-spontaneous regime, the radicand is positive and we have two real solutions at $k=0$ -- the two light pseudo-phonons.
    \item $4 \omega_o^2=(\Gamma-\bar\Omega)^2$: Increasing the quotient EXB/SSB, the square root eventually vanishes where the two poles collide.
    \item $4 \omega_o^2<(\Gamma-\bar\Omega)^2$: After the collision, the square root is purely imaginary yielding two purely imaginary solutions at $k=0$.
\end{enumerate}
In the section about explicit breaking, we verified $\Gamma=m^2(V_X+2V_Z)/(4\pi T)\sim m^2/T$ for potentials corresponding to explicit breaking within our holographic model. In fact, generalizing this to potentials containing explicit and spontaneous translational breaking, we find ~\cite{Amoretti:2018tzw,Davison:2013jba,Ammon:2019apj,Ammon:2019wci}
\begin{equation}
    \Gamma+\frac{\omega_o^2}{\bar\Omega}=\frac{m^2\,(V_X+2V_Z)}{4\pi\,T}=\frac{m^2}{2\pi\,T}\,\alpha+c\,\frac{m^2}{\pi\,T}\,\beta\sim \langle\text{EXB}\rangle^2+\langle\text{SSB}^2\rangle,
\end{equation}
where $c=5/2$ for the solid potential in \eqref{eq:potentials2} and $c=1$ for the fluid.
In the pseudo-spontaneous limit, the EXB scale is much smaller than the SSB scale implying
\begin{equation}
 \langle\text{EXB}\rangle\sim   \sqrt{\Gamma}\sim m\,\sqrt{\alpha/T},\quad\quad\quad \langle\text{SSB}\rangle\sim   \sqrt{\omega_o^2/\bar\Omega}\sim m\,\sqrt{\beta/T}.
\end{equation}
 The QNMs in the pseudo-spontaneous limit at $k=0$ encode $\bar\Omega$ and $\omega_o^2$ according to eq. \eqref{eq:k0}, where we approximate $\Gamma$ by eq. \eqref{d1}. The results are depicted in figure \ref{pic:GMOR}.
In the pseudo-spontaneous limit, the results indicate the following scaling behaviors\footnote{In contrast to \cite{Ammon:2019apj,Baggioli:2019abx}, we conclude that the dimensionless quantity $\bar\Omega/T$ is proportional to the fraction $\frac{\langle\text{EXB}\rangle}{\langle\text{SSB}\rangle}$, which differs by a factor of $T$.}
\begin{equation}
    \bar\Omega/T\sim\sqrt{\frac{\alpha}{\beta}}\sim\frac{\langle\text{EXB}\rangle}{\langle\text{SSB}\rangle},\quad\omega_o^2/T^2\sim m^2/T^2\,\sqrt{\alpha\,\beta}\sim\langle\text{EXB}\rangle\,\langle\text{SSB}\rangle/T^2.
\end{equation}
The scaling behavior signifies the following results
\begin{enumerate}[label=(\roman*)]
    \item The mass of the pseudo-phonons satisfies the GMOR relations, i.e. the mass is proportional to the symmetry breaking scales EXB and SSB.
    \item The phase relaxation parameter $\bar\Omega/T$ is proportional to the fraction of EXB and SSB. The phase relaxation parameter vanishes in the spontaneous limit (EXB$=0$) on the one hand. On the other hand, for mostly explicit symmetry breaking the mode is overdamped and does not influence the low energy physics. 
    \item In the fluid model, we did not observe any propagating phonons for purely spontaneous symmetry breaking since the elastic shear modulus is zero. Yet, by switching on a small explicit symmetry breaking term, we observe light  and underdamped modes -- the pseudo-phonons. 
\end{enumerate}
The authors of \cite{Ammon:2019wci}, also observed the pseudo-phonons in the solid model. Yet our observation of propagating shear waves is even more surprising, since the transverse phonons are non-propagation in the spontaneous regime. 
In \cite{Amoretti:2018tzw,Andrade:2018gqk}, the authors suggested a universal relation between the phase relaxation $\bar\Omega$, the pinning frequency $\omega_o^2$ and the Goldstone diffusion $\xi_\perp$
\begin{equation}\label{universe}
    \bar\Omega/T\sim M\,\xi_\perp/T\sim\omega_o^2\,\chi_{\pi\pi}\,\xi_\perp/T,
\end{equation}
where $M$ is the mass of the pseudo-phonons and~\cite{Amoretti:2019kuf,Ammon:2019wci,Ammon:2019apj} \begin{equation}\xi_\perp=\lim\limits_{\omega\to 0}\omega\,\lim\limits_{k\to 0}\, G^R_{\phi_\perp \phi_\perp}(\omega,k)=\frac{4\,\pi\,s\,T^2}{2\,m^2\,\chi_{\pi\pi}^2\,\left(V_X(1,1)\,+\,2\,V_Z(1,1)\right)}. \label{eq:formulaxi}\end{equation}
In the right side of figure \ref{pic:GMOR}, we checked the validity of this formula in our fluid model.

To conclude our discussion of the $k=0$ behavior in the transverse sector, we investigate the lowest modes at $k=0$ in dependence of the SSB and EXB scale and connect the behavior to hydrodynamic arguments.

In the left side of figure \ref{pic:kzeropseudo}, we depict the modes at $k=0$ extracted from the QNMs in the holographic model. We keep the SSB scale fixed and slowly increase the EXB scale. For small EXB, we observe two modes with finite real part. Initially, the mass of the pseudo-phonons, i.e. the real part of the modes, increases for increasing EXB. After reaching a maximum the modes are bound towards one another and eventually collide for increasing the EXB further. After the collision the modes are sitting on the imaginary axis moving away from one another. 

In the right side of figure \ref{pic:kzeropseudo}, we depict the limit in the opposite direction. This time, we hold the EXB scale constant. Increasing the SSB scale brings us deeper in the pseudo-spontaneous regime and we observe two modes on the imaginary axis colliding with one another and moving away from the axis.

For the remainder of the section, we focus on the momentum dependence of the dispersion relations in the three different regimes.
For dispersion relations of the form of eq. \eqref{disp2}, we find indeed three different regimes, depending on the values of $\Gamma,\,\bar\Omega,\,\omega_0$. Similar to the $k=0$ case, we discuss the cases in the order $4\, \omega_o^2>(\Gamma-\bar\Omega)^2$, $4\, \omega_o^2=(\Gamma-\bar\Omega)^2$, $4\, \omega_o^2<(\Gamma-\bar\Omega)^2$.
\begin{enumerate}[label=(\roman*)]
    \item   $\omega= i\,c_1\pm c_2+(i\,c_3\pm c_4)\,k^2$. For vanishing momentum, the modes have a finite real part at $k=0$. The real part scales quadratic in the momentum and either increases or decreases depending on the sign of $c_4$. For the negative sign, the two modes will collide eventually at $k^\star$.
\item  $\omega=-i\, c_1\pm c_2\,k-i c_3\,k^2$.  In this limit, the modes have vanishing real part for $k=0$. For sufficiently small momenta, the real part increases linearly. 
\item $\omega=-i\, (c_1\pm c_2)-i (c_3\pm c_4)\,k^2$. The modes are purely imaginary around $k=0$ and we observe the $k$-gap phenomenon. Furthermore, we observe two non-propagating diffusive modes since the dispersion relation has two branches. For larger $k$, the modes eventually collide.
\end{enumerate}
\begin{figure}[H]
    \centering
   \includegraphics[width=5cm]{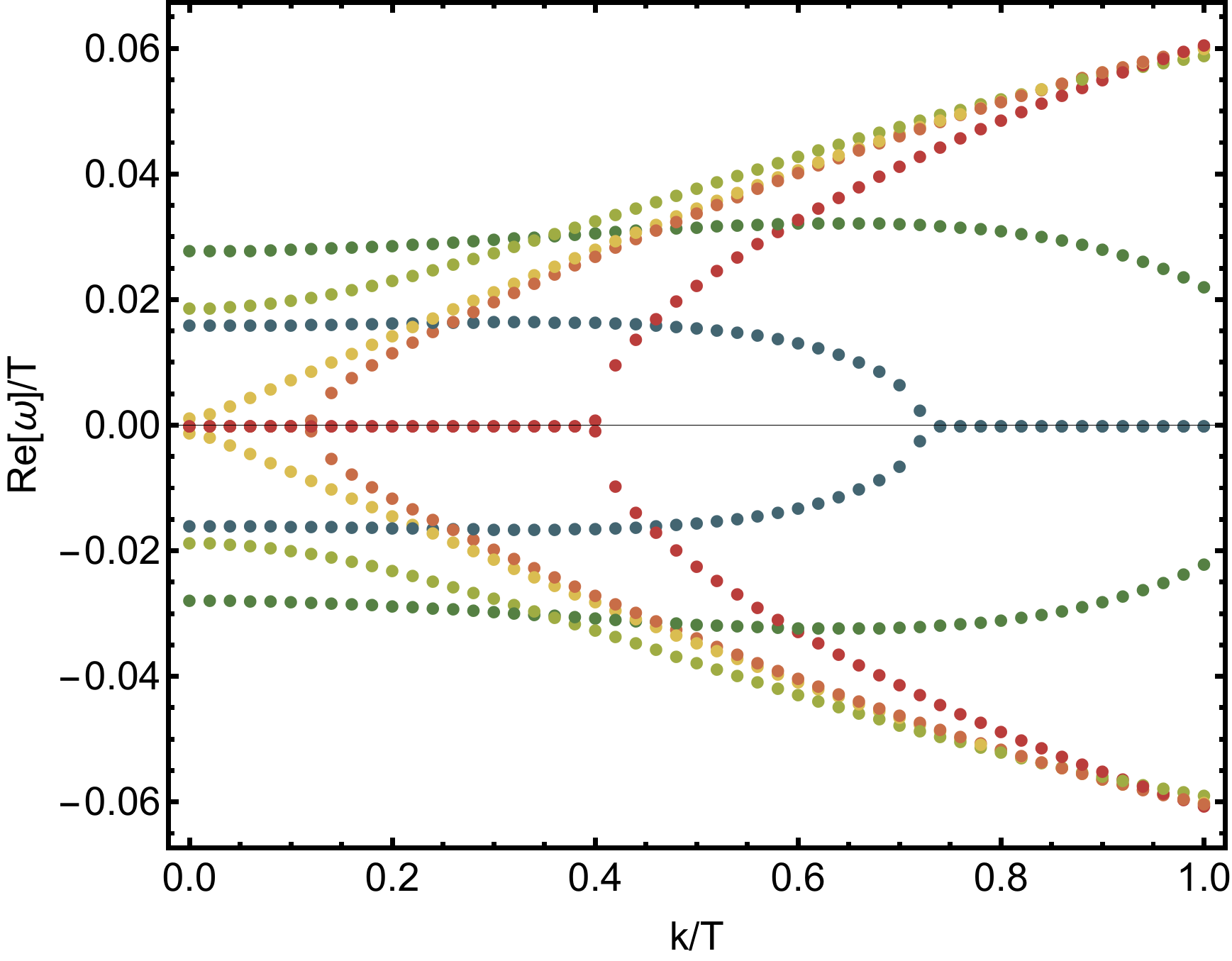}\quad  \includegraphics[width=5cm]{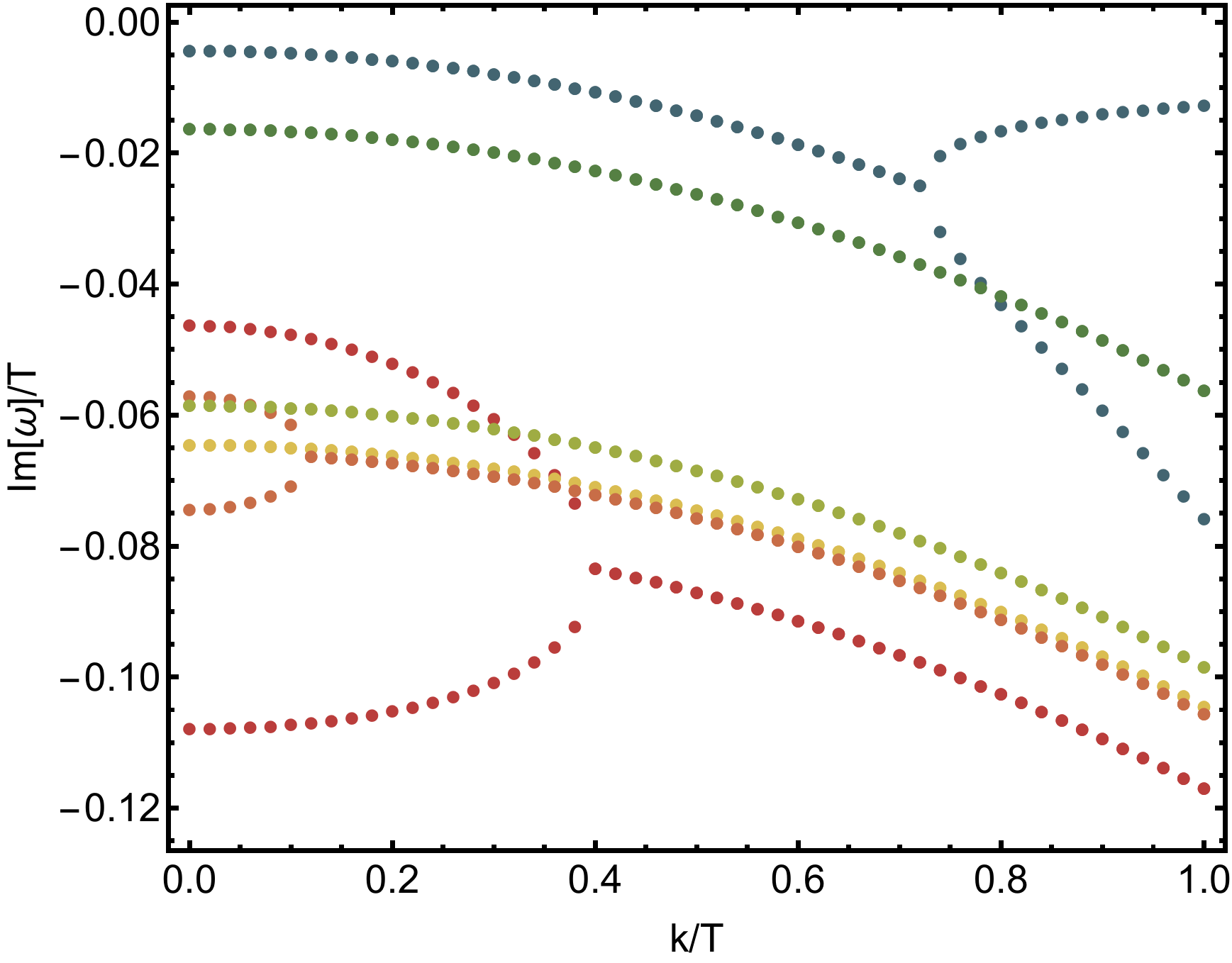}
        \caption{The lowest modes in the transverse sector for $m/T=0.1,\,\beta=1, \,\alpha\in\{0.001,0.151\}$ (blue to red). At small EXB, \textit{i.e.} small $\alpha$, a gapped phonon is present in the spectrum. Going to large EXB, such mode is destroyed and the $k-$gap, typical of the pure EXB case, appears.}
      \label{pic:kdeptr} 
\end{figure}
In figure \ref{pic:kdeptr}, we depict the dispersion relations obtained from QNMs in our holographic model in the three different regimes. The dispersion relation depicted in red corresponds to regime (iii) where we observe two purely imaginary modes for small $k$. Regime (ii) is depicted in yellow; we find two sound-like modes with gapped imaginary part. Even deeper in the pseudo-spontaneous regime, we find the dispersion relations depicted in green. The modes have a finite real part at $k=0$ indicating regime (i).

\subsection*{Longitudinal sector}
After we discussed the transverse sector in depth, we shift our focus on the longitudinal sector of the fluid and solid model. The longitudinal sector is significantly more complicated in hydrodynamics and holography. For simplicity, we will focus on the results obtained from holography in this chapter. Furthermore, since some of the results of this section are very similar to one another or to previous section we move the respective graphics to the appendix for the sake of a compact presentation. In addition to the two sound modes in the longitudinal sector, the system also includes a third diffusive mode in the hydrodynamic regime. Similar to the transverse sector, we will outline the $k=0$ behavior of the modes first. 

\begin{figure}[h]
    \centering
    \includegraphics[width=5cm]{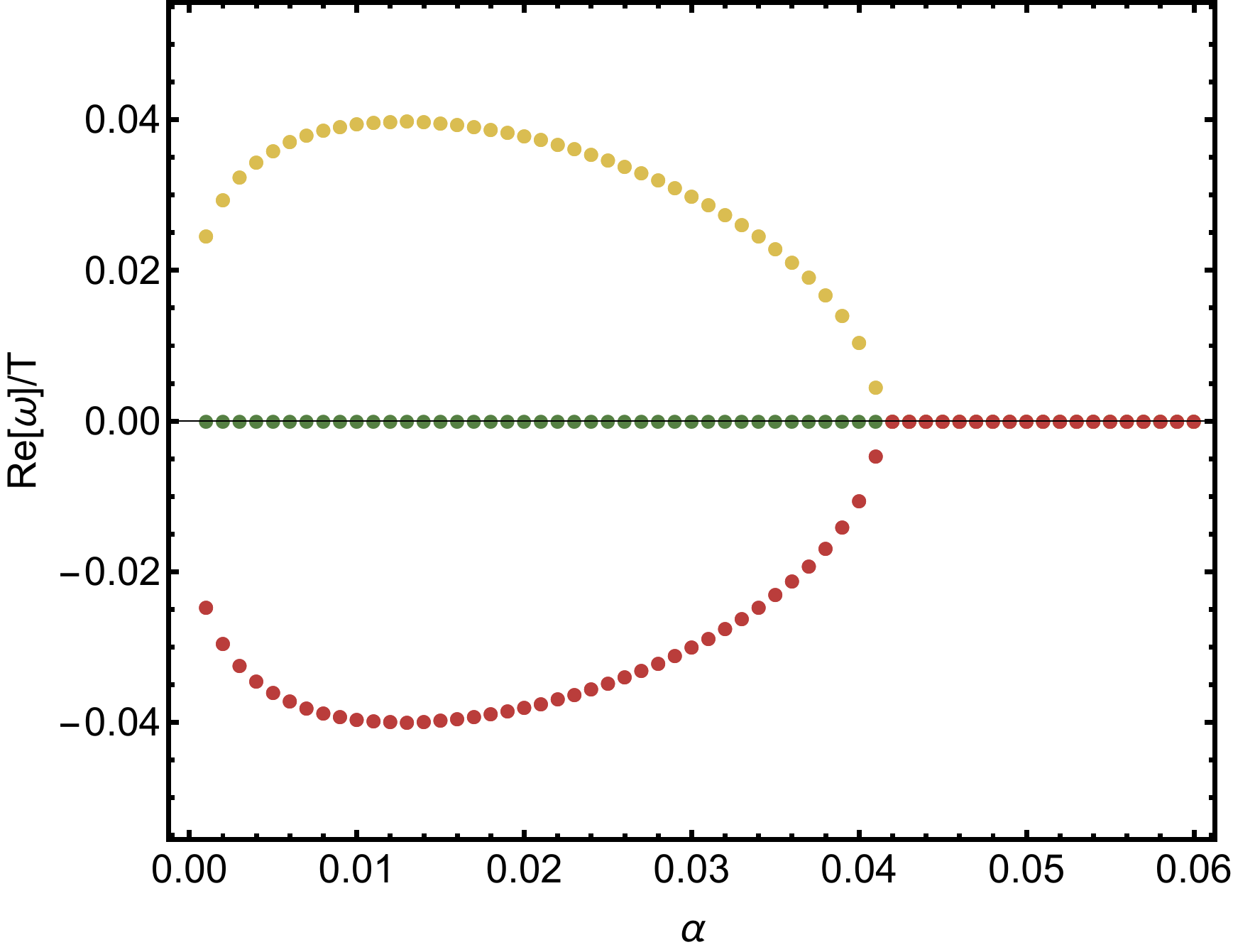}\quad \includegraphics[width=5cm]{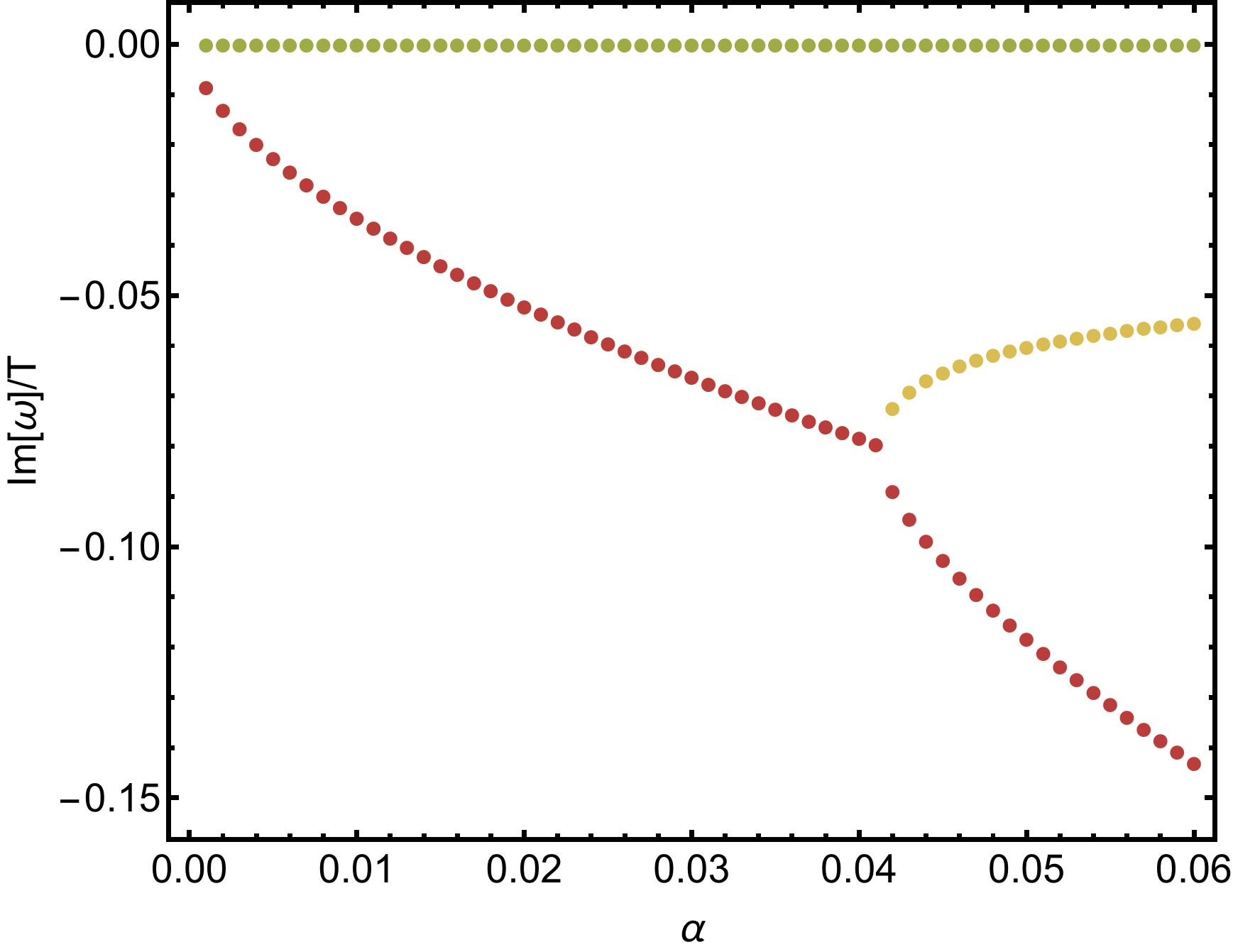}
    \caption{The dynamics of the lowest modes in the longitudinal spectrum of the solid model $V(X)=\alpha\,X+\beta\,X^5$ increasing the EXB scale. The parameters are fixed to $m/T = 0.1, \beta = 5$.\label{fig:otherincrease}}
    \end{figure}
\begin{figure}[h]
    \centering
    \includegraphics[width=5cm]{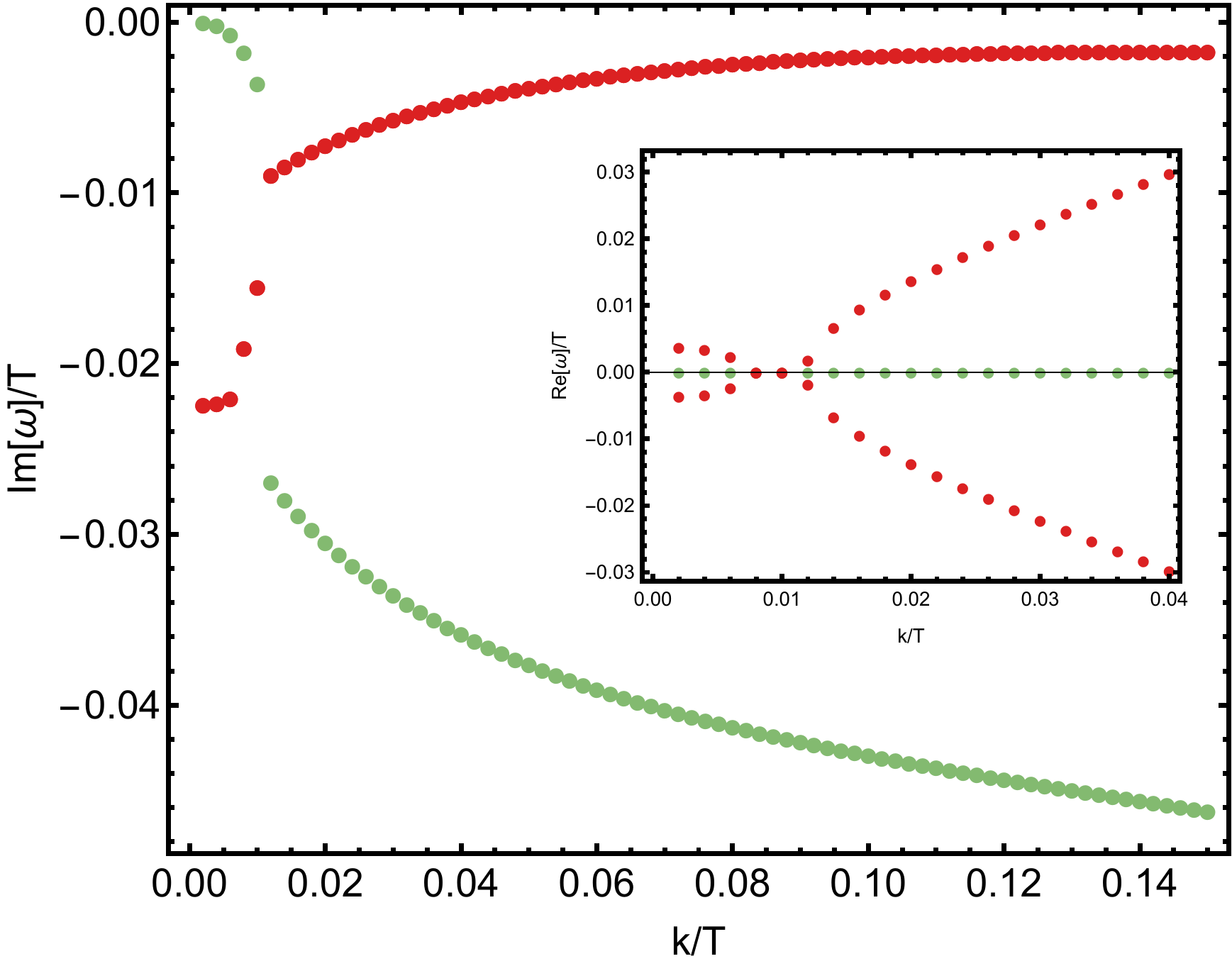}\ \includegraphics[width=5cm]{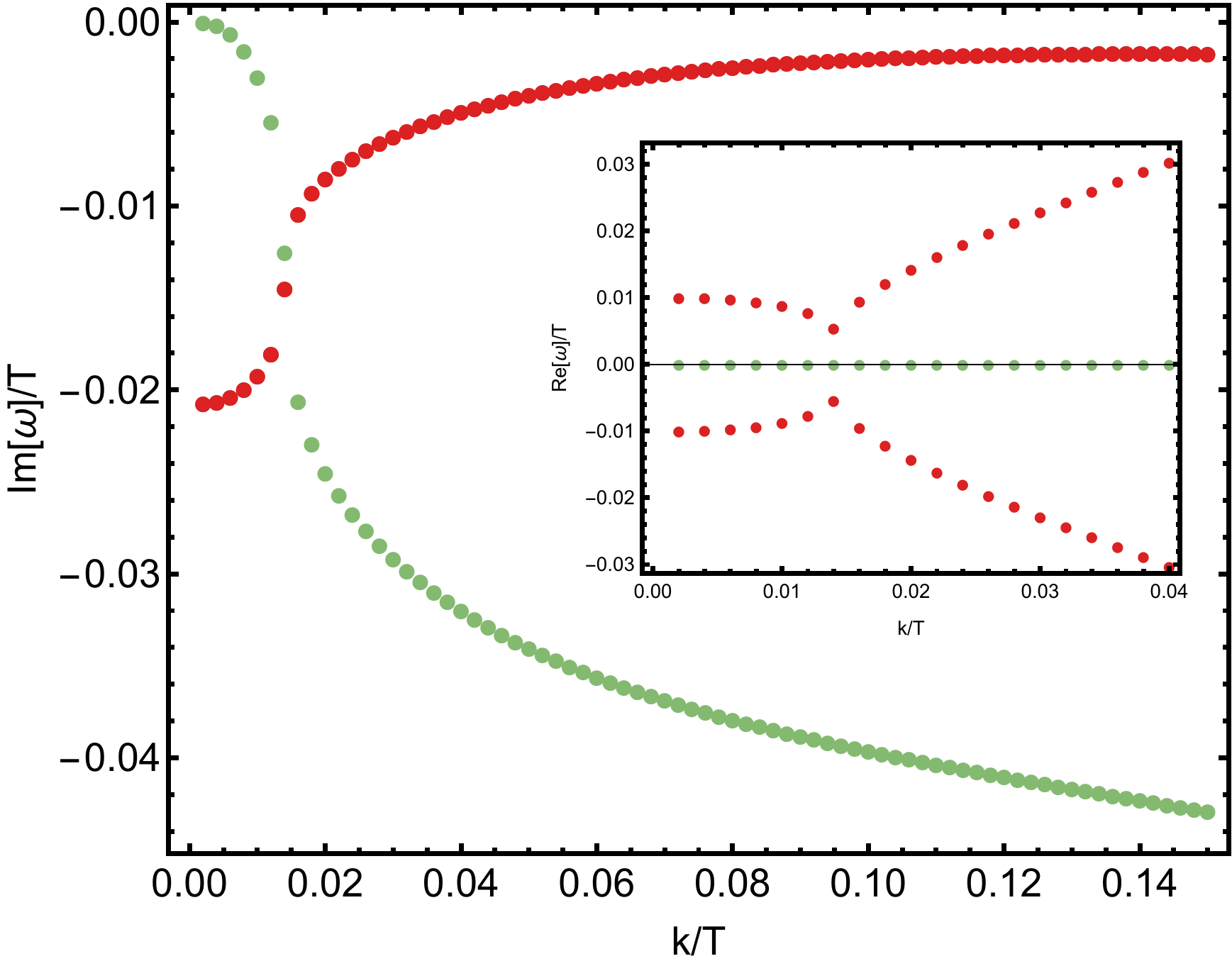}
      \ \includegraphics[width=5cm]{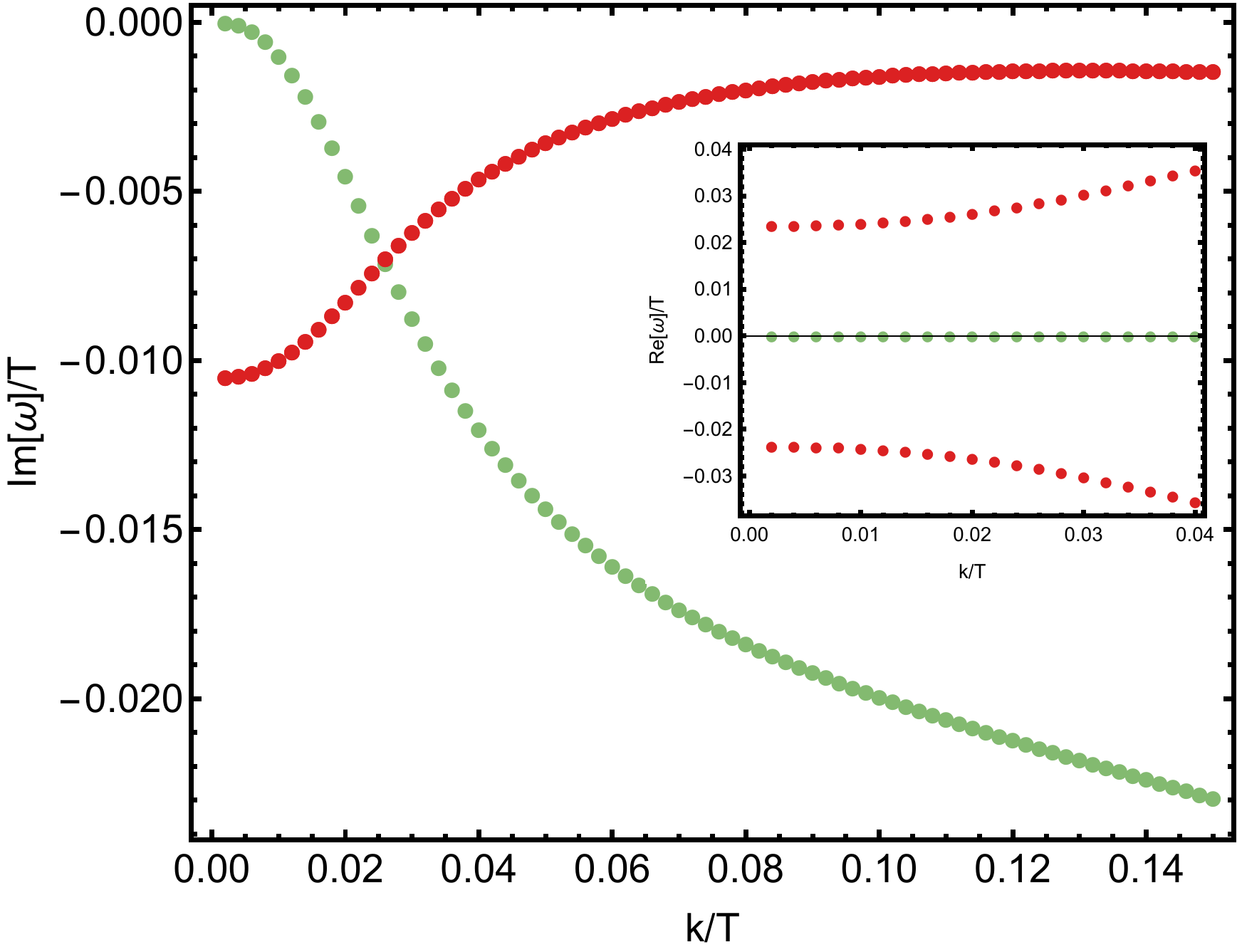}
    \caption{The lowest modes in the longitudinal sector for the fluid model $V(Z)=\alpha Z+\beta Z^2$.
We fix $\alpha = 0.01, m/T = 0.1$ and $\beta\in[1.8, 5]$ (left to right).}
    \label{pic:phononemergence}
\end{figure}

In figure \ref{fig:otherincrease}, we are increasing the EXB scale in the longitudinal sector of the solid model while keeping the SSB scale fixed. Note that the limit is the same limit we took in the left side of figure \ref{pic:kzeropseudo} in the transverse sector of the fluid model, even though we chose to a different graphic presentation. In this limit, we start deep in the pseudo-spontaneous regime with two modes displaying a finite real part at $k=0$. In contrast to the transverse sector, we observe an additional mode depicted in green which stays at the origin for $k=0$.
In figure \ref{pic:phononemergence}, we depict the emergence of the light pseudo-phonons with finite real part at $k=0$ in more detail. Note that the cartoon is similar to our observations for (mostly) explicit symmetry breaking. For explicit symmetry breaking, the purely imaginary modes collide for a certain $k^\star$. With increasing SSB strength, $k^\star$ decreases to the point where the collision already happened at $k=0$. 
\begin{figure}[h]
    \centering
    \includegraphics[width=5.8cm]{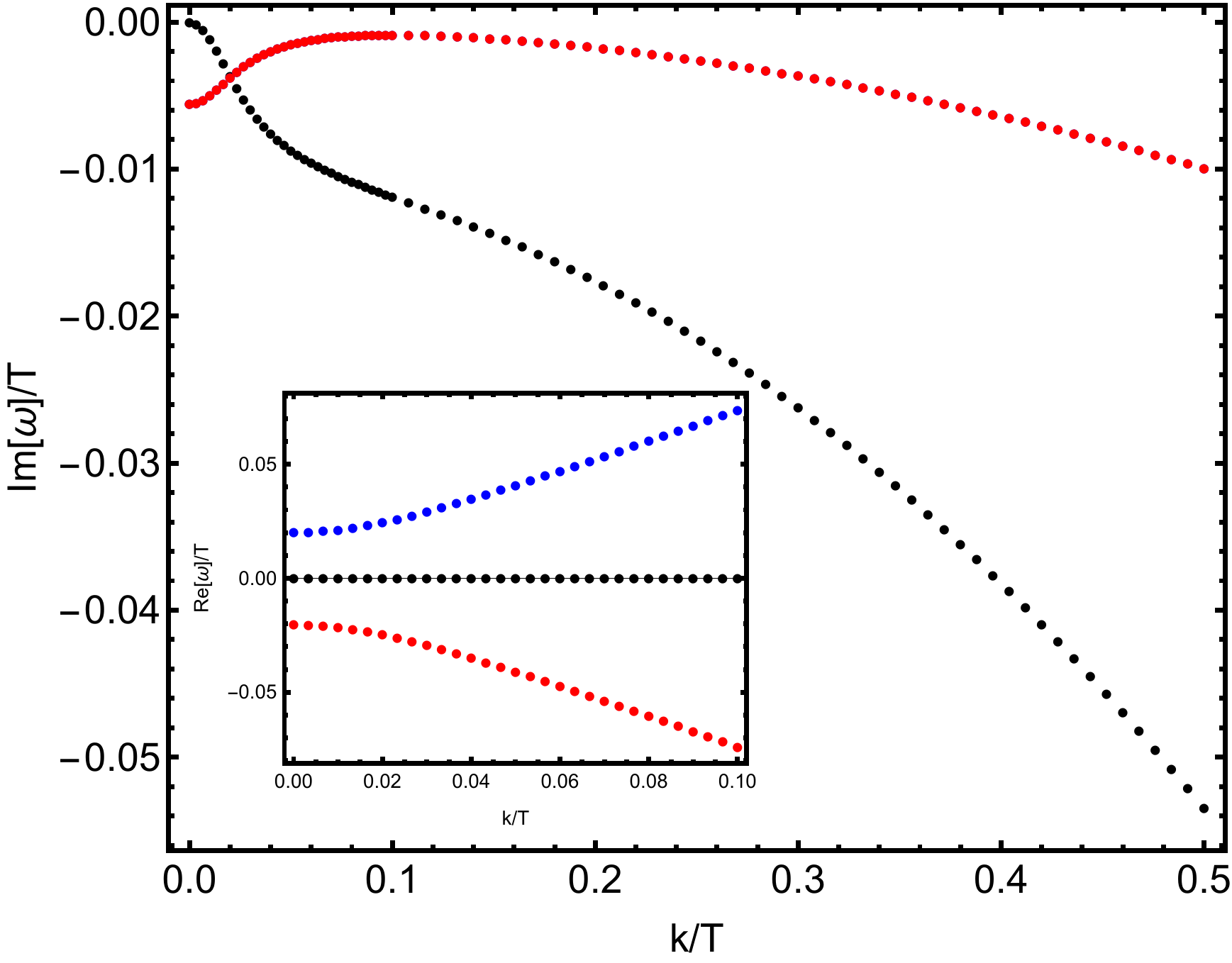}\quad \includegraphics[width=5.8cm]{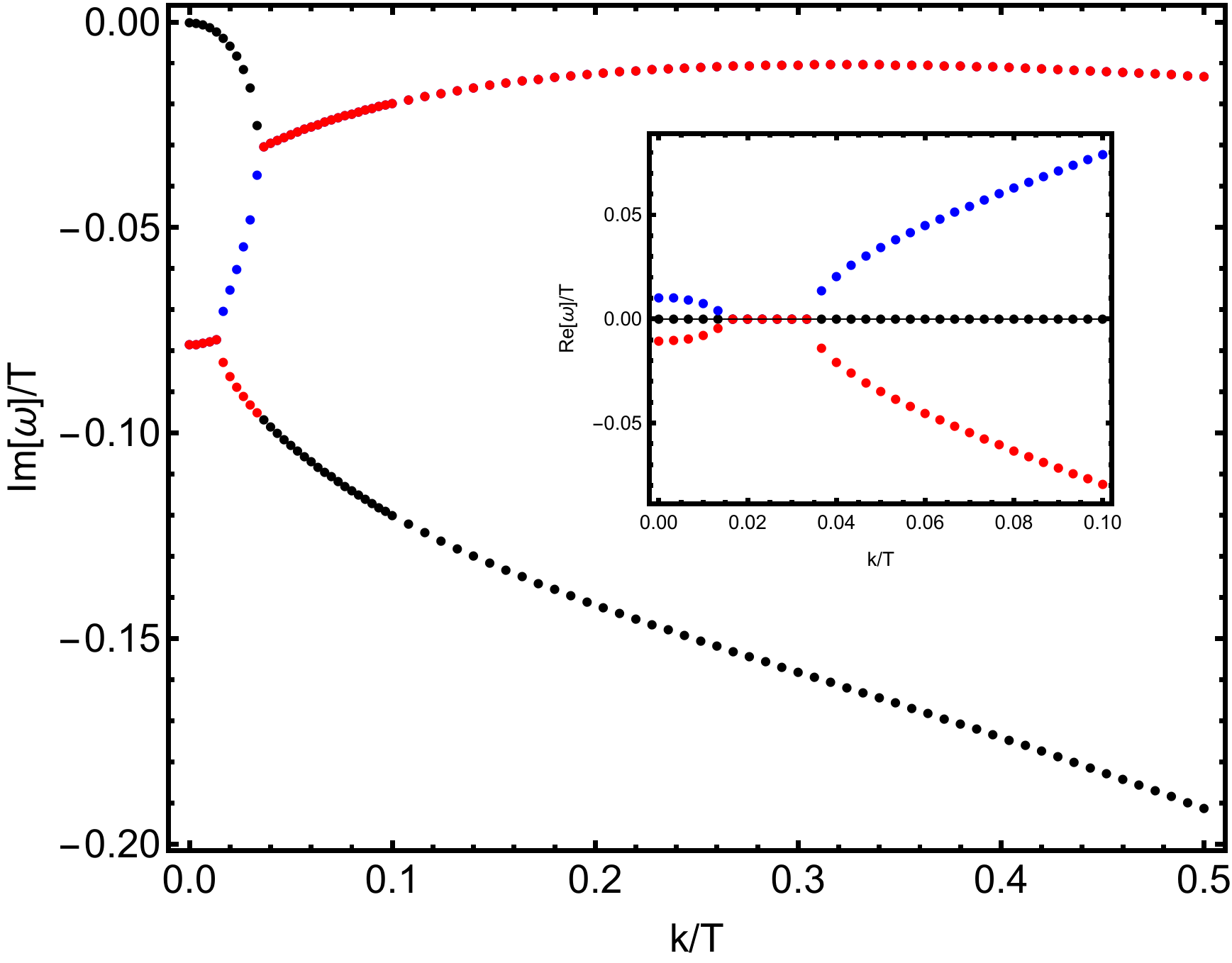}\\
      \includegraphics[width=5.8cm]{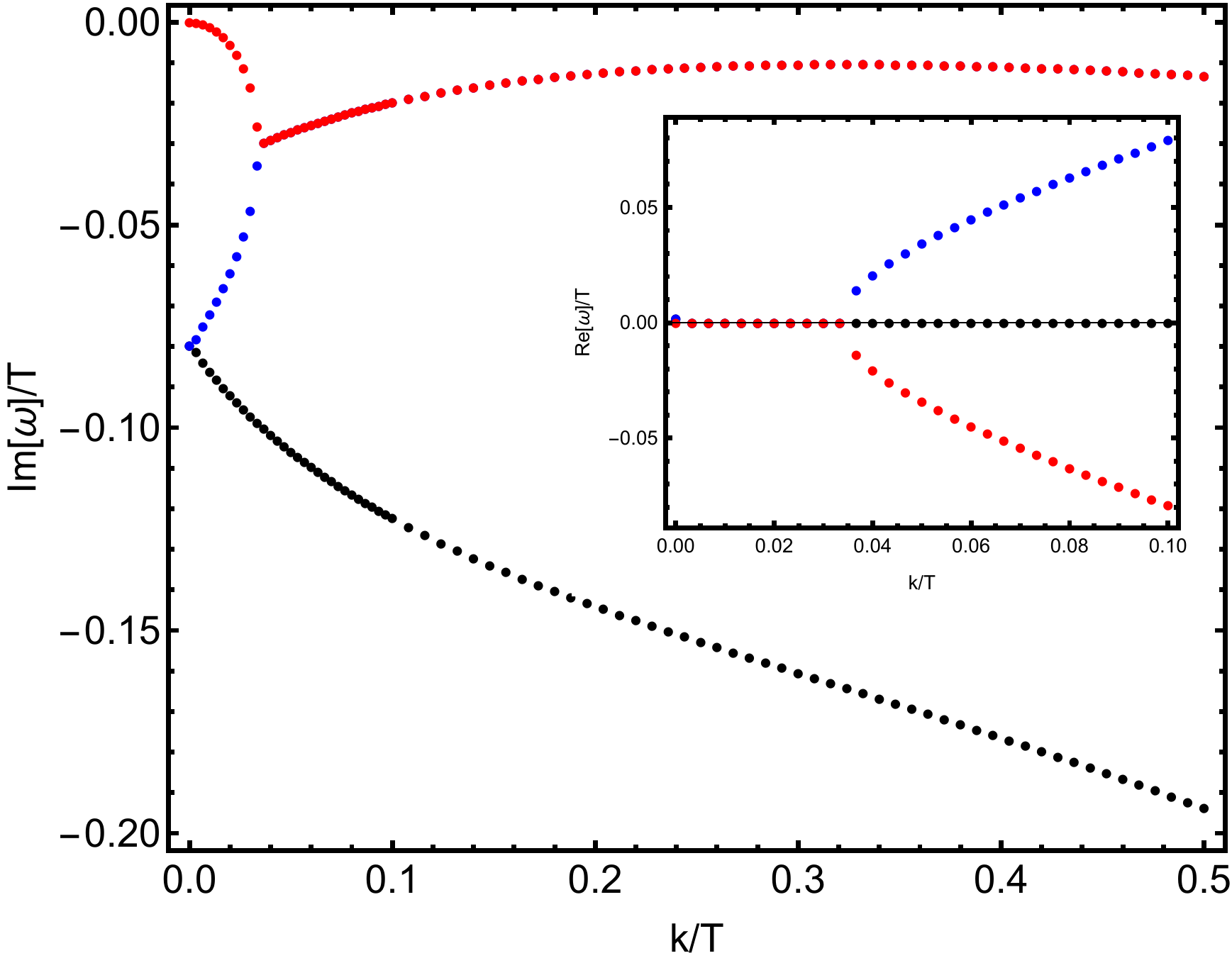}\quad \includegraphics[width=5.8cm]{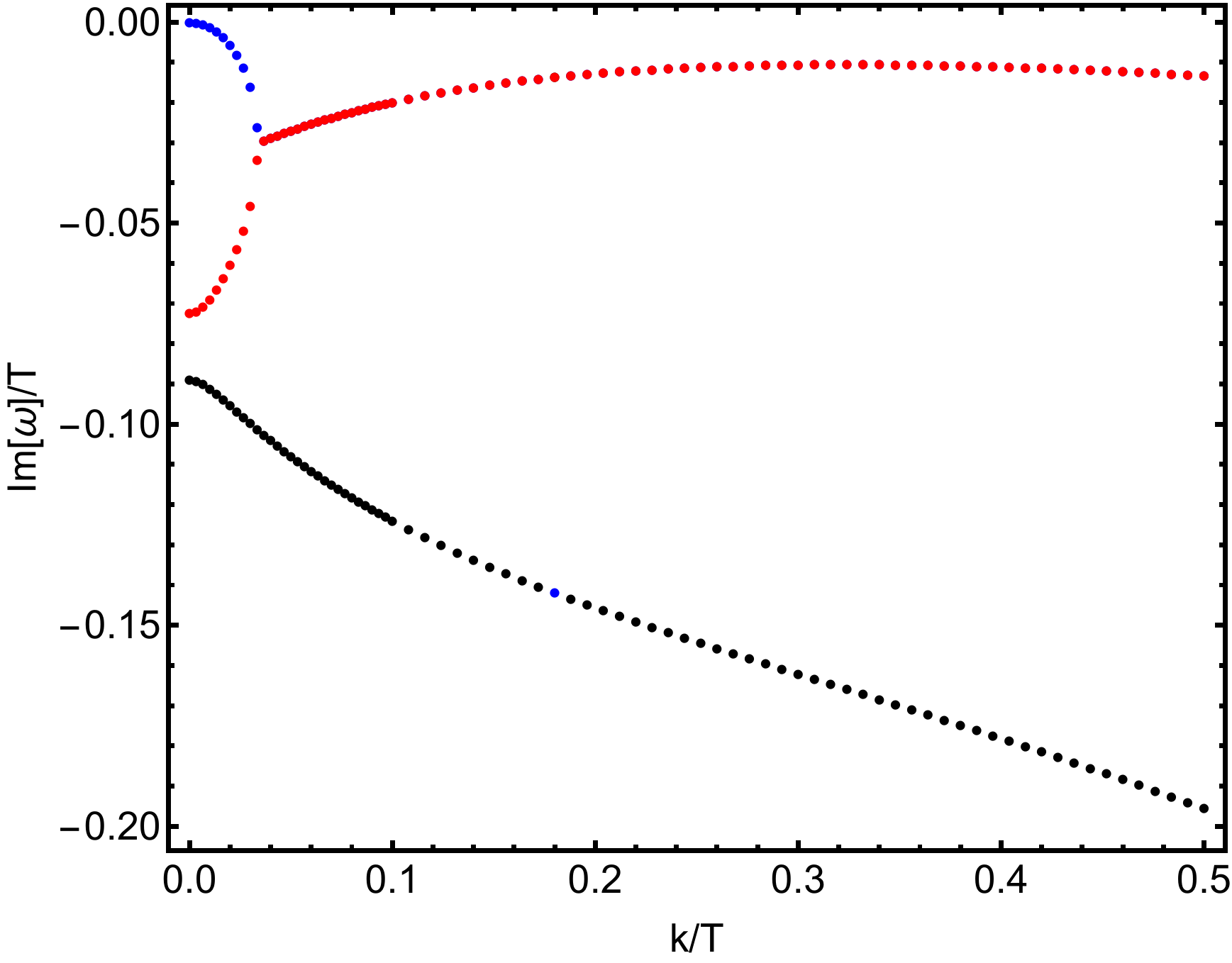}
    \caption{The lowest modes in the longitudinal sector of the solid model $V(X)=\alpha X+\beta X^5$ for $\beta=5, m/T=0.1$ and $\alpha \in [0.0005,0.0412]$ (top to bottom).} 
    \label{finalf}
\end{figure}
However, the dynamics is not as simple as it looks like as may be seen from figure \ref{fig:otherincrease}. In the spontaneous case, both modes are located at the origin for $k=0$ (see right side of figure \ref{pic:moving}). Introducing a small explicit breaking, renders the sound modes massive yielding the light pseudo-phonons. As shown in figure \ref{finalf}, the purely diffusive mode still vanishes for $k=0$ and the sound-like and diffusive mode are crossing one another at a certain momentum. Dialing up the explicit breaking further, a complex dynamic is unfolding; the diffusive mode and the sound modes collide. For large enough explicit symmetry breaking, the diffusive mode decouples, explaining the splitting of the modes we saw at $k=0$. In this case, the two sound modes collide and sound is restored after the collision. In this regime, we observe the $k$-gap phenomenon as we observed for explicit symmetry breaking in figure\ref{figdueb}.
The numerical results show that the purely diffusive mode does not simply acquire a damping term and the dynamics of the full system can not be understood simply by adding up the different contributions to the phase relaxation as suggested in \cite{Delacretaz:2017zxd,Amoretti:2018tzw,Andrade:2018gqk}. The relaxation mechanism is more complicated and highly entangled with the longitudinal sound. For the sake of completeness, we also checked that the GMOR relations are satisfied in the longitudinal sector.

\section{Nonlinear dynamics and broken translations}
In the preceding sections, we studied the properties of systems with broken translations in the linear response regime. This is particularly valuable to strengthen our understanding of the effects of broken spacetime symmetries on the transport behavior of strongly coupled field theories. Within the linear response regime, we have some access to physics in terms of effective field theories and hydrodynamic methods. Beyond linear response, however, we still lack a satisfying theoretical field theoretic framework. The AdS/CFT correspondence is the appropriate framework to study time-dependent problems in strongly coupled field theories from first principles. In this section, we study periodically driven QFTs with (spontaneously) broken translational invariance.
Notably, we drive the system periodically for various amplitudes and driving frequencies of the applied strain as initiated for various periodically driven systems in~\cite{Rangamani:2015sha,Biasi:2017kkn,Biasi:2019eap,Li:2013fhw,Auzzi:2013pca,Hashimoto:2016ize,Kinoshita:2017uch,Ishii:2018ucz,Andrade:2019rpn,Carracedo:2016qrf}.

\subsection*{The nonlinear holographic setup}
In this section, we employ the same setup as in the previous sections given by eq. \ref{eq:model} with potentials of the form $V(X)=X^3$ corresponding to spontaneous breaking of the translational symmetry in the dual field theory. 
In contrast to the last sections, we consider the full nonlinear time-dependent behavior in this section instead of small fluctuations about the equilibrium state. Far from equilibrium, the configuration is no longer described by a black brane and we make the ansatz
\begin{equation*}
    \dd s^2\!=\!-A(u,t)\,\dd t^2-\frac{2}{u^2}\,\dd u\,\dd t+S(u,t)^2\left(\!\cosh(H(u,t))\,(\dd x^2\!+\!\dd y^2)+2\sinh(H(u,t))\,\dd x\dd y\!\right)\!,
\end{equation*}
based on the characteristic formulation adapted to AdS by Chesler and Yaffe \cite{Chesler:2008hg,Chesler:2009cy,Chesler:2010bi,Chesler:2013lia}.
At the boundary, we require asymptotically AdS boundary conditions for the unperturbed state and the metric functions behave near $u=0$ as
\begin{eqnarray}
&&A=\frac{1}{u^2}+\frac{2(s_1-\dot{s_0})}{s_0 \ u}+\left( \frac{s_1^2}{s_0^2}-\frac{2\dot{s_1}}{s_0}-\frac{3\dot{h_0}^2}{4} \right)+a_3 u+\mathcal{O}(u^2),\nonumber\\
&&S=\frac{s_0}{u}+s_1-\frac{s_0\dot{h_0}^2}{8}u+\frac{s_1 \dot{h_0}^2}{8} u^2+\mathcal{O}(u^3),\label{bdry-exp}\\
&&H=h_0+\dot{h_0}u-\frac{s_1 \dot{h_0}}{s_0}u^2+h_3 u^3+\mathcal{O}(u^4).\nonumber
\end{eqnarray}
We focus on driving the $xy$ component of the boundary metric periodically with the applied shear
\begin{equation}
    h_{xy}(t)=\gamma(t)=\gamma_0\,\sin(2\pi\,\omega\,t),
\end{equation}
where $\gamma_0$ is the strain amplitude and $\omega$ the characteristic frequency.\footnote{We keep the length scales of the spatial dimensions fixed, i.e. $h_{xx}=h_{yy}=1$.} In terms of our metric ansatz, this translates to 
\begin{equation}
    h_0(t)=\arcsinh\left(\frac{\gamma(t)}{(1-\gamma(t)^2)^{1/2}}\right),\ s_0(t)=(1-\gamma(t)^2)^{1/4}.\label{eq:strainten}
\end{equation}
In eq. \eqref{eq:strainten}, we restrict the amplitudes to be in the range $\gamma_0\in[0,1]$, where $\gamma_0=1$ is the extremal amplitude. The response to the applied strain function may be read off in terms of the shear stress $\sigma(t)$ encoded in the $T_{xy}(t)$ component of the boundary stress tensor.\footnote{For details on the numerics and how to read of the shear stress see appendix \ref{app:HMGtime}, in particular eq. \eqref{eq:readoffTXY}.} 
There is one caveat, however. Imposing a strain at $t=0$ violates the boundary conditions at the beginning of the time evolution since time derivatives of the strain are non-zero. Following~\cite{Ammon:2016fru,Grieninger:2017jxz}, we turn on the amplitude smoothly according to
\mbox{$\gamma(t)= \frac{\gamma_0}{2}\left(1+\tanh\left(\frac{t-t_c}{w_c}\right)\right)\sin\left(2\pi\omega t\right)$}, where parameters $t_c, w_c$ control the width and abruptness.
\begin{figure}[!t]
    \centering\includegraphics[width=9.3cm]{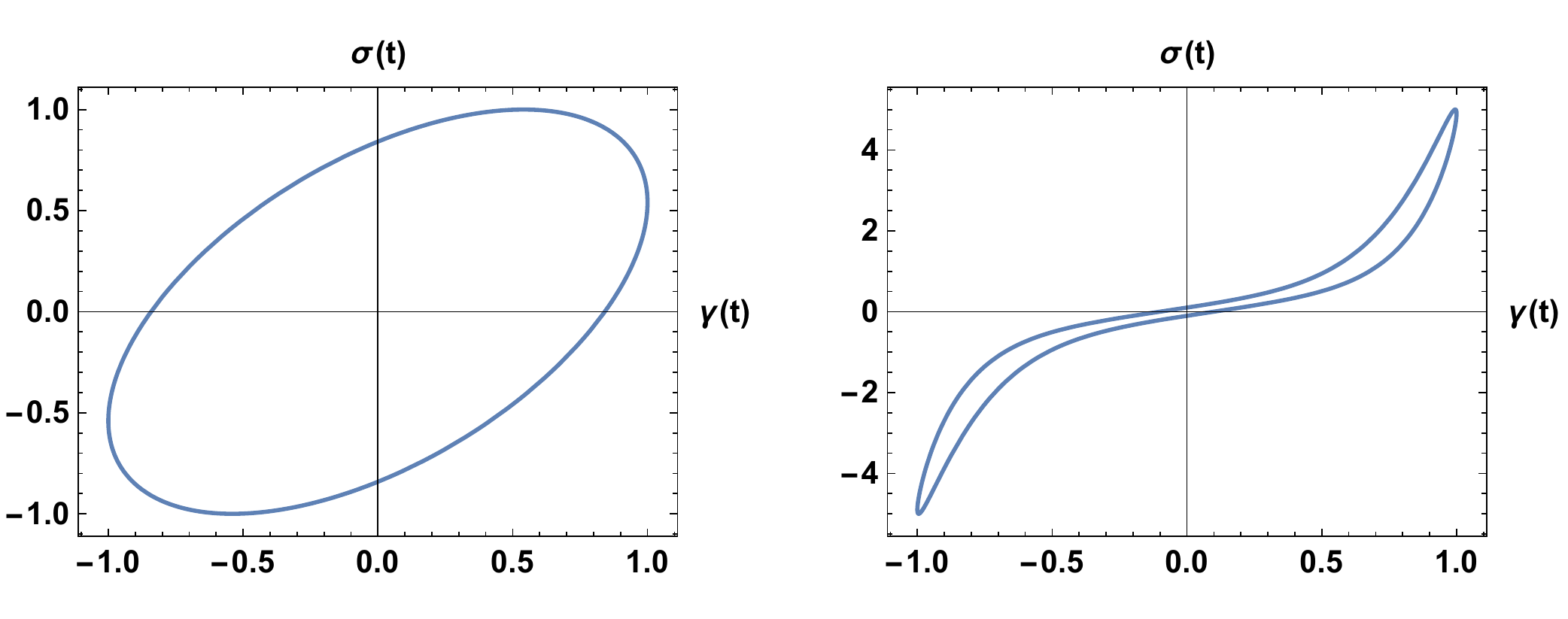}
    \caption{Demonstration of nonlinearities. \textbf{Left: } $\gamma(t)=\sin(t-1)$ over $\sigma(t)=\sin(t)$. \textbf{Right: } $\gamma(t)=\sin(t-1)$ over $\sigma(t)=\sin(t-0.9)+\sin(t-0.9)^3+2\,\sin(t-0.9)^5+\sin(t-0.9)^7$}.\label{fig:linvsnonline}
    \end{figure}
Recall that in the linear response limit, the shear response is given in terms of the two point function $ G^R_{T_{xy}T_{xy}}(\omega, k=0)$.
Beyond linear response, we may expand the stress in the sinusoidal strain by
\begin{equation}
    \sigma(t)=\sum\limits_{n\text{,odd}}\sum\limits_{p\text{,odd}}^n\gamma_0^n\,(G'_{np}\,\sin(2\pi\,p\,\omega\,t)+G''_{np}\,\cos(2\pi\,p\,\omega\,t)), \label{eq:stresssinus}
\end{equation}
where linear response theory is captured by $n=1$.

\begin{figure}[h!]
    \centering\includegraphics[width=6.35cm]{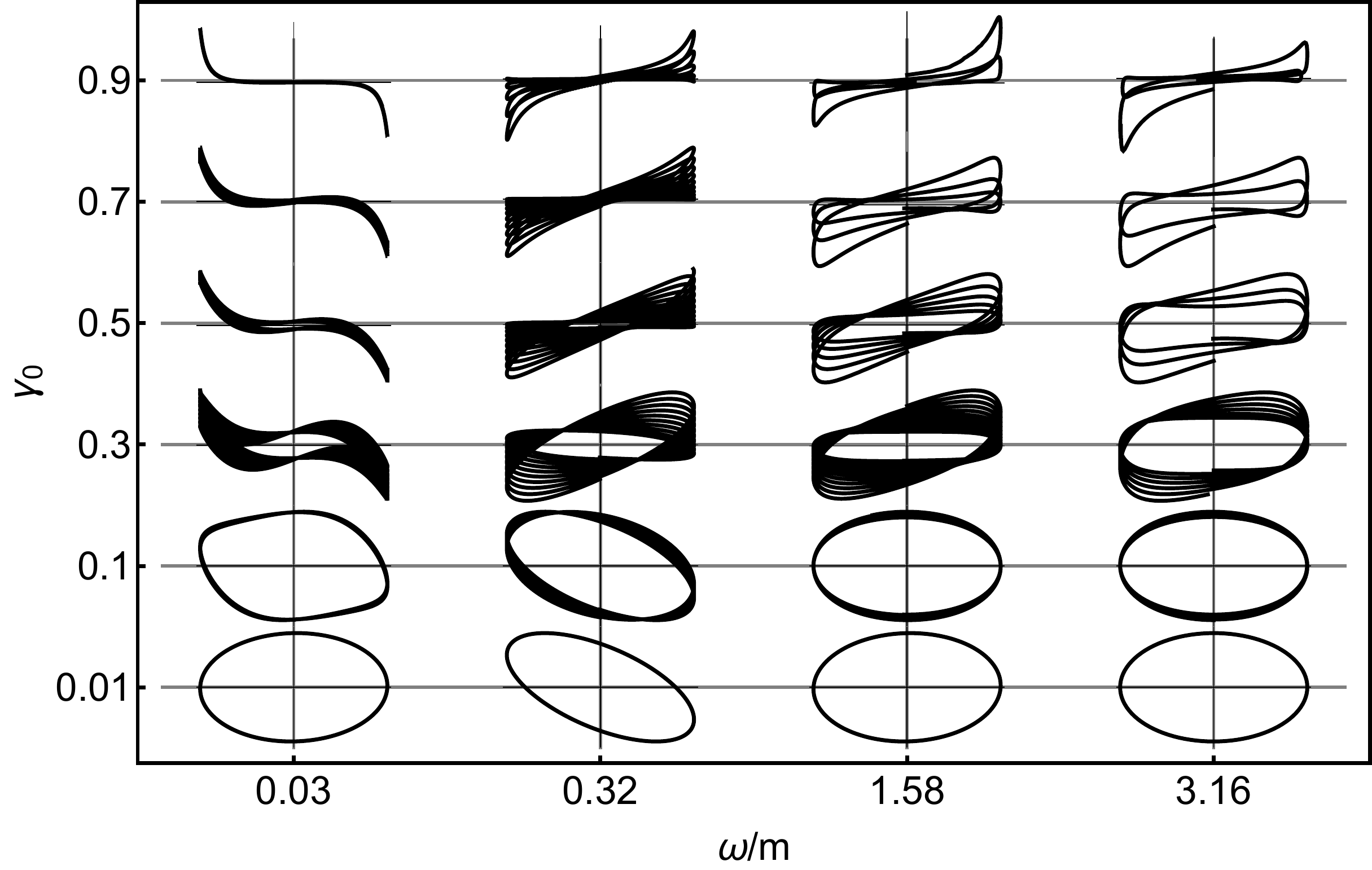}\quad 
    \includegraphics[width=8cm]{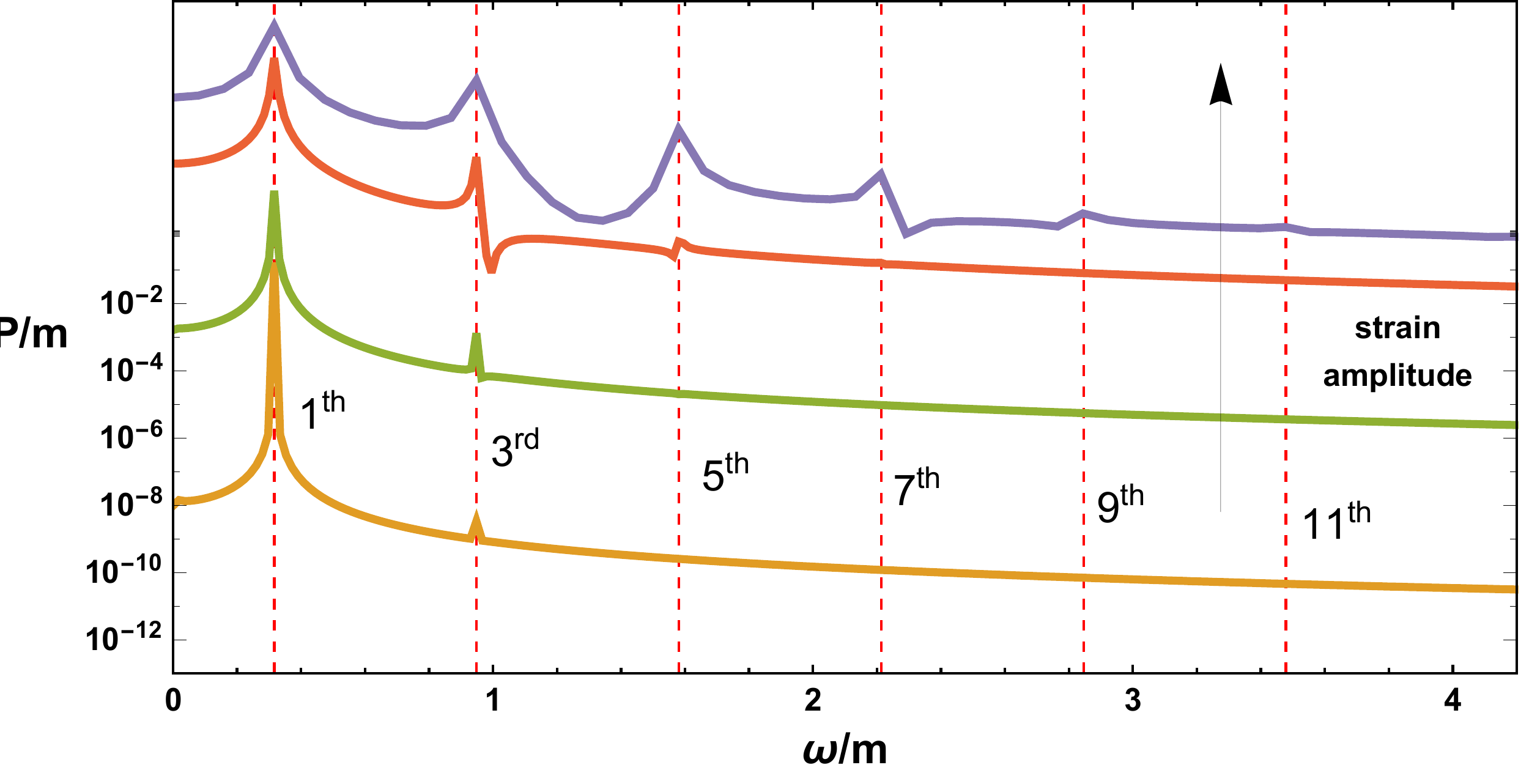}
    \caption{\textbf{Left: }Pipkin diagram: the Lissajous figures as function of $\gamma_0$ and $\omega$ for $m/\color{black}T_{in}\color{black}=1.81$. \textbf{Right: }Fourier spectrum $\mathcal{P}\equiv |I|^2$ of the stress for increasing $\gamma_0=\{0.01,0.1,0.4,0.75\}$ (orange to black).}
    \label{fig:3b}
\end{figure}

First, we work out the so-called Pipkin diagram for the dual field theory in order to get an intuition about the holographic model at hand. In the Pipkin diagram, we depict the Lissajous figures of the response, i.e. we plot $\sigma(t)$ over $\gamma(t)$. As shown in figure \ref{fig:linvsnonline}, the Lissajous figures give insight into the the linearity and nonlinearity, respectively, of the system  \cite{doi:10.1122/1.2970095}. In the left side, we see that $\sigma(t)$ is described by a simple sinus. The graphics on the right side includes higher order corrections to eq. \eqref{eq:stresssinus}, indicating the nonlinear regime. The Lissajous figures of our system are depicted in figure \ref{fig:3b} for various frequencies and amplitudes. In the linear response regime ($\gamma_0\ll1$), the Lissajous figures are perfectly elliptic shaped. By increasing the amplitude, we see higher harmonics contributing to $\sigma(t)$ and the Lissajous figures are more and more elongated. 

In the Pipkin diagram, we also see the dissipation of our system. On one hand, if the system has more time to react to the strain and dissipate (small frequencies), the viscous properties are dominant. On short time scales (large frequencies), on the other hand, the elastic properties dominate the response. In the intermediate regime, we observe both properties; for small amplitudes, we are in the regime of linear viscoelasticity, which is described by linear response theory for spontaneously broken translations. For large amplitudes, linear response theory breaks down and we are in the regime of large amplitude oscillatory shear tests. The inclination of the curves -- either bound to the left like in the first column or bound to the right -- gives insights how the system with spontaneously broken translations dissipates energy; we observe viscous and elastic nonlinearities. The latter are not present in systems with translational invariance. Note that when the Lissajous figures changes over the different cycles, we are also breaking the discrete time translation symmetry of our periodically driven system. 

 The onset of the nonlinearity is also apparent in the Fourier spectrum in dependence of the strain amplitude as depicted in the right side of figure \ref{fig:3b}. For small strain amplitudes (orange, $\gamma_0=0.01$), the response is completely captured in terms of the first frequency ($p=1$) which is described by linear response. For larger $\gamma_0$, we see the onset of higher harmonics which contribute to the stress eq. \eqref{eq:stresssinus} indicating the onset of nonlinearity. The bigger $\gamma_0$, the more higher harmonics appear in the spectrum.
 
 \begin{figure}[h]
    \centering
    \includegraphics[width=5.5cm]{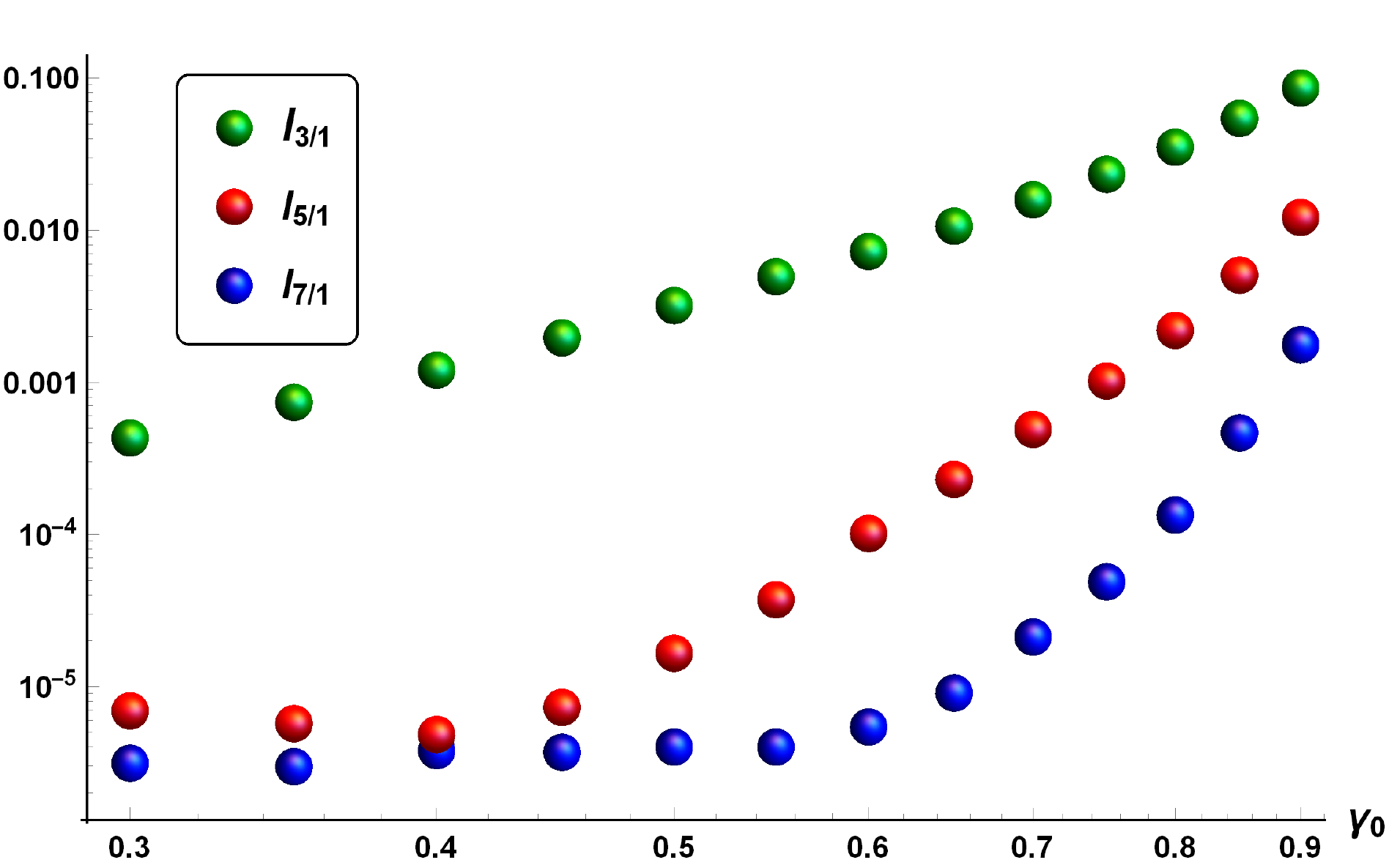}\quad\quad\quad
    \includegraphics[width=5.5cm]{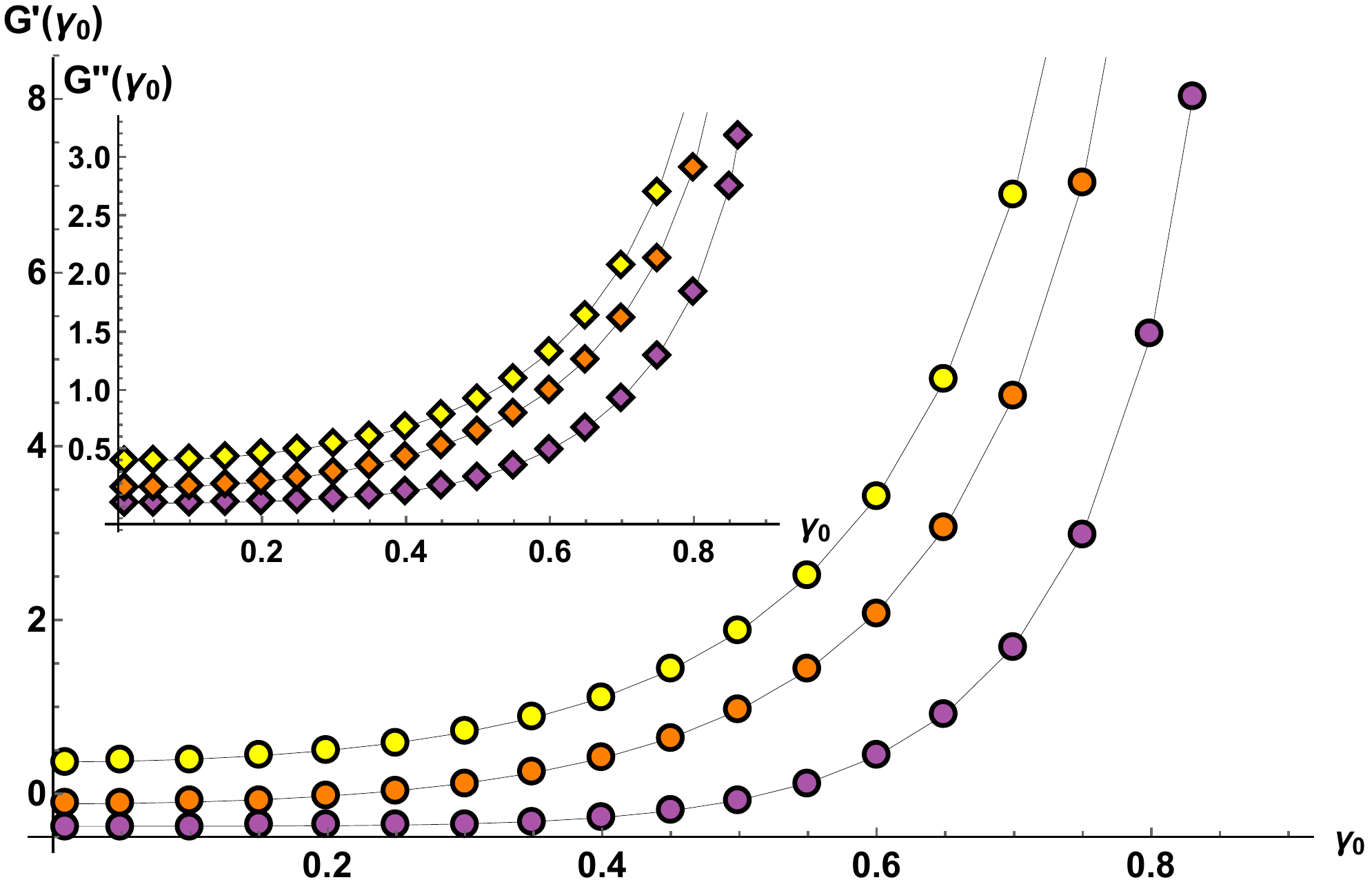}
   \caption{\textbf{Left:} Normalized intensity $I_{n/1}\equiv I_n/I_1$ as function of $\gamma_0$.  \textbf{Right:} First complex moduli $G_1'(\omega,\gamma_0),G_1''(\omega,\gamma_0)$ at fixed frequency, as function of the strain amplitude for $m/T_\text{in}=0.01,1.81,30$ (yellow, red, purple).}
    \label{figbb}
\end{figure}

In figure \ref{figbb}, we investigate the normalized intensity of the first three higher harmonics in dependence of the amplitude $\gamma_0$ of the applied strain. In the range $0.3\le\gamma_0\le0.5$, only the first higher harmonic $I_{3/1}$ contributes. For larger values of $\gamma_0$, the higher harmonics $I_{5/1}$ and $I_{7/1}$ are also present in the spectrum. Furthermore, the higher harmonics $I_{5/1}$ and $I_{7/1}$ follow a power law behavior for large amplitudes as found in~\cite{doi:10.1146/annurev.fl.27.010195.001125}. Another indicator to investigate the nonlinearity are the first nonlinear complex moduli $G'_1(\omega,\gamma_0), G_1''(\omega,\gamma_0)$, defined as contributions of the first harmonics ($p=1$) in eq. \eqref{eq:stresssinus}. In the limit $\gamma_0\ll 1$, the complex moduli are in the linear response limit and thus independent of $\gamma_0$ as shown in the right side of figure \ref{figbb}. For small $\gamma_0$, the response does not change for increasing the amplitude. At around $\gamma_0=0.25$, the first complex moduli start increasing with the amplitude, thus indicating the onset of the nonlinear regime. Note that this happens independently of the value of $m/T$. In the nonlinear regime, the complex moduli grow faster than linearly indicating a strain stiffening mechanism as exhibited by hyperelastic materials such as rubber-like systems or complex polymers.

\begin{figure}[h]
    \centering
   \includegraphics[width=0.3 \linewidth]{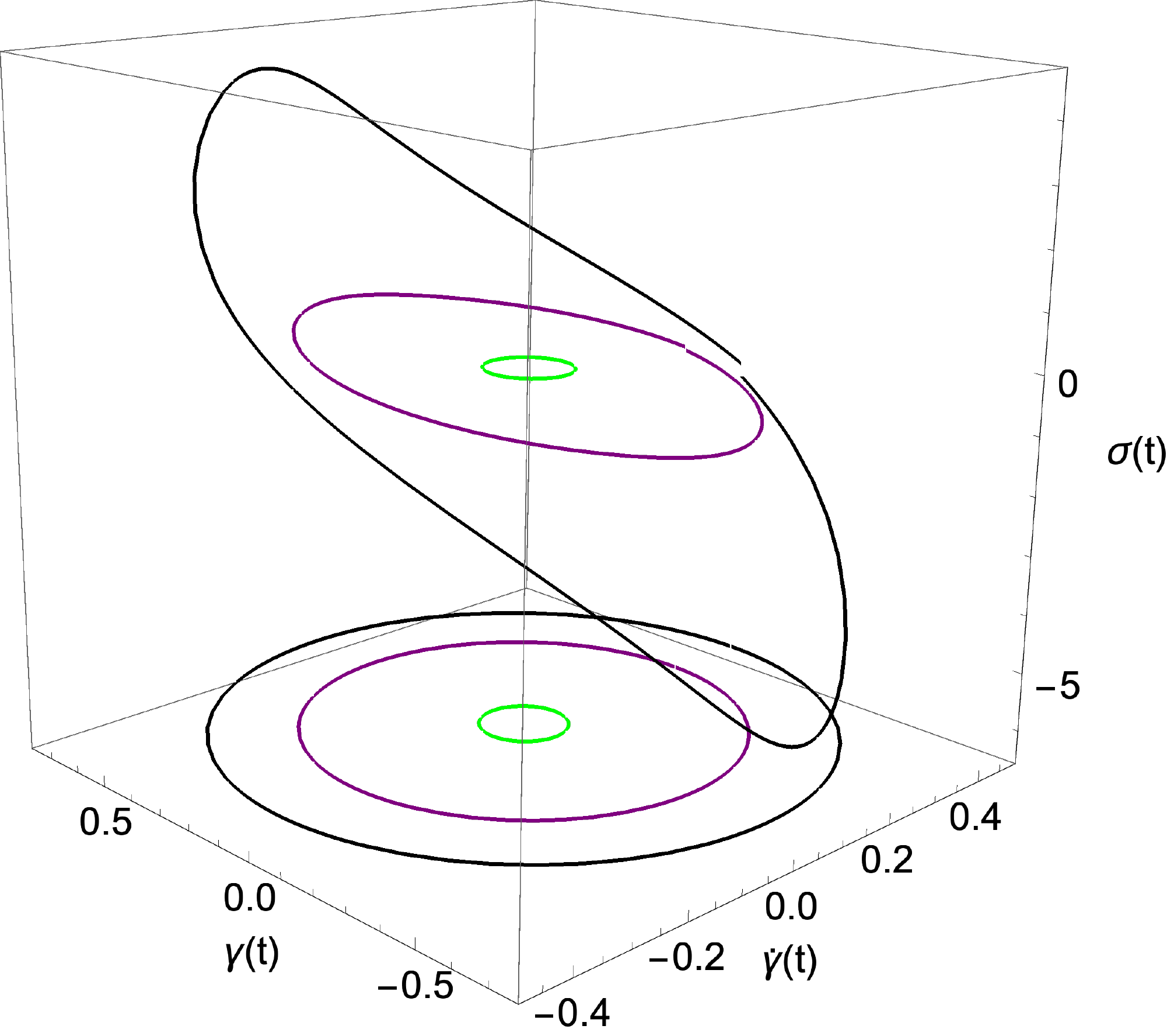}\hspace{0.7cm}\includegraphics[width=0.6 \linewidth]{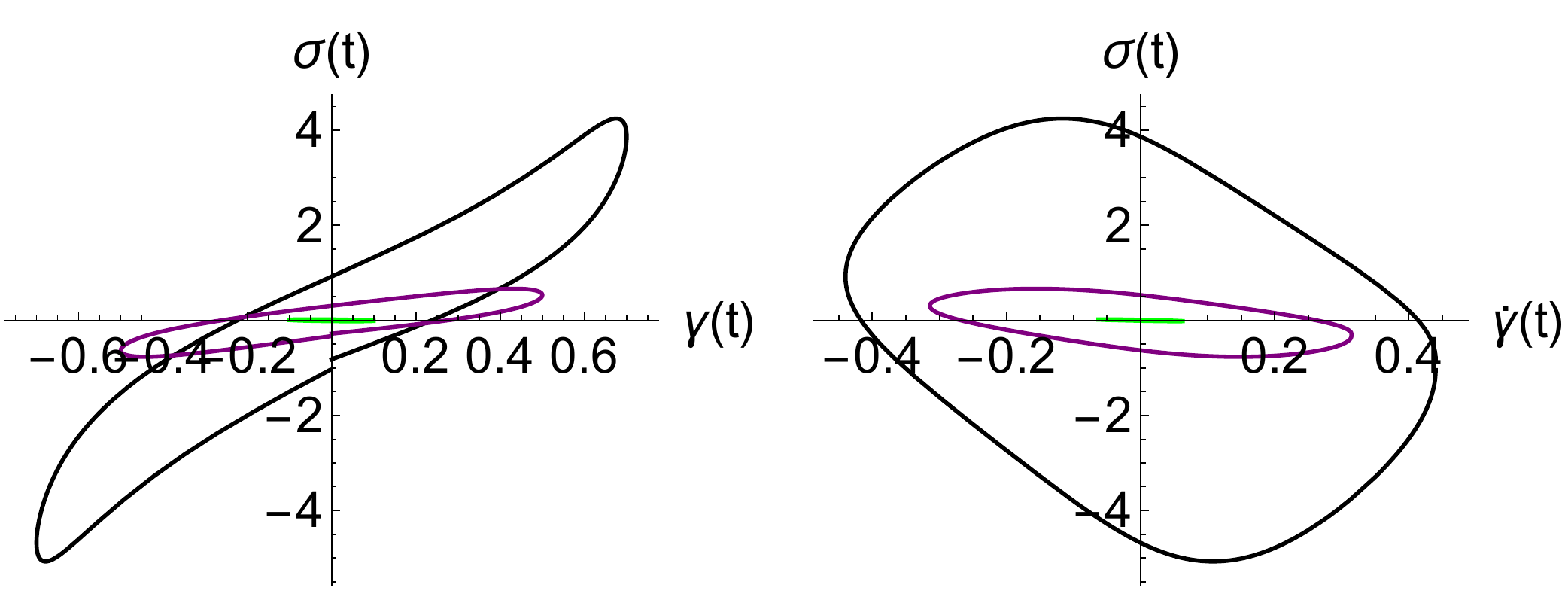}
    \caption{Three dimensional phase space $\{ \sigma(t),\gamma(t),\dot{\gamma}(t)\}$ for $\gamma(t)=\gamma_0\,\sin (2\pi\omega t)$ with $\gamma_0=0.1,0.5,0.7$ (from green to black) at fixed $m/T_\text{in}= 1.81$, $\omega= 0.1$.}
    \label{fig:3qq}
\end{figure}

The three dimensional response curves encode two perspectives: the elastic perspective and the viscous perspective. In figure \ref{fig:3qq}, we show both perspective for increasing the amplitude of the applied strain (green to black). In the elastic perspective, we project the response onto the $\{\sigma(t),\gamma(t)\}$ plane. Since the response of the black curve clearly deviates from an ellipsis, we nonlinearitites in the elasticity are present in the response of the system.
The viscous perspective consists of projecting the three-dimensional curve onto the  $\{\sigma(t),\dot{\gamma}(t)\}$ plane. Recall that the elastic properties of the system are caused by the spontaneously broken translations

\begin{figure}[h]
    \centering\includegraphics[width=4.4cm]{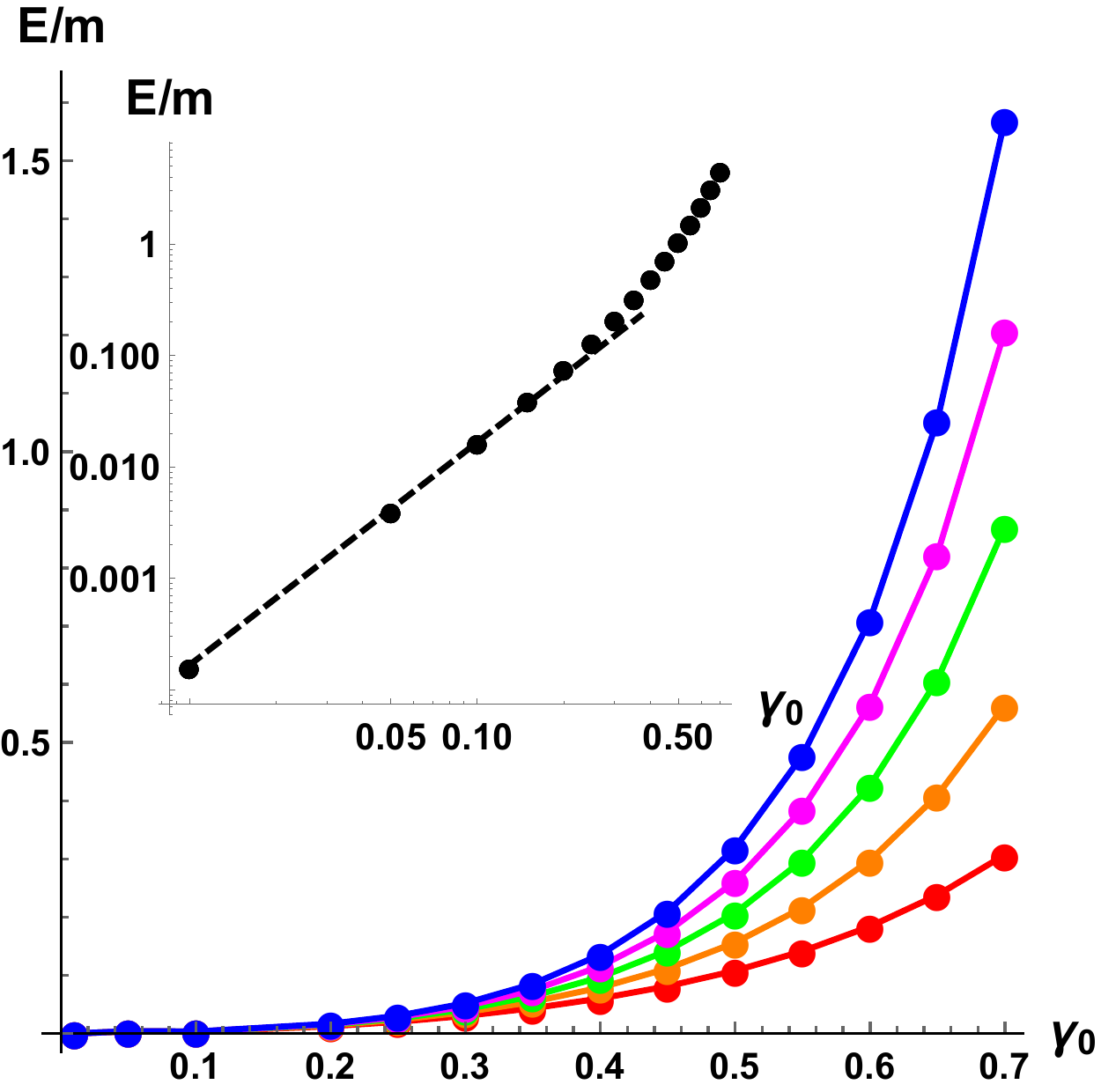}\caption{
\textbf{Right:} The dissipated energy $\mathrm{E}$ over the first $5$ cycles (red to blue). \textbf{Inset:} Average over the first $5$ cycles; low amplitude scaling $\sim \gamma_0^2$.}
    \label{fig:3disse}
\end{figure}

The area of the Lissajous curves measures the energy dissipated in each cycle $\mathcal C$ of the periodic driving, $E\equiv\int_{\mathcal C}\sigma(\gamma)\,\dd\gamma.$ In figure \ref{fig:3disse}, we depict the dissipated energy for the first 5 cycles. At fixed frequency, the energy dissipation grows in response to increasing the amplitude. For small strains, we observe a quadratic increase of the averaged dissipated energy $E\sim\gamma_0^2$.

\chapter{Anomaly induced transport in the presence strong magnetic fields}\label{sec:anotranspo}
In this chapter, we study the influence of chiral anomalies on the transport properties of systems subject to strong magnetic fields. So far, only weak magnetic fields, which are of order 1 in the derivatives, were considered in the literature. We derive the anomalous hydrodynamics for strong external magnetic fields for the first time. We characterize the transport properties of the system in terms of thermodynamic susceptibilities and the so-called shear and bulk viscosities. The magnetic field breaks the rotational invariance down to an $SO(2)$ symmetry and hence the viscosities are different depending on the orientation with respect to the magnetic field. For a magnetic field pointing in the $z$-direction of the field theory, we focus specially on the shear viscosities \mbox{$\eta_\perp\equiv\eta_{xyxy},$} \mbox{$\tilde \eta_\perp\equiv1/2(\eta_{xy xx}-\eta_{xy yy})$},  (perpendicular to the magnetic field, \mbox{helicity-2} sector), \mbox{$\eta_\|\equiv\eta_{xzxz}=\eta_{yzyz}$}, \mbox{$\tilde \eta_\|\equiv\eta_{yz xz} =\eta_{xz yz}$} (parallel to the magnetic field, \mbox{helicity-1} sector) and the bulk viscosities $\zeta_1,\,\zeta_2,\,\eta_1,\,\eta_2$. 
Shear viscosities in strong magnetic field have previously been investigated in~\cite{Hernandez:2017mch,Finazzo:2016mhm,Critelli:2014kra,Huang:2011dc,Hattori:2016cnt,Hattori:2017qih}. This chapter is based on my work in collaboration with Martin Ammon, Juan Hernandez, Matthias Kaminski, Roshan Koirala, Julian Leiber, and Jackson Wu which will appear shortly in~\cite{Ammon:2020rvg}.

\subsubsection{Thermodynamics}
Following \cite{Ammon:2020rvg,Hernandez:2017mch,Banerjee:2012iz,Jensen:2012jh,Kovtun:2016lfw,Kovtun:2018dvd}, we consider the equilibrium generating functional in order to study the anomalous thermodynamic system subject to a strong axial magnetic field \mbox{($F\sim B\sim \mathcal O(1)$)}
\begin{equation}
W_s = \int d^4x \sqrt{-g}\left(p(T,\mu,B^2) + \sum_{n=1}^{5} M_n(T,\mu,B^2)\, s_n + O(\partial^2)\right)\,,\label{eq:thermgenfunc}
\end{equation} 
where we count $T,\mu,u^\mu, B^\mu\sim\mathcal O(1)$.
In this equation, $\mu$ refers to the chemical potential, $B$ to the magnetic field, $u^\mu$ the velocity, $T$ is the temperature, $g$ the external metric, $s_n$ the first order equilibrium scalars\footnote{In particular, we have $ s_1=B^\mu \partial_\mu (\frac{B^2}{T^4}),\, s_2=\epsilon^{\mu\nu\rho\sigma} u_\mu B_\nu \nabla_{\!\rho} B_\sigma,\,s_3=B{\cdot}a,\,s_4= B{\cdot}E,\,s_5=B{\cdot}\Omega$, with $\Omega^\mu=\epsilon^{\mu\nu\alpha\beta}u_\nu\nabla_\alpha u_\beta$ and $a^\mu=u^
\lambda\,\nabla_\lambda u^\mu$. For more details see~\cite{Ammon:2020rvg,Hernandez:2017mch}.} and the associated coefficients $M_n$ are determined by the underlying theory.
$M_5$, for example, is the magneto-vortical susceptibility. In parity preserving theories, $M_5$ is the only coefficient $M_n$ which may be non-zero. However, in the presence of an axial chemical potential (or a chiral anomaly) parity is broken and all other $M_n$ might be non-vanishing and will contribute in the thermodynamic constitutive relations.

By the method of Kadanoff and Martin as explained in section \ref{sec:hydro} and \cite{Hernandez:2017mch}, we are able to derive the thermodynamic transport coefficients in terms of static correlation functions at small momentum. We will focus on $M_2$ and $M_5$ which are given by varying the generating functional \eqref{eq:thermgenfunc} with respect the the metric $g_{\mu\nu}$
\begin{align}\label{eq:thermo_25}
\frac{1}{k_z} {\rm Im}\, G_{T^{xz} T^{yz}}(\omega = 0, k_z \hat{k}) = -2\, B_0^2\, M_2\,,  \quad  \quad
\frac{1}{k_z} {\rm Im}\, G_{T^{tx} T^{yz}}(\omega = 0, k_z \hat{k}) = - B_0 \, M_5,
\end{align}
    where $B_0$ refers to the external axial magnetic field.

\subsubsection{Hydrodynamics}
The next step after considering the equilibrium coefficients is to discuss the non-equilibrium transport properties. 
In the presence of chiral anomalies and a strong axial background magnetic field, the generating functional is no longer gauge invariant. 
The consistent generating functional of a 3+1 dimensional theory is given by 
\begin{equation}
\label{eq:WA}
W_\text{cons}\! = W_s + \int\! d^4 x\sqrt{-g} \left( c_1 T^2\, \Omega{\cdot}A + c_2\, T\left(B{\cdot}A + \mu \,\Omega{\cdot}A\right) + \frac{C}{3}\mu \left( B{\cdot}A+\coeff{1}{2} \mu\, \Omega{\cdot}A \right)\!\right),
\end{equation}
where $W_s$ is the generating functional for the theory without anomalies~\eqref{eq:thermgenfunc}. The consistent generating functional gives rise to the consistent current $J^{\text{cons}}_\mu$ and the energy-momentum tensor $T_{\mu\nu}$. Note that $W_\text{cons}$ is no longer gauge invariant; under a gauge transformation it transforms as
\begin{equation}
\label{eq:WA-gauge}
\delta_\alpha W_{cons} =  \frac{C}{24} \int d^4x \sqrt{-g}\, \alpha \, \epsilon^{\mu\nu\rho\sigma}F_{\mu\nu}F_{\rho\sigma}  \equiv {\cal A}  \,.
\end{equation}
The $c_i$ are transport coefficients. Note that $c_1$ is related to the mixed gauge-gravitational anomaly~\cite{Chen:2012ca,Jensen:2012kj} and will not be present in our holographic model since we do not take higher curvature terms into account. The coefficient $c_2$ breaks the CPT symmetry and is hence not allowed in Lorentz invariant theories~\cite{Jensen:2013vta}.  

The associated conservation laws read
\begin{subequations}
\label{eq:ConsViol}
\begin{align}
& \nabla_\mu T^{\mu\nu}_A = F^{\mu\nu}J^{cons}_\nu -A^\nu \nabla_\mu J^\mu_{cons} \,, &
& \nabla_\mu J_{cons}^\mu = - \frac{C}{24}\epsilon^{\mu\nu\rho\sigma}F_{\mu\nu} F_{\rho\sigma} = \frac{C}{3} \,E{\cdot}B\,,
\end{align}
where we see that the (consistent) axial current is anomalous.
\end{subequations}
In equilibrium, the decomposition of the energy-momentum tensor $T^{\mu\nu}_\text{eq.}$ and current $J^\mu_\text{eq.}$ is given by eq. \eqref{eq:hydrodecompositionequilibrium}, with $\cal{E}=\cal{E}_{\rm eq.}, {\cal P} ={\cal P}_{\rm eq.}, {\cal Q}^\mu={\cal Q}_{\rm eq.}^\mu, {\cal T}^{\mu\nu}  ={\cal T}^{\mu\nu}_{\rm eq.}, {\cal J}^\mu={\cal J}^\mu_{\rm eq.}$, and ${\cal N} = {\cal N}_{\rm eq.}$. We do not give explicit expressions for the equilibrium quantities since they are very complicated expressions in terms of the thermodynamic variables. The deviations from equilibrium are given in terms of the constitutive relations
\label{eq:TTF}
\begin{align}
  {\cal P} &={\cal P}_{\rm eq.} -\zeta_1 \nabla{\cdot}u - \zeta_2 {b^\mu b^\nu}\nabla_{\!\mu} u_\nu + c_3 b\cdot V \,,\\%[5pt]
  {\cal T}^{\mu\nu} & ={\cal T}^{\mu\nu}_{\rm eq.} - \eta_\perp \sigma^{\mu\nu}_\perp
  -\eta_\parallel (b^\mu \Sigma^\nu + b^\nu \Sigma^\mu) 
  - b^{\langle \mu}b^{\nu\rangle} \left(\eta_1 \nabla{\cdot}u + \eta_2 b^\alpha b^\beta \nabla_\alpha u_\beta - c_{14} b\cdot V \right)\nonumber\\%[5pt]
  & \ \ \ \,  
  - \tilde\eta_\perp \tilde\sigma^{\mu\nu}_\perp 
  - \tilde\eta_\parallel (b^\mu \tilde\Sigma^\nu + b^\nu \tilde\Sigma^\mu) + c_{15} (b^{\mu} V_\perp^{\nu} + b^\nu V^\mu) + c_{17}( b^{\mu}\tilde{V}^{\nu} +b^\nu \tilde{V}^\mu) \,,
\end{align}
where $\sigma^{\mu\nu}_\perp \equiv \coeff12 \left(\mathbb{B}^{\mu\alpha} \mathbb{B}^{\nu\beta} + \mathbb{B}^{\nu\alpha} \mathbb{B}^{\mu\beta} - \mathbb{B}^{\mu\nu} \mathbb{B}^{\alpha\beta}\right) \sigma_{\alpha\beta}$ is the part of the shear viscosity transverse to the magnetic field and $\tilde{\sigma}_\perp ^{\mu\nu} = \frac{1}{2} \left( \epsilon^{\mu\lambda\alpha\beta} u_\lambda b_\alpha \sigma_{\perp\beta}^{\ \ \ \nu} + \epsilon^{\nu\lambda\alpha\beta} u_\lambda b_\alpha \sigma_{\perp\beta}^{\ \ \ \mu}\right)$. 
Furthermore, the current is given by
\begin{align*}
  {\cal J}^\mu & = {\cal J}^\mu_{\rm eq.} + \sigma_\perp V^\mu_\perp +  \tilde\sigma_\perp\, \tilde V^\mu + b^\mu (\sigma_{||}b\cdot V + c_4 \nabla \cdot u + c_5 b^\mu b^\nu \nabla_\mu u_\nu) + c_8 \Sigma^\mu + c_{10} \tilde{\Sigma}^\mu\,.
\end{align*}

The two-point correlation functions of the energy-momentum tensor and conserved currents may be obtained by varying the one-point functions in the presence of external sources with respect to those external sources. In particular, we may vary the on-shell expressions $T^{\mu\nu}[A,g]$ and $J^\mu[A,g]$ with respect to $g_{\mu\nu}$ and $A_\mu$ to determine the retarded hydrodynamic correlation functions~\cite{Ammon:2020rvg,Hernandez:2017mch}
\begin{subequations}
\label{eq:corr-funcs}
\begin{align}
  & G^R_{T^{\mu\nu} T^{\alpha\beta}} = \frac{2}{\sqrt{-g}} \frac{\delta}{\delta g_{\alpha\beta}} \left( \sqrt{-g}\, T^{\mu\nu}_\textrm{on-shell}[A,g] \right)\,,
  &&  G^R_{J^\mu T^{\alpha\beta}} = \frac{2}{\sqrt{-g}} \frac{\delta}{\delta g_{\alpha\beta}} \left( \sqrt{-g}\, J^{\mu}_\textrm{on-shell}[A,g] \right)\,,\\[5pt]
  & G^R_{T^{\mu\nu} J^\alpha} = \frac{\delta}{\delta A_{\alpha}}   T^{\mu\nu}_\textrm{on-shell}[A,g]  \,,
  &&  G^R_{J^\mu J^\alpha} = \frac{\delta}{\delta A_{\alpha}}  J^{\mu}_\textrm{on-shell}[A,g] \,,
\end{align}
\end{subequations}
and the source perturbations $\delta g$ and $\delta A$ are set to zero after the variation. In this way, we may directly derive the Kubo formulas for transport coefficients as well as Onsager constraints imposed on the transport coefficients. The retarded Green's functions are constrained due to Onsager relations. Consider an anti-unitary operator $\Theta$, which is in our case time-reversal ($\Theta={\cal T}$) or a combination of parity and time reversal ($\Theta={\cal PT}$); then the retarded Green's function has to satisfy~\cite{Ammon:2020rvg,Hernandez:2017mch,Onsager1,Onsager2} 
\begin{equation}
\label{eq:Ons}
    G^R_{\varphi_a \varphi_b}(\omega,\mathbf{k};\chi) = \eta_{\varphi_a} \eta_{\varphi_b} G^R_{\varphi_b^\dagger \varphi_a^\dagger}(\omega,-\mathbf{k};-\chi)\,,
\end{equation}
where $\eta_{\varphi_i}$ is the $\Theta$ eigenvalue of $\varphi_i$.
In case of $\Theta = {\cal T}$, we have to reverse $B$, i.e. $\chi = B$ while for $\Theta = {\cal PT}$, $\chi = \mu$, we have to reverse the chemical potential, $\chi=\mu$. All two-point functions must satisfy the Onsager constraints which is an important check of our numerical methods.

\section{The holographic setup}
In the following, we determine the hydrostatic and hydrodynamic transport coefficients in a particular \textit{top-down} holographic model. In particular, we consider an Einstein-Maxwell-Chern-Simons theory with negative cosmological constant in five spacetime dimensions, which is a consistent \textit{top-down} truncation of IIb supergravity compactified on the S$^5$. The action reads
\begin{equation}
    S_\text{grav}=\frac{1}{2\kappa^{2}}
\left[\int_{\mathcal{M}}\! d^{5}x\, 
\sqrt{-g}\left(
R+\frac{12}{L^{2}}-\frac{1}{4}F_{mn}F^{mn}
\right)
-\frac{\gamma}{6}\int_{\mathcal{M}} A\wedge F\wedge F\right],\label{eq:EMCS1}
\end{equation}
where we have to set $\gamma=2/\sqrt{3}$ in order to obtain the bosonic part of minimal gauged supergravity in five dimensions~\cite{Buchel:2006gb,Gauntlett:2006ai,Gauntlett:2007ma,Colgain:2014pha,Hartnoll:2016apf}.\footnote{The coupling strength of the anomaly in our holographic model $\gamma$ is related to the hydrodynamic description \eqref{eq:WA} by $\gamma=-C$.} The additional CS term in the bulk corresponds to a chiral anomaly in the dual field theory. 

 Coupling the axial gauge field to gravity has dramatic consequences for the four dimensional energy momentum tensor; its trace is anomalous since it is proportional to the field strength squared. To cure the resulting divergences caused by the trace anomaly of the energy momentum tensor $T_\mu^\mu\sim F^2$, we have to include a logarithmic term to the boundary terms. This term is not diffeomorphism invariant since it explicitly depends on the radial coordinate. In addition to the logarithmic term, we have to supplement the action by the usual Gibbons-Hawking term in order to have a well defined variational principle~\footnote{Throughout this chapter, we will refer to the radial direction as $\varrho$ and to the induced metric as $h$. For simplicity, we set $L=1$ and $2\,\kappa^2=16\pi\,G_5=1$.}
\begin{equation}\label{eq:actionSbdy}
S_\text{bdy}= \frac{1}{\kappa^2} \int_{\partial\mathcal{M}}\! d^4x \, \sqrt{-h} \left( K - \frac{3}{L} + \frac{L}{4} R(h) + \frac{L}{8} \ln\left( \frac{\varrho}{L} \right) F_{\mu\nu} F^{\mu\nu} \right) \, .
\end{equation}
The equations of motion associated with the action \eqref{eq:actionSbdy} read
\begin{equation}\label{eq:EOM1}
R_{mn}=-4\,g_{mn}+\frac{1}{2}\left(F_{mo}F_{n}{}^{o}-\frac{1}{6}g_{mn}F_{op}F^{op}\right),\quad\quad\quad
d\star F+\frac{\gamma}{2}F\wedge F=0.
\end{equation}
In addition to the equation of motion \eqref{eq:EOM1}, the field strength also has to satisfy the Bianchi
identity $\dd F=0$. In the following, we want to consider the simplest possible static ansatz for the a CFT in the presence of a homogeneous external magnetic field pointing in the $z$-direction~(note that this is the 3-direction in the field theory) and a finite charge density $\rho$. For our holographic setup \eqref{eq:EMCS1}, we require translational invariance in $(t,x,y,z)$ and rotational invariance in the $x-y$ plane. The most general ansatz respecting these symmetries was worked out by D'Hoker and Kraus~\cite{DHoker:2009ixq}
\begin{eqnarray}
\label{eq:ansatzF} 
F  &=&E(\varrho)\,\dd  t\wedge \dd\varrho +  B\,\dd x\wedge \dd y+ \mathcal P(\varrho)\,  \dd z\wedge\dd \varrho
\end{eqnarray}
for the field strength and
\begin{eqnarray}\label{eq:ansatzmetric}
d s^{2}  &=&  \frac{1}{\varrho^{2}}\left[\left(-u(\varrho)+c(\varrho)^2\,w(\varrho)^2\right)\, \dd t^{2} -2 \,\dd t \, \dd \varrho + 2\, c(\varrho)\,w(\varrho)^2 \, \dd z\, \dd t \right. \nonumber \\
 &&\left. + v(\varrho)^{2}\, \left( \dd x^{2} +\dd y^{2}\right)  + w(\varrho)^{2}\,\dd z^{2}\right] 
\end{eqnarray}
for the metric in Eddington-Finkelstein coordinates. The temporal of the field strength \eqref{eq:ansatzF}, corresponds to a radial gauge field. In addition, the radial derivative of the $\varrho z$-component $\mathcal P'(\varrho)$ corresponds to a current induced by the magnetic fields and the chiral anomaly. Furthermore, the Bianchi identities require that the magnetic field $B$ is constant.
The field strength eq. \eqref{eq:ansatzF} may be induced by a gauge field of the form
\begin{equation}
A= A_t(\varrho)\, \dd t + \frac{B}{2} \left(- y \, \dd x + x \, \dd y \right) +P(\varrho)\, \dd z\,,\label{eq:gaugefield}
\end{equation}
where $ P'(\varrho)=-\mathcal P(\varrho)$ and $A'_t(\varrho)=-E(\varrho)$. By fixing the radial shifts, we set the event horizon to $\varrho=1$, where $u(\varrho)=0=c(\varrho)$ which fixes all residual symmetries.\footnote{There is an additional shift symmetry discussed in \cite{DHoker:2009ixq} which we may fix by setting $c(1)=0$.}

The conformal boundary is located at $\varrho=0$ in our coordinates. At the conformal boundary, the metric functions asymptote to AdS which fixes the leading coefficients in the asymptotic expansion. The subleading powers are affected by the trace anomaly of the energy momentum tensor which induces logarithmic divergences scaling with the squared of the magnetic field
\begin{eqnarray}\label{eq:boundaryExpansionBackground}
&& u(\varrho) = 1 + \varrho^4\left[  u_4 +{\cal O}(\varrho^2) \right]+ \varrho^4\ln(\varrho)\left[ \frac{B^2}{6}+ {\cal O}(\varrho^2) \right]  \nonumber \, ,\\
&& v(\varrho) = 1 + \varrho^4\left[- \frac{w_4}{2} +{\cal O}(\varrho^2)\right] + \varrho^4\ln(\varrho)\left[- \frac{B^2}{24}+ {\cal O}(\varrho^2) \right] \, , \nonumber\\
&& w(\varrho) = 1 + \varrho^4\left[  w_4 +{\cal O}(\varrho^2)\right] + \varrho^4\ln(\varrho)\left[ \frac{B^2}{12}+ {\cal O}(\varrho^2) \right] \, , \nonumber\\
&& c(\varrho) =\varrho^4 \left[ c_4 + {\cal O}(\varrho^2) \right] + \varrho^8\ln(\varrho) \left[ - \frac{B^2}{12}  {c_4} + {\cal O}(\varrho^2)\right]  \, ,\nonumber  \\	
&&A_t(\varrho)  =  \mu -  \frac{\rho}{2} \varrho^2  -  \frac{\gamma B  p_1}{8} \varrho^4  +  {\cal O}(\varrho^6)\, ,\nonumber\\
&&P(\varrho) = \varrho^2\left( \frac{p_1}{2}  + \frac{\gamma B  \rho}{8}  \varrho^2 + {\cal O}(\varrho^4 ) \right) \,,
\end{eqnarray}
where $u_4,\, w_4,\, c_4, \, \rho, \, p_1$ are the undetermined coefficients in the asymptotic expansions.

\subsubsection{Thermodynamics of charged magnetic branes}
Charged magnetic branes were previously discussed in~\cite{DHoker:2009ixq,DHoker:2009mmn,Ammon:2016szz,Ammon:2017ded}. In this section, we explain the fundamental thermodynamic quantities which are important for characterizing the transport behavior. 

Without matter content, the one-point function of the energy-momentum tensor is given by eq. \eqref{EMtensor} and \eqref{EMcounter}. It is straightforward to include the matter contributions by varying the matter action with respect to the induced metric according to~\cite{Balasubramanian:1999re}
\begin{equation}
\left\langle T_{\mu\nu} \right\rangle=\lim\limits_{\varrho\rightarrow 0}\frac{1}{\varrho^{2}}\left(-2K_{\mu\nu}+2(K-3) \, h_{\mu\nu}+\ln(\varrho)\left(F_{\mu}^{\ \alpha}F_{\nu\alpha}
-\frac{1}{4} \, h_{\mu\nu}F^{\alpha\beta}F_{\alpha\beta}\right)\right).\label{eq:onepointback}
\end{equation}
Similarly, the covariant current follows from varying the action with respect to the gauge field~\cite{DHoker:2009ixq}
\begin{equation}\label{eq:one-current}
\left\langle J_{\text{cov}}^\mu  \right\rangle= \lim\limits_{\varrho\rightarrow 0}\frac{1}{\varrho^{3}}
h^{\mu\alpha}\partial_{\varrho}A_{\alpha}
\,.
\end{equation}
In our specific ansatz, the one-point function of the energy-momentum tensor \eqref{eq:onepointback} reads
\begin{equation}
\label{eq:EnergyMomTensor}
\langle T^{\mu\nu}\rangle =\left( \begin{array}{cccc}
-3 \, u_4 & 0 & 0 &  
-4 \, c_4 \\
0 & -\frac{B^{2}}{4}-u_4-4\,w_4  & 0 & 0\\
0 & 0 & -\frac{B^{2}}{4}-u_4-4\,w_4  & 0\\
-4 \, c_4  & 0 & 0 & 8 \, w_4 - u_4 \end{array} \right) \, .
\end{equation} 
Similarly, the one-point function of the covariant current reads
$\langle J^\mu \rangle = \left(\rho,\,0,\,0,\,p_1\right).$ By taking the trace of the energy-momentum tensor one-point function \eqref{eq:EnergyMomTensor}, the trace anomaly of the energy-momentum tensor is clearly evident $\langle T_\mu{}^\mu \rangle = -B^2/2$. The entropy density $s$ is encoded in the horizon values of the metric functions $v(\varrho)$ and $w(\varrho)$ reads evaluated at $\varrho=1,$
 \begin{equation}
     s=4\pi \, v(1)^2 \, w(1).\label{eq:entropyTTT}
 \end{equation}
The pressure $p$ is defined in terms of the variables and their respective thermodynamic conjugate 
\begin{equation}
    p = s\, T - \epsilon + \mu \langle J^t \rangle = u'(1)\, v(1)^2 \, w(1) + 3 u_4 + \mu \rho
\end{equation} 
where we introduced the energy density $\epsilon=\langle T^{tt}\rangle$.
Finally, the enthalpy $w$ is given by $w = \epsilon + p = s T + \mu \langle J^t \rangle.$

\subsection*{Transport properties from holography}
In order to probe the transport properties of the system within linear response theory, we have to consider fluctuations of the metric and gauge field ($\tilde h_{nm},\tilde a_m$) about the numerically constructed equilibrium state ($\bar g_{nm},\bar A_m$). Expanding the fields to first order for $\varepsilon\ll 1$, the fields read
\begin{equation}\label{eq:ansatzqnm}
g_{mn} = \bar{g}_{mn} + \varepsilon \,  \tilde{h}_{mn} \,, \qquad A_{m} = \bar{A}_{m} + \varepsilon \,\tilde{a}_m \,,
\end{equation}
where the zeroth order $(\bar g_{nm},\bar A_m)$ corresponds to the equilibrium state and the first order ($\tilde h_{nm},\tilde a_m$) to the linear fluctuations 
\begin{equation}
    \tilde{h}_{mn}(x, \varrho) = \int \frac{\dd^4k}{(2\pi)^4} \, e^{i k_\mu x^\mu} \, h_{mn}(k,\varrho) \, , \qquad     \tilde{a}_{m}(x, \varrho) = \int \frac{\dd^4k}{(2\pi)^4}
    \, e^{i k_\mu x^\mu} \, a_{m}(k,\varrho) \, , 
\end{equation}
with $k_\mu x^\mu = - \omega t + \bm{k} \cdot \bm{x}.$ In the following, we align the momentum with the direction of the magnetic field, i.e. $\bm{k}=(0,0,k)$. By classifying the fluctuations with respect to the momentum pointing in the $z-$direction, we find that the fluctuations decouple in three different sectors -- helicity-2, helicity-1, and helicity-0:
\begin{center}
\begin{tabular}{c|c}
Helicity \hspace{1mm} & Fluctuation modes \\ \hline
2\, &$ \, h_{xy},\,h_{xx}-h_{yy}  $\\
1\, &$ \, h_{tx},\,h_{xz},\,a_{x},\,h_{\varrho x}  $\\
    &$ \,h_{ty},\,h_{yz},\,a_{y},\,h_{\varrho y} $\\
0\, &$ \, h_{tt},\,h_{tz},\,h_{zz},\,h_{xx}+h_{yy},\,h_{\varrho t},\,h_{\varrho z},\,h_{\varrho \varrho},\,a_t,\,a_z,\,a_\varrho $
\end{tabular}
\end{center}
In particular, we may treat the three sectors separately since the equations of motions of different helicities are decoupled. First, we have to fix the gauge. We implement radial gauge for the gauge field $a_\varrho=0$ and for the metric fluctuations, we choose a gauge where $h_{m\varrho}=0, m\not= t$ and $h_{t\varrho}=1/2\,h_{tt}$. Fixing the gauge in this way leads to constraint equations consisting of the equations of motions of the modes we set to zero. In terms of helicity, we will find constraint equations from the following modes:
\begin{center}
\begin{tabular}{c|c}
Helicity \hspace{1mm} & \hspace{1mm} Constraint modes\\
\hline
2\,& \,  none\\
1\,& \, $h_{\varrho x},\,h_{\varrho y}$ \\
0\,& \, $h_{\varrho t},h_{\varrho z},\,h_{\varrho\varrho},a_{\varrho}$
\end{tabular}
\end{center}

In order to calculate the expectation values of the fluctuations and extract the one-point function of the energy momentum tensor and the current, respectively, we have to perform the holographic renormalization procedure for the fluctuations. By evaluating eq. \eqref{eq:onepointback} and eq. \eqref{eq:one-current} for the metric fluctuations, we obtain for the one-point functions
\begin{equation}
\left\langle T_{\mu\nu} \right\rangle = \frac{1}{6} \frac{\partial^4}{\partial \varrho^4} h_{\mu\nu}(\varrho)\Big|_{\varrho=0},\quad\text{ and }\quad
\left\langle J_{\mu} \right\rangle = \frac{\partial^2}{\partial \varrho^2} a_{\mu}(\varrho)\Big|_{\varrho=0}.
\end{equation}
Equipped with this, we are now able to compute the two-point functions in our setup corresponding to the thermodynamic and hydrodynamic transport coefficients. 

Throughout this chapter, we will normalize all quantities by means of the temperature, i.e. the respective dimensionless quantities in our holographic setup are
\begin{equation}
    \tilde B\equiv B/T^2, \ \tilde \mu\equiv \mu/T,\ k/T,\ \omega/T.
\end{equation}
We solved all appearing equations of motions by means of so-called pseudo-spectral methods (see \ref{ref:appnumrel} and \ref{app:sectionconv}).
\section{Results}\label{sec:2pointsep}
To compute the Kubo formulas, for example \eqref{eq:thermo_25}, we have to compute two-point functions of the form 
\begin{equation}
  \coeff{1}{k_z}{\rm Im}\, G_{T^{ab} T^{cd}}(\omega = 0, k_z \hat{k})\quad \text{or}\quad  \coeff{1}{\omega}{\rm Im}\, G_{ T^{ab}T^{cd}}(\omega,{\bf k}{=}0)
\end{equation}
and similarly $JT$ and $TJ$ correlators. Note that we implicitly assume the $\omega$ and $k_z$ in the pre-factors to be small. Given we have a solution for the retarded Green's function, there are two ways to extract the coefficient linear in $\omega$ and $k_z$, respectively. On the one hand, we can extract the coefficient by a linear fit in $\omega$ and $k^z$. On the other hand, we can recast the problem by introducing the auxiliary metric functions
\begin{equation}
h_{mn}(\varrho,\omega)= h_{mn}^{(0)}(\varrho) + \omega\, h_{mn}^{(1)}(\varrho)\quad\text{and}\quad h_{mn}(\varrho,\bm{k})= h_{mn}^{(0)}(\varrho) + k_z \, h_{mn}^{(1)}(\varrho),
\end{equation} 
and the same combinations for the gauge fluctuations. By expanding the functions to linear order we may solve the coupled equations of motion to order for order in $\omega$ or $k$. In the two-point functions of the form $\coeff{1}{\omega}{\rm Im}\, G_{T^{ab}T^{cd}}(\omega,{\bf k}{=}0)$, for example, we have to source the fluctuation $h_{cd}^{(0)}$ and read of the vacuum expectation value of the fluctuation $h_{ab}^{(1)}$ in order to get the order linear in $\omega$.

\begin{figure}[h]
    \centering
    \includegraphics[width=6cm]{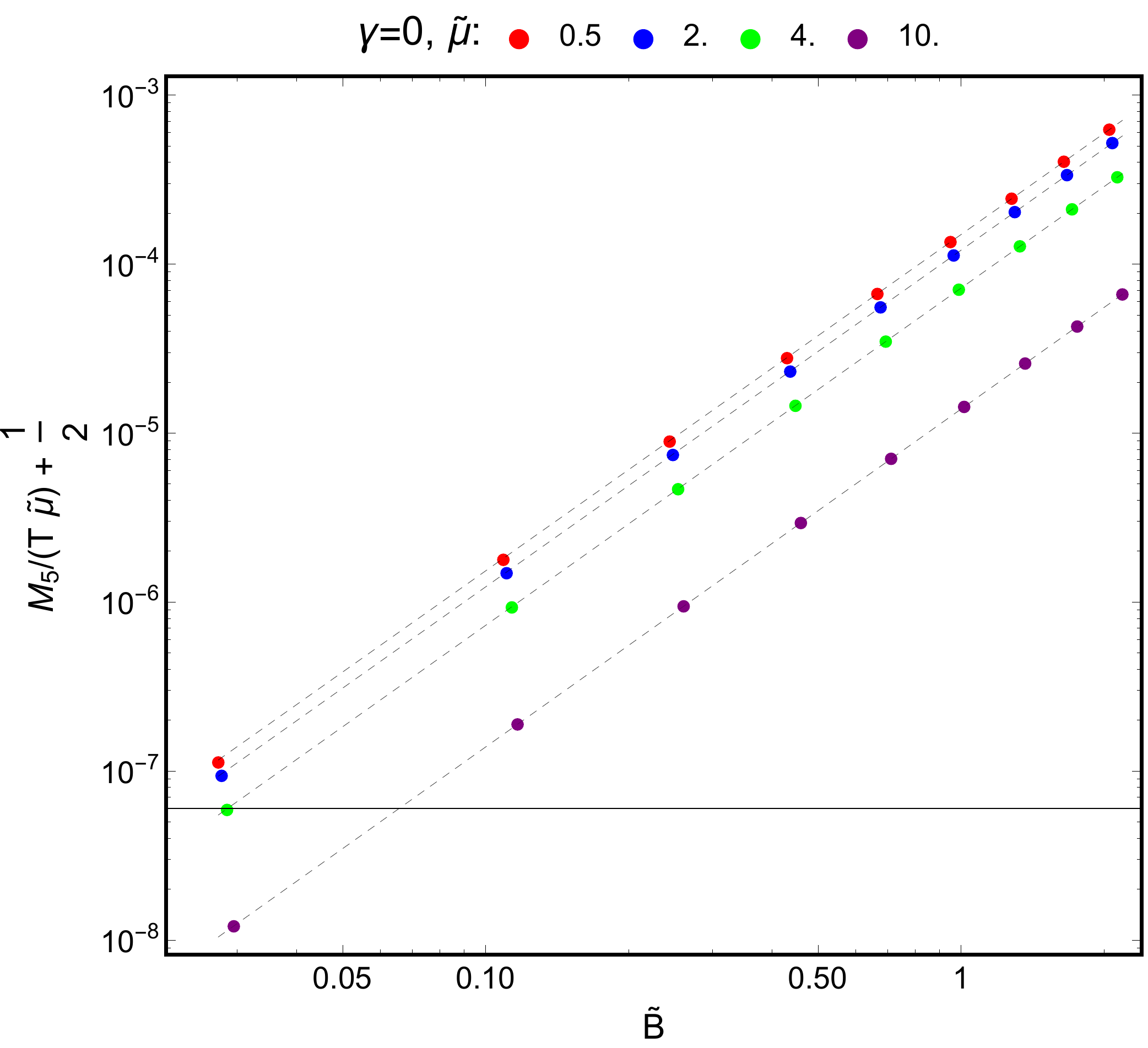}
    \caption{Double logarithmic plot of the dimensionless magneto-vortical susceptibility $M_5/T$ divided by the dimensionless chemical potential $\tilde \mu$ in dependence of $\tilde B$. The dashed lines are quadratic fits.}
    \label{fig:my_label5}
\end{figure}
\subsection{Thermodynamics}
In this section, we apply the just outlined procedure to compute two-point functions and study the thermodynamic properties of our system subject to a strong axial magnetic field in presence (and absence) of chiral anomalies by means of our holographic model. In particular, we will focus on the impact of the chiral anomaly on the transport properties of the system.
\begin{figure}[!t]
    \centering
    \includegraphics[width=6.2cm]{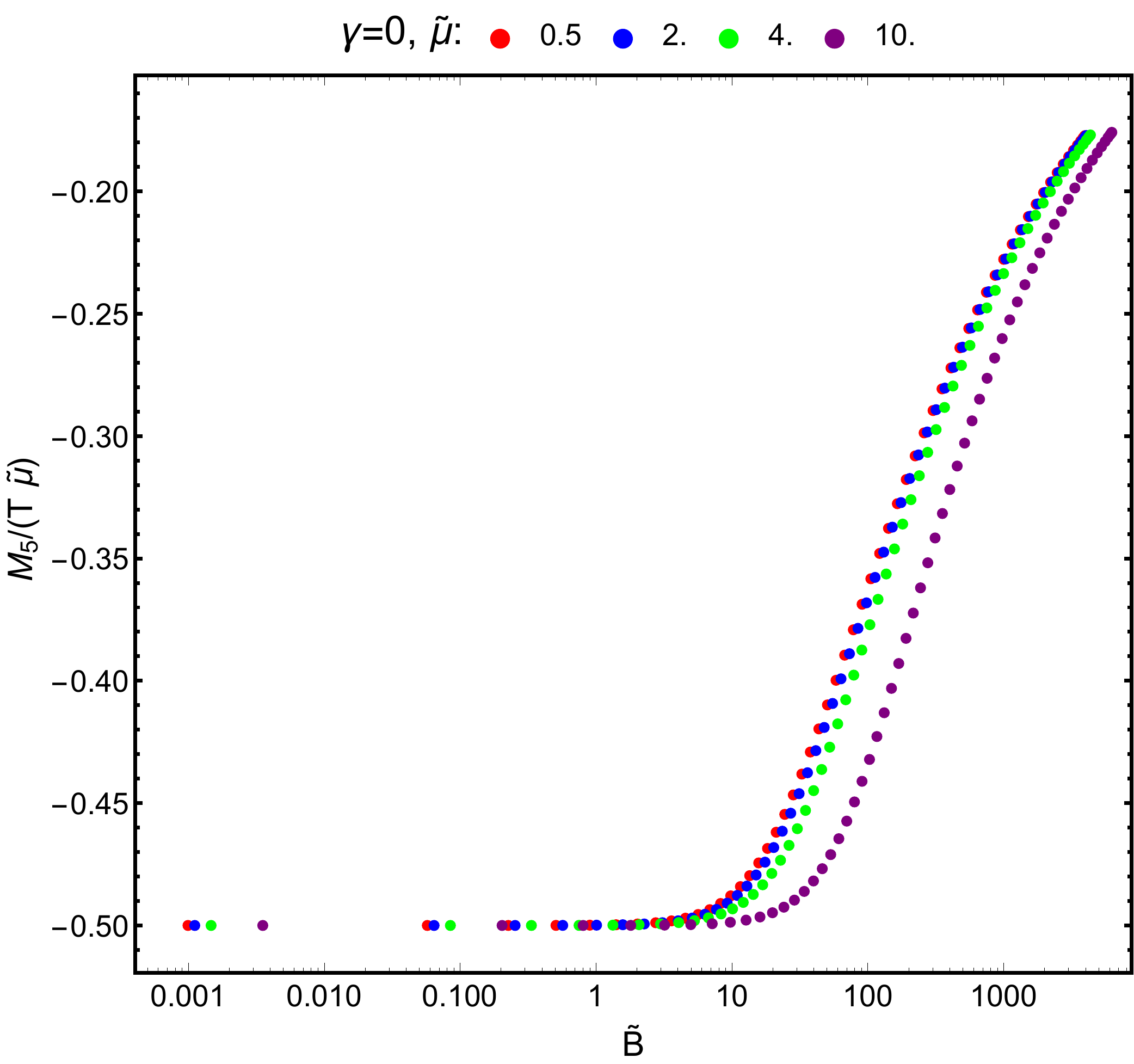}  \hspace{0.5cm} \includegraphics[width=6.2cm]{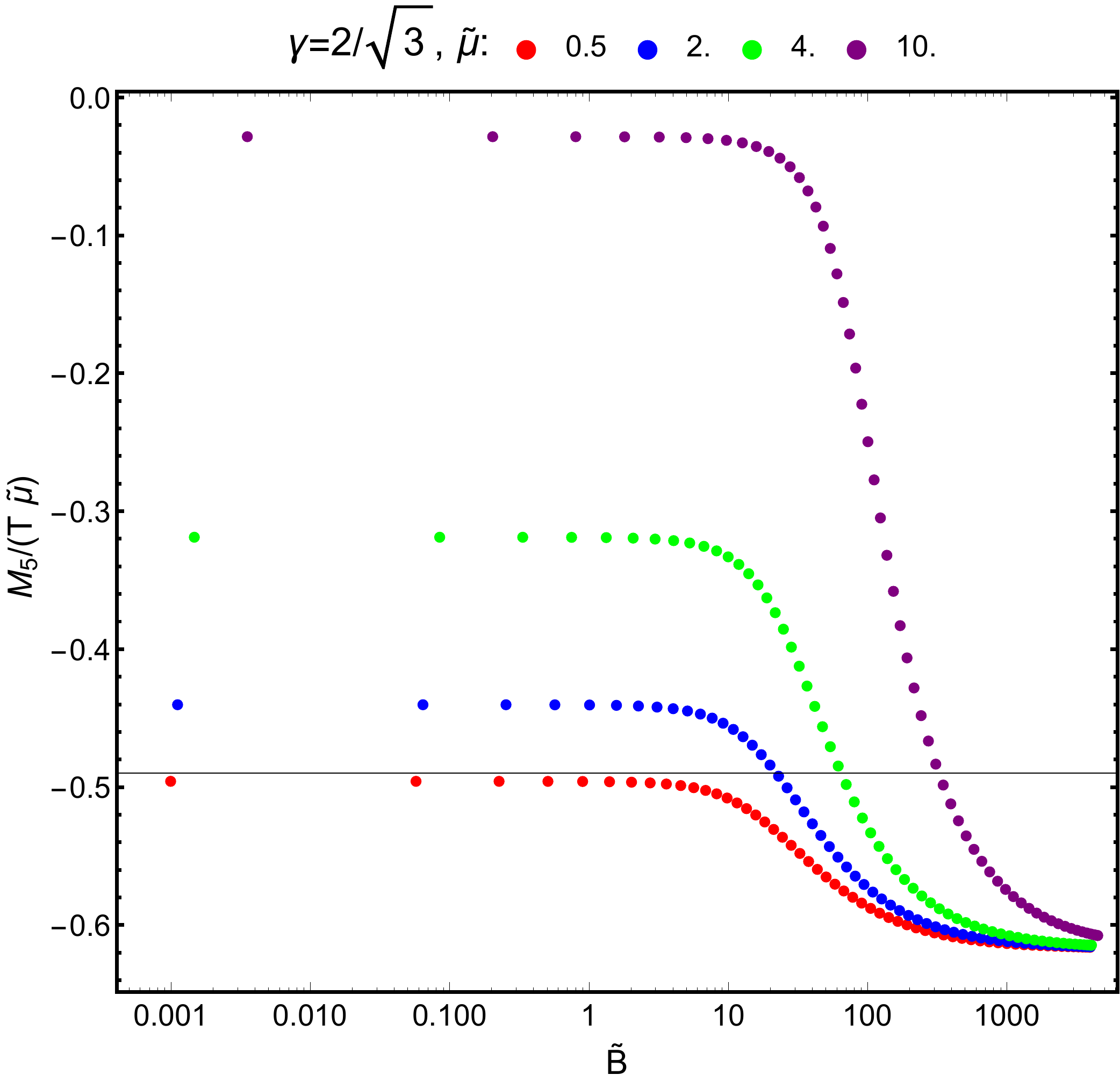}
    \caption{Logarithmic plot of the dimensionless magneto-vortical susceptibility $M_5/T$ divided by the dimensionless chemical potential $\tilde \mu$ in dependence of $\tilde B$ without (left) and in presence (right) of the chiral anomaly for various $\tilde\mu$.}
    \label{fig:my_label5_gamma}
\end{figure}
 
 We start with the magneto-vortical susceptibility, given by the right equation in eq. \eqref{eq:thermo_25}. The authors of~\cite{Bu:2019qmd} derived the magneto-vortical susceptibility for a non-anomalous system in the limit $B\to 0$ and found $M_5=-1/2\, \mu$. We may compute the magneto-vortical susceptibility according to eq. \eqref{eq:thermo_25}. In order to make $M_5$ dimensionless, we have to multiply it by $T$. In figure \ref{fig:my_label5}, we present the the magneto-vortical susceptibility extracted by the novel method developed in section \ref{sec:2pointsep}. We see that the curves for all considered $\tilde \mu$ all correctly tend to $M_5/T=-\tilde \mu/2$ in the $\tilde B\to 0 $ limit. Furthermore, for increasing $\tilde B$ the curves all grow quadratic in $\tilde B$ for small enough $\tilde B$.

In the right side of figure \ref{fig:my_label5_gamma}, we consider the same scenario in presence of the chiral anomaly, i.e. $\gamma=2/\sqrt{3}$. Most notably, for increasing chemical potential $\tilde \mu$ the \mbox{} increasingly deviates from the value observed for $\gamma =0$ and the ratio $M_5/(T\,\tilde \mu)$ is approaching 0. Without anomaly, the absolute value of $M_5/(T\,\tilde \mu)$ decreases for increasing $\tilde B$ but it behaves the opposite for the theory with chiral anomaly; after a plateau in the small $\tilde B$ regime, the absolute value of $M_5/(T\,\tilde \mu)$ increases significantly in the large $\tilde B$ regime.

\begin{figure}[h]
    \centering
    \includegraphics[width=6.2cm]{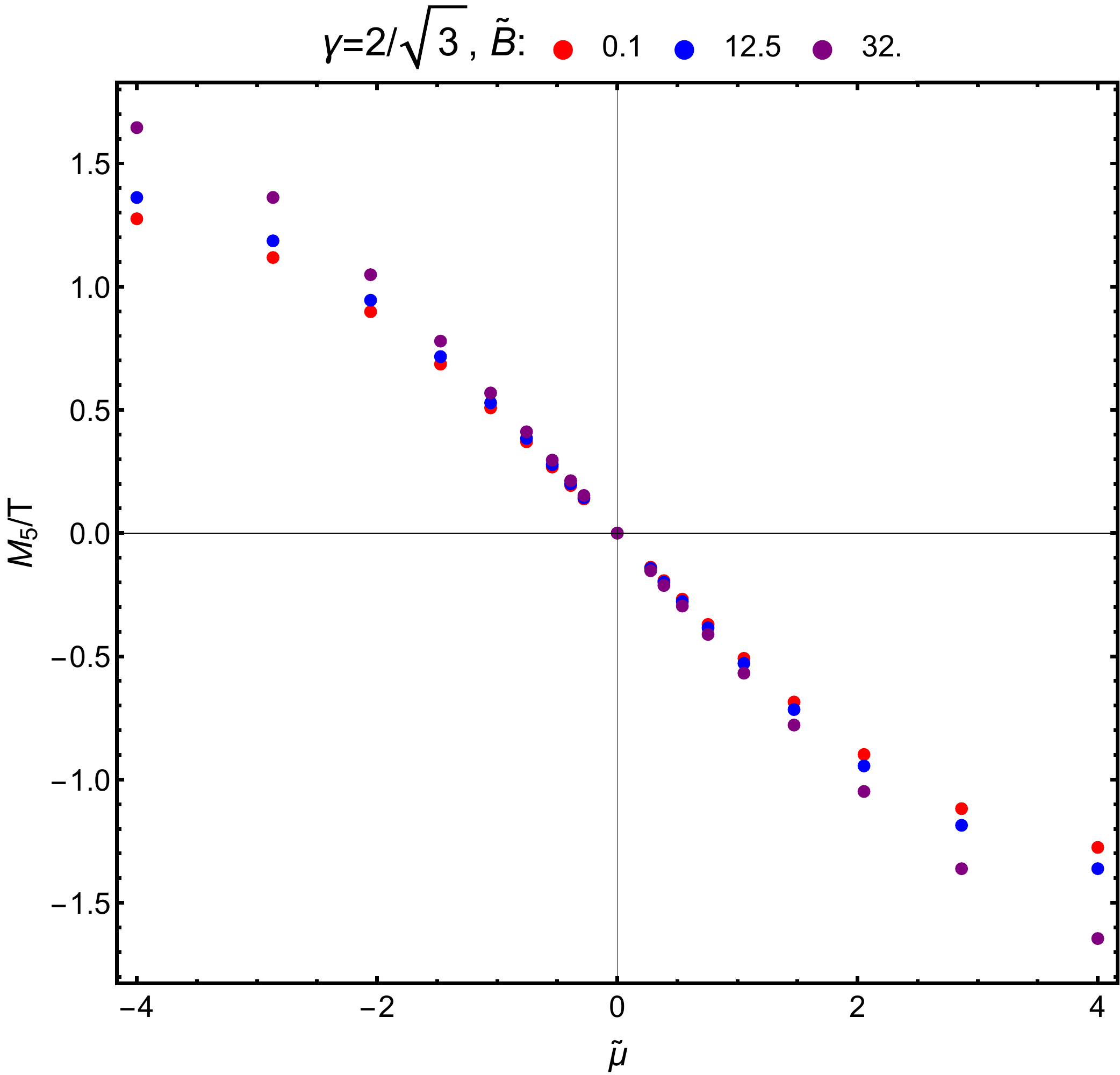}
    \hspace{0.5cm} \includegraphics[width=6.2cm]{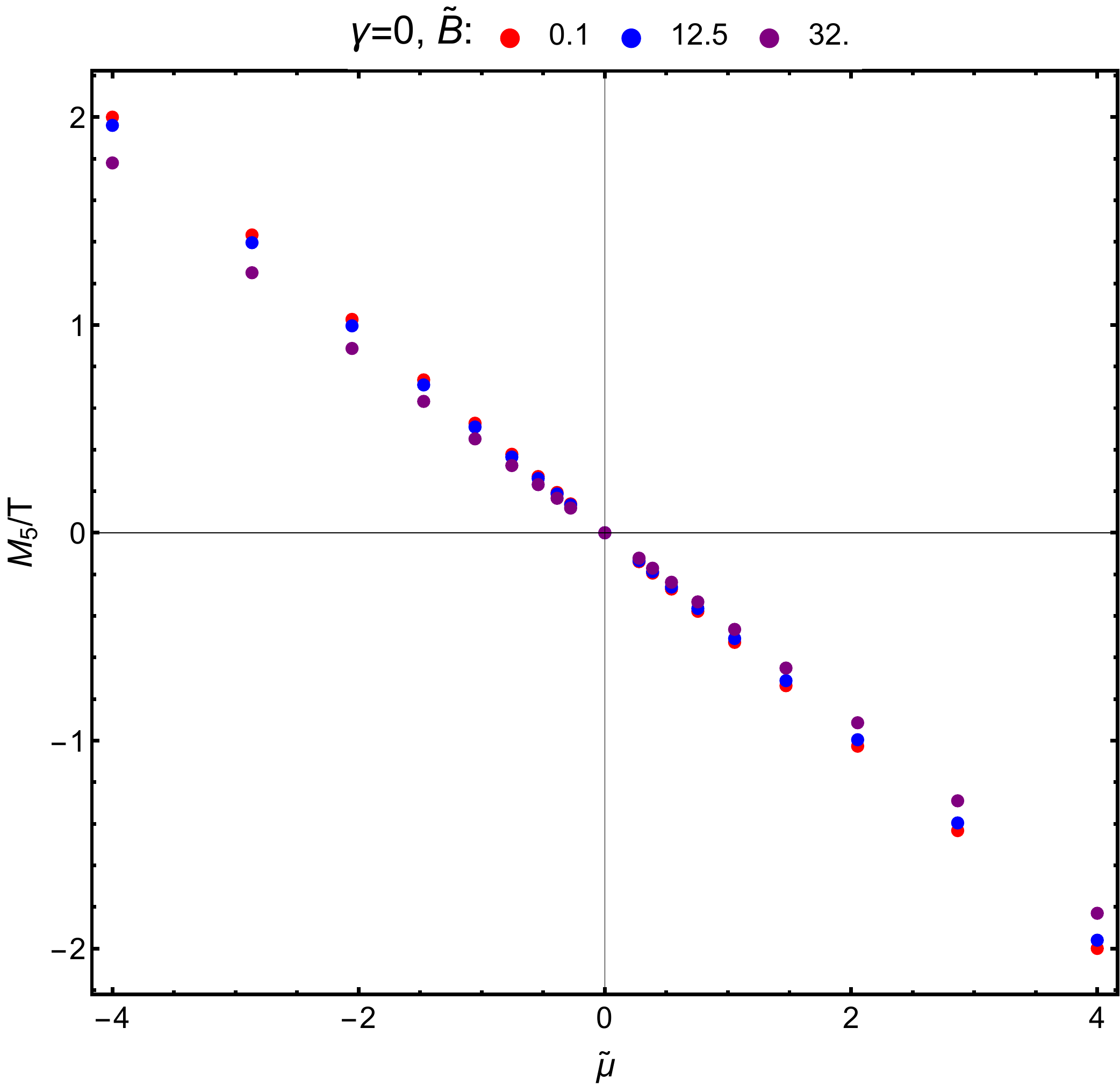}
    \caption{Dependence of the dimensionless magneto-vortical susceptibility $M_5/T$ on $\tilde\mu$ for fixed $\tilde B$. We show $\tilde B=\{0.1,12.5,32\}$ (red, blue purple) in presence of the chiral anomaly (left) and without anomaly (right).}
    \label{fig:my_label5_3_Bval}
\end{figure}

We quantified the dependence on the chemical potential $\tilde \mu$ in figure \ref{fig:my_label5_3_Bval} for three different values of the magnetic field $\tilde B$. In the anomalous case with $\gamma=2/\sqrt{3}$, the curve clearly deviates from a straight line, specially for the smallest considered value of the magnetic field $\tilde B=0.1$. The $\gamma=0$ case on the right side, however, is clearly linear in $\tilde\mu$ for $\tilde B=0.1$. In both cases we notice that $M_5$ is odd under parity transformations, i.e. changes sign for $\tilde\mu\to-\tilde\mu$.

\begin{figure}[h]
    \centering
    \includegraphics[width=6.2cm]{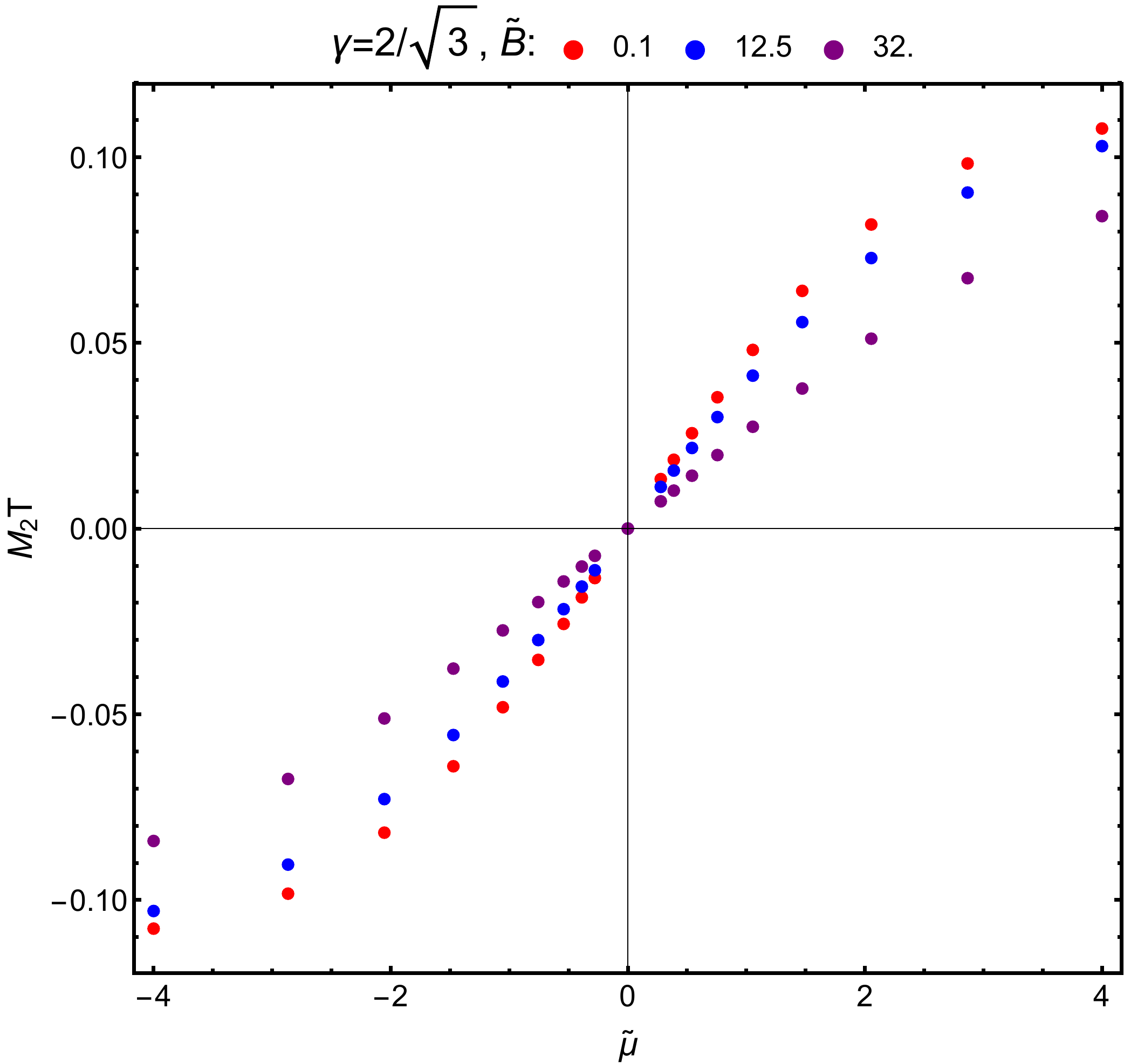}
    \hspace{0.5cm} \includegraphics[width=6.2cm]{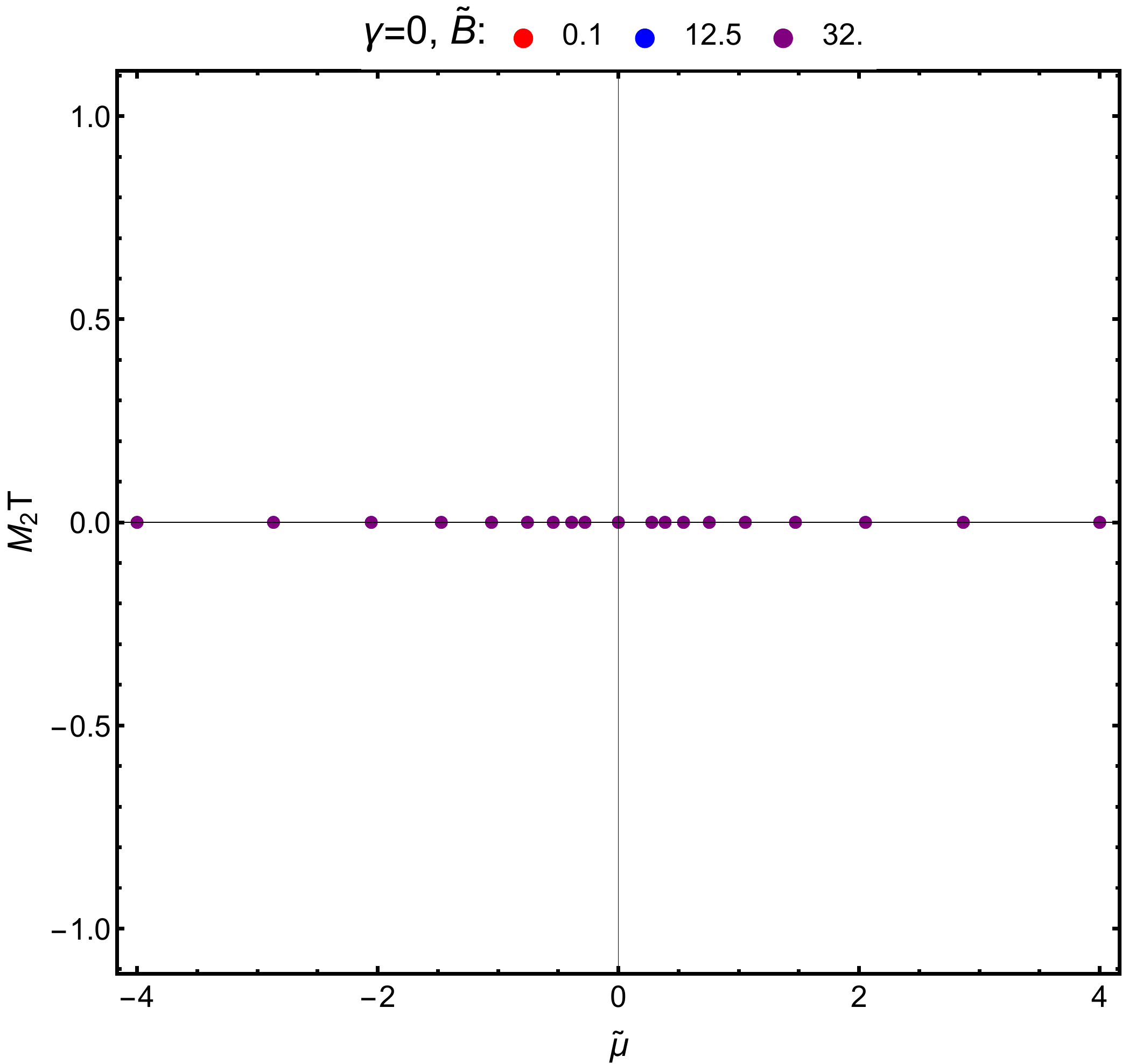}
    \caption{Dependence of the dimensionless thermodynamic susceptibility $M_2/T$ on $\tilde\mu$ for fixed $\tilde B$. We show $\tilde B=\{0.1,12.5,32\}$ (red, blue purple) in presence of the chiral anomaly (left) and without anomaly (right).}
    \label{fig:my_label5_M2}
\end{figure}

The imprint of the chiral anomaly is even more clear in terms of $M_2$. Similar to $M_5$, we have to multiply the thermodynamic susceptibility $M_2$ by $T$ to render it dimensionless. In figure \ref{fig:my_label5_M2}, we see that the thermodynamic susceptibility $M_2$ is only non-zero in presence of the anomaly (and non-vanishing chemical potential) and we show for the first time in the literature that this coefficient is present and indeed contributes in an anomalous QFT. For increasing the value of magnetic field $\tilde B$, the absolute value $M_2\,T$ decreases in the range of the considered magnetic fields and chemical potentials. In a similar fashion as we already observed for the magneto-vortical susceptibility, $M_2\, T$ is odd under parity transformations $\tilde \mu\to -\tilde\mu$. We furthermore numerically verified that $M_1=M_3=M_4=0$ within our holographic model.

\subsection{Hydrodynamic transport coefficients in the helicity-two sector}\label{sec:hydrotrans}
After discussing the thermodynamical susceptibilities, we will now examine the hydrostatic transport properties of the system. We start with the helicity-two sector, where we have analytical control of the transport coefficients.
The helicity-two sector contains only two decoupled fields, $h_{xy}$ and $h_{xx}-h_{yy}$, both formally satisfying the same differential equation. In particular, the transport coefficient of interest in this sector is the so-called transverse shear viscosities $\eta_\perp$ and $\tilde\eta_\perp$ given by
\begin{align}
  & \coeff{1}{\omega}{\rm Im}\, G_{T^{xy} T^{xy}}(\omega,{\bf k}{=}0) =  \eta_\perp \,,\label{eq:s2:1}&  & \coeff{1}{\omega}{\rm Im}\, G_{T^{xy} O_3}(\omega,{\bf k}{=}0) =  \tilde\eta_\perp\, {\rm sign}(B_0) \,,
\end{align}
where we introduced $O_3=\frac12 \,\left(T^{xx}-T^{yy}\right)$. Since both two-point functions are evaluated at $\bm{k}=0$, we can set it to zero throughout this calculation.
Both transport coefficients in eqs. \eqref{eq:s2:1} correspond to the coefficient linear in $\omega$. We can extract this coefficient easily by introducing auxiliary fields in terms of a $\omega$ expansion
\begin{equation}
    h^x{}_y{} (z) = h_0(z) + \omega h_1(z) + \mathcal{O}(\omega^2) \, .\label{eq:wsplits2}
\end{equation}
The split in powers of $\omega$ (and also $\bm{k}$) is an important concept on which we will rely heavily when we compute the transport coefficients in the helicity-1 and helicity-0 sector. 
In the new variables, we may solve the equation of motion analytically and read off the expectation value in presence of the source $h_{xy}(0)=1$ according to eq. \eqref{eq:one-current}
\begin{equation}
\frac{\eta_\perp}{s}\! =\! \frac{1}{s\omega}{\rm Im}\, G_{T^{xy} T^{xy}}(\omega,{\bf k}{=}0) = \frac{1}{6 \,s}\left. \frac{\partial^4}{\partial \varrho^4} h^x_{\ y}(\varrho)\right|_{ \varrho=0}\!\!\!\!\!= \frac{1}{6\, s} \left. \frac{\partial^4}{\partial \varrho^4}\, \mbox{Im} \, h_1(\varrho)\right|_{ \varrho=0}\!\!\!\!
= \frac{1}{4\pi },
\end{equation}
where we used eq. \eqref{eq:entropyTTT} for the entropy. The second transport coefficient $\tilde\eta_\perp$ given by the two-point function in \eqref{eq:s2:1} is zero since the differential equations are decoupled and the metric fluctuation $h_{xy}$ is not influenced by a source term for $O_3$.

\subsection{Hydrodynamic transport coefficients in the Helicity-one sector}
In this section, we focus on the effects of the chiral anomaly on the helicity-1 transport coefficients. Notably, we discuss the parallel shear viscosities $\eta_\|,\,\tilde\eta_\|$ and the perpendicular resistivity $\rho_\perp$ as well as the perpendicular Hall resistivity $\tilde\rho_\perp$. 

\subsubsection*{Kubo formulas in the presence of anomalies in the helicity-one sector}
In \cite{Hernandez:2017mch}, the authors classified the transport coefficients in strong magnetic fields without chiral anomaly. In contrast to~\cite{Hernandez:2017mch}, we \cite{Ammon:2020rvg} find that in presence of the chiral anomaly the parallel shear viscosities acquire extra contributions in terms of novel transport coefficients, for example $c_8, c_{15},\tilde c_{10},\tilde c_{17}$
\begin{align}\label{eq:KuboTT1an1}
     &\coeff{1}{\omega}{\rm Im}\, G_{T^{xz} T^{xz}}(\omega,{\bf k}{=}0) =  \eta_\parallel +  (c_8 c_{15} - \tilde{c}_{10} \tilde{c}_{17}) \rho_\perp  - ( c_8 \tilde{c}_{17} + \tilde{c}_{10} c_{15})  \tilde{\rho}_\perp \,, \\ 
   &\coeff{1}{\omega}{\rm Im}\, G_{T^{yz} T^{xz}}(\omega,{\bf k}{=}0) = \left( \tilde\eta_\parallel  + ( c_8 \tilde{c}_{17} + \tilde{c}_{10} c_{15}) \rho_\perp +  ( c_8 c_{15} - \tilde{c}_{10} \tilde{c}_{17} ) \tilde{\rho}_\perp \right) {\rm sign}(B_0)\,, \label{eq:KuboTT1an2}  
\end{align}
where $\tilde{c}_{10} = c_{10} -\xi_{TB},\,\tilde{c}_{17} = \bar{c}_{17} +\xi_{TB} = c_{17} + B_0^2 M_{2,\mu}+\xi_{TB} \,$
with $\xi_{TB} = \frac12C\mu^2 +c_1T^2+2c_2 T\mu $. The Kubo formulas for parity violating non-equilibrium coefficients are
\begin{align}\label{eq:KuboTT2an1}
  & \coeff{1}{\omega}{\rm Im}\, G_{T^{tx} T^{xz}}(\omega,{\bf k}{=}0) = - \frac{w_0-\MO_{,\mu}B_0^2}{B_0}(c_8 \tilde{\rho}_\perp + \tilde{c}_{10}\rho_\perp) \,,\\[5pt]  \label{eq:KuboTT2an2}
   & \coeff{1}{\omega}{\rm Im}\, G_{T^{tx} T^{yz}}(\omega,{\bf k}{=}0) = - \frac{w_0-\MO_{,\mu}B_0^2}{|B_0|}(c_8 \rho_\perp - \tilde{c}_{10}\tilde{\rho}_\perp) \,,\\[5pt] \label{eq:KuboTT2an3}
  & \coeff{1}{\omega}{\rm Im}\, G_{ T^{xz}T^{tx}}(\omega,{\bf k}{=}0) = \frac{w_0}{B_0}(c_{15} \tilde{\rho}_\perp + \tilde{c}_{17}\rho_\perp) \,,\\[5pt] \label{eq:KuboTT2an4}
  & \coeff{1}{\omega}{\rm Im}\, G_{T^{yz}T^{tx}}(\omega,{\bf k}{=}0) = - \frac{w_0}{|B_0|}(c_{15} \rho_\perp - \tilde{c}_{17}\tilde{\rho}_\perp) \,.
\end{align}
\begin{figure}[h]
    \centering
    \includegraphics[width=6.2cm]{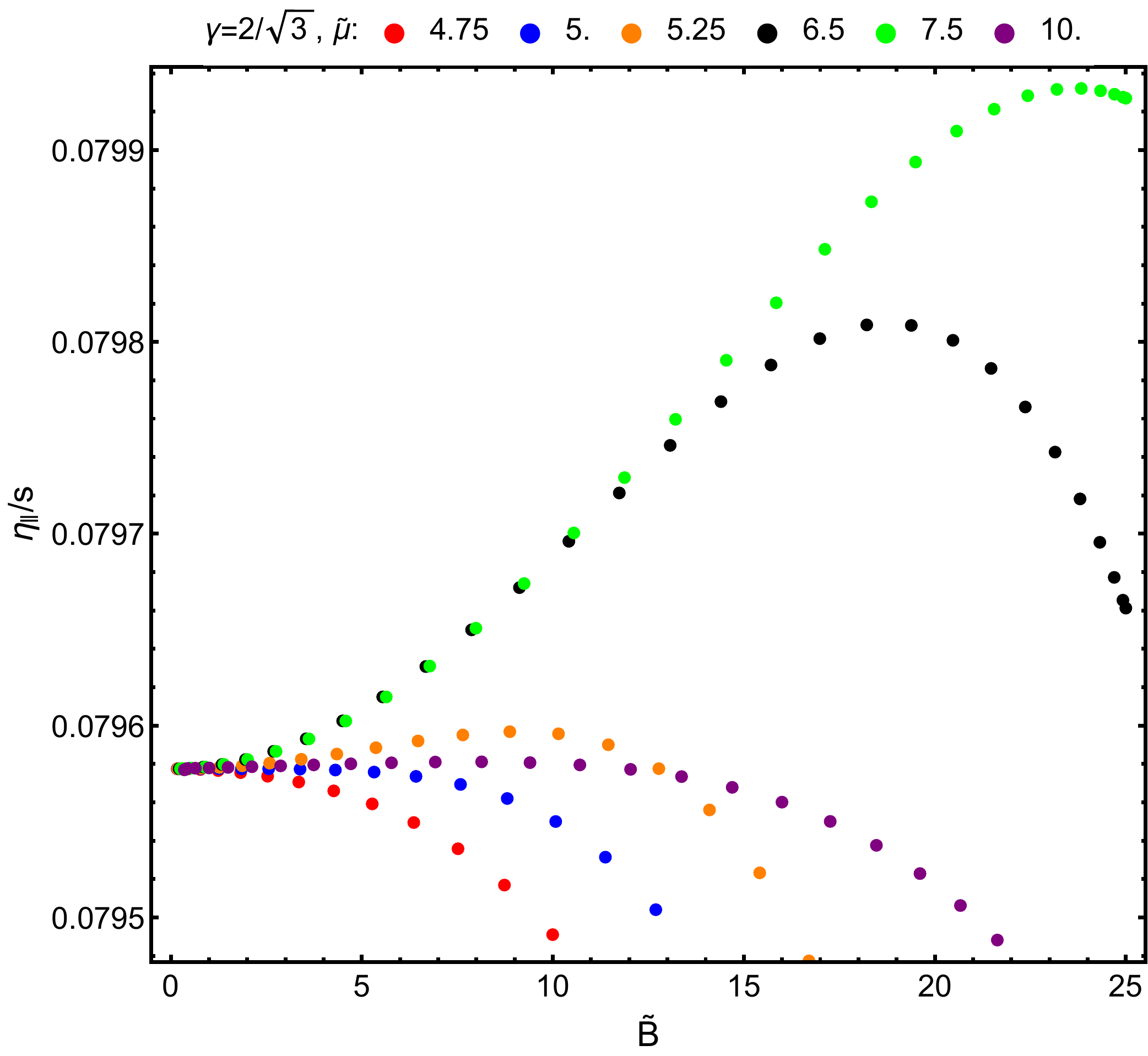}\hspace{0.5cm} \includegraphics[width=6.2cm]{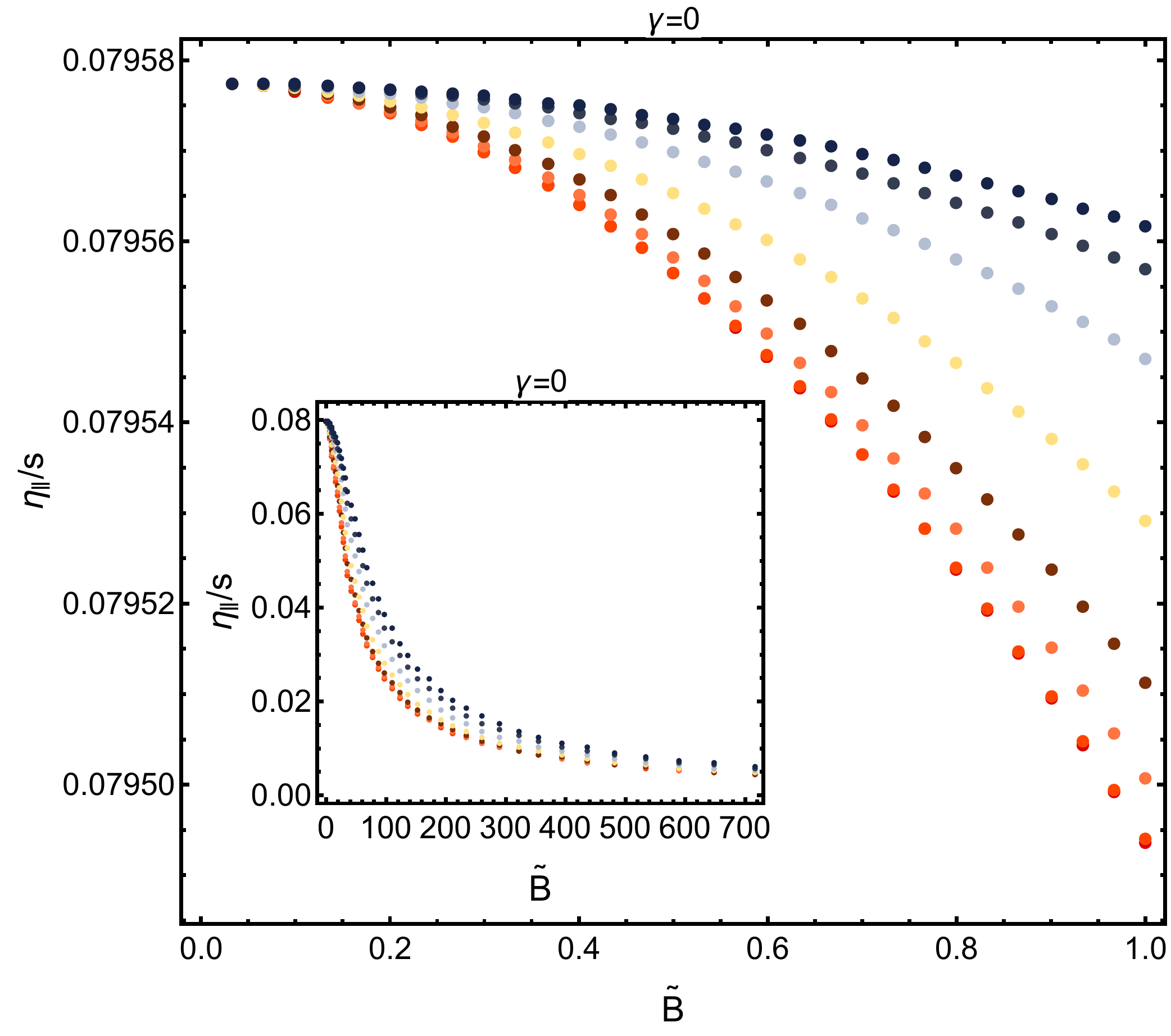}
    \caption{Dimensionless ratio of parallel shear viscosity and entropy density for \mbox{$\tilde\mu=\{0., 0.253, 1.224, 2.078, 3.404, 5.105, 6.521, 7.5\}$} (red to dark blue). \textbf{Left: }In presence of the chiral anomaly, the ratio increases for certain values of the chemical potential and small $\tilde B$. \textbf{Right: } Without the anomaly, all curves tend downwards.}
    \label{fig:my_label_eta_para_h1}
\end{figure}
\begin{figure}[h]
    \centering
    \includegraphics[width=6.5cm]{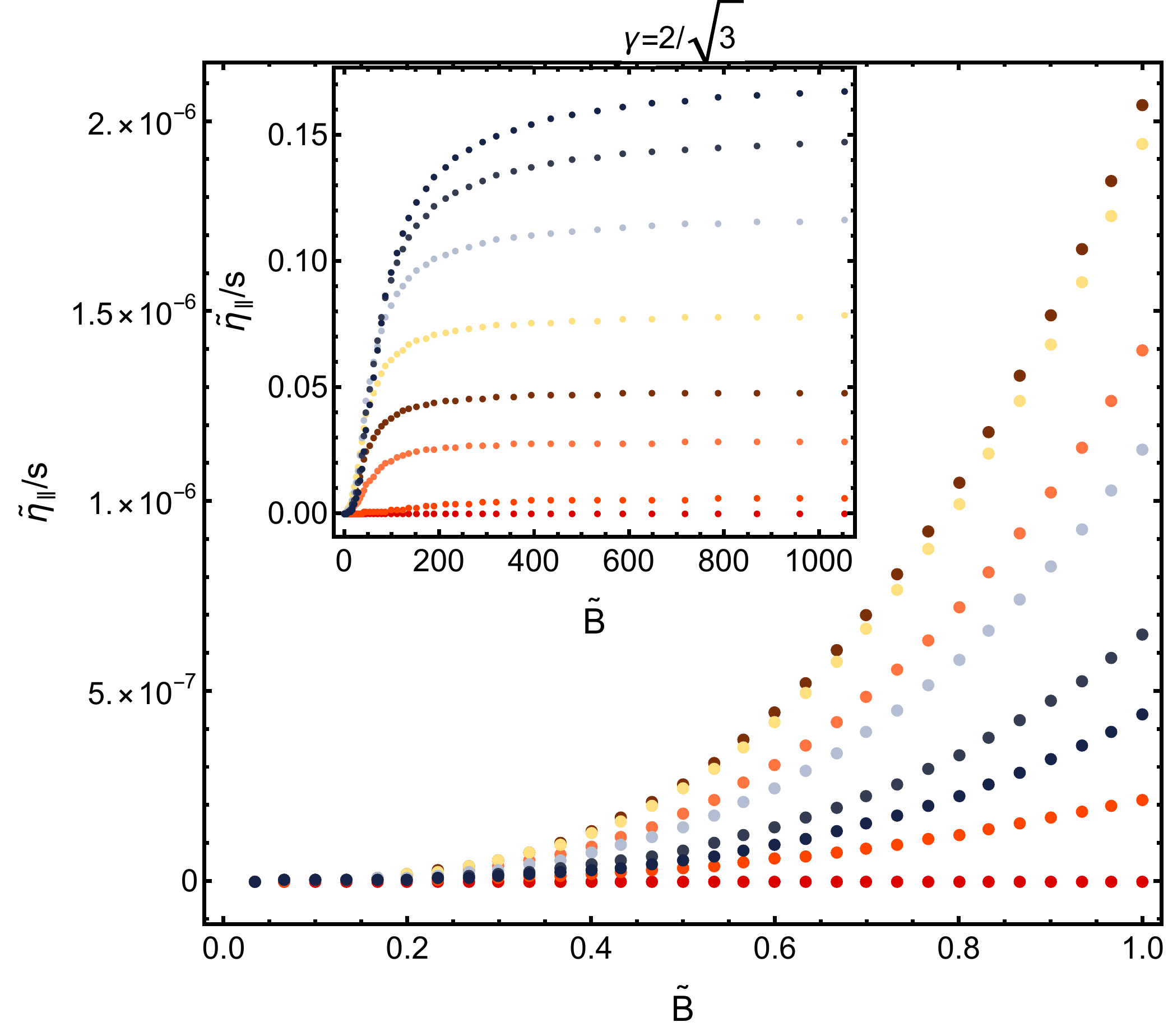}\hspace{0.5cm} \includegraphics[width=6.2cm]{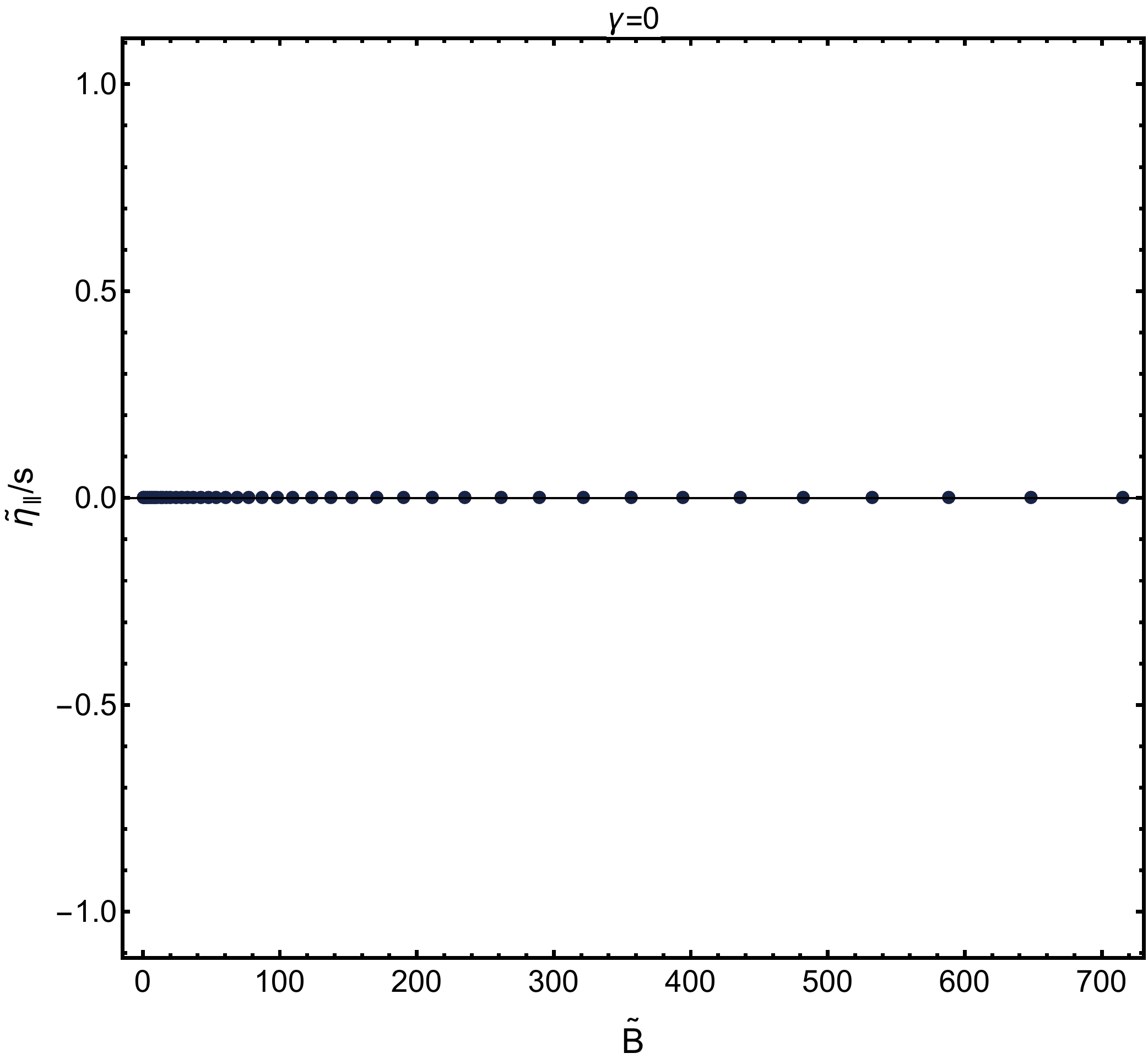}
    \caption{Dimensionless ratio of parallel Hall viscosity and entropy density for \mbox{$\tilde\mu=\{0., 0.253, 1.224, 2.078, 3.404, 5.105, 6.521, 7.5\}$} (red to dark blue). In presence of the chiral anomaly (left), all curves increase with $\tilde B$ (except $\tilde\mu=0$). Without chiral anomaly, the viscosity is zero for all values $\tilde\mu$ (right).}
    \label{fig:my_label_eta_para_h12}
\end{figure}
In the following, we discuss the parallel shear viscosity $\eta_\|$ in a strong external magnetic field.
In figure \ref{fig:my_label_eta_para_h1}, we depicted the parallel shear viscosities $\eta_\|$ and $\tilde\eta_\|$ with and without chiral anomaly. As usual, the parallel shear viscosity is encoded in the $T^{xz}T^{xz}$ correlators as indicated in eq. \eqref{eq:KuboTT1an1}. In the left side of figure \ref{fig:my_label_eta_para_h1}, we depict the dimensionless ratio of the parallel shear viscosity and the entropy in dependence of the magnetic field divided by the temperature squared. We see that for a certain range of the chemical potential $5\lesssim\tilde\mu\lesssim10$ the parallel shear viscosity initially increases for small magnetic fields with the increase being maximal for $\tilde\mu\approx 7.5$. We thus observe either an increase or decrease in the parallel shear viscosity depending on the considered value of the chemical potential. Since the magnetic field breaks the rotational invariance to a $SO(2)$ subgroup, it is not surprising that the dimensionless ratio of parallel shear viscosity and entropy deviates from the universal value of $1/(4\pi)$. Indeed, this was first noticed by the authors of \cite{Erdmenger:2010xm} in the context of spontaneously broken rotations in an anisotropic p-wave superfluid.
The increase in the parallel shear viscosity for small values of the magnetic field is solely an effect of the chiral anomaly which is obvious if we compare to the system without chiral anomaly. In the right side of figure \ref{fig:my_label_eta_para_h1}, we present the ratio of parallel shear viscosity over entropy in a strong external magnetic field without chiral anomaly in dependence of $\tilde B$. As it can be seen from the inset of the graphics, the shear viscosity always decreases with increasing magnetic field independent of the value of the chemical potential. This is in strong contrast to the behavior in presence of the chiral anomaly.

Even more remarkable are the results for the parallel Hall viscosity depicted in figure \ref{fig:my_label_eta_para_h12}. The parallel Hall viscosity may be extracted from the $T^{yz}T^{xz}$  according to eq. \eqref{eq:KuboTT1an2}. From the left side in figure \ref{fig:my_label_eta_para_h12}, we can see that the parallel Hall viscosity is non-zero in the anomalous system when a finite chemical potential is present. We furthermore notice that the Hall viscosity increases for increasing the magnetic field. Similar to the parallel shear viscosity, the behavior for small $\tilde B$ depends on the considered chemical potential. There is a certain critical $\tilde\mu_0$ for which the parallel Hall viscosity increases the fastest for small $\tilde B$. The inset indicates, however, that the behavior for very large $\tilde B$ is monotonous in $\tilde\mu$. Note that larger values of the chemical potential yield bigger values of $\tilde\eta_\|/s$ at large $\tilde B$. Without anomaly, the situation changes drastically. In the right side of figure \ref{fig:my_label_eta_para_h12}, we see the parallel Hall viscosity in strong magnetic fields without chiral anomaly. It vanishes for all values of the chemical potential and not only for vanishing chemical potential as in the anomalous case. The parallel Hall viscosity is a novel transport coefficient which is only present in the anomalous system and vanishes without chiral anomaly, similar to the thermodynamic susceptibility $M_2$.

To conclude the discussion of the transport coefficients in the helicity-one sector, we compute the perpendicular resistivity $\rho_\perp$ and the perpendicular Hall resistivity $\tilde \rho_\perp$ with formula \eqref{eq:KuboTT2an1}-\eqref{eq:KuboTT2an4}. Note that in order to evaluate eq. \eqref{eq:KuboTT2an1} and \eqref{eq:KuboTT2an2}, we have to  (numerically) calculate the derivative of $M_5$ \eqref{eq:thermo_25} with respect to $\tilde\mu$ at fixed $\tilde B$. In order to make the results for the resistivities dimensionless, we have to multiply them by $T$. 

\begin{figure}[!t]
    \centering
    \includegraphics[width=6.2cm]{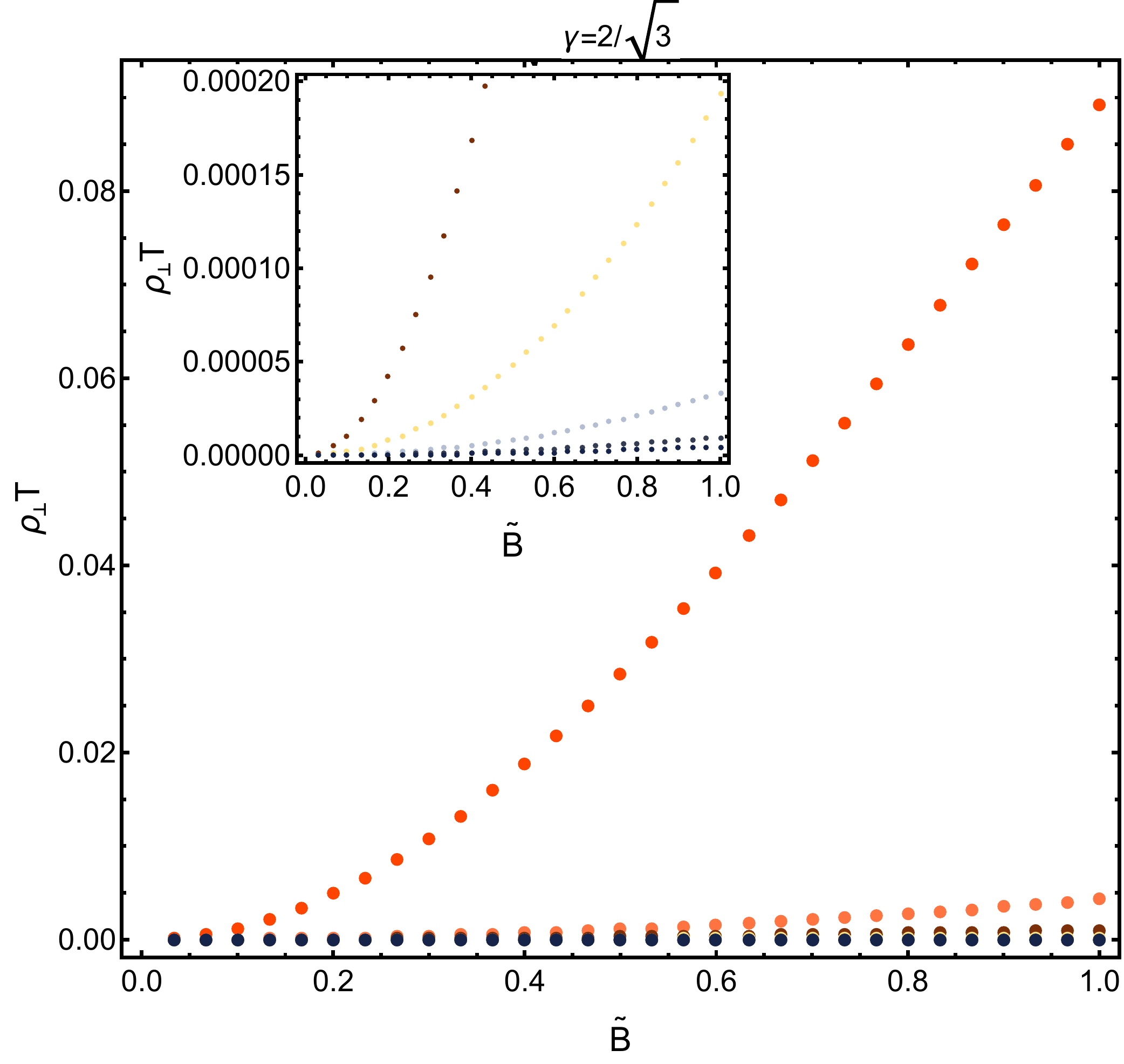}\hspace{0.5cm} \includegraphics[width=6.2cm]{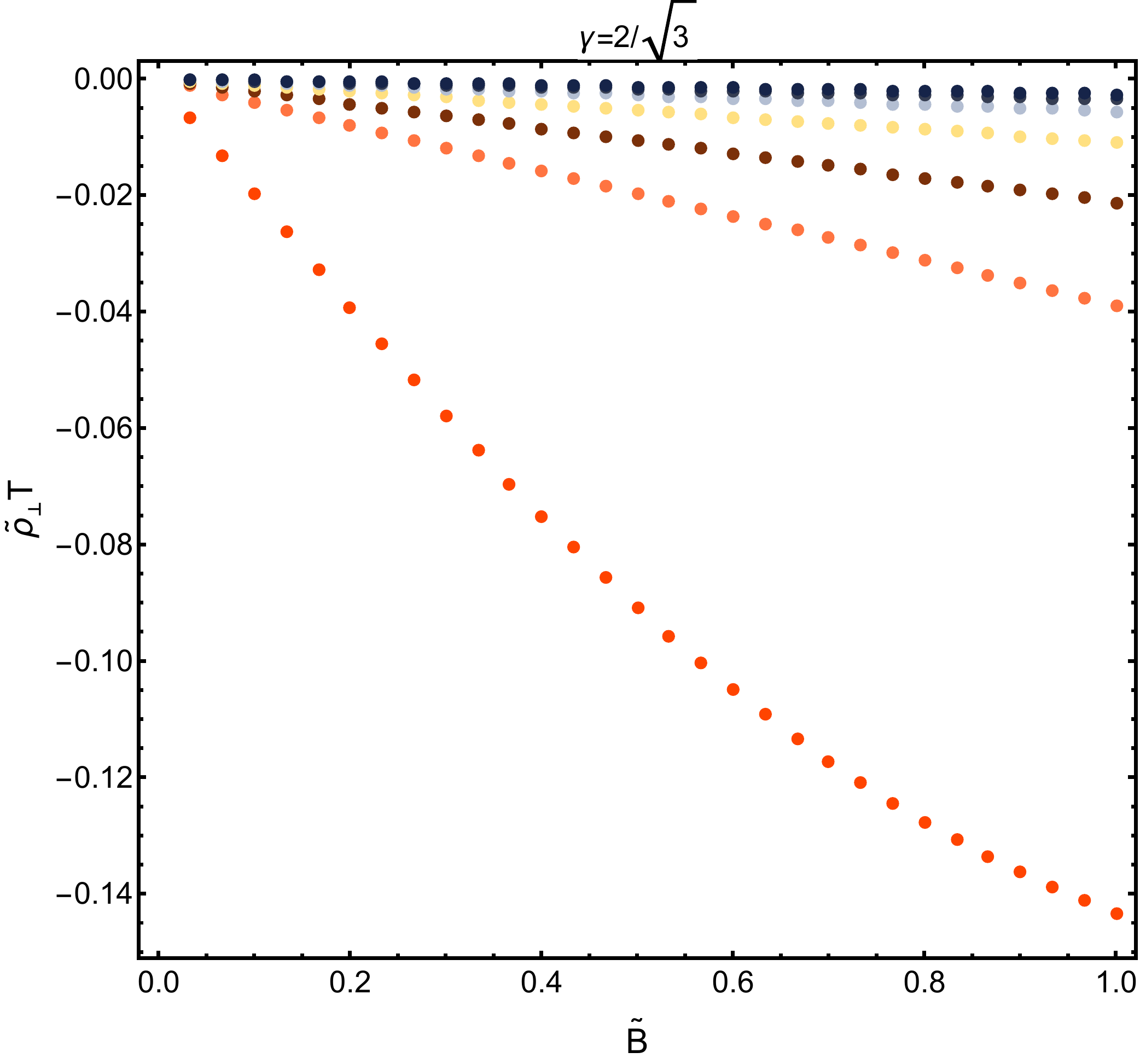}
    \caption{Perpendicular resistivity in presence of the chiral anomaly (left) and perpendicular Hall resistivity (right) for \mbox{$\tilde\mu=\{0.253, 1.224, 2.078, 3.404, 5.105, 6.521, 7.5\}$} (from red to dark blue). The absolute values decrease with increasing chemical potential.}
    \label{fig:my_label_rho}
\end{figure}

In the left side of figure \ref{fig:my_label_rho}, we show a cartoon of the dimensionless perpendicular resistivity $\rho_\perp\,T$ in dependence of the magnetic field. With increasing magnetic field, the resistivity also increases while it decreases for increasing chemical potential. We furthermore note that the behavior is qualitatively not effected by the presence of the chiral anomaly and we do not display the results for $\gamma=0$ for the sake of a compact presentation.

The results for the perpendicular Hall resistivity are similar and depicted in right side of figure \ref{fig:my_label_rho}. The perpendicular Hall resistivity increases for increasing magnetic field and the absolute value decreases for increasing chemical potential. Similarly to the perpendicular resistivity the perpendicular Hall resistivity is not qualitatively affected by the presence of the chiral anomaly and we thus do not present a cartoon of it.  

\subsection{\mbox{Hydrodynamic transport coefficients in the Helicity-zero sector}}
 In this sector, we consider the transport coefficients in the helicity-zero sector. For the helicity-zero sector, the parity-violating terms do not yield new contributions to the bulk viscosities which are thus given by the expressions derived in~\cite{Hernandez:2017mch}
\begin{align}\label{eq:Kubo-Tmne}
  & \coeff{1}{\omega}\delta_{ij} {\rm Im}\, G_{T^{ij} O_1}(\omega,{\bf k}{=}0) = 3 \zeta_1+\cdots\,,\\ \label{eq:Kubo-Tmnf}
  & \coeff{1}{3\omega} \delta_{ij} \delta_{kl}\, {\rm Im}\, G_{T^{ij} T^{kl}}(\omega,{\bf k}{=}0) = 3\zeta_1 + \zeta_2 +\cdots\,,\\ \label{eq:Kubo-Tmng}
  & \coeff{1}{\omega} {\rm Im}\, G_{O_1 O_1}(\omega,{\bf k}{=}0) = \zeta_1 - \coeff23 \eta_1+\cdots\,,\\ \label{eq:Kubo-Tmnh}
  & \coeff{1}{\omega} {\rm Im}\, G_{O_2 O_2}(\omega,{\bf k}{=}0) = 2\eta_2+\cdots\,,
\end{align}
where $O_1 = \frac12(T^{xx}+T^{yy})$ and $O_2 = T^{zz} - \frac12(T^{xx}+T^{yy})$. The $\delta^{ij}$ is the projector onto the spatial coordinates, i.e. $i=x,y,z$. The ellipsis denote terms which are zero when $M_1=M_3=M_4=0$, or when $B_0 \ll T_0^2$ as we already confirmed numerically. 

By computing the correlators in eq. \eqref{eq:Kubo-Tmnf} and eq. \eqref{eq:Kubo-Tmnf} numerically, we are able to extract the bulk viscositites $\zeta_1 $ and $\zeta_2$. Both viscosities $\zeta_1$ and $\zeta_2$ vanish in both cases, with and without chiral anomaly and we thus do not show them in a graphic. 

We may extract bulk viscosities by computing the corresponding 2-point functions in our holographic model according to eq. \eqref{eq:Kubo-Tmng} and \eqref{eq:Kubo-Tmnh} since $\xi_1=0.$ In order to make the viscosities dimensionless, we divide them by the entropy density.

\begin{figure}[!t]
    \centering
    \includegraphics[width=6.2cm]{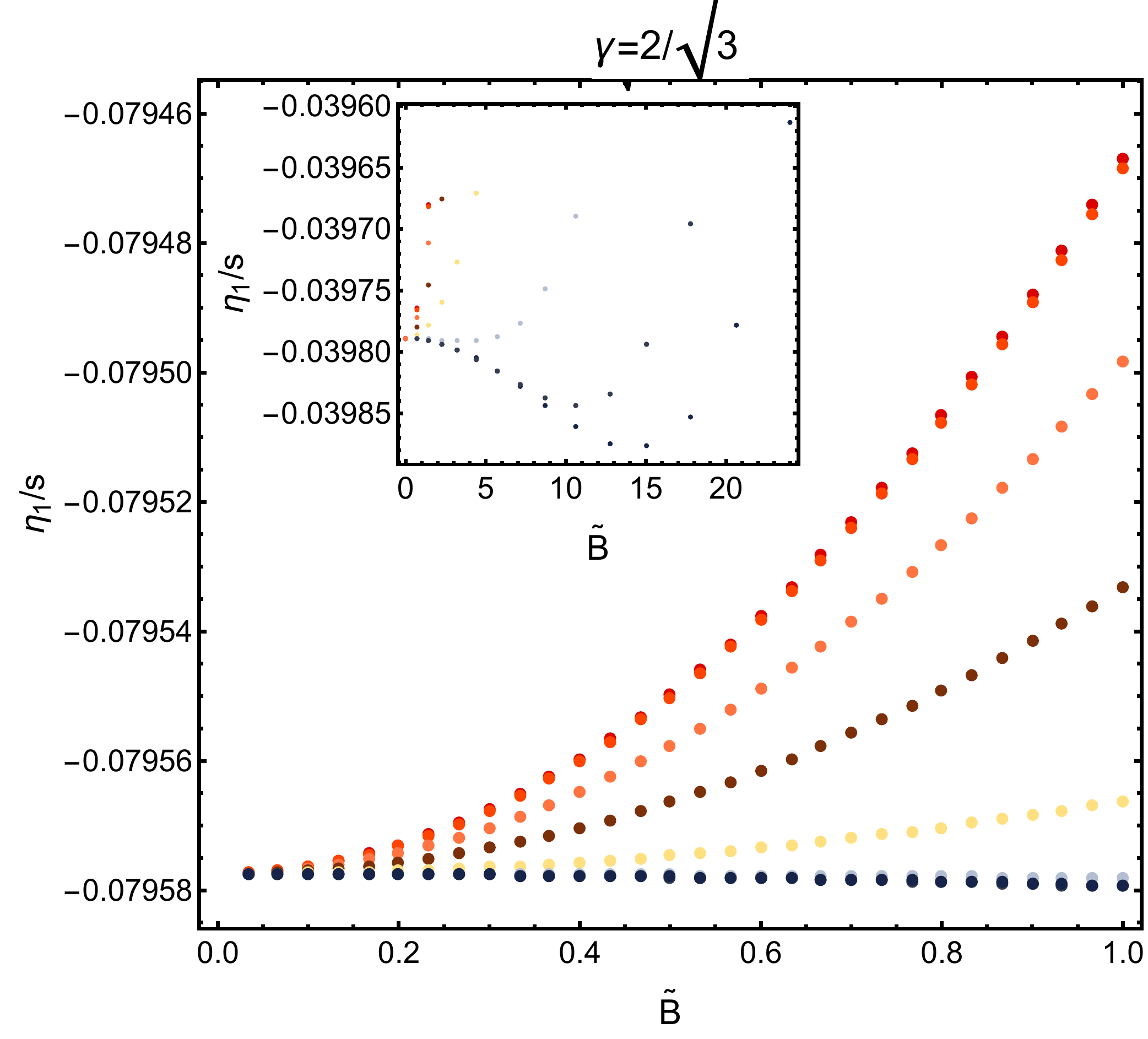}\hspace{0.5cm} \includegraphics[width=6.2cm]{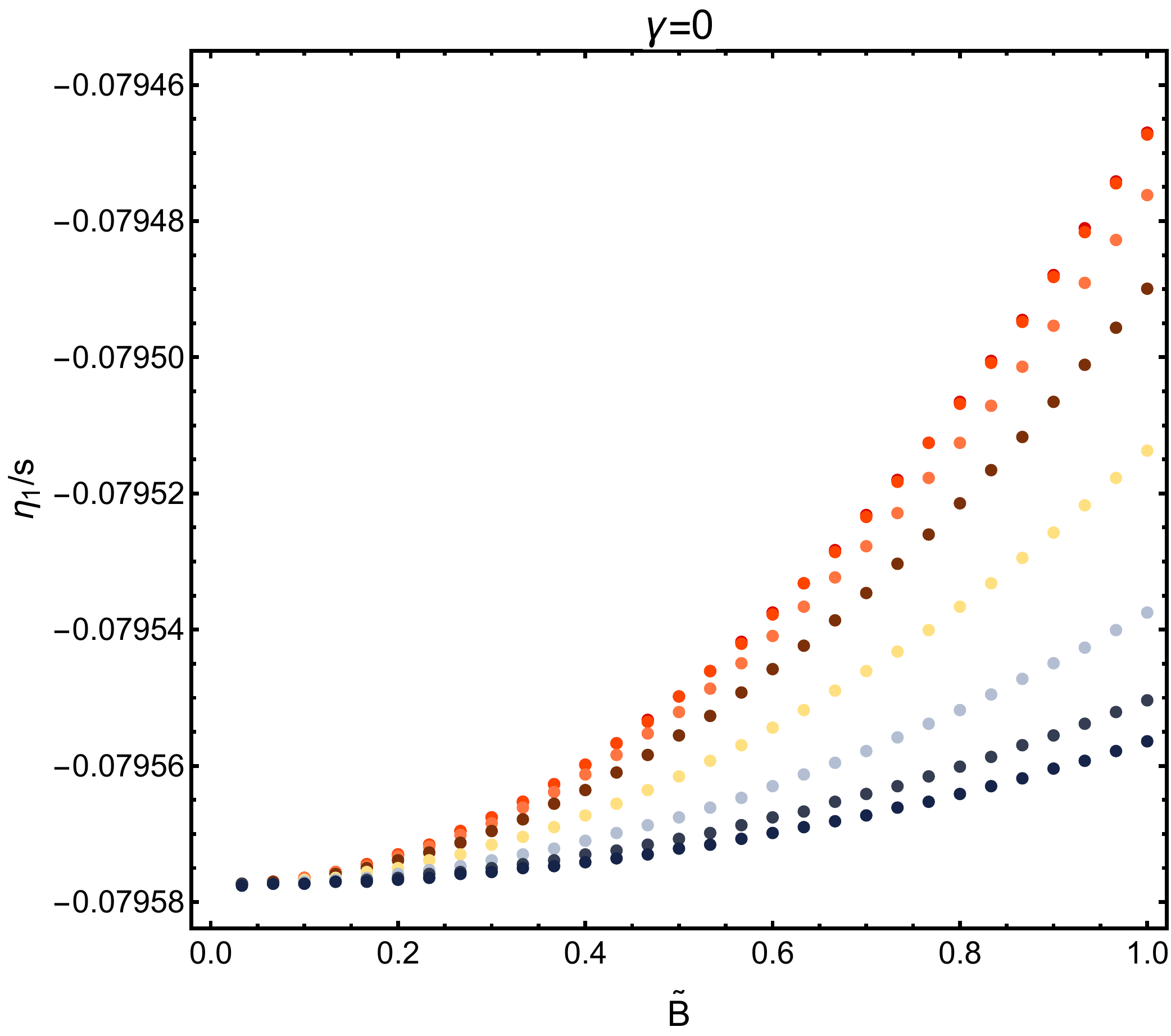}\vspace{-0.1cm}
    \caption{Dimensionless ratio of bulk viscosity $\eta_1$ and entropy density for \mbox{$\tilde\mu=\{0., 0.253, 1.224, 2.078, 3.404, 5.105, 6.521, 7.5\}$}. In presence of chiral anomaly (left) the absolute values increase for certain $\tilde\mu$ and small $\tilde B$. Without anomaly (right) all curves decrease in absolute value for increasing $\tilde B$.}
  \label{fig:mysp0eta1}
\end{figure}

In figure \ref{fig:mysp0eta1}, we depict the numerical results for the bulk viscosity $\eta_1$ with and without chiral anomaly. In presence of the chiral anomaly (left side of figure \ref{fig:mysp0eta1}). In particular, we observe that the anomaly influences the bulk viscosity $\eta_1$ in the left side of the figure in a way that they change slope at $\tilde B\to 0$ in a certain range of the chemical potential. As we already observed for the parallel shear viscosity in the helicity-1 sector, the absolute value of viscosity $\eta_1$ increases  for a certain range of values of the chemical potential as may be seen from the inset in figure \ref{fig:mysp0eta1}. This increase is not present in the non-anomalous theory (right side of figure \ref{fig:mysp0eta1}), where we notice that the absolute value of $\eta_1/s$ decreases for all values of the chemical potential. We hence conclude that the presence of the chiral anomaly changes -- similar to the helicity-1 case -- the slope of $\eta_1/s$ in the small $\tilde B$ regime for certain values of the chemical potential.

\begin{figure}[!t]
    \centering
    \includegraphics[width=6.2cm]{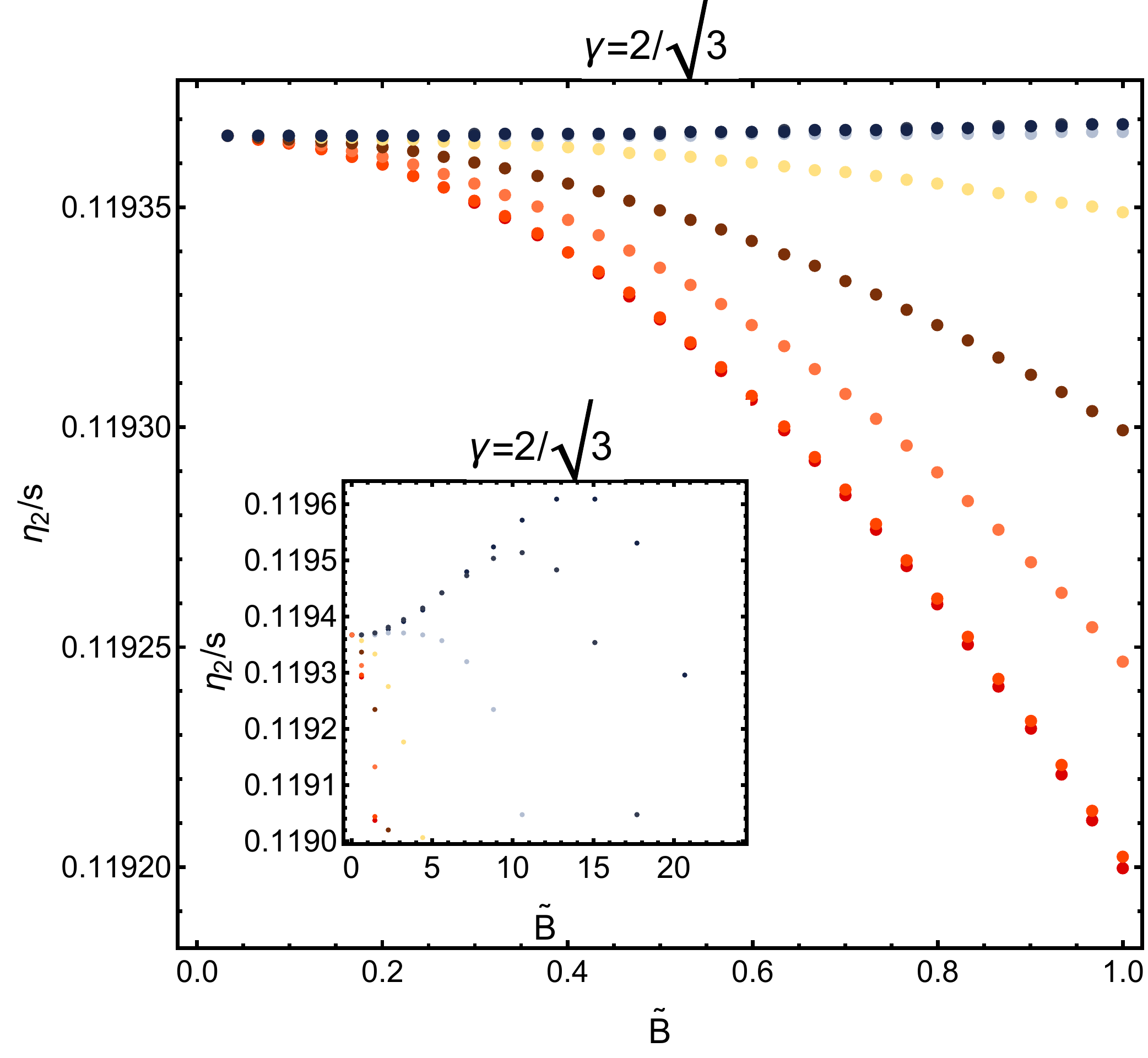}\hspace{0.5cm} \includegraphics[width=6.2cm]{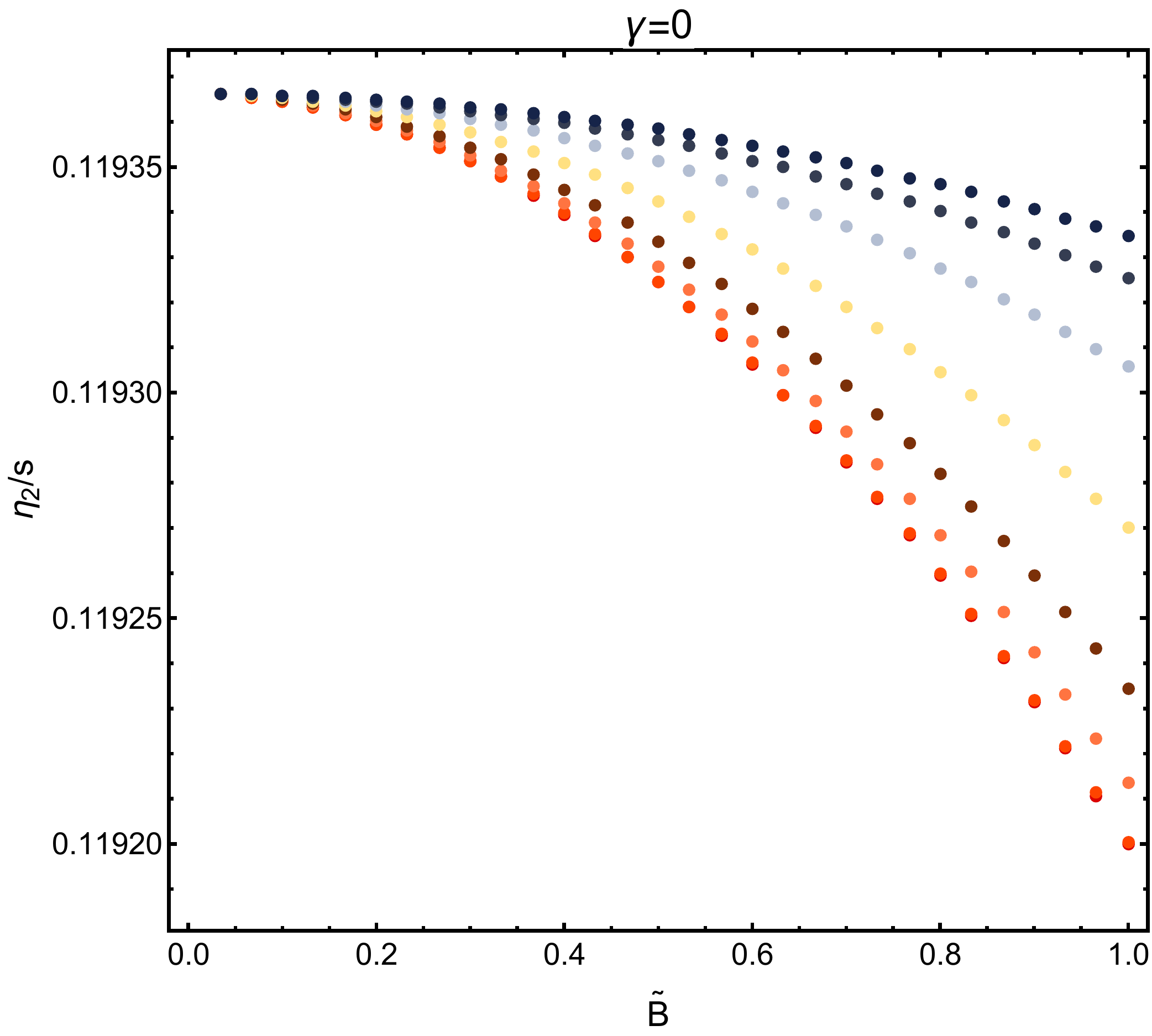}\vspace{-0.1cm}
    \caption{Dimensionless ratio of bulk viscosity $\eta_2$ and entropy density for \mbox{$\tilde\mu=\{0., 0.253, 1.224, 2.078, 3.404, 5.105, 6.521, 7.5\}$}. In presence of chiral anomaly (left) the absolute values increase for certain $\tilde\mu$ and small $\tilde B$. Without anomaly (right) all curves decrease in absolute value for increasing $\tilde B$.}
  \label{fig:mysp0eta2}
\end{figure}

We observe qualitatively the same behavior for $\eta_2$. In figure \ref{fig:mysp0eta2}, we show that the presence of the chiral anomaly changes the slope of $\eta_2/s$ for small values of $\tilde B$. All curves eventually decrease for larger $\tilde B$. In case of vanishing chiral anomaly, the curves always decrease with increasing the magnetic field independent of the value of the chemical potential.

\chapter{Entanglement entropy, Dirichlet walls and the swampland}\label{section:EEandtheswampland}
So far, two types of Ryu-Takayanagi surfaces have been known in de Sitter: One measures the entanglement entropy between the two CFTs on the UV slice \cite{Dong:2018cuv} while the other measures the entanglement entropy across the horizon on the UV slice in the static patch. The natural question to ask is: what happens with the entangling surfaces between the horizon and the UV slice? Do they interpolate between the two already known entanglement entropies? Furthermore, can we use the notion of entanglement entropy to establish possible consistency requirements on matter fields in dS gravity (swampland bounds)?

This chapter is based on my work with Hao Geng and Andreas Karch, published in \cite{Geng:2019bnn}.
\section{Entanglement entropy in DS/dS: A one parameter family}
\begin{figure}[H]
    \centering
    \includegraphics[width=6cm]{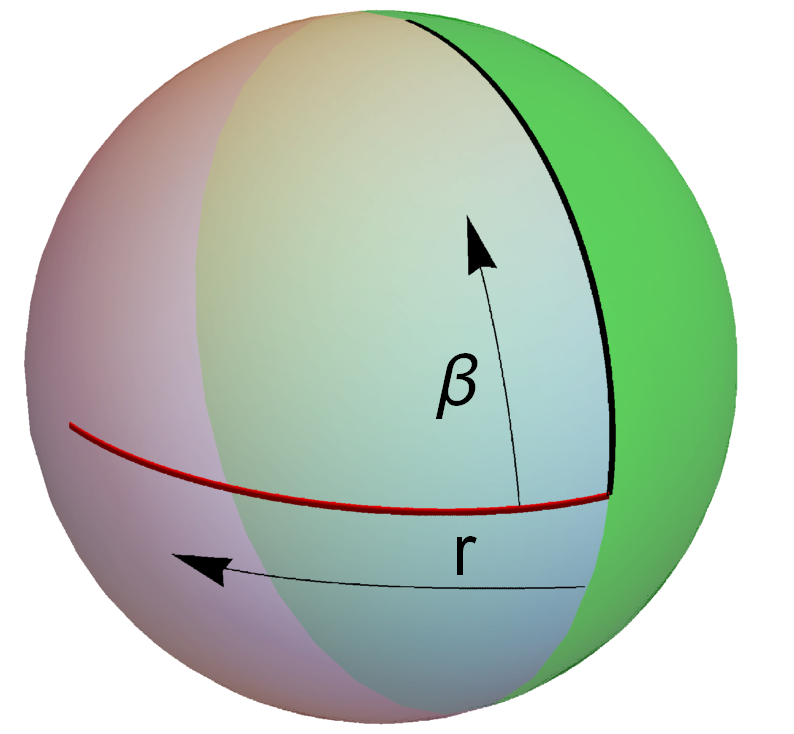}\hspace{0.5cm}
    \includegraphics[width=6cm]{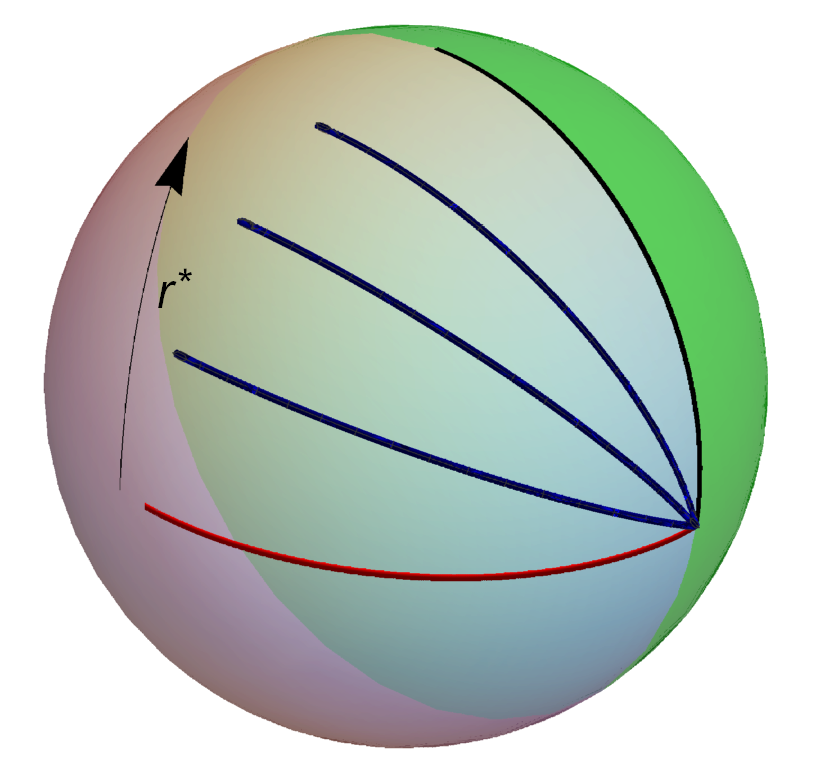}
    \caption{Euclidean dS$_3$ in the static patch at $\tau=0$. \textbf{Left: }The class U surface is depicted in red and the class D surface in black, respectively. The UV slice separates the left half (purple) from the right (green). \textbf{Right: }The one-parameter family of entangling surfaces interpolates between class U and class D.}
    \label{fig:ClassUaD}
\end{figure}

First, we consider spherical entangling surfaces in the static patch of dS given by the metric
\begin{equation}\label{eq:metricdS}
    \dd s^2_{\text{dS}_{d+1}}\!=\dd r^2+L^2\sin(r/L)\,\dd s^2_{\text{dS}_d}, \ \ \text{with }\ \dd s^2_{\text{dS}_d}=-(1-\rho^2)\,\dd\tau^2+\frac{\dd\rho^2}{1-\rho^2}+\rho^2\,\dd\Omega^2_{d-2},
\end{equation} where $\rho=\cos(\beta)$. 
We choose our observer to sit at the origin of the static patch
\begin{equation}
    \rho=\rho_0\in(0,1],\quad \tau=0.
\end{equation}
 Note that the cosmological horizon is located at $\rho=1$. The two known classes of entangling surfaces are the ``U"-shaped entangling surfaces (class U) which are hanging down from the central UV slice to the IR along the cosmological horizon. The second class (class D) -- constructed in \cite{Gorbenko:2018oov} -- are living on the UV slice and separate the left CFT from the right CFT. In order to illustrate the difference between the two classes, we depict both cases in the left side of figure \ref{fig:ClassUaD}.
\subsection{Spatial entanglement in dS (class U)}
The class U surfaces compute the entanglement entropy between the left and the right CFT. In the gravity picture, this corresponds to a double sided surface, which stretches from the left side to the right side. The easiest way to obtain this solution is to write down the Lagrangian for the minimal surfaces parametrized as $\beta(r)$
\begin{equation}
\mathcal L_\text{U}=L^{d-2}\,\cos^{d-2}(\beta)\,\sin^{d-2}\left(\frac rL\right)\,\sqrt{1+L^2\,\sin^2\left(\frac rL\right)\,(\beta'(r))^2}.\label{eq:lagrangian}
\end{equation}
For $\beta_0=0\ (\rho_0=1)$, we find the simple solution $\beta(r)=0$, or
\begin{equation}
 \rho(r)=1,
\end{equation}
since $\beta'$ vanishes for constant $\beta$ and $(\cos(\beta))'$ vanishes for $\beta=0$. This implies the class U surfaces (depicted in red in figure \ref{fig:ClassUaD}) just stretch along the cosmological horizon which is located at $\beta=0$ and the entanglement entropy associated with this type of surfaces is simply given by the entropy of the de-Sitter space
\begin{equation}
    S_\text{EE,U}=S_\text{dS}.\label{eq:classU}
\end{equation}
For entangling surfaces with $\rho_0<1$, we naturally expect to end up with surfaces which end at $\rho(r)<1$ and thus yield a smaller entanglement entropy as $S_\text{dS}$ (as it is the case in AdS). First, we discuss the other known type of entangling surfaces -- the class D.

\subsection{Integrating out one CFT (class D)}
In order to see the solution for the other class of entangling surfaces, we consider yet another parametrization namely $r(\beta)$ (instead of $\beta(r)$)
\begin{equation}
\mathcal L_\text{D}=L^{d-1}\,\cos^{d-2}(\beta)\,\sin^{d-2}\left(\frac {r(\beta)}{L}\right)\,\sqrt{(r'(\beta)^2+L^2\,\sin^2\left(\frac {r(\beta)}{L}\right)}.
\end{equation}
Similar to the preceding subsection, 
we note that the Euler-Lagrange equations vanish for 
\begin{equation}
    r(\beta)/L=\frac\pi 2,
\end{equation}
since $r'=0$ and $(\sin(r(\beta)/L))'$ vanishes for $r(\beta)/L=\pi/2$. Moreover, since $r(\beta)/L=\pi/2$ for all $\beta$, the entangling surface lives completely on the UV slice for generic $\beta_0$. In particular, the enclosed volume is given by the $\beta_0=0$ surface which yield the entanglement entropy
\begin{equation}
    S_\text{EE,D}=\frac{\Omega_{d-2}\,\int_{0}^{\pi/2}\cos^{d-2}(\beta)\,\dd\beta}{4\,G_N}=\frac 12\,S_\text{dS}.
\end{equation}
The authors of \cite{Gorbenko:2018oov} argued that the class D surfaces integrate out the degrees of freedom of one CFT and we are thus left with half of the original de-Sitter entropy. Intuitively this makes sense, since the entangling surface separates the left and the right side. Moreover, if we are tracing out the entire spatial part of one CFT, the Ryu-Takayanagi surface is simply the whole spatial region of the dS$_d$ localized at $r/L=\pi/2$.

The volume inside the static patch is given by half the volume of the unit sphere, while the spatial volume of the dS$_d$ is the volume of a unit sphere. Hence, the ``global" or ``full" entropy is twice the entropy of the static patch and we find
\begin{equation}
    S_\text{global,D}=S_\text{dS}.
\end{equation}
To summarize, we calculated the de-Sitter entropy in two very different ways: first, we calculated the entanglement entropy on the gravity side captured by the entangling surface along the horizon. In the second, we traced out the degrees of freedom of one of the CFTs on an entire spatial slice yielding the same result.

\subsection{A one parameter family of entangling surfaces}
Interestingly, it is possible to combine both concepts in terms of a one parameter family of entangling surfaces interpolating between both concepts. In this section, we consider entangling surfaces of class U but with $\rho_0<1$ and turning point at $r^\star$ in the bulk. The already known class U surface corresponds to the case $r^\star/L=0$ and $r^\star/L=\pi$. 

The Lagrangian for the minimal surface is given in eq. \eqref{eq:classU}. It is straightforward to check that the corresponding equations of motions are solved by\footnote{Note that our $r$ is shifted by $\pi L/2$ compared to the original work~\cite{Geng:2019bnn}.}
    \begin{equation}
    \beta(r)=\arcsin\left[\tan\left(r^\star/L\right)/\tan\left(r/L\right)\right].\label{eq:analytical}
    \end{equation}
By evaluating the Lagrangian eq. \eqref{eq:classU} on the analytical solution eq. \eqref{eq:analytical} and integrating over $r$ (where $\Omega_{d-2}$ is the volume of the sphere), it is straightforward to obtain the entanglement entropy
\begin{equation}
    S_\text{EE}=\frac{\Omega_{d-2}}{2\,G_N}\,\int_0^{\pi/2}\dd r\,\mathcal L(\beta(r))=\frac{L^{d-1}\sqrt{\pi}\,\Gamma((d-1)/2)}{4\,G_N\Gamma(d/2)}=S_\text{dS}.
\end{equation}
Surprisingly, all entangling surfaces yield the same entropy -- the de-Sitter entropy -- independently of their turning point $r^\star$. The exact field theoretic interpretation of these entangling surfaces is not clear. Since they interpolate between the two extreme cases -- the class U and D surfaces -- a possible explanation is that they are all representing slightly different ways of how to trace out degrees of freedom. From a geometric point of view, the picture in the bulk is a lot clearer. In the right side of figure \ref{fig:ClassUaD}, we depicted five different values of $r^\star$ of the one-parameter family: the two extreme cases $r^\star/L=\pi/2$ (class D) and $r^\star=0$ (class U), as well as three intermediate values. The entangling surfaces all correspond to great circles on the sphere which inherently have the same area. This also implies, however, that the existence of the one-parameter family of solutions is closely related to the symmetry of de-Sitter. In geometries slightly deviating from de-Sitter, for example, by introducing a black hole or matter fields, the family of solutions is no longer present. We will see an interesting application of the one-parameter family of solutions in the next chapter.
\section{Warped de-Sitter}
In order to probe our notion of entanglement entropy in DS/dS, we switch on sources in the dual field theory. The sources deform the spacetime and we have to allow for a general warpfactor in the metric
\begin{equation}
    \dd s_{\text{dS}_{d+1}}^2 = \dd r^2+e^{2\,A(r)}\,\dd s^2_{\text{dS}_{d}}.
\end{equation}
We have to impose certain conditions on the warpfactor, since we still want to apply the framework of DS/dS
\begin{itemize}
    \item The spacetime should still asymptote to the two de-Sitter horizons. This requires that the warpfactor $A(r)$ exhibits two asymptotic horizon regions $r_\text{min}$ and $r_\text{max}$ indistinguishable from the ones in dS$_d$ sliced dS$_{d+1}$.
    \item We assume that the deformations act symmetrical on both de-Sitter halves. However, our final conclusions are independent of this assumption, as argued later.
    \item The warpfactor $A(r)$ has a maximum $r_\text{m}$ between the two asymptotic regions $r_\text{min}$ and $r_\text{max}$. Without loss of generality, we can redefine our coordinates system in a way that the maximum $r_\text{m}$ is located at $r_\text{m}=0$. By doing so, we allow $r_\text{max}=\pi\,L/2-r_\text{min}$ to have values different than $-\pi L/2$ and $\pi L/2$, respectively, due to stretching and contracting of the spacetime in response to the deformation.
\end{itemize}
We can declare the surface at the maximum of the warpfactor to be our UV brane with curvature radius given by $L_\star=e^{A(0)}$ since the graviton is localized at $r_m=0$. The Newton's constant on the slice dS$_d$, $g_N$, and in the bulk dS$_{d+1}$, $G_N$, are related by
\begin{equation}
    g_N^{-1}=G^{-1}_N\,\int\dd r\, e^{(d-2)\,A(r)}.\label{eq:newton}
\end{equation}
The class U and class D surfaces with general warpfactor read
\begin{align}
    \mathcal L_\text{U}&=\cos^{d-2}(\beta(r))\,e^{(d-2)\,A(r)}\sqrt{1+e^{2A(r)}(\partial_r\beta(r))^2}\\ \mathcal L_\text{D}&=\cos^{d-2}(\beta)\,e^{(d-2)\,A(r(\beta))}\sqrt{(\partial_\beta r(\beta))^2+e^{2A(r(\beta))}},
\end{align}
where the two special solutions (note that we shifted the middle slice by $\pi/2$ to $r_\text{m}=0$ instead of $r_\text{m}/L=\pi/2$ as considered in the last section)
\begin{equation}
    \beta(r)=0\quad\text{and}\quad r(\beta)=0
\end{equation}
still solve the equations of motion with general warp factor. The former solves the equations of motion since $(\cos(\beta))'$ vanishes at the horizon of each patch. The latter relies on the fact that the warp factor has a maximum on the UV slice, namely \mbox{$A'(r_m)=A'(0)=0$}.
With eq. \eqref{eq:newton}, the class U surface simply yields the entanglement entropy
\begin{equation}
    S_\text{EE,U}=\frac{A}{4\,G_N}=\frac{\Omega_{d-2}\,\int_{-r_\text{max}}^{r_\text{max}}\dd r\,e^{(d-2)A(r)}}{4\,G_N}=\frac{\Omega_{d-2}}{4\,g_N}=S_\text{dS}.
\end{equation}
We conclude that the class U surfaces always reproduce the correct de-Sitter entropy, since the Newton's constant picks up exactly the contribution of the warp factor. Even for general deformations, the class U surface reproduces the correct de-Sitter entropy, as we already expected from general considerations in~\cite{Emparan:2006ni,Myers:2013lva} since the class U surface corresponds to a horizon in the bulk.

For the class D surfaces, however, the situation is entirely different. Even if we account for the entire spatial volume, hence integrating out one CFT, the entanglement entropy does not correspond to the entropy of dS$_{d+1}$. On the UV slice, the volume is totally determined by the maximum of the warpfactor $e^{A(r_m)}$ which sets the curvature radius $L_\star$. 
It would be natural to assume that the entropy of dS$_d$ is an upper bound for any entanglement entropy which we can obtain for a density matrix by tracing out degrees of freedom or in other words an upper bound on the class D entanglement entropy. This assumptions leads to constraints on the warp factor $A(r)$ and thus on the matter content which ensure that the obtained entanglement entropy is never larger than the dS$_d$ entropy. Recall that we focus on the class D entangling surface associated with tracing out the degrees of freedom of one of the CFTs. In the undeformed case, the class D entangling surfaces correctly reproduce the dS$_d$ entropy. For a general warp factor the entanglement entropy associated with two copies of this particular class D surface reads
\begin{equation}
    S_\text{global,D}=\frac{L_\star^{d-1}\,V_{d-1}}{2\,G_N}=\frac{\Omega_{d-2}\,L_\star^{d-1}}{4\,G_N}\int_{-\pi/2}^{\pi/2}\dd\beta\,\cos^{d-2}(\beta).
\end{equation}
Our consistency requirement thus imposes the constraint
\begin{equation}
    1\ge \frac{S_\text{global,D}}{S_\text{dS}}=e^{(d-1)\,A(r_m)}\frac{\int_{-\pi/2}^{\pi/2}\dd\beta\,\cos^{d-2}(\beta)}{\int_{-r_\text{max}}^{r_\text{max}}\dd r\,e^{(d-2)\,A(r)}}.\label{eq:swamplandbound}
\end{equation}
At first sight, this consistency constraint seems to be really challenging to evaluate since we have to compare a local quantity $\sim e^{A(0)}$ to an integrated one $\int\dd r\, e^{(d-1)\,A}$. As we will see in the next section though, we are able to derive quite general statements from this bound.

\subsection*{Matter fields in warped de-Sitter}
In order to understand the implications of our consistency requirement, we consider the backreaction of matter fields on the de-Sitter space. We study a generic, homogeneous matter content in $d>2$ by considering the action
\begin{equation}
    S=\frac{1}{16\pi\,G_N}\,\int\dd^{d+1}x\,\sqrt{-g}\left(R-2\Lambda\right)+S_\text{matter},
\end{equation}
with positive cosmological constant $\Lambda=d\,(d-1)/(2\,L^2)$ and energy-momentum tensor $T_{ij}=-2/\sqrt{-g}\,\delta S/\delta g^{ij}$.
\newpage\noindent
The Einstein's equations for a massive scalar field in de-Sitter read
\begin{align}
    &\frac{d\,(d-1)}{2}\left(A'^{\,2}-\frac{e^{-2A}}{L^2}+\frac{1}{L^2}\right)=8\pi\,G_N\,T^r_r\label{eq:efg1}\\
 &(d-1)\left(A''+\frac d2\,A'^{\,2}-e^{-2A}\,\frac{d-2}{2\,L^2}+\frac{d}{2\,L^2}\right)=8\pi\,G_N\,T^t_t   \label{eq:efg2}.
\end{align}
Equipped with the Einstein's equations, we may rewrite the integral of the warp factor as
\begin{equation}
    \int_{-r_\text{max}}^{r_\text{max}}\dd r\,e^{(d-2)\,A}=2\int_{-\infty}^{A(0)}\,\frac{\dd A}{A'}\,e^{(d-2)\,A}=2\int_{-\infty}^{A(0)}\,\frac{\dd A}{\sqrt{\frac{16\pi\,G_N\,T^r_r}{d\,(d-1)}-\frac{1}{L^2}+\frac{e^{-2A}}{L^2}}}\,e^{(d-2)\,A},\label{eq:warpint}
\end{equation}
where we replaced the derivative of $A'$ using eq. \eqref{eq:efg1}. The change of variables from $r$ to $A$ is valid because of the monotonicity of $A$. The warp factor $A$ increases monotonous from $r_\text{min}$, reaches a maximum at $0$ and decreases monotonous until it eventually reaches its second minimum at $r_\text{max}$. The monotonicity tells us even more about the integral. We know that the integrand vanishes at the maximum and we thus have
\begin{equation}
    \frac{16\pi\,G_N}{d\,(d-1)}T_r^r(0)=\frac{1}{L^2}-\frac{e^{-2A(0)}}{L^2}.\label{eq:condma}
\end{equation}
We furthermore note that $A'(r)>0$ for $-r_\text{max}<r<0$ and $A'(r)<0$ for $0<r<r_\text{max}$. Combining these two observations, we may rewrite the integral eq. \eqref{eq:warpint} with a shift of the integration variable by its maximum value $A(r)=F(r)+A(0)$ as
\begin{equation}
    \int_{-r_\text{max}}^{r_\text{max}}\dd r\,e^{(d-2)A}=2L\,e^{(d-1)\,A(0)}\int_{-\infty}^{0}\frac{e^{(d-2)\,F}\,\dd F}{\sqrt{e^{-2\,F}-\frac{16\pi\,G_N\,L^2\,T^r_r-d(d-1)}{16\pi\,G_N\,L^2\,T^r_r(0)-d(d-1)}}}
    .\label{eq:toin}
\end{equation}
In order to find an inequality for this expression, we have to take a look at the monotonicity of $T^r_r$. We note that
\begin{equation}
    \partial_rT^r_r=\frac{d\,(d-1)}{8\pi\,G_N}\,A'\left(A''+e^{-2A}/L^2\right)=d\,A'\left(T^t_t-T^r_r\right),
\end{equation}
where we used Einstein's equation in the last step. Since we know, when $A'$ is greater or smaller than zero, the sign of the derivative of $T_r^r$ depends on the sign of the difference ($T^t_t-T_r^r$). Equation \eqref{eq:condma} tells us that the left hand side is bounded by $1/L^2$ since the exponential function is always greater than zero. We conclude that the fraction in the denominator of eq. \eqref{eq:toin} is smaller or equal than one if and only if
\begin{equation}
    -T_t^t+T^r_r>0.\label{eq:neq}
\end{equation}
The inequality in eq. \eqref{eq:neq} has to be satisfied for any reasonable form of matter since it is exactly the Null-Energy Condition.
In this case, we may approximate the integral by
\begin{equation}
    \int_{-r_\text{max}}^{r_\text{max}}\dd r\,e^{(d-2)A}\le2\,L\,e^{(d-1)A(0)}\int_{-\infty}^{0}\frac{\dd F\,e^{(d-2)\,F}}{\sqrt{e^{-2F}-1}}=L\,e^{(d-1)A(0)}\int_{-\pi/2}^{\pi/2}\dd \beta\,\cos^{d-2}(\beta).
\end{equation}
However, this is exactly equal to the nominator in eq. \eqref{eq:swamplandbound} leading to
\begin{equation}
     \frac{S_\text{global,D}}{S_\text{dS}}\ge 1.\label{eq:bounv}
\end{equation}
This is a rather surprising result; eq. \eqref{eq:bounv} states that any form of matter satisfying the Null-Energy condition violates our proposed consistency requirement.

\chapter{$T\bar T$ deformations and cutoff (A)dS}\label{sec:TTbarandcutoff}
Irrelevant deformations within QFTs are still very uncharted territory. 
This is particularly the case because we cannot follow the normal renormalization group procedure; consider a theory where we switch on an (IR) relevant coupling in the UV. In the UV the coupling does not contribute and we can follow the trajectory to the IR. For irrelevant operators, we have to start in the IR and flow towards the UV. The so-called $T\bar T$ deformation is unlike the generic irrelevant deformations exactly solvable~\cite{Smirnov:2016lqw,Cavaglia:2016oda,Zamolodchikov:2004ce}. In the field theory space for a generic seed QFT, we may define a trajectory from the IR towards the ultraviolet (UV) by triggering the flow with a $T\bar T$ deformation. Even though we flow towards the UV, we may extract valuable information about the theory such as the finite volume spectrum, the S-matrix, and the deformed Lagrangian exactly.

In this chapter, we derive entanglement entropies in the $T\bar T$ deformed theories from field theory and holography in general spacetime dimensions. Since it challenging to compute entanglement entropies in higher dimensions -- even for free theories -- we employ the trick introduced in \cite{Donnelly:2018bef} and consider a QFT on a sphere. For spheres, we may compute the entanglement entropy in terms of the sphere partition function. Within holography, the QFT on a sphere is realized by choosing dS slicing in the radial direction (in contrast to Poincar\'e/flat slicing for example). The results in this chapter are based on my work published in~\cite{Grieninger:2019zts}.

\section{$T\bar{T}$ deformations in $2d$ field theory}
We can view the $T\bar T$ flow from a renormalization group perspective, where $t$ parametrizes the trajectory in the field theory space. We start with the undeformed Lagrangian $\mathcal L$ and trigger the flow by deforming the theory with the operator irrelevant operator $\det\left(T_{\mu\nu}^{(t)}\right))$
\begin{equation}
    \mathcal L^{(t+\delta t)}=\mathcal{L}^{(t)}+\delta t\,\det(T^{(t)}_{\mu\nu})=\mathcal L^{(t)}-\delta t \,T\bar T^{(t)},
\end{equation}
where $T\bar T=\frac18\left(T^{\mu\nu}T_{\mu\nu}-(T_\mu^\mu)^2\right)$ for a 2d QFT.
In this equation, $T\bar T$ denotes a composite operator of dimension 4.\newpage

Starting from a conformal field theory, the stress tensor of the $T\bar T$ deformed theory is no longer traceless 
\begin{equation}
    S_\text{QFT}=S_\text{CFT}+2\pi\lambda\,\int\dd^2x\,\sqrt{\gamma}\,T\bar T
\end{equation}
where the deformed action obeys $\dd S_\text{QFT}/\dd\lambda=2\pi\int\dd^2x\,\sqrt{\gamma}\,T\bar T$. 
Assuming the theory exhibits only a single mass scale $\mu$, we may perform a dimensional analysis and find
\begin{equation}
    \mu\frac{\dd S_\text{QFT}}{\dd\mu}=\int\dd^2x\,\sqrt{\gamma}\, T^\mu_\mu.
\end{equation}
With $\mu=1/\sqrt{\lambda}$, the trace of the energy momentum tensor yields
\begin{equation}\label{eq:fieldtheorytrajectory}
    T^i_i=-\frac{c}{24\pi}\,\tilde R-4\pi\,\lambda\,T\bar T+\ldots,
\end{equation}
where the dots denote higher orders in $\lambda$; we also introduced the two dimensional Ricci tensor $\tilde R$ which accounts for the curvature on a sphere and vanishes for flat space.

\section{$T\bar T$ deformations in holography}
To understand the holographic proposal by McGough, Mezei, and Verlinde~\cite{McGough:2016lol}, we consider AdS$_3$ restricted to~\cite{Hartman:2018tkw,Kraus:2018xrn, Gorbenko:2018oov}
\begin{align}
    \dd s^2=\dd r^2+g_{\mu\nu}(r,x)\,\dd x^\mu\,\dd x^\nu, && \text{with } r<r_c,\label{eq:AdSsub}
\end{align}
where the radial coordinate $r$ is related to the ``standard'' radial coordinate in eq. \eqref{eq:poincareM} by $1/u=e^{r/L}$.
The CFT is located at the boundary as usually. Throughout this section, the boundary of the space \eqref{eq:AdSsub} is no longer at conformal infinity but rather at a finite radial distance $r=r_c$. The central charge is given by~\cite{Brown:1986nw}
\begin{equation}\label{eq:TTbarcentralcharge}
   \mu=\frac{16\pi G_N}{r_c^2}=\frac{24\pi\,L}{c\,r_c^2}.
\end{equation}
At large central charge $c$, we can employ the weak form of the AdS/CFT correspondence and equate the generating functionals at $r=r_c$ (with notation $g^0_{\mu\nu}=g_{\mu\nu}(r_c)$)
\begin{equation}
    Z_\text{QFT}(\gamma_{\mu\nu},J)=\text{exp}\left(-S_\text{grav}(g_{\mu\nu}^0=e^{2\,r_c/L}\,\gamma_{\mu\nu})\right),
\end{equation}
where $S_\text{grav}$ is the action of the classical gravitational theory with metric eq. \eqref{eq:AdSsub} subject to Dirichlet boundary conditions $\dd s^2|_{r=r_c}=g_{\mu\nu}\,\dd x^\mu\,\dd x^\nu$. 
In the following, we set $e^{r_c/L}=1$ for convenience.\newpage \noindent
To understand the trajectory within the holographic duality, we consider a gravity theory in (A)dS$_3$
\begin{equation}
    S=-\frac{1}{16\pi\,G_N}\int_{\mathcal M}\dd^3x\,\sqrt{g}\,\left(R-2\,\frac{\eta}{L^2}\right)-\frac{1}{8\pi\,G}\int_{\partial M}\dd^2x\,\sqrt{\gamma}\,(K-L^{-1})\label{eq:actionTTbar3}
\end{equation}
with the holographic stress tensor (see eq. \eqref{EMtensor})
\begin{equation}
    T_{\mu\nu}=\frac{1}{8\pi\,G_N}\left(K_{\mu\nu}-K\,\gamma_{\mu\nu}+\frac{\gamma_{\mu\nu}}{L}\right).\label{eq:emtensorttbarK}
\end{equation}
In expression \eqref{eq:actionTTbar3}, we introduced $\eta$ which is 1 for AdS and $-1$ for dS and hence reflecting the sign change in the cosmological constant. 
The metric for dS$_d$ sliced (A)dS$_{d+1}$ as introduced in eq. \eqref{Adsds} reads in a unified notation
\begin{equation}
    \dd s^2_{\text{(A)dS}_{d+1}}=\dd r^2+L^2\,\sinhh(r/L)\,\dd s^2_{\text{dS}_d},
\end{equation}
where the hyperbolic sine corresponds to the AdS$_{d+1}$ case and the sine to dS$_d$. 
In $d=2$, the trace of the radial Einstein equation reads
\begin{equation}\label{eq:radialEinstein}
    G^r_r=\frac 12\,\left(K^2-K^{\mu\nu}K_{\mu\nu}\right)-\tilde R-\frac{\eta}{L^2}=0,
\end{equation}
where we denote the $d$ dimensional Ricci scalar on the cutoff slice by $\tilde R$.
By evaluating \mbox{$T\bar T=(T^{\mu\nu}T_{\mu\nu}-(T_\mu^\mu)^2)$} for the energy-momentum tensor \eqref{eq:emtensorttbarK} and using \eqref{eq:radialEinstein} to remove the extrinsic curvature from the expressions, we find the more generalized trace flow equation
\begin{equation}
    T_\mu^\mu=\!-\frac{L\,\tilde R}{16\,\pi\,G_N}-4\pi\,G_NL\,(T^{\mu\nu}T_{\mu\nu}-(T_\mu^\mu)^2)-\frac{\eta-1}{8\pi\,G_N\,L}=-\frac{c\,\tilde R}{24\,\pi}-4\pi\,T\bar T+\frac{\eta-1}{8\pi\,G_NL}.\label{eq:traceEMTT}
\end{equation}
In eq. \eqref{eq:traceEMTT}, the trajectory for dS picks up and additional contribution from the cosmological constant compared to AdS and the ``usual'' field theory trajectory \eqref{eq:fieldtheorytrajectory}. 
For the sphere (or in other words Euclidean dS), we can immediately solve for the ground state; on the sphere, the energy-momentum tensor is proportional to the metric of the sphere $\langle T_{\mu\nu}\rangle=\omega_2(R)\,\gamma_{\mu\nu}$, where $R$ denotes the radius of the sphere. Using this expression for the first equality in \eqref{eq:traceEMTT} yields a quadratic equation in $\omega_2(R)$ 
\begin{equation}
    8\pi\,G_N\,\omega_2^2-\frac{2}{L}\omega_2-\frac{\eta-1}{8\pi\,G_N\,L^2}-\frac{\tilde R}{16\pi\,G_N}=0
\end{equation}
with solutions
\begin{equation}\label{eq:omega2eq}
    \omega_2=\frac{1}{8\pi\,G_N\,L}\left(1\pm \sqrt{\frac{\tilde R}{2}+\frac{\eta}{4\,L^2}}\ \right).
\end{equation}
For dS$_2$ sliced (A)dS$_3$, we simplify this further by evaluating the Ricci tensor $\tilde R$ for the two-sphere of radius $R=\sinhh(r_c/L)$.

Note that the previous statements implicitly assumed that we are taking the expectation values of the operator valued quantities, i.e. we should write
\begin{equation}
    \langle T\bar T\rangle=\frac 18\,\left(\langle T^{\mu\nu}\rangle\langle T_{\mu\nu}\rangle-\langle T_\mu^\mu\rangle^2\right),
\end{equation}
where we assumed a factorization property. In general, this does not hold for a higher dimensional CFT. However, if we work in the large $N$ limit the factorization of the expectation values holds even in $d>2$.

\section{Entanglement entropy and $T\bar T$ deformations}\label{sec::EEandfieldtheorystu}
In this section, we derive the entanglement entropy for antipodal points on a sphere in $T\bar T$ deformed theories from field theory and holography. The calculation was first performed in \cite{Donnelly:2018bef} and we follow it closely.

The partition function of a two dimensional CFT on a sphere parametrized by \newline\mbox{$\dd s^2=R^2\,(\dd \beta^2+\sin(\beta)^2\,\dd \phi^2)$}, with $\phi\in[0,2\pi]$ and $\beta\in [-\pi/2,\pi/2]$ is given by
\begin{equation}
    \frac{\dd}{\dd R}\log(Z)=-\frac 1R\,\int\dd^2x\,\sqrt{\gamma}\,T^\mu_\mu.\label{eq:derPartition}
\end{equation}
Note that the trace of the energy momentum tensor, deformed by the $T\bar T$ deformation, is given in terms of eq. \eqref{eq:traceEMTT}.
As we explained in \eqref{eq:omega2eq}, the symmetries on the sphere dictate that the energy momentum tensor is proportional to the metric on the sphere $\langle T_{\mu\nu}\rangle=\omega_d(R)\,\gamma_{\mu\nu}.$ Re-writing eq. \eqref{eq:omega2eq} in terms of field theory quantities according to eq. \eqref{eq:TTbarcentralcharge}, we find for the proportionality function $\omega_2(R)$
\begin{equation}
    \omega_2(R)=\frac 2\mu\left(1-\sqrt{1+\frac{c\,\mu}{24\pi\,R^2}}\right).\label{eq:propconst}
\end{equation}
The partition function is given by integrating eq. \eqref{eq:derPartition} with \eqref{eq:propconst}
\begin{equation}
    \log Z=\frac c3\,\sinh^{-1}\left(\sqrt{\frac{24\pi}{c\,\mu}}\,R\right)+\frac{8\pi}{\mu}\,\left(R\,\sqrt{\frac{c\,\mu}{24\,\pi}+R^2}-R^2\right), 
\end{equation}
where we fixed the integration constant by imposing $\log Z|_{R=0}=0$ as argued in \cite{Donnelly:2018bef}. Throughout this thesis, we follow this convention of fixing the integration constant which is based on the argument that it leads to a free theory in the UV even though we have concerns if this is consistent with the AdS/CFT machinery.
To compute the entanglement entropy from the partition function, we employ the replica trick which works in our case as follows; we consider the $n$-sheeted cover of the sphere with radius $R$
\begin{equation}
    \dd s^2=R^2\,(\dd \beta^2+n^2\,\sin(\beta)^2\,\dd \phi^2).
\end{equation}
The energy-momentum tensor is isotropic on the sphere and the sphere partition function picks up for varying $n$
\begin{equation}
    \left.\frac{\dd \log Z}{\dd n}\right|_{n=1}=-\int\dd^2x\,\sqrt{\gamma}\,T^\phi_\phi=-\frac{1}{2}\int\dd^2x\,\sqrt{\gamma}\,T^\mu_\mu.
\end{equation}
For antipodal points on the sphere, we can rotate in the $\phi$ direction around the axis connecting the two points. In this case, the entanglement follows from the partition function by taking the limit $n\to1$ by
\begin{equation}
    S_\text{EE}=\left.\left(1-n\frac{\dd}{\dd n}\right)\log Z\right|_{n=1}=\frac c3\,\sinh^{-1}\left(\sqrt{\frac{24\pi}{c\mu}}\,R\right).
\end{equation}

On the other side of the duality, it is straightforward to derive the entangling surfaces which correspond to antipodal points. In the coordinates of eq. \eqref{eq:metricdS}, the entangling surfaces are given by the solutions to
\begin{equation}
    \mathcal L=\sqrt{1+L^2\,\sinhh\left(\frac rL\right)\,\left(\frac{\dd \beta(r)}{\dd r}\right)^2}
\end{equation}
which are anchored to the cutoff surface at $r=r_c$. For antipodal points, the straightforward solution is $r=0,\dd\beta/\dd r=0$ with the associated entanglement entropy given by
\begin{equation}
    S_\text{EE}=\frac{1}{4\,G_N}\int_0^{r_c}\dd r=\frac{L}{2\,G_N}\,\arcsinhh(R/L),
\end{equation}
where we introduced the radius of the circle $R$ in terms of the cutoff $R=L\,\sinhh(r_c/L)$.

\section{Dirichlet walls in (A)dS}
In this section, we work out the impact of a Dirichlet wall at a fixed radial position on the entanglement entropy associated with spherical entangling surfaces. We first derive the results for the one-parameter family of entangling surfaces in dS which we constructed in the last chapter. Then we proceed to generalize the results to (A)dS by considering spherical entangling surfaces in AdS.

\subsection{Entanglement entropy for subintervals on the sphere in dS}
In the last chapter, we found a novel one-parameter family of entangling surfaces in dS~\cite{Geng:2019bnn}. All entangling surfaces correctly reproduce the entanglement entropy of dS and are given in terms of the standard ``U"-shaped surfaces that are hanging down towards the IR and are parametrized in terms of $\beta(r)$
\beq
{\cal L}_{I} = L^{d-2}\, \cos^{d-2}(\beta) \sin^{d-2} \left ( \frac{r}{L} \right ) \sqrt{1 +L^2\, \sin^2 \left ( \frac{r}{L} \right )
 (\beta')^2}.\label{lagr}
\eeq
The equations of motions associated with the Lagrangian are solved by~\cite{Geng:2019bnn}
\beq
\label{analyticsolution}
 \beta(r)=\arcsin\left[\tan\left(r^\star/L\right)/\tan\left(r/L\right)\right],
\eeq
where $r^\star$ is the turning point of the entangling surface.

For pedagogical reasons, we consider the solutions \eqref{analyticsolution} in the parametrization $r(\beta)$
\beq
{\cal L}_{II} =L^{d-2}\,  \cos^{d-2}(\beta) \sin^{d-2} \left ( \frac{r}{L} \right ) \sqrt{(r')^2 +L^2\, \sin^2 \left ( \frac{r}{L} \right )\label{para2}
},
\eeq
where the entangling surfaces are given by
\beq
r(\beta)=L\,\arccot\left(\sin(\beta)/\tan(r^\star/L)\right).\label{solpara2}
\eeq
The one-parameter family of entangling surfaces all reach the cosmological horizon with vanishing first derivative. The second derivatives, however, are different and the entangling surfaces yield different entanglement entropies in presence of a bulk cutoff
\begin{equation}
    r_c/L=\eps/L.
\end{equation}
The Dirichlet wall starts eating up the entangling surfaces when we move it deeper in the bulk 
since the entangling surfaces have to satisfy $r^\star/L>\varepsilon/L$ (figure \ref{pic:geo}).

To find the entanglement entropy, we have to evaluate the Lagrangian \eqref{para2} on the analytical solution \eqref{solpara2} and integrate $\beta$ from the cutoff surface at $\beta=\beta_\eps$ to the turning point of the entangling surfaces $\beta=\pi/2$
\begin{align}
\int\limits_{\beta_\eps}^{\pi/2}\!\dd\beta \mathcal L(r(\beta))&=\frac{L^{d-1}\sqrt{\pi}\,\Gamma\!\left(\frac{d-1}{2}\right)\!}{2\,\Gamma(d/2)}\!-\!\frac{L^{d-1}\text{}_2F_1[\frac12,\frac32-\frac d2,\frac 32,\frac{\sin(\beta_\eps)^2}{\sin(r^\star/L)^2+\cos(r^\star/L)^2\sin(\beta_\eps)^2}]}{\sqrt{\cos(r^\star/L)^2+\sin(r^\star/L)^2/\sin(\beta_\eps)^2}}\nonumber\\&=
4\,G_N\,\text{EE}_\text{dS}-\Delta(\eps,r^\star/L),\label{genEEdS}
\end{align}
where $\text{}_2F_1$ is the hypergeometric function $\text{}_2F_1(a,b;c;z)$.
We denote the de-Sitter entropy by EE$_\text{dS}$ which we recover if we either push the turning point to $r^\star=0$ (studied in \cite{Gorbenko:2018oov}) or remove the cutoff $\epsilon=0$ (studied in \cite{Geng:2019bnn}).
Since the quantity $\Delta(\epsilon,r^\star)$ is strictly positive for $\epsilon>0$ and $r^\star>0$, the entanglement entropies in presence of the cutoff are smaller than the de-Sitter entropy. The bulk variable $r^\star$ does not have an obvious field theory interpretation and we hence rewrite it in terms of the boundary coordinate $\beta$ by introducing $\beta_\eps$. The angle $\beta_\eps$ is related to the turning point in terms of the analytical solution
$r^\star=L\,\arctan\left(\frac{R\,\sin(\beta_\eps)}{\sqrt{L^2-R^2}}\right)$, where we also introduced the radius $R=L\,\sin(\eps/L)$ on the slice given in terms of the warpfactor at $r_c$. With this, the quantity $\Delta$ reads in field theory variables
\begin{equation}
\Delta(\eps,\beta_\eps)=2\,L^{d-2}\,\sqrt{L^2-R^2\,\cos(\beta_\eps)^2}\,\text{}_2F_1\left(1/2,3/2-d/2,3/2,1-\frac{R^2\,\cos(\beta_\eps)^2}{L^2}\right).
\end{equation}

\subsection{Entanglement entropy for subintervals on the sphere in AdS}
We can use the analytical solutions of the last section in order to perform an analog calculation in AdS$_{d+1}$. In particular, dS$_d$ sliced AdS$_{d+1}$ follows from dS$_{d+1}$ by Wick rotating the $d+1$-dimensional curvature constant. We therefore change the warpfactor in the Lagrangian from $\sin(r/L)$ to $\sinh(r/L)$
\begin{equation}
\mathcal L_I=L^{d-2}\cos^{d-2}(\beta)\,\sinh^{d-2}\left(\frac rL\right)\,\sqrt{1+L^2\,\sinh^{2}\left(\frac rL\right)\,(\beta')^2},   
\end{equation}
which is solved by replacing $\tan(r/L)$ with $\tanh(r/L)$ compared to the last section
\begin{equation}
   \beta(r)=\arcsin(\tanh (r^\star/L)/ \tanh (r/L)).\label{eq:anaads}
\end{equation}
The cutoff $r/L=\eps/L$ implies a radius of $R=L\,\sinh(\eps/L)$ in the AdS case and the turning point of the entangling surfaces reads in terms of field theory coordinates $\beta_\eps=\arcsin(\tanh(r^\star/L)/\tanh(\eps/L))$.

The minimal area associated with the entangling surfaces is given by integrating the Lagrangian evaluated on the analytical solution \eqref{eq:anaads}
\begin{equation}
A=2\,L^{D-3}\int_{r^\star}^{\eps}dr\,\frac{\sinh(r/L)}{\cosh(r^\star/L)}\,\left(-1+\frac{\cosh(r/L)^2}{\cosh(r^\star/L)^2}\right)^{d/2-3/2}. \label{OSLag}   
\end{equation}
The integral is straightforward to solve by switching coordinates according to\newline \mbox{$y^2=-1+\cosh(r/L)^2/\cosh(r^\star/L)^2$}
\begin{equation}
    A=2\,L^{d-1}\,\int_{y(r^\star)}^{y(\eps)}dy\,\frac{y^{d-2}}{\sqrt{1+y^2}}=\frac{(R \cos (\beta_\eps))^{d-1}}{d-1} \, _2F_1\left(\frac{1}{2},\frac{d-1}{2};\frac{d+1}{2};-\frac{R^2 \cos ^2(\beta_\eps)}{L^2}\right).\label{genEEAdS}
\end{equation}

\subsection{Entanglement entropy for generic intervals: Summary}
\begin{figure}
    \centering
  \includegraphics[width=6.1cm]{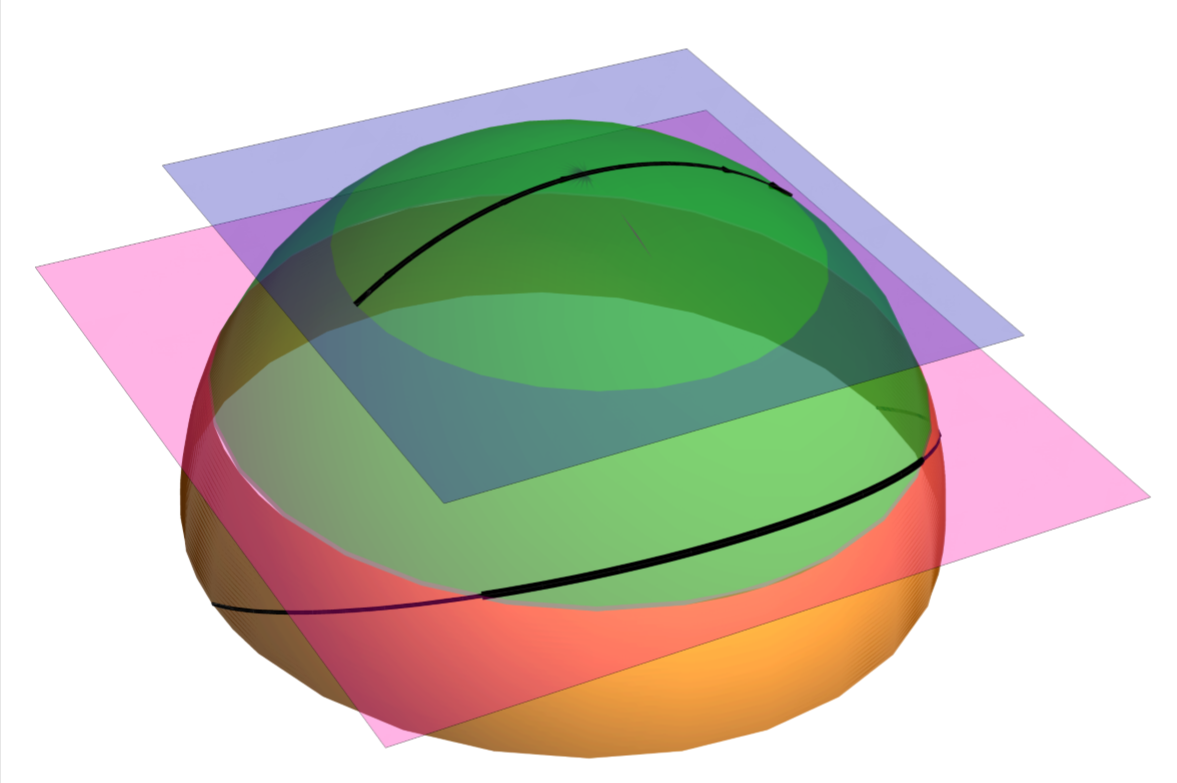}\hspace{0.5cm}  \includegraphics[width=3.8cm]{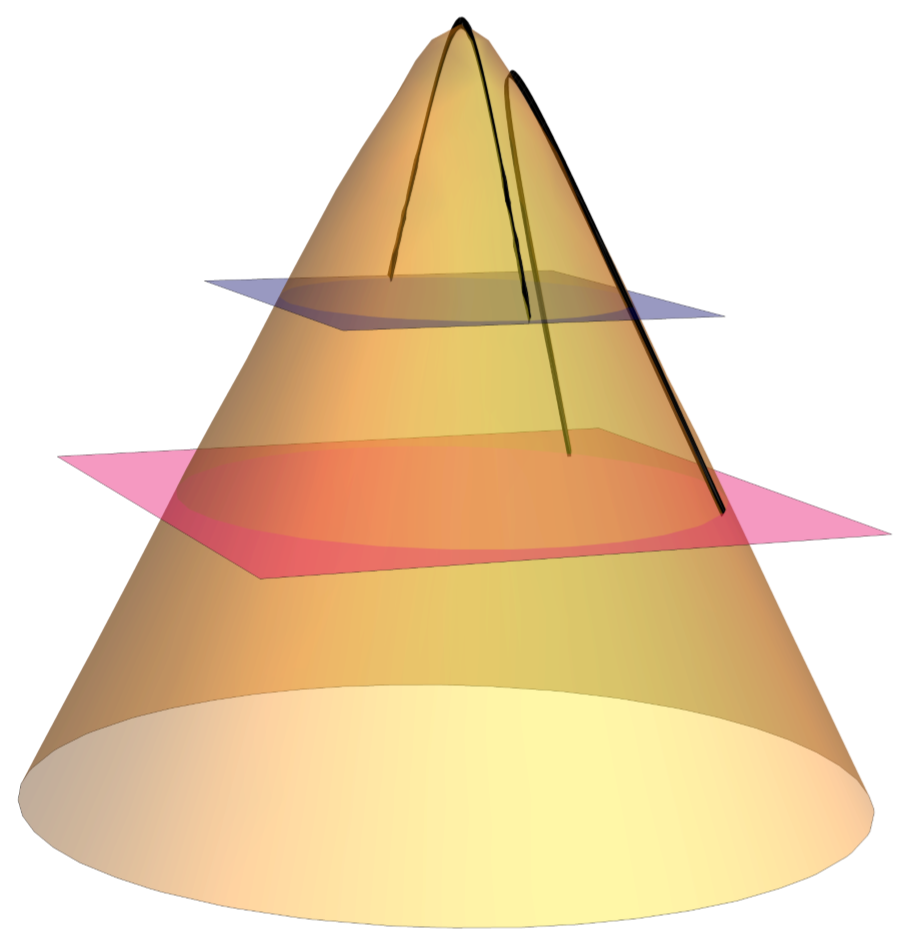}
       \caption{\textbf{Left:} The entangling surface for $r^\star/L=\pi/3$ -- $(\theta,r)$ are the polar and azimuthal angles, respectively, in the static patch of Euclidean dS$_3$ in presence of a cutoff $\epsilon$ (magenta surface) restricting the entangling surface to the bolder line. We may rotate this surface by $\theta_0=\pi/3$ to bring it to the top of the sphere corresponding exactly to a cutoff surface for antipodal point with radius $R_\text{eff}=R\,\cos(\beta_\epsilon(r^\star))$ (blue). \textbf{Right:} Analogous picture for Euclidean AdS$_3$. In AdS$_3$, the transformation consists of a spacetime rotation and a special conformal transformation.}\label{pic:geo}
\end{figure}

The entanglement entropies for cutoff (A)dS are given in eq. \eqref{genEEdS} and \eqref{genEEAdS}, respectively. By varying the turning point in the bulk, we are able to probe arbitrary subinvervals of the entanglement entropy on the cutoff sphere. The radius $R$ of the sphere on the cutoff slice appears in the expressions for the entanglemenet entropy only in combination with the cosine of the ending point on the cutoff surface $R\,\cos(\beta_\eps)$; we can simplify the expressions by introducing an effective radius $R_\text{eff}(\beta_\eps)=R\cos(\beta_\eps)$. Interestingly, in terms of the effective radius, the entanglement entropies match formally the expressions for the entanglement entropies of antipodal points on the sphere
 \begin{align}
d=2:\ & S_\text{EE}(\beta_\eps)=\frac{L}{2\,G_N}\,\arcsinhh\left(\frac{R_\text{eff}}{L}\right)\label{holoEEd2}\\
d=3:\ & S_\text{EE}(\beta_\eps)=\frac{L\,\pi}{2\,G_N}\,\eta\left(-L+\sqrt{L^2+\eta\,R_\text{eff}^2}\right)\label{holoEEd3}\\
d=4:\ & S_\text{EE}(\beta_\eps)=\frac{\pi\,L}{2\,G_N}\,\eta\left(R_\text{eff}\sqrt{\eta\,R_\text{eff}^2+L^2}-L^2\,\arcsinhh\left(\frac{R_\text{eff}}{L}\right)\right)\label{holoEEd4}\\
d=5:\ & S_\text{EE}(\beta_\eps)=\frac{\pi^2\,L}{6\,G_N}\left(2L^3+(\eta\,R_\text{eff}^2-2\,L^2)\,\sqrt{L^2+\eta\,R_\text{eff}^2}\right)\label{holoEEd5}\\
d=6:\ & S_\text{EE}(\beta_\eps)=\frac{\pi^2\,L}{12\,G_N}\!\left(\!R_\text{eff}\,\sqrt{L^2+\eta\,R_\text{eff}^2}\,(2\,\eta\,R_\text{eff}^2-\!3L^2)\!+\!3\,L^4\arcsinhh\!\left(\frac{R_\text{eff}}{L}\right)\!\right)\label{holoEEd6}\!.
\end{align}
\begin{figure}
    \centering
  \includegraphics[width=6cm]{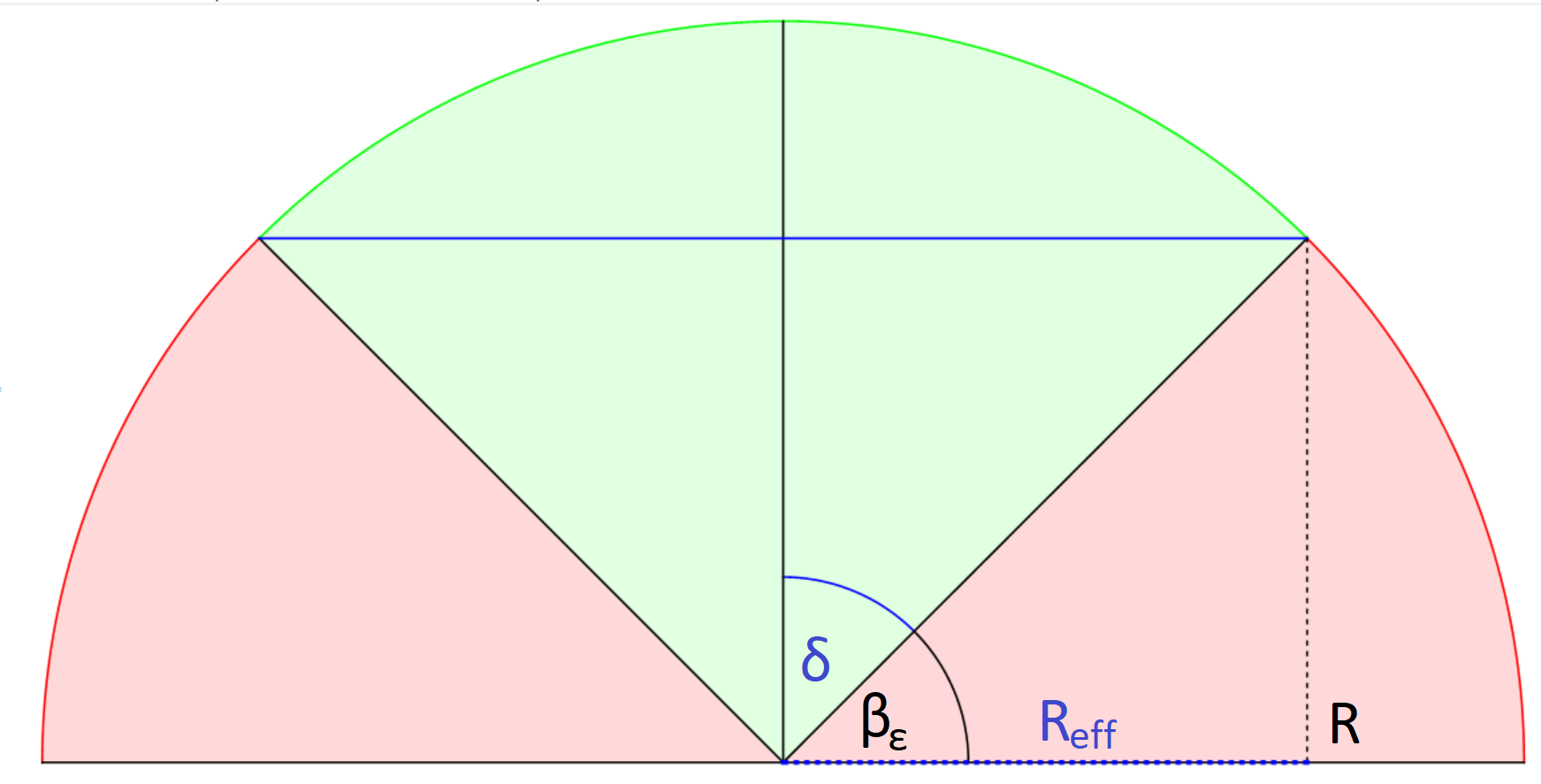}
       \caption{The effective radius $R_\text{eff}=R\,\cos(\beta_\epsilon)$ (blue) corresponds to the radius, where the points of the interval are antipodal by the basic definition of the cosine.}\label{pic:geo2}
\end{figure}
The results are easier to interpret from a geometric perspective in $d=2$ ~(see figure \ref{pic:geo2} and \ref{pic:geo}). 
The half circle in figure~\ref{pic:geo2} corresponds to antipodal points on a sphere with radius $R$. If the entangling surfaces end on the sphere with angle $\beta_\eps$, the interval corresponds to the region highlighted in green. With the basic definition of cosine, we see that the points are the endpoints of a smaller circle with radius $R_\text{eff}$.
In figure~\ref{pic:geo}, we notice that the entangling surfaces without cutoff correspond to great circles on the sphere (in the dS case). By introducing a cutoff surface (magenta), we cut away part of the entangling surface. With a rotation along the sphere, we may bring the entangling surface to the top of the sphere. On the top, the entangling surface correspond to antipodal points on the cutoff surface with smaller radius, $R_\text{eff}$, depicted in blue. A similar result holds for the AdS case. Instead of rotating the entangling surface to the top of the sphere, we have to perform a boost which brings the entangling surface on the top of the cone. The endpoints correspond to antipodal points for a boundary sphere with radius $R_\text{eff}$.

In the previous expressions, the angle $\beta_\eps$ indicates how much smaller the subintervall is compared to antipodal points on the sphere. It therefore makes sense to introduce the angle $\delta=\pi/2-\beta$, with $R_\text{eff}=R\cos\theta_\eps=R\sin\delta$. 
In the AdS case, the results for pushing the cutoff surface to the boundary have asymptote to the results without cutoff surface. Introducing the cutoff $\Lambda$, the entanglement entropies for AdS read
\begin{align}
 S^{d=2}_\text{EE}(\delta)&\!=\!\frac{L}{2\,G_N}\,\left(\log\!\left(\frac{2\,\Lambda \sin(\delta)}{L}\right)+\frac{L^2}{4 \Lambda^2\,\sin(\delta)^2}+\mathcal O\left(\frac{1}{\Lambda^3}\right)\right)\label{holoEEd2a}\\
 S^{d=3}_\text{EE}(\delta)&\!=\!\frac{L\,\pi}{2\,G_N}\!\left(\Lambda\,\sin(\delta)-L+\frac{L^2}{2\Lambda\,\sin(\delta)}+\mathcal O\left(\frac{1}{\Lambda^3}\right)\right)\label{holoEEd3a}\\
 S^{d=4}_\text{EE}(\delta)&\!=\!\frac{L\,\pi}{2\,G_N}\!\left(\Lambda^2 \sin(\delta)^2-\frac 12 L^2-L^2\log\!\left(\frac{2\,\Lambda \sin(\delta)}{L}\right)+\frac{L^2}{4 \Lambda^2\,\sin(\delta)^2}\!+\!\mathcal O\!\left(\frac{1}{\Lambda^3}\!\right)\!\right)\label{holoEEd4a}\\
 S^{d=5}_\text{EE}(\delta)&\!=\!\frac{\pi^2\,L}{6\,G_N}\!\left(\Lambda^3\,\sin(\delta)^3-\frac 32 \Lambda L^3\,\sin(\delta)+2L^3-\frac{9 L^4}{8\,\Lambda\,\sin(\delta)}+\mathcal O\left(\frac{1}{\Lambda^3}\right)\right)\label{holoEEd5a}\\
  S^{d=6}_\text{EE}(\delta)&\!=\!\frac{\pi^2\,L^3}{12\,G_N}\!\left(\!\frac{2\Lambda^4\sin(\delta)^4}{L^2}-2\Lambda^2\sin(\delta)^2-\!\frac{7\,L^2}{4}+3L^2\log\!\left(\!\frac{2\Lambda \sin(\delta)}{L}\!\right)\!+\!\mathcal O\!\left(\frac{1}{\Lambda^2}\!\right)\!\right)\!,\label{holoEEd6a}
\end{align}
which agrees with the result of Casini, Huerta and Myers \cite{Casini:2011kv}. In $d=2$, we can compare our result to the well known field theory result for a subsystem $\ell$ in a system of length $L$ \cite{Holzhey:1994we,Calabrese:2004eu,Calabrese:2005zw}
\begin{equation}
S=\frac{c}{3}\,\log\left(\frac{L}{\pi\, a}\,\sin\left(\frac{\pi \ell}{L}\right)\right),
\end{equation}
with the cutoff $a$ ($a\to 0$).
In the dS case, the quantity $\Delta$ vanishes for pushing the cutoff to the UV slice and we recover the full de-Sitter entropy as observed in \cite{Geng:2019bnn}.

\subsection{Renormalization and Generalized entanglement entropies}
On the one hand, in computations of the entanglement entropies in field theory or holography, we expect the result per s\'e to be divergent. The leading coefficients are sensitive to the cutoff, and the leading term is the so-called area term~\cite{Nishioka:2009un,Casini:2003ix,Plenio:2004he,Cramer:2005mx,Das:2005ah}. Subleading terms, in particular the logarithmic divergences, encode universal information characterizing the properties of the underlying QFT. On the other hand, in CFT calculations we consider renormalized quantities since they are universally well defined and make sense in the continuum limit. As we have seen in the holographic picture, the $T\bar T$ deformation acts as UV regulator which renders all quantities inherently finite. In principle, we are allowed to add an arbitrary amount of counterterms to the dual effective action describing the field theory but we are restricting ourselves to the standard counterterms obtained within the canonical holographic renormalization procedure~\cite{deHaro:2000vlm,Skenderis:2002wp}. If we now na\"ively compare the holographic entanglement entropies to entanglement entropies from field theory obtained from the renormalized action, we will find a seeming mismatch as the authors of \cite{Banerjee:2019ewu}. The counterterms of the field theory action on the cutoff slice have an impact on the entanglement entropy and thus must also be taken into account in the calculation of entanglement entropies by means of Ryu-Takayanagi surfaces (see for example \cite{Murdia:2019fax} for a discussion about this) \cite{Taylor:2016aoi,Cooperman:2013iqr,Emparan:1999pm,Faulkner:2013ana,Faulkner:2013ica}. They contribute to the entanglement entropy precisely on the points where the entangling surface reaches the cutoff surface.

The renormalized action for the gravity theory which also leads to the renormalized holographic energy-momentum tensor is given by eq. \eqref{actionS}. The counterterms living on the cutoff slice are higher curvature terms and we may compute the contributions to the holographic entanglement by calculating the Wald entropy associated with the counterterms.\footnote{An alternative approach is considering the counterterms outlined in~\cite{Anastasiou:2018mfk,Anastasiou:2018rla}.} The Wald entropy \cite{Wald:1993nt} is given by \cite{Jacobson:1993vj,Jacobson:1994qe,Brustein:2007jj}
\begin{equation}
    S_\text{Wald}=-2\pi\oint\dd^dx\,\frac{\delta\mathcal L}{\delta\tilde R_{abcd}}\,\hat\epsilon_{ab}\,\hat\epsilon_{cd},\label{wald}
\end{equation}
where $\hat\epsilon_{ab}$ are the binormals to the horizon. 
It is easier to rewrite the metric in terms of eq. \eqref{eq:metricdS} with $R(r)=L\,\sinhh(r/L), R\equiv R(r_c)$ and $\rho=\cos\phi $
\begin{equation}
    \dd s^2_\text{(A)dS}=\dd r^2+ R^2(r)\,\left(-(1-\rho^2)\,\dd \tau^2+\frac{\dd \rho^2}{1-\rho^2}+\rho^2\,\dd\Omega_{d-2}\right).
\end{equation}
In these coordinates the radial coordinate in the cutoff slice is $\rho$ and we can parametrize inward pointing quantities in terms of it. On the cutoff slice $r/L=r_c/L$ in the static patch, the $\hat\epsilon_{\tau\rho}$, are the binormals and we have to vary the Lagrangian in eq. \eqref{wald} with respect to $\tilde R_{\tau\rho\tau\rho}$ in order to find the Wald entropy associated with the co-dimension 2 entangling surface ending on the cutoff slice.

For the counterterms given in eq. \eqref{counter}, we find on the slice $r/L=r_c/L$ for an entangling surface ending on the surface at $\rho_\epsilon=\cos(\beta_\epsilon)$ with the effective radius \mbox{$R_\text{eff}=R(r_c)\,\rho_\epsilon$}
\begin{align}
    S_\text{W,ct}&=-\frac{1}{4\,G_N}R_\text{eff}^{d-2}\,\oint \sqrt{h}\,\left(\frac{c_2\,L}{d-2}+\frac{c_3\,L^3}{(d-4)(d-2)^2}\left(\tilde R-h^{ab}\tilde R_{ab}-\frac{2d}{4\,(d-1)}\tilde R\right)\right)\nonumber\\
    &=-\frac{\pi^{(d-1)/2} \,R_\text{eff}^{d-2}}{2\,(d-2) \,G_N\,\Gamma((d-1)/2)} \left(c_2\,  L-\frac{c_3\, L^3}{2\,(d-4)\, R_\text{eff}^2}\left(d-2\right)\right),\label{ctwald}
\end{align}
where $h_{ab}$ is the induced metric on the unit boundary sphere and where we used that on the cutoff slice $\tilde R=d\,(d-1)/R_\text{eff}^2$ and $R_{ab}=(d-1)/R_\text{eff}^2\,h_{ab}$. 
Evaluating the expression in eq. \eqref{ctwald} for $3\le d\le 6$, we find
\begin{align}
    S^{d=3}_\text{W,ct}&=-\frac{\pi\,L\,R_\text{eff}}{2\,G_N},\label{wald1}\quad &
    S^{d=4}_\text{W,ct}&=-\frac{\pi\,L\,R_\text{eff}^2}{2\,G_N}, \\
    S^{d=5}_\text{W,ct}&=\frac{\pi^2\,L}{8\,G_N}\left(-\frac{4\,R_\text{eff}^3}{5}+2\,R_\text{eff}\,L^2\right), \quad &
    S^{d=6}_\text{W,ct}&=\frac{\pi^2\,L}{8\,G_N}\left(-\frac{4\,R_\text{eff}^4}{3 }+\frac{4\,R_\text{eff}^2\,L^2}{3}\right)\label{wald4}.
\end{align}
In $d=2$, the contributions of the counterterms to the entanglement entropy vanish since the counterterm acts as a boundary cosmological constant.

\section{$T\bar T$ deformations from field theory in DS/dS}
In this section, we derive the corresponding entanglement entropies from field theory in general dimensions. So far the results for field theories living on the cutoff slice of (A)dS$_3$ were found in the literature in \cite{Donnelly:2018bef} and we reviewed them in section \eqref{sec::EEandfieldtheorystu}. For higher dimensions, the entanglement entropies for $T\bar T$ deformed theories dual to AdS$_{d+1}$ were derived in \cite{Banerjee:2019ewu}. However, as already alluded to in the last section, the authors only found a match up to missing area terms. In this section, we generalize the calculation for the field theory dual to dS$_{d+1}$ to general dimensions and demonstrate that we are able to find an exact match between the results from cutoff (A)dS$_{d+1}$ and the corresponding field theories.

\subsection{The $d$ dimensional deforming operator}
In section \ref{section:stresstensor}, we gave the explicit expressions for the holographic energy momentum tensor $T_{ij}$ \eqref{EMtensor} with curvature counter terms $C_{ij}$ \eqref{EMcounter} for spacetimes up to $6+1$ dimensions. The trajectory of the $T\bar T$ deformation may be expressed in terms of the trace of the energy momentum tensor and we can define a deforming operator according to
\begin{equation}
    X_d=-\frac{1}{d\,\lambda_d}\langle T^\mu_\mu\rangle.
\end{equation}
We drop all $\langle \dots\rangle$ from now on and implicitly assume to take the expectation values of the operator valued quantities since we are working in the large $N$ limit of the field theory which exhibits a factorization property. In order to eliminate the extrinsic curvature from the expression for the energy momentum tensor eq. \eqref{EMtensor}, we consider the trace of the radial Einstein equation of the $(d+1)-$dimensional gravitational theory in with metric \eqref{Adsds} 
\begin{equation}
K^2-K^{\mu\nu}K_{\mu\nu}-\eta\,\frac{d(d-1)}{L^2}-\tilde R=0,\label{flow}
\end{equation}
where $\tilde R$ is the $d$ dimensional Ricci scalar, describing the curvature on the cutoff slice $r=r_c$.
Analogous to the preceding section, we introduced $\eta=1$ for AdS and $\eta=-1$ for dS. Using eq. \eqref{flow}, to eliminate the extrinsic curvature we find for the trace of the energy momentum tensor
\begin{equation}
    X_d=\!\left(T_{\mu\nu}\!+\!\frac{\alpha_d}{\lambda_d^{\frac{d-2}{d}}}C_{\mu\nu}\!\right)^{\!2}\!\!\!-\!\frac{1}{d-1}\left(T^\mu_\mu\!+\!\frac{\alpha_d}{\lambda_d^{\frac{d-2}{d}}}C_\mu^\mu\!\right)^{\!2}\!\!+\!\frac1d \frac{\alpha_d}{\lambda_d^{\frac{2(d-1)}{d}}}\!\left(\frac{d-2}{2}\tilde R\!+\!C^\mu_\mu\!\right)\!+\!\frac{(d-1)(\eta-1)}{4\,d\,\lambda_d^2}.\label{eq:ddimedefo}
\end{equation}
For $d=2$, the $C_{\mu\nu}$ vanish and we find the results derived in eq. \eqref{eq:traceEMTT}. Analogous to the $d=2$ case, the deforming operator for the field theory dual to dS$_{d+1}$ exhibits an extra term compared to the AdS case.

The field theory parameters are related to the parameters on the gravity side by~\cite{Banerjee:2019ewu}
\begin{equation}
    \lambda_d=\frac{4\pi\,G_N\,L}{d},\quad \alpha_d=\frac{L^{2(d-1)/d}}{(2d)^{\frac{d-2}{d}}\,(d-2)\,(8\pi\,G_N)^{2/d}},\quad L^2=2d\,(d-2)\,\alpha_d\,\lambda^{2/d}_d.\label{gravityfieldtheory}
\end{equation}
In particular, in a two-dimensional CFT, the central charge $c$ is related to the gravity theory quantities by eq. \eqref{eq:TTbarcentralcharge}.

\newpage
\subsection{From sphere partition functions to entanglement entropy}
Analogous to the case $d=2$ explained in section \ref{sec::EEandfieldtheorystu}, we compute the entanglement entropy in the field theory by means of the sphere partition function. The energy momentum tensor is proportional to the metric on the sphere $T_{\mu\nu}=\omega_d(R)\,\gamma_{\mu\nu}$, where the function $\omega(R)$ depends on the radius $R$ of the sphere. Taking the trace of this expression, the sphere partition function reads
\begin{equation}
   R\frac{\partial}{\partial R}\,\log Z_{S^d}=-d\,\int d^dx\,\sqrt{\gamma}\,\omega_d(R),\label{partf}
\end{equation}
from which we may compute the entanglement entropy for antipodal points on the sphere using the replica trick by
\begin{equation}
S_\text{d,EE}=\left(1-\frac{R}{d}\frac{d}{dR}\right)\log Z_{S^d}.\label{EEreplica}
\end{equation}
The $n-$folded cover of the sphere needed for the replica trick is given by~\cite{Donnelly:2018bef,Banerjee:2019ewu}
\begin{equation}
    \dd s^2=R^2\left(\dd\beta_1^2+\sum_{i=2}^{d-1}\prod_{j=1}^{i-1}\cos(\beta_j)^2\,\dd\beta_i^2+n^2\,\prod_{j=1}^{d-1}\cos(\beta_j)^2\,\dd\beta_d^2 \right),\label{nfoldcover}
\end{equation}
with $\beta_j\in[-\pi/2,\pi/2]$ for $j=1,\ldots,d-1$ and $\beta_d\in[0,2\pi]$. By evaluating the deforming operator for the metric of the boundary sphere and using the flow equation \mbox{$d\,\omega_d(R)=T_\mu^\mu=-d\,\lambda_d\,X_d$}, we find a quadratic equation in the proportionality function $\omega_d(R)$. In the following, we denote the two signs of the solution by $\bm{s}$
\begin{align}
     \omega_2(R)\ =&\ \frac{1+\bm{s}\sqrt{\eta+\frac{c\,\lambda_2}{3\pi\,R^2}}}{4\,\lambda_2},\\
     \omega_{d>2}(R)=&\ \frac{(-1+d)}{4\,d\,R^4\,\lambda_d}\left(\!2\,(d-2)\,d\,R^2\,\lambda^{2/d}\,c_2\,\alpha_d-(d-2)^2\,d^2\,\lambda^{4/d}\,c_3\,\alpha_d^2\right.\nonumber\\ 
&\left.+ 2R^3\,\left(R+\bm{s}\,\sqrt{\eta\,R^2+2\,(d-2)\,d\,\lambda^{2/d}\alpha_d}\right)\right).\label{omegaeq}
\end{align}
From \eqref{omegaeq}, it is easy to obtain the entanglement entropies. The recipe goes as follows: Use the proportionality function $\omega_d(R)$ eq. \eqref{omegaeq} to compute the partition function according to eq. \eqref{partf}. Fix the integration constant by demanding $\left.\log Z\right|_{R=0}=0,$ leading to a trivial theory in the UV. For antipodal points on the sphere, the entanglement entropy follows with eq. \eqref{EEreplica}.\newpage
\section{Entanglement Entropy from field theory in general dimensions}
With the recipe outlined in the last section it is straightforward to compute the entanglement entropies. The entanglement entropies and their holographic counter-parts obtained from the gravitational theory read (for more details see appendix \ref{app:EETT}).
    \begin{align}
& S^{d=2,}_{\text{QFT}}=S^{d=2}_{\text{holo}}=
    \frac{c}{3}\,\arcsinhh\left(\frac{\sqrt{3\,\pi}\,R}{\sqrt{c\,\lambda_2}}\right)\label{EEd233}\\
& S^{d=3,}_{\text{QFT}}=S^{d=3}_{\text{holo}}=\frac{4\,\pi^2\,t_3}{\lambda_3}\left(- R-\eta\,\sqrt{6t_3}+\eta\,\sqrt{\eta\,R^2+6\,t_3}\right) \left(\sqrt{6}-\frac{6 \sqrt{t_3}}{\sqrt{\eta\, R^2+6\, t_3}}\right)\\
 &S^{d=4}_{\text{QFT}}\!=S^{d=4}_{\text{holo}}\!=\frac{8\pi^2\,t_4}{\lambda_4}\left(R\left(-R+\eta\,\sqrt{\eta\, R^2+16\,t_4}\right)-16\,\eta\,t_4\,\arcsin(h)\left(\frac{R}{4\,\sqrt{t_4}}\right)\right)\\
 &S^{d=5}_{\text{QFT}}\!=S_{\text{holo}}^{d=5}\!=\!\frac{4\pi^3\,t_5}{\lambda_5}\! \left(\!-R^3\!+45 R\,t_5\!+\eta R^2\sqrt{\eta R^2+30 t_5}+\!60t_5\!\left(\!\sqrt{30\,t_5}\!-\!\sqrt{\eta R^2+30 t_5}\right)\!\right) \\
& S_{\text{QFT}}^{d=6}\!=S_{\text{holo}}^{d=6}\!= -\frac{16\,\pi^3\,t_6}{3\lambda_6}\left(R\left(R^3-48\,R\,t_6-\eta\,R^2\sqrt{\eta\,R^2+48\,t_6}+72\,t_6\,\sqrt{\eta\,R^2+48\,t_6}\right)\right.\nonumber\\
    &\ \quad\quad\quad\quad\quad\quad\left.-3456\,t_6^2\,\arcsinhh\left(\frac{R}{4\,\sqrt{3\,t_6}}\right)\right),\label{EEholoEF}
   \end{align}
   where we translated the results we obtained in the gravitational theory into field theory language by using eq. \eqref{gravityfieldtheory} and added the contributions of the counterterms on the slice encoded in the Wald entropy.
 For the interested reader, we provide some additional steps of the calculation in appendix \ref{app:EETT}. We see that the entanglement entropies in both sides of the proposed duality -- cutoff (A)dS and $T\bar T$ deformed field theories -- match perfectly. Since the counter terms are finite for the $T\bar T$ deformed theory, our results depend on the cutoff and hence on the regularization scheme. We can remove the dependence on the cutoff and consider only cutoff independent quantities by taking daking derivatives of the entanglement entropy according to~\cite{Liu:2012eea,Liu:2013una}
\begin{equation}
    \mathcal S_\text{R,EE}(R)=\begin{cases}
    \frac{R}{(d-2)!!}\,R\,\frac{d}{dR}(R\,\frac{d}{dR}-2)\ldots(R\,\frac{d}{dR}-(d-2))\,S_\text{EE}\quad\quad\quad\quad\text{d even},
    \\ \frac{R}{(d-2)!!}(R\,\frac{d}{dR}-1)(R\,\frac{d}{dR}-3)\ldots(R\,\frac{d}{dR}-(d-2))\,S_\text{EE}\quad\quad\text{d odd.}\end{cases}\label{runningC}
\end{equation}
The results for $\mathcal S_\text{R,EE}(R)$ are also outlined in appendix \ref{app:EETT}.

\chapter{Conclusions \& Outlook}\vspace{-0.2cm}
In this thesis, we investigated the non-equilibrium properties of strongly coupled field theories within the framework of holography. Specifically, we investigated the impact of broken spacetime symmetries and chiral anomalies on the transport behavior of strongly coupled field theories and examined the entanglement entropy within non-equilibrium scenarios in de-Sitter and deformed QFTs. In the following, we give a detailed summary of the central results and discuss the further directions in the outlook. 
\vspace{-0.18cm}
\subsection*{Broken spacetime translations}
The main long term goal in considering broken spacetime symmetries in hydrodynamic is to extend the hydrodynamic prescription to explicitly broken symmetries. Within holography, we can directly compute the dispersion relations and transport coefficients and thus gain useful insights into how to construct the prescription. First, we considered the case of spontaneously broken translational symmetry. Even though the hydrodynamic description for spontaneously broken translations is considered in textbooks~\cite{chaikin2000principles,PhysRevA.6.2401} and was recently reviewed in \cite{Delacretaz:2017zxd,Ammon:2019apj}, we set our sight on verifying this prescription in our holographic model in the longitudinal sector of the solid model and the transverse and longitudinal sector of the fluid model~\cite{Ammon:2019apj,Baggioli:2019abx}. A match in the transverse sector of the solid model was established in~\cite{Alberte:2017oqx}.
On the field theory side, we established a comprehensive formulation of the hydrodynamic dispersion relations in the spontaneous case. On the gravity side, we were able to check the dispersion relations obtained from hydrodynamics and to compute the corresponding transport coefficients. We find excellent agreement with the predictions from hydrodynamic in the transverse sector as well as reasonable agreement for the sound modes the longitudinal sector of the model. To our surprise, we find a significant disagreement for the purely diffusive mode which appears to be shifted. By considering the decoupling limit analytically, we were able to verify that this disagreement between the hydrodynamic description and the numerical prediction from holography is indeed systematic and not due to an error in the computation. In the final stages of this thesis, we were able to resolve the disagreement. The authors of~\cite{Donos:2019hpp,Armas:2019sbe} showed that we have to take a novel transport coefficient referred to as \textit{configuration pressure} into account and hence the textbook treatment of spontaneously broken translations in hydrodynamics has to be revised. In \cite{Ammon:2020xyv}, we demonstrated that the holographic results derived for~\cite{Ammon:2019apj,Baggioli:2019abx} perfectly match the hydrodynamic description of~\cite{Donos:2019hpp,Armas:2019sbe,Armas:2020bmo} and the mismatch is no longer present. The novel transport coefficient takes into account that the background configuration might be strained, as it is indeed the cases for the holographic models we investigated in~\cite{Ammon:2019apj,Baggioli:2019abx}. However, its temperature derivative contributes even in unstrained states.

In the second part of the discussion of chapter \ref{chapter:brokenspacetime}, we laid out important steps towards a realization of a hydrodynamic prescription including momentum dissipation and phase relaxation. In particular, we considered explicitly and pseudo-spontaneously broken translation in the longitudinal and transverse sector. For this kind of symmetry breaking, we observe momentum dissipation as well as a phase relaxation mechanism for the phonons. In the pseudo-spontaneous limit, the phonons acquire a mass term similar to the pion in chiral symmetry breaking. We verified that the light (pseudo)-phonons satisfy the so-called Gell-Mann-Oakes-Renner relations~\cite{PhysRev.175.2195}. We demonstrate that the holographic results agree with an heuristically extended hydrodynamic prescription where we put in the momentum dissipation and phase relaxation by hand.

Within linear response -- for example, in the hydrodynamic regime -- we have some access to the physics of broken translations in terms of effective field theories methods. Beyond linear response, however, this intuition completely breaks down since we have to study the full nonlinear real time dynamics of the strongly coupled field theory in order to characterize the out-of-equilibrium behavior. We force the system out of its equilibrium state by driving it periodically for different strengths and frequencies of the periodic driving. We may isolate the imprint of the translational symmetry breaking by comparing to the results for periodically driven holographic CFTs with translational invariance (for example \cite{Rangamani:2015sha}). Within the linear response regime, the spontaneously broken translational invariance leads to a tilted elliptic shape of the Lissajous figures. The Lissajous figures include the information about the dissipation mechanisms in the CFT with broken translations. Beyond the linear regime, we also break the discrete time translations of the system. Due to the broken translations, the system shows viscoelastic properties. In particular, we observe nonlinear elasticity in the nonlinear regime in contrast to periodically driven CFTs with translational symmetry. We found a strain hardening mechanism as typically found for rubber-like systems, contradicting the claim that the homogeneous holographic massive gravity models describe strange metals.

The main motivation in this thesis was to address fundamental questions like how do broken symmetries affect the hydrodynamic description and how to extend standard textbook hydrodynamics in this regards. Some of the results, such as the broken translational invariance, may also be interpreted within the AdS/CMT program. For example, the far-from-equilibrium behavior in periodically driven CFTs in presence of spontaneously broken translations describes so-called large oscillatory shear tests on solids. Not much is known about the theoretical description of these processes and our holographic results might help to get a better intuition about the underlying physical processes. 
\vspace{-0.18cm}
\subsection*{Anomaly induced transport in strong magnetic fields}
 In chapter \ref{sec:anotranspo}, we considered the transport coefficients of a strongly coupled anomalous QFT subject to a strong (axial) magnetic field. In particular, we demonstrated that in the presence of the chiral anomaly we have to take into account several novel transport coefficients constrained by the chiral anomaly. The direct computation of those transport coefficients in Einstein-Maxwell-Chern-Simons theory showed that these transport coefficients are non-zero and thus contribute to the transport properties of the system under consideration. Within this thesis, we focused on the computation of the shear and bulk viscosities as well as perpendicular resistivities in strong magnetic fields and stressed the effects of the chiral anomaly. This work is an important step within the holographic community since the underlying Einstein-Maxwell-Chern-Simons theory is a simple top-down construction from string theory and we thus have control over the dual CFT. The AdS/CFT correspondence is particularly well suited to study transport in strongly coupled anomalous QFTs since the physics is universal in a sense that the coefficients in the anomalous currents for example are totally determined by the anomaly. Note for example that the parallel Hall viscosity is a non-dissipative transport coefficient and may be topologically protected.
\vspace{-0.18cm}
\subsection*{Entanglement Entropy, (A)dS, and $T\bar T$ deformations}
So far, two sorts of entangling surfaces have been known in de Sitter: One measures the entanglement entropy (EE) between the two CFTs on the UV slice \cite{Dong:2018cuv}, the other one measures the EE across the horizon on the UV slice in the static patch. In chapter \ref{section:EEandtheswampland}, we constructed a one-parameter family of solutions which interpolates between the two concepts. Surprisingly, all entangling surfaces of the one parameter family reproduce the dS entropy correctly. We used the notion of EE to investigate possible consistency requirements on matter fields in dS gravity (swampland bound). Starting from the assumptions that the dS entropy is the largest possible EE, we explored the parameters in which matter fields obey this bound. Remarkably, we found that any matter obeying the Null Energy condition violates our proposal.

Starting from the one-parameter family of entangling surfaces, we derived the EE for generic subintervals in the presence of a hard radial cutoff in (A)dS. In the field theory, this scenario corresponds to deforming the field theory with a so-called $T\bar T$ deformation which is an exactly solvable irrelevant deformation. Concretely, we computed the EE for $T\bar T$ deformed CFTs on a sphere for generic intervals in general dimensions dual to (A)dS. Intriguingly, we found that we may express the results for generic subintervals in terms of the result for antipodal points \cite{Donnelly:2018bef} by introducing an effective radius for the sphere. Additionally, we generalized the field theory calculation for a CFT dual to dS to general dimensions. Moreover, we solved the confusion in the literature about a seeming mismatch between the EE on the field theory side and the holographic result obtained from the gravitational theory and were able to demonstrate that both sides match perfectly.
\vspace{-0.08cm}
\section*{Outlook}
In this thesis, we advanced our knowledge about non-equilibrium aspects in several regards. 
In the last section, we want to discuss the future direction building upon the central results of this thesis.
\vspace{-0.18cm}
\subsection*{Hydrodynamic description for broken spacetime translations}
In particular, we established a consistent framework to study spontaneously broken translational invariance within hydrodynamics and holography. This is first basic steps towards a consistent ``hydrodynamic description'' to model explicitly and pseudo-spontaneously broken spacetime symmetries. The road map to this description is already laid out in the holographic results obtained within this thesis. Within holography, we worked out the response in terms of QNMs and compared to heuristic extensions of hydrodynamics which include phase relaxation and momentum dissipation ``put in by hand''. The consistent framework may help to answer some problems important for the AdS/CMT program. One natural extension is to study the effects of spontaneous translational symmetry breaking at finite density within an external magnetic field in terms of so-called magneto-phonons and magneto-plasmons. Beyond linear response, the spectrum of questions is even richer. In periodically driven systems exist non-equilibrium states which are possibly analytically tractable -- the non-equilibrium steady states. These special states in the non-equilibrium dynamics are states in which the system is quasi-static but the continuous driving is necessary to keep the system in this state; the system dissipates all energy introduced by the driving at the same rate. The long-term goal is to classify the non-equilibrium steady states. It would be interesting to investigate the existence and properties of the steady states within the non-equilibrium dynamics of systems with broken translations. In the context of holography it would be especially interesting to apply the procedure of computing the spatial collective modes as outlined in \cite{Sonner:2017jcf}. 
Another intriguing, albeit a more technically challenging direction, is to consider quenches in the strength of the spontaneous symmetry breaking and investigate the resulting response with regards to universal behavior. 
\thispagestyle{empty}
\vspace{-0.18cm}
\subsection*{Chiral Anomalies in strongly coupled QFTs}
The full classification of the hydrodynamic transport coefficients for a four dimensional anomalous QFT subject to a strong magnetic field opens a wide range of interesting follow up questions. The next step towards the full nonlinear response is to consider the frequency and momentum dependence of the transport coefficients. The frequency dependent coefficients allow us to quantify the transport behavior in oscillating magnetic and electric fields.

The derivation of an upper bound on the isotropization time in supersymmetric Yang-Mills plasma from holography was a big success \cite{Chesler:2008hg}. However, the authors chose a simple bottom-up model consisting of the 5D Einstein-Hilbert action. Follow-up works included an external magnetic field and finite charge density \cite{Fuini:2015hba,Cartwright:2019opv} but still are lacking a Chern-Simons term which is unavoidable in consistent top-down constructions giving rise to (supersymmetric) four-dimensional QFTs. 
By studying the full time-dependence in the Einstein-Maxwell-Chern-Simons setup, we may study the equilibration of the strongly coupled anomalous QFT within a well-defined truncation from top-down models in string theory. It is interesting, to study whether and how the holographic magnetic quantum critical system will thermalize, in particular close to the quantum critical point and how the anomaly coefficients affects the equilibration time scales. 

Weyl semimetals may be realized by applying a circular polarised laser on a Dirac semimetal~\cite{H_bener_2017}. In holography, we may mimick this by considering the time dependence of Maxwell-Chern Simons theory in presence of a rotating electric field.
Quantum states in such a time periodic driven system can be described by Floquet theory which is restricted to small amplitudes and frequencies. However, by studying the full time dependence within holographic setup of~\cite{Landsteiner:2015lsa}, we overcome the usual restrictions of Floquet theory. 
Within the holographic mode, we can study non-equilibrium steady states in periodically driven anomalous QFTs, specially their stability in dependence on the applied electric fields. It is tempting to investigate, whether we observer chaotic behavior in the regime of large amplitudes.
\vspace{-0.18cm}
\subsection*{Quantum gravity and matter in de-Sitter and the $T\bar T$ deformation}
In chapter \ref{section:EEandtheswampland}, we discussed a framework to study quantum gravity in dS -- the DS/dS correspondence. Although the conjecture has been put on more solid footing over the last years, there are plenty of open questions. For example, it would be interesting to understand the exact physical interpretation of the one-parameter family of entangling surfaces in the dual CFTs. We have some intuition about the special cases, where the entangling surface corresponds to the cosmological horizon or is located on the UV brane~\cite{Dong:2018cuv}. However, the precise interpretation of the one-parameter family of entangling surfaces which interpolates between those two concepts remains an open question. With the notion of EE in DS/dS, we set out to derive consistency conditions on the matter content in dS$_{d+1}$ but surprisingly any matter obeying the null energy condition violates this bound! There are several ways out of this dilemma. For one, the framework may not be well defined. Either it is not possible to define a theory of quantum gravity with matter on dS or DS/dS is not the correct framework to model it. On a more confident note, we could conclude that the bound was simply too strong.

The EE in the nominator, $S_{d+1,\text{global}}$, accounts for the EE on the entire spatial volume. This EE is twice as large as the EE in the dS static patch. For a single observer living on the UV brane of dS$_d$ the maximum entropy available corresponds to $S_\text{dS}$ and this observer thus does not have sufficient information to fully reconstruct the higher dimensional dS$_{d+1}$ geometry. To do so, we require some information from beyond the horizon. This is a surprising fact but must not necessarily be an inconsistency but may rather be another fascinating property of the DS/dS correspondence.

For future investigations, we may postulate a less stringent bound
\begin{equation*}
    \label{futurebound}
\frac{S_{d+1,\text{static}}}{S_\text{dS} } \leq 1 \quad \Leftrightarrow \quad 
\frac{S_{d+1,\text{global}}}{S_\text{dS} } \leq 2,
\end{equation*}
which is, however, the minimum consistency requirement we have to demand under any circumstances. Within this bound, we are requiring the EE between the left and right CFT within the static patch to be less than the dS$_d$ entropy. It would be interesting, to quantify the impact of this consistency requirement in the context of constraints on the potential of matter in dS. The one-parameter family of entangling surfaces which all yield the same EE is intimately connected to the symmetries of dS and thus not present if we deform the geometry away from dS. For a Dirichlet wall in dS corresponding to a hard radial cutoff, the single member of the one-parameter family yield different EE depending on their turning point in the bulk. This was in part topic of the EE in the context of $T\bar T$ deformed QFTs. 

We derived the holographic EE for general subintervals for $T\bar T$ deformed QFTs. It would be very interesting, to also extend the field theory calculation to general geometries. To realize this, we have to evaluate the associated partition functions on manifolds with conical defects. Furthermore, we alluded to the construction of \cite{Gorbenko:2018oov}, relating the QFTs in terms of the DS/dS correspondence to the AdS/CFT correspondence by $T\bar T$ deformations. Working out this construction explicitly is an important future direction for understanding quantum gravity in dS within the framework of holography. 
\appendix

\chapter{Numerical methods}\label{ref:appnumrel}
\pagenumbering{roman}
\setcounter{page}{1}
In this chapter, we review the numerical methods we applied throughout this thesis in order to solve the various differential equations and eigenvalue problems efficiently and accurately following~\cite{Boyd1989ChebyshevAF} closely. For problems in numerical holography, spectral methods have the big advantage that they solve the equations of motions globally, i.e. we can simultaneously demand boundary conditions on both ends of the domain. This is a big advantage compared to Finite-difference or shooting methods where we have to vary the initial conditions on one side in order to find the appropriate function values on the other side of the domain. Another point in favor of applying a spectral method is their fast convergence rate; we have a very robust and accurate numerical technique at hand which is not necessarily a lot more complicated than using a shooting method. Lastly, using spectral methods provides us with very good control of about the quality and accuracy of the numerical solution which is essential for solving differential equations numerically. To quote Boyd himself: ``One must watch the convergence of a numerical code as carefully as a father watching his four year old play near a busy road.''
\section{(Pseudo)-spectral methods}
In order to get an intuition about how spectral methods work, we consider the differential equation
 
where $\mathcal L$ is a differential operator and $u(x)$ the solution satisfying $\mathcal L\,u(x)-f(x)=0$. We can expand the solution in terms of a basis $\{\phi_n(x)\}$ by formally writing $u(x)=\sum_{n=0}^\infty\, c_n\phi_n(x)$. The basic idea of spectral methods is to approximate the exact solution $u(x)$ by a finite number $N$ of basis polynomials $\phi_n(x)$
\begin{equation}
    u(x)\approx u_N(x)=\sum\limits_{n=0}^N\,c_n\,\phi_n(x)\label{eq:spectralrep}
\end{equation}
and find the coefficients $\{c_n\}_{n=0,\dots,N}$ which minimize the residuum $R(x;c_n)$
\begin{equation}
    R(x;c_0,c_1,\dots,c_N)=\mathcal L\,u_N(x)-f(x)\label{eq:res}
\end{equation}
The first important step is to choose the appropriate set of basis function and to discretize the coordinates $x_i$ in order to make it tractable for computer algorithms.

Since we know the expressions of the basis functions analytically, we can re-write their first and second derivative by introducing derivative matrices $\hat D$. In the given basis, the derivative of the basis functions is given as a linear combination of the basis functions i.e. $\phi'_m(x)=\sum_{n=0}^N\hat D_{mn}\phi_n(x), \ \phi''_m(x)=\sum_{n,l=0}^N\hat D_{mn}\hat D_{nl}\,\phi_l(x)$.  With the differentiation matrices, we can rewrite derivatives so that they act on the coefficients 
\begin{align}
    &u'(x)\, \approx\sum\limits_{n=0}^N c_n\,\phi_j'(x)\,=\sum\limits_{n,m=0}^Nc_n\,\hat D_{nm}\phi_m(x)=\sum\limits_{n=0}^Nc_n'\,\phi_n(x),\\
    &u''(x)\approx\sum\limits_{n=0}^N c_n\,\phi_j''(x)=\sum\limits_{n,m,l=0}^Nc_n\,\hat D_{nm}\hat D_{ml}\,\phi_l(x)=\sum\limits_{n=0}^Nc_n''\,\phi_n(x).
\end{align}
The difference between spectral- and pseudo-spectral methods is that the former uses the coefficients in order to minimize the residuum \eqref{eq:res} while the latter uses the function values $u_i$ evaluated on the gridpoints $x_i$. In contrast to finite differences, the gridpoints are not necessarily equidistant but given in terms of the zeros and extrema of the basis functions $\phi_n$. The relation of the function values $u_i$ and the coefficients $c_i$ is $u_i=\sum_{j=0}^Nc_j\,\phi_j(x_i)$. 

The best choice for the basis functions in the homogeneous case are the so-called Chebychev polynomials given by
\begin{equation}
    T_k(x)=\cos(k\,\arccos(x)).
\end{equation}
In the Chebychev basis, we can choose between different grids in order to discretize the coordinate. There are the Chebychev-Radau grids which do not include either the left or the right boundary of the domain, the Chebychev-Gau\ss grid where both boundaries of the domain are excluded and the Chebychev-Lobatto which includes boundaries of the domain. In AdS/CFT it is very convenient when we can impose boundary conditions at conformal boundary and know the numerical solution there. Note that the boundary and the horizon are regular singular points to the equations of motion in AdS/CFT and the differential equation is degenerate there. This means that we technically do not have to impose boundary conditions there and we can use the Chebychev-Gau\ss and Chebychev-Radau grids. Even though these grids are interesting, we restrict our discussion to the Chebychev-Lobatto grid, in particular since we want the conformal boundary to be part of the integration domain. The Chebychev-Lobatto gridpoints are given by 
\begin{equation}
    x_i=\cos\varphi_i=\cos\frac{\pi \,i}{N}
\end{equation}
with the differentiation matrix given by
\begin{equation}
    \hat D_{mj}\begin{cases}-\frac{2N^2+1}{6}, & m=j=N,\\
    \frac{2N^2+1}{6}, & m=j=0,\\
    -\frac{x_j}{2\,(1-x_j^2)}, &0<m=j<N,\\
    \frac{\kappa_m}{\kappa_j}\frac{(-1)^{m-j}}{x_m-x_j},& m\not =j, 
    \end{cases}
\label{eq:diffmatrix}
\end{equation}
where we defined
\begin{equation}
    \kappa_j=\begin{cases}2,& \text{for }j=0, j=N\\
    1,& \text{for }j=1,\ldots N-1.\end{cases}
\end{equation}
The differentiation matrix for the second derivative is simply given by applying the matrix for the first derivative \eqref{eq:diffmatrix} twice. Note that by choosing non-equidistant gridpoints, we avoid the Runge-phenomenon which appears in interpolations on equidistant grids. We demand that the residuum vanishes exactly on the collocation points $x_i$ and the numerical solution is thus exact on the gridpoints (note that the conformal boundary is one of the gridpoints).
For periodic problems it is best to use a Fourier grid. For spherical symmetrical problems, we recommend using spherical harmonics as basis functions for latitude and longitude. In these cases, we refer the interested reader to~\cite{Boyd1989ChebyshevAF,Grandclement:2007sb,canuto2007spectral,trefethen2000spectral}. The attentive reader might have noticed that the Chebychev-Lobatto gridpoints are defined on $x\in[-1,1]$ while the radial coordinate of AdS is usually in the interval $u\in[0,1]$ where the conformal boundary is located at $z=0$. In practice, it is thus convenient to map the gridpoints and differentiation matrices according to 
\begin{equation}
    z_i=\frac{x_i+1}{2}\in[0,1], \quad \tilde D=\left(\partial_x\left(\frac{x+1}{2}\right)\right)^{-1}\hat D=2\,\hat D, \quad\tilde D^2=\tilde D\,\tilde D=4\,\hat D\,\hat D.\label{eq:diffmatr2}
\end{equation}
The corresponding Chebychev polynomials are given by $T_k(z)=\cos(k\,\arccos(2 z-1))$.

\section{Pseudospectral solutions to Boundary Value Problems}
After discussing the discretization of differential equations in terms of spectral methods, we now work out how to solve linear boundary value problems in practice. The step to solving nonlinear boundary values is formally not more complicated since the nonlinear are solved by iteratively solving the linear problem. We discuss them in section~\ref{sec:nonlinear}.
\subsection*{Boundary conditions}\label{app:bc}
Before we impose boundary conditions on the differential equation, we have to understand the difference between behavioral and numerical boundary conditions. Periodicity, for example, is a behavioral boundary condition. We require that the solution satisfies $u(x)=u(x+2\pi)$ but that does not impose any specific value on the function $u(x)$ or its derivative. For periodic functions, we choose sines and cosines as basis functions which obviously satisfy the periodicity condition. By expanding the functions in terms of this basis, our numerical solution is automatically periodic and we do not have to impose any explicit boundary conditions since the periodicity requirement is inherently satisfied by the basis functions. 
Even more interestingly, the differential equation
\begin{equation}
    z\,(1-z)\,\partial^2_z u(z)-(z+1)\,\partial_z u+5\,u(z)=3, \ z\in[0,1],\label{eq:sssss}
\end{equation}
is singular on both endpoints. However, the Chebychev polynomials are individually analytic at $z=\{0,1\}\ (x=\pm1)$ and so are linear combinations of them. This leads to the important insight that we can solve singular differential equations with an exponential rate of convergence (we get to the meaning of that in the next section) without imposing any additional constraints!
This is extremely convenient in the context of AdS/CFT. As we learned, the equations of motions we come across in the context of the AdS/CFT correspondence have regular singular points on both ends of the intervals. By solving these equations in terms of a Chebychev basis, we cure those singularities in a very elegant way and are able to construct numerical solutions converging exponentially despite the singular endpoints.
In contrast, numerical boundary conditions are of the form $u(1)=\left.\dd u/\dd x\right|_{x=0}=5$ and have to be imposed explicitly.

To impose boundary conditions in terms of spectral methods, we usually follow the following strategies: 
\begin{enumerate}
    \item[(i)] \textit{boundary bordering} In the $N\times N$ pseudo-spectral matrix, we replace the lines corresponding to the two endpoints by the explicit boundary condition. The remaining $(N-2)\times N$ block corresponds to the collocation conditions in the inner of the interval.
    \item[(ii)] \textit{basis recombination} We re-formulate the boundary value problem so that the new boundary conditions in the modified boundary value boundary problem are homogeneous. In this strategy, we modify the set of basis functions so that they satisfy the constraints individually.
\end{enumerate}
In practice, we usually have to use (or have the luxury of using) both strategies. To see how this works in context of the AdS/CFT correspondence, we can consider a very simple setup: the famous holographic superconductor in a planar AdS$_4$ Schwarzschild background~\cite{Hartnoll:2008vx}. The resulting equations of motion are given by
\begin{align}
   & z^2 f(z) \Psi''(z)+z \left(z f'(z)-2 f(z)\right) \Psi'(z)+\Psi(z) \left(\frac{z^2 \Phi(z)^2}{f(z)}+2\right)=0\label{eq:sc1}\\
   & z^4\, \Phi''(z)-\frac{2\, z^2\, \Phi(z) \Psi(z)^2}{f(z)}=0,\label{eq:sc2}
\end{align}
where $f(z)=1-z^3$. Furthermore, the function $\Phi(z)$ behaves as $\Phi(z)\sim (1-z)\,\Phi_h+\dots$ at the horizon. We see that both of the coupled (nonlinear) differential equations are singular at the endpoints $z=0$ and $z=1$ (since $f(1)=0$). Additionally, we know the solution at the boundary in terms of a power series
\begin{align}
    &\Psi(z)=\Psi_{(s)}\,z+\Psi_{(v)}\,z^2+\dots,\label{eq:psifrob}\\&\Phi(z)=\mu+\rho\,z+\dots,
\end{align}
where the leading and subleading mode are identified with the field theory quantities according to the AdS/CFT dictionary. The precise physical interpretation is not important for now. Let us formulate the following boundary value boundary problem:
We want to find the solutions of $\Psi(z),\Phi(z)$ to the differential equations \eqref{eq:sc1} and \eqref{eq:sc2} subject to the boundary conditions: $\Psi(0)=0,\,\Psi'(0)=0, \,\Phi(0)=\mu,\,\Phi(1)=0$ (and also where $\Psi'' \not = c$)\footnote{The ambiguity of multiple possible solution is due the nonlinearity of system of equations}. 

Na\"ively, we could go forward and simply implement the boundary conditions and try to solve the system. However, there is a more elegant way. Since we know the solution in terms of a power series at $z=0$ \eqref{eq:psifrob}, we notice that the re-defined function $\Psi(z)=z^2\,\tilde\Psi(z)$ automatically satisfies the two conditions $\Psi(0)=0,\,\Psi'(0)=0$ and we do not have to impose them explicitly anymore. Similarly, the new function $\Phi(z)=(1-z)\,\tilde\Phi(z)$ automatically implements $\Phi(1)=0$. In terms of the new functions, the system reads
\begin{align}
   & \tilde\Psi \left(2\, z\, f'+\frac{(z-1)^2\, z^2\, \tilde\Phi^2}{f}-2 f+2\right)+z \left(\left(z\, f'+2\, f\right) \tilde\Psi+z\, f\, \tilde\Psi''\right)=0\label{eq:sc11}\\
   & -(z-1) \,\tilde\Phi''-2 \tilde\Phi'+\frac{2\, (z-1)\, z^2\, \tilde\Phi\, \tilde\Psi^2}{f}=0,\label{eq:sc22}
\end{align}
where the boundary condition $\tilde\Phi(0)=\mu$ is left (and we want to pick the solutions with $\tilde\Psi(0)\not =0$). The new boundary value problem is still a system of second order differential equations but instead of four boundary conditions, we have only one boundary condition left! So where do we get the missing boundary conditions from? The answer is simple and we already stated it in this chapter: in case the equation of motion degenerates (in terms of singular points) we do not have to impose any boundary condition since the choice of Chebychev polynomials which are regular at the singular points we already imposed the missing boundary conditions in terms of behavioral boundary conditions. Hence it is sufficient to require that the residuum vanishes at the respective points (which is a fancy way of saying, "Do not do anything" and just impose the equation of motion at the endpoint).

Note that it also possible to introduce the auxiliary function $\tilde \Phi(z)$ as $\Phi(z)=(1-z)\,(\mu+z\,\tilde\Phi(z))$. In this way, $\mu$ will inevitably appear in the equations of motion. This is no problem as long as we have a determining equation for $\mu$ (for example, a constraint equation). In this case, we can add the $\mu$ to the unknown parameters and solve for the corresponding $(N+1)\times(N+1)$ dimensional matrix (in case of one unknown function).

All in all, we can write this in a compact way (for a (non)linear problem depending on one coordinate). Let $n$ be the number of unknown functions with $I\in\{1,\dots,n\}$ and $\xi$ denote additional $m$ additional parameters with $J\in\{1,\dots,m\}$. Then we find the following set of equations (see eq. \eqref{eq:res} for the definition of the residuum) for a given resolution $N$
\begin{align}
    R_0^I(u'^I,u^I;\,\xi^J)=0&\quad \text{for }z=0,\\
    R^I(u''^I,u'^I,u^I,z;\,\xi^J)=0&\quad \text{for }0<z<1,\\
    R^I_1(u''^I,u'^I,u^I,z;\,\xi^J)=0&\quad \text{for }z=1,\\
    \xi^J(u''^I,u'^I,u^I;\,\xi^J)=0&\quad \text{for }0<z<1.
\end{align}
The coordinate is discretized according to \eqref{eq:diffmatr2} and all derivatives are replaced by the differentiation matrices \eqref{eq:diffmatr2}.
The resulting spectral matrix is a $(n\cdot N+m)\times (n\cdot N+m)$ matrix. For a pseudo-spectral methods, we can then determine the numerical values of the solution by solving $(R,\xi)(u)=0$ for the components $u$. The spectral coefficients may be determined by inverting \eqref{eq:spectralrep}.

\section{Convergence, Accuracy \& All that}
\begin{figure}[h]
    \centering
 \includegraphics[width=7cm]{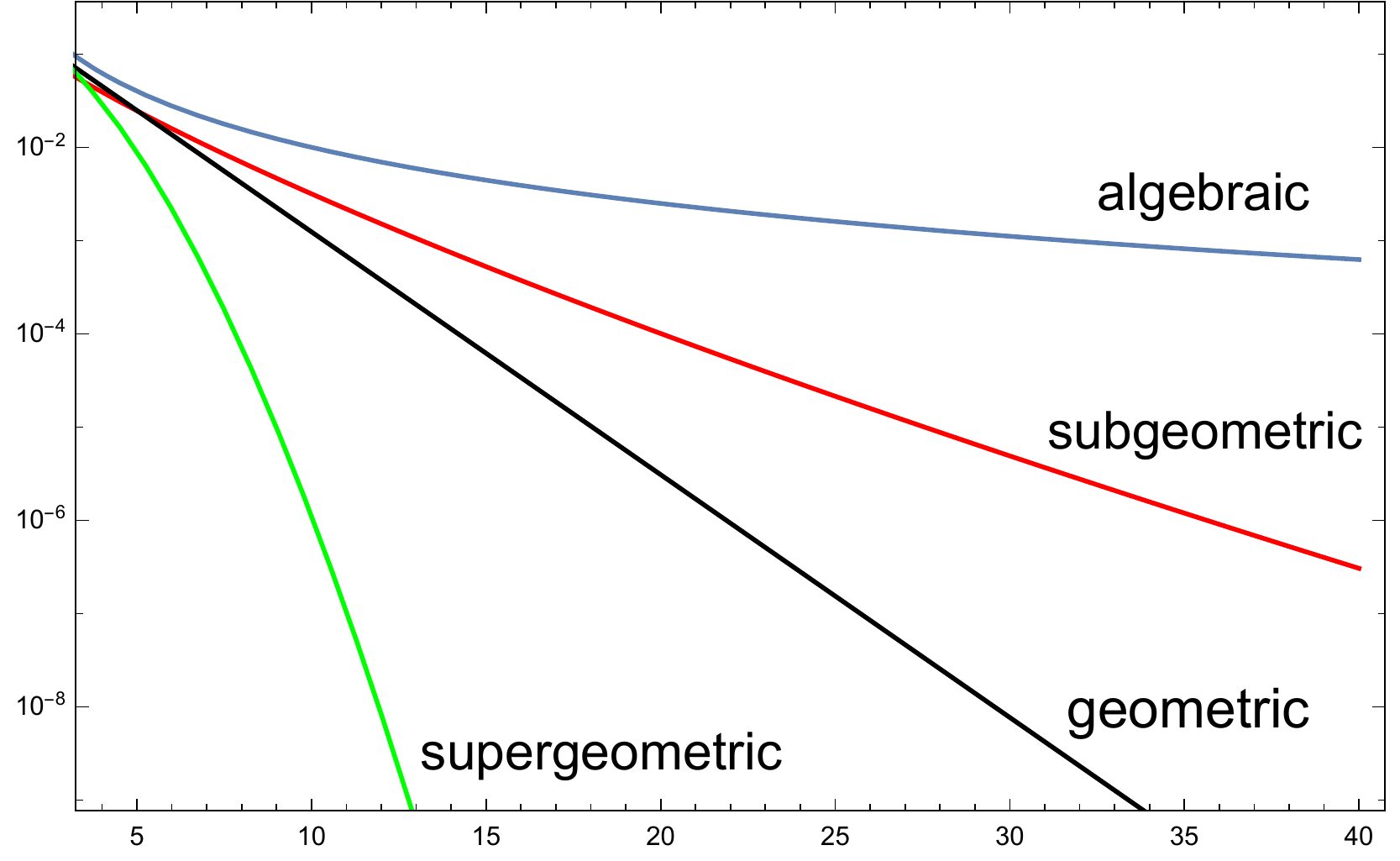}
        \caption{Different rates of convergence.}\label{fig:convergence_rate}
\end{figure}

The approximation by $N$ basis polynomials in \eqref{eq:spectralrep} is only a good approximation if the coefficients beyond the truncation do not contribute significantly. Furthermore, we can only make statements about higher coefficients if the coefficients of our solution consistently fall off for larger indices. There are various rates of convergence. Instead of a rigorous abstract mathematical treatment, we will keep the discussion about convergence on a graphical level. This is more than enough for using spectral methods on a day-to-day basis. The different rates of convergence are depicted in graphic \ref{fig:convergence_rate}.
Spectral convergence corresponds to exponential decay of the coefficients. In a log-linear plot where we plot the absolute value of the coefficients against the number of basis polynomials a purely exponential decay corresponds to a straight line with negative slope (geometric, black line). The coefficients decay according $|a_n|\sim \mathcal O(c_1\,\text{exp}(-\,c_2\,N))$. If the coefficients decay even faster, we speak of supergeometric convergence ($|a_n|\sim\mathcal O(c_1\,\text{exp}(-(N/c_2)\,\log(N)))$ depicted as green line in figure \ref{fig:convergence_rate}. In the log-linear plot, the slope of supergeometric convergence rate gets more and more negative instead of converging to a straight line. On the contrary, in the case of $|a_n|\sim \mathcal O(c_1\,\text{exp}(-\,c_2\,N^r))$ with $r<1$, the coefficients are decaying slower than the straight line in a log-linear plot and we refer to the decay rate as subgeometric (depicted in red in figure \ref{fig:convergence_rate}). Lastly, if the coefficients decay as a polynomial in the number of gridpoints, the convergence is algebraic. In this case, the curve is bend upwards in a log-linear plot with slope tending to zero. There is one caveat; the notions of convergence we just discussed are asymptotic quantities and in practise the effective rate of convergence might differ from the asymptotic ones.

\subsection*{Coordinate mappings}
As explained in~\cite{Boyd1989ChebyshevAF}, logarithmic singularities render the asymptotic form of the spectral coefficients algebraic $(\sim N^{-1}\,p^N)$. The spectral solutions, we can construct in presence of those singularities have very poor convergence. In AdS/CFT, it is very common that logarithms appear in the asymptotic expansions of the functions at the conformal boundary. One example is the trace anomaly of the energy momentum tensor in $4+1$ dimensions due to the presence of a magnetic field. As so often when dealing with problems in spectral methods, there is a solution by Boyd for it~\cite{Boyd1989TheAC}.
For functions with endpoint singularities (remember that $x\in[-1,1]$ while $z\in[0,1]$), where the expansion reads
\begin{equation}
    f(x)=c_1\,(1-x)^k\,\log(1-x)+c_2\,(1+x)^k\,\log(1+x),
\end{equation}
the algebraic coordinate mapping
\begin{equation}
    x=\sin\left(\frac\pi 2\,y\right)\label{eq:boyd:sin}
\end{equation}
leads to a much higher resolution near the singularity an may restore the degraded rate of convergence. In our case, the functions have endpoint singularities only at the conformal boundary for example
\begin{equation}
    f(z)=1+ z^4\,f_4+ c_1\,\log(z)\,z^4,
\end{equation}
we propose the Boyd inspired mapping
\begin{equation}\label{eq:mappingz2}
    z\mapsto z^2
\end{equation}
which was put forward in my master thesis~\cite{Grieninger:2017jxz,Ammon:2016fru}. This mapping is obtained by transforming eq. \eqref{eq:boyd:sin} to $z\in[0,1]$ and expand the expression around $z=0$. The mapping moves the logarithmic divergences to higher order according to
\begin{equation}
    z^n\,\log(z)\mapsto 2\,z^{2n}\,\log(z)
\end{equation}
and we find a much improved rate of convergence, as shown in figure \ref{voncergencecm}.

\section{Nonlinear Ordinary differential equations}\label{sec:nonlinear}
Applying (pseudo)-spectral methods to nonlinear equations instead of linear ones is conceptually not significantly more difficult. In terms of Newton's method, the equations we have to solve numerically are still linear. Due to the ambiguity of the solutions to nonlinear equations, however, we have to start with a good initial guess of the solution so that the Newton method converges to the solution we are interested in.
\subsubsection{Newton-Raphson method}
The Newton method is an iterative method in order to find the roots of a nonlinear equations. It works as follows; for an initial guess close to the true solution, the Newton method is based on using the derivative to do a step in direction of the true solution. Let $x_i$ be the current approximation of the solution (for the first step it is the initial guess). Then we may find derive an expression for the next best solution by using the Taylor expansion
\begin{equation}
    y=f'(x_i)(x-x_i)+f(x_i)
\end{equation}
The tangent intersects the $x$-axis at $x_{i+1}$ (where $y=0$)
\begin{equation}
    x_{i+1}=x_i-\frac{f(x_i)}{f'(x_i)}.\label{eq:neWT}
\end{equation}
We generalize this to a nonlinear system of algebraic equations $F(\bm{X})=0$. $\bm{X}$ is the vector of all fields evaluated on the collocation points \newline $\bm{X}^T=(X_1^1,\ldots,X_N^1,\ldots,X_N^n,\xi^1,\ldots,\xi^m)$, where $N$ is the number of collocation points, $n$ the number of fields and $m$ the number of additional parameters.
Since $\bm{X}$ is a vector the generalization of the derivative $f'(x_i)$ in eq. \eqref{eq:neWT} will be matrix valued and we introduce the Jacobian $\hat J(\bm{X})=\partial F(\bm{X})\partial \bm{X}$. With this, the Newton step reads
\begin{equation}
    \bm{X}_{i+1}=\bm{X}_i-[\hat J(\bm{X}_i)]^{-1}\,F(\bm{X}_i),
\end{equation}
where we assume that the Jacobian is invertible. For more complicated problems, it can be very time consuming to evaluate (and store) the Jacobian. In theses cases, we recommend introducing auxiliary variables for the derivatives (instead of evaluating them on the gridpoints with the differentiation matrices) and compute the Jacobian according to the chain rule. In this way, we separate zeroth, first and second derivatives acting on the fields. For a discussion see the appendices in~\cite{dbt_mods_00033484,Ammon:2017ded}.
\section{Eigenvalue problems and Quasi-Normal modes}
Computing the Quasi-Normal modes within the AdS/CFT correspondence is a very important task since it gives us insight into the linear response regime in strongly coupled field theories. In general, we have to find solutions to the linearized equations of motion, subject to infalling conditions at the horizon and Dirichlet conditions at the conformal boundary. One elegant approach to impose the infalling condition at the horizon is by means of the coordinate choice in the gravity theory. If we use infalling Eddington-Finkelstein coordinates instead of Poinca\'e coordinates, all fields automatically satisfy infalling conditions at the horizon. We may collect the system of equations of motion in terms of powers in $\omega$ (where $\omega$ the QNM frequency)
\begin{equation}
    \bm{\alpha}[\bm{X}]+\omega\,\bm{\beta}[\bm{X}]+\omega^2\,\bm{\gamma}[\bm{X}]=0.\label{eq:quadre}
\end{equation}
The vector $\bm{X}$ includes all fields and $\bm{\alpha},\,\bm{\beta},$ and ${\gamma}$ are differential operators. The eq. \eqref{eq:quadre} is a quadratic eigenvalue problem with the eigenvalues given by the QNM frequencies.
In most cases we are lucky and $\bm{\gamma}[\bm{X}]$ is zero. Nevertheless, it is always possible to reduce the quadratic eigenvalue problem \eqref{eq:quadre} to a linear one by introducing auxiliary fields. For example, $ \bm{\alpha}f+\omega\,\bm{\beta}f+\omega^2\,\bm{\gamma}f=0$ may be reduced to the system
\begin{equation}
    \bm{\alpha} f+\omega\,(\bm{\beta}f+\bm{\gamma}\tilde f)=0, \ \ \tilde f=\omega f
\end{equation}
This obviously increases the spectral matrix by $N$ for each auxiliary field since we have to add the new fields to $\bm{X}$. The resulting linear eigenvalue problem reads
\begin{equation}
    \bm{\alpha}[\bm{X}]+\omega\,\bm{\beta}[\bm{X}]=0.\label{eq:quadre2}
\end{equation}
    In order to solve this problem numerically, we discretize the (radial) coordinate with eq. \eqref{eq:diffmatr2} and replace all derivatives by the differentiation matrices \eqref{eq:diffmatr2}.
The Dirichlet boundary conditions may be implemented by using the methods explained in \ref{app:bc}. We redefine the functions by factorizing out an appropriate power $k$ in z, $f=z^k\,\tilde f$. In practical applications, we have  constraint equations in addition to the equations of motion\footnote{Note that constraint equations are not present if one works in so-called gauge invariant variables}. Implementing these constraint equations is the biggest difficulty in obtaining the QNMs (or solving the (non)linear problems described in the previous section). Usually, it is enough to demand constraint equations at one of the endpoints and we can replace one of the boundary condition on the fields with the constraint equation. There is no general way to determine which boundary conditions may be replaced by the constraint equations. One big advantage of this approach is that we are computing not only one but up to $n\cdot N+m$ QNMs at once! ($n$ is the number of fields, $m$ the number of extra conditions and $N$ the number of gridpoints)

\section{Time-dependent problems}
The pioneering work of studying time-dependent problems in the AdS/CFT correspondence was done by Chesler and Yaffe \cite{Chesler:2008hg,Chesler:2009cy,Chesler:2010bi} (see \cite{Chesler:2013lia} for a detailed review). The authors developed the standard approach which is the so-called characteristic formulation by Bondi and Sachs. The characteristic formulation has the advantage that the coupled set of partial differential equations decouples and exhibits a nested structure. In the nested structure, we can solve the equations successively which simplifies the hard task of solving coupled partial differential equations -- a numerically very costly task -- and hence allows for fast and efficient codes.

\subsubsection{The numerical routine for nonlinear oscillatory shear tests in AdS}
In this section, we display the numerical routine to solve the time-dependent partial differential equations in chapter \ref{chapter:brokenspacetime}. This subsection is taken from the appendix of my publication~\cite{Baggioli:2019mck}. We make the following re-definitions of the metric functions
\begin{eqnarray}
&&A=\frac{1+\tilde{A}\ u}{u^2},\quad S=\frac{s_0+\tilde{S}\ u}{u},\nonumber \\
&&H=h_0+\tilde{H}\ u,\quad d_+S=\frac{\widetilde{d_+S}}{u^2},\\
&&d_+H=\frac{\dot{h_0}}{2}+\widetilde{d_+H}\ u.\nonumber
\end{eqnarray}
In our numerical calculation, we use the Chebyshev discretization with 50-100 grid points for the integration  along the radial coordinate.

Following the characteristic feature of the bulk equations of motion in EF coordinates \cite{Chesler:2013lia},  the numerical recipe to solve the equations of motion is: \eqref{constraint}-\eqref{eq-dppS}:
\begin{enumerate}
\item We start with a static black hole solution with $\tilde{H}=0, \tilde{S}=s_1(t_0)$ as an initial configuration and choose the strain function $\gamma(t)$.
\item We check the accuracy of the numerical calculation, by plugging $\tilde{H}, \tilde{S}$ in the constraint equation \eqref{constraint}. 
\item We use the definition of apparent horizon, $d_+S(u=1)=0$,  as a boundary condition to calculate $\widetilde{d_+S}$ by solving eq. \eqref{eq-dpS}.
\item Then we solve eq. \eqref{eq-dpH} with one boundary condition for $\widetilde{d_+H}$ at asymptotic region, $\widetilde{d_+H}(u=0)=\tfrac{\dot{h_0}\dot{s_0}}{s_0}+\ddot{h_0}$.
\item Now we can solve eq. \eqref{eq-A} to find $\tilde{A}$ with two boundary conditions: the first is $\tilde{A}(u=0)=2(s_1-\dot{s_0})/s_0$ and the second we can find by expanding eq. \eqref{eq-dppS} near the horizon which leads to
\begin{equation}
A =-\frac{1 }{3} \left({d_+H}\right)^2\bigg|_{u=1}.
\end{equation}
\item By using the definition of operator $d_+$,  we find $\dot{\tilde{S}}, \dot{\tilde{H}}$. Then we integrate in time using a fourth-order Runge-Kutta method for the first three time steps and then the fourth order Adams-Bashforth method, to compute $\tilde{H}(u, t_0+\delta t)$ and $\tilde{S}(t_0+\delta t)$ and repeat the same routine from step 2.
\end{enumerate}

\begin{figure}[h!]
    \centering
\includegraphics[width=0.47\linewidth]{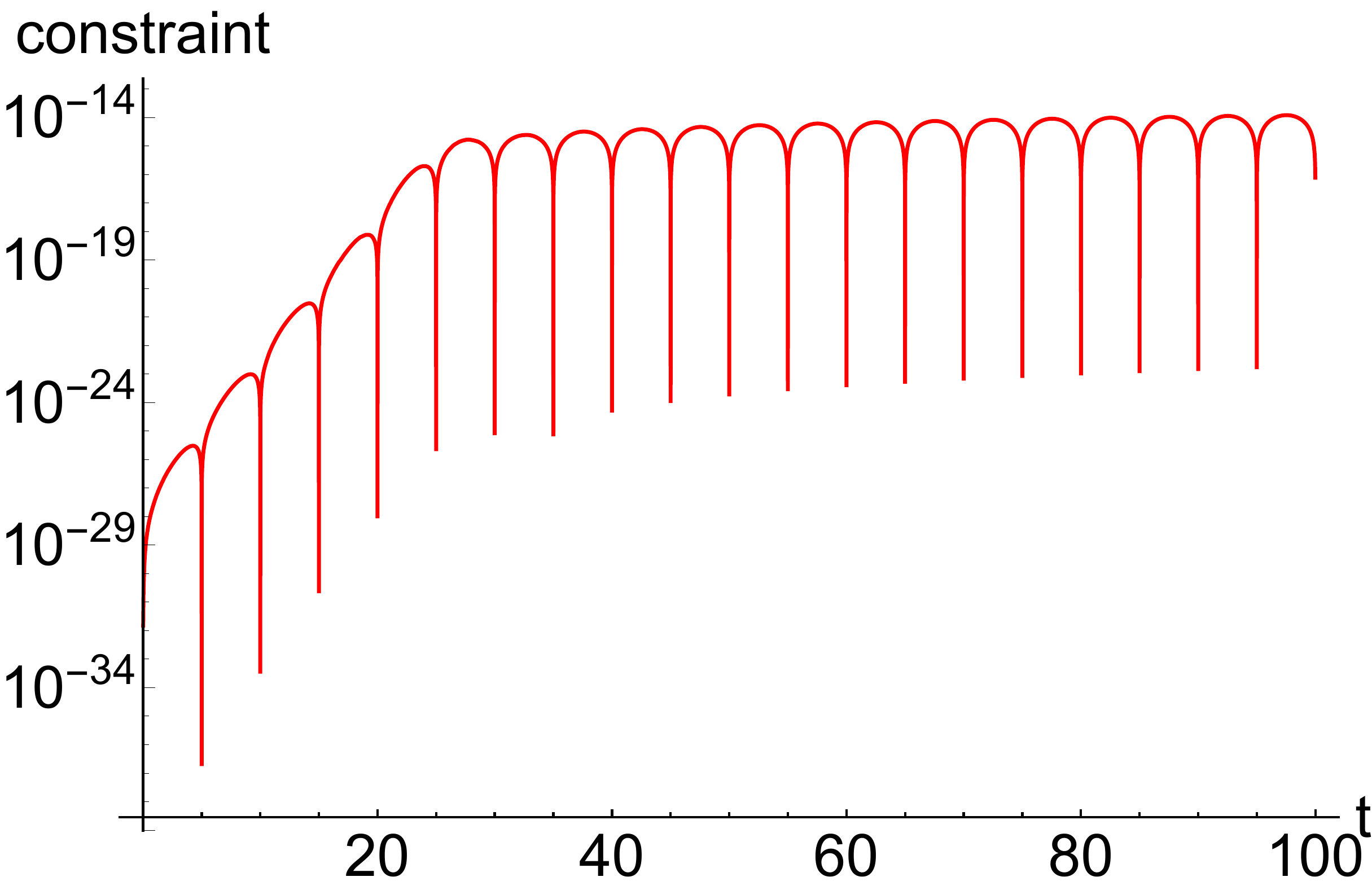}
   \caption{Constraint equation \eqref{constraint} for $s_1|_{t=0}=0, \gamma_0=0.1, w_c=2, t_c=25$ for \color{black} $\omega/m=0.32$. \color{black}}
    \label{example}
\end{figure}

To impose the sinusoidal strain, we turn on the amplitude smoothly (in the spirit of \cite{Ammon:2016fru,Grieninger:2017jxz}) as
\begin{equation}
    \gamma(t)= \frac{\gamma_0}{2}\left(1+\tanh\left(\frac{t-t_c}{w_c}\right)\right)\sin\left(2\pi\omega t\right),
\end{equation}
where parameters $t_c, w_c$ control how accurate the initial configuration satisfies the constraint equation \eqref{constraint} and how fast the maximum strain is reached respectively. In figure \eqref{example}, we show a concrete example of the constraint equation \eqref{constraint}.
\section{Convergence plots}\label{app:sectionconv}
\subsection*{Chiral transport in strong magnetic fields}
\begin{figure}[h]
    \centering
 \includegraphics[width=7cm]{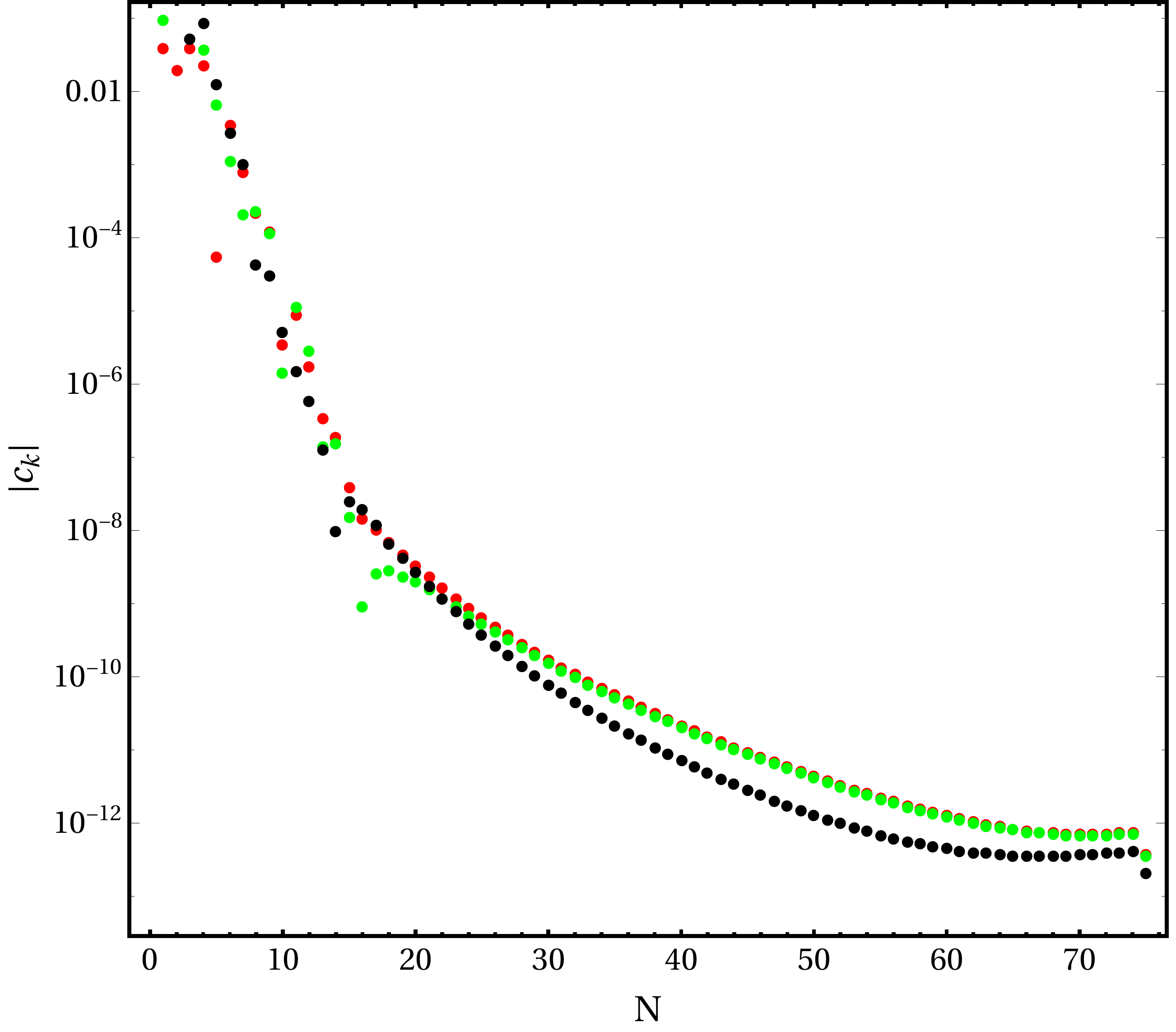}\hspace{0.5cm} \includegraphics[width=7cm]{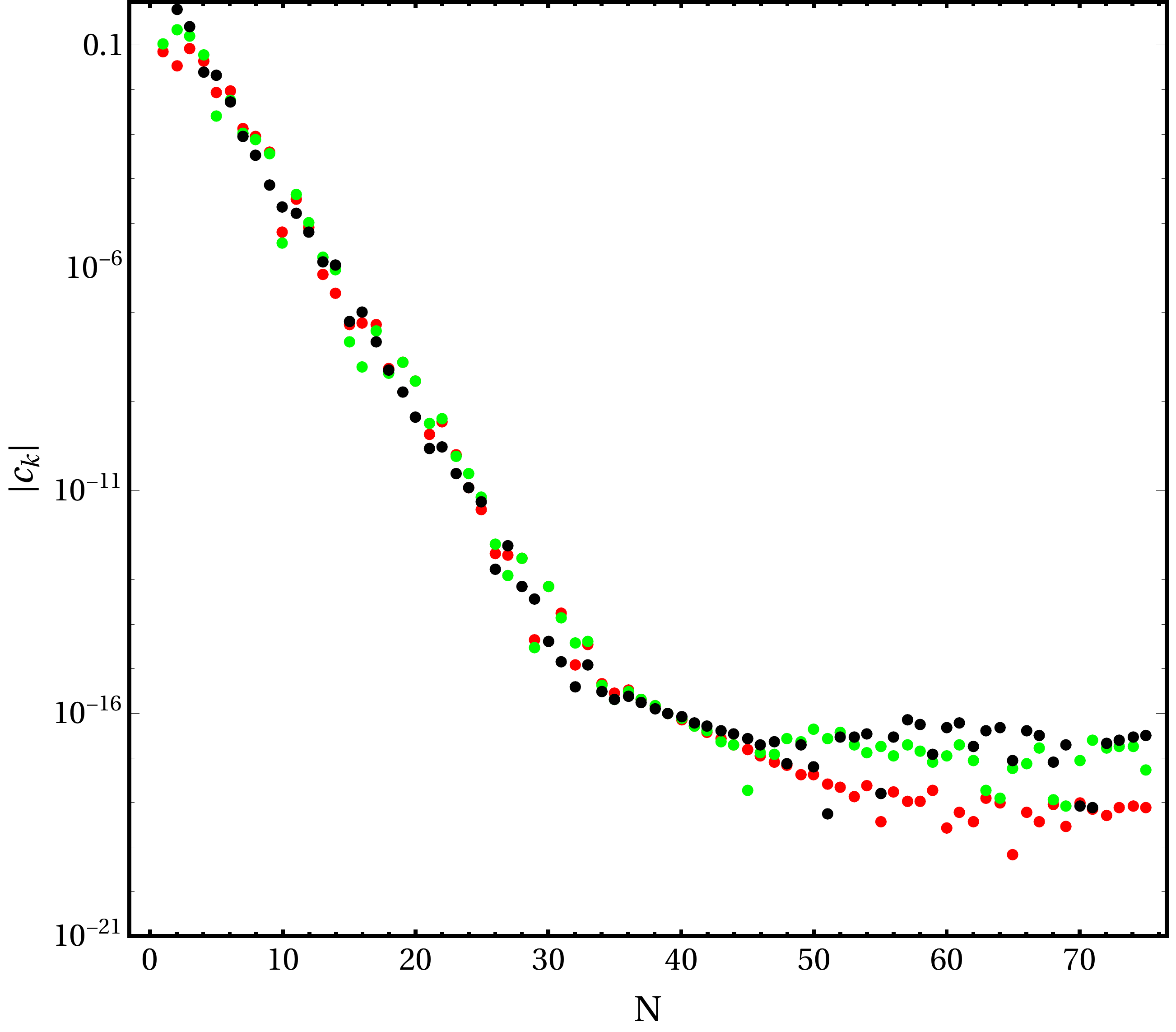}
        \caption{Convergence in the helicity-1 sector without (left) and with (right) the coordinate mapping \eqref{eq:mappingz2}. The parameters are $\gamma=2/\sqrt{3},\, \tilde B=109.658,\,\tilde\mu=7.5$ and the fluctuations are $h_{t+},\,h_{z+},\,a_+$ (red, green, black).}\label{voncergencecm}
\end{figure}

In this subsection, we exemplarily present a convergence plot for chapter \ref{sec:anotranspo}. In the helicity-1 sector, it is convenient to decouple the fluctuation equations further by introducing $h_{t\pm} = h_{tx} \pm i h_{ty},$ $h_{z\pm} = h_{xz} \pm i h_{yz},$ and $a_{\pm} = a_{x} \pm i a_y.$ In figure \ref{voncergencecm}, we depict the convergence for $h_{t+},\,h_{z+},\,a_+$ without coordinate mapping (left) and with coordinate mapping (right). We note that with coordinate mapping, the coefficients fall off geometrically to machine precision before they reach a plateau.

\subsection*{Holographic massive gravity}
\begin{figure}
\centering
\includegraphics[width=7cm]{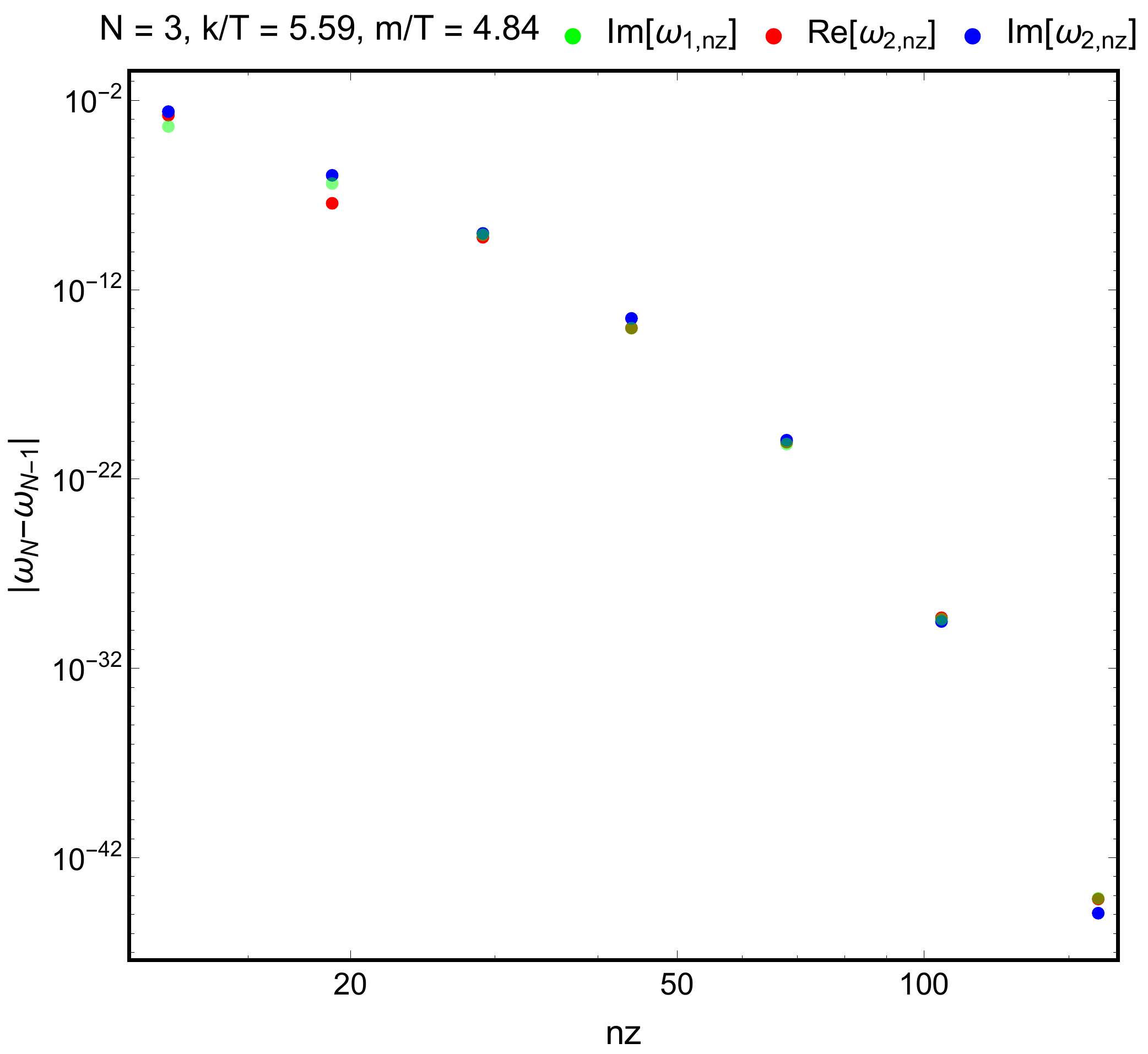}
\quad
\includegraphics[width=7cm]{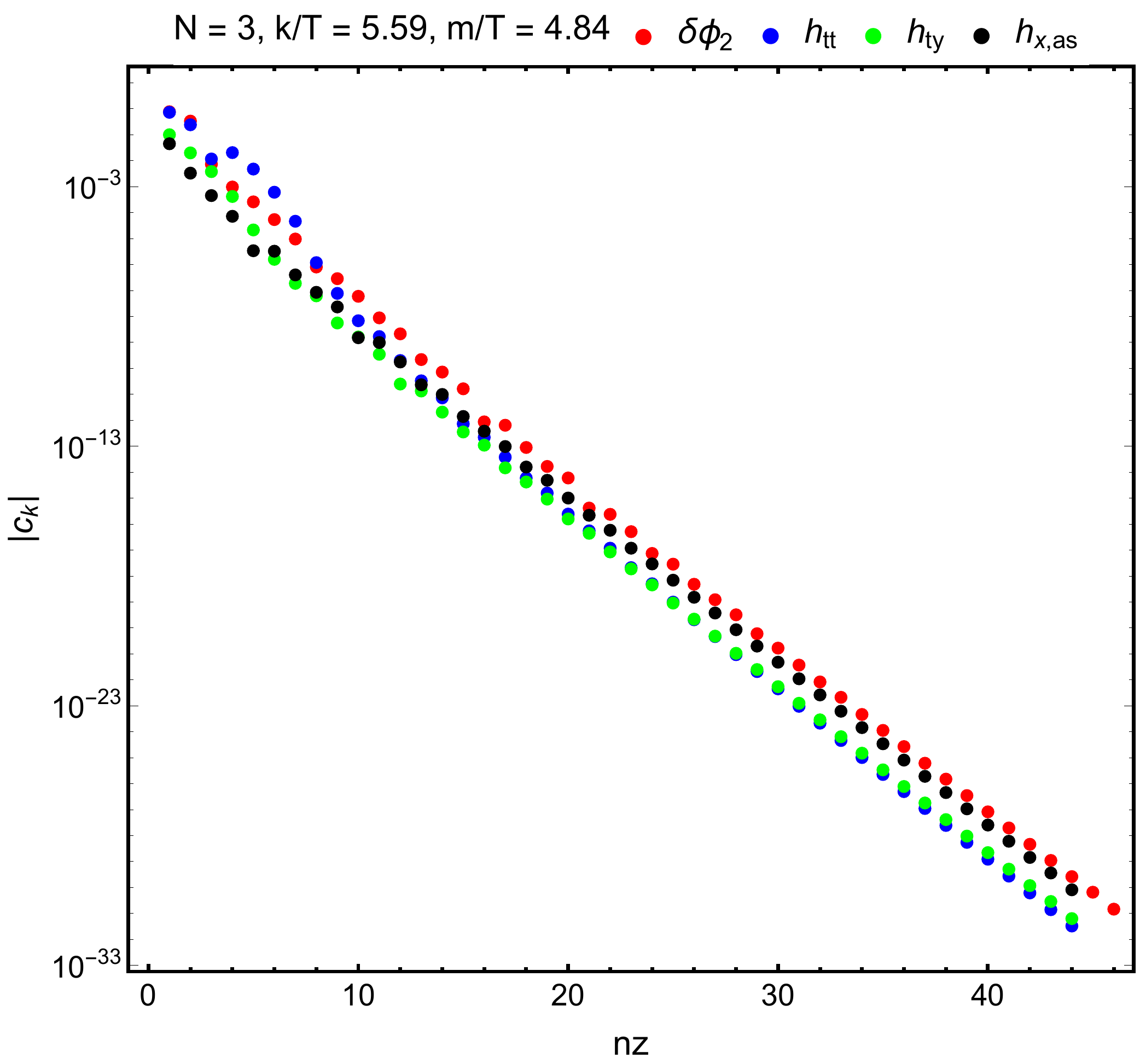}
\caption{\textbf{Left: }Moving of the lowest three QNMs with increasing grid size. \textbf{Right: }Chebychev-coefficients of the eigenfunctions corresponding to the second lowest QNM. \label{pic:conv}}
\end{figure}

We checked that our solutions satisfy the equations of motions and all constraint equations. To check the convergence of the numerical solution, we monitor the change of the solution for finer discretizations. As depicted in the l.h.s. of figure \ref{pic:conv}, the change of the quasi-normal mode frequency decays exponentially with a growing number of grid points; the same is valid for the corresponding eigenfunctions. Another check for the numerical method are the Chebychev-coefficients of the solution, displayed in the r.h.s. of figure \ref{pic:conv}; the coefficients decay exponentially, indicating exponential accuracy of our numerical method.  

\chapter{Holographic massive gravity}
\section*{Equations of motions}\label{app1}
In this appendix, we provide the equations of motions for the fluctuations (taken from my publications \cite{Baggioli:2019abx,Baggioli:2019mck}). We consider the momentum aligned in the $y$ direction, $\vec{k}=(0,0,k)$. 
Throughout this appendix, derivatives of the potential always denote derivatives with respect to the argument, i.e.  $V'(Z)\equiv d V(Z)/dZ,\,V''(Z) \equiv d^2 V(Z)/dZ^2$ and $V'(X)\equiv d V(X)/dX,\,V''(X) \equiv d^2 V(X)/dX^2$. To simplify the notion, we suppress all arguments of the functions. 
\subsection*{Transverse sector}
In the transverse sector we consider the following set of metric and scalar fluctuations $\{\delta \phi_x,\,h_{tx},\,h_{xy}\}$, where we assumed radial gauge $h_{xu}=0$~\cite{Baggioli:2019abx}.\\[0.2cm]
\textbf{V(Z) case}\\[0.15cm]
\begin{align}
   & V'\,\left( \delta\phi_x'\, \left(f'+2\, i\, \omega \right)+f \delta\phi_x''+h_{tx}'\right)+4 u^3\, V'' \left(f\, \delta\phi_x'+h_{tx}+i\, \omega\,  \delta\phi_x\right)=0\\
    & h_{tx} \left(2\, u\, f'-6 f+k^2 u^2+4 m^2 u^4\, V'-2 m^2\, V+6\right)\nonumber\\&+u \left(-u\, f\, h_{tx}''+2 f\, h_{tx}'+k\, u\, \omega\,  h_{xy}+4\, i\, m^2\, u^3\, \omega\,  \delta\phi_x'\, V'-i\, u\, \omega\, h_{tx}'\right)=0\\
 &u \left(h_{xy}' \left(-u\, f'+2\, f-2\, i\, u\, \omega \right)-u f h_{xy}''-i\, k\, u\,h_{tx}'\right)\nonumber\\&+2 h_{xy} \,\left(u\, f'-3 f-m^2\, V+i\, u\, \omega +3\right)+2\, i\, k\, u\, h_{tx}=0\\
 &i\, k\, u\, h_{xy}'-u \left(4\, m^2\, u^2\, \delta \phi_x'\, V'+h_{tx}''\right)+2 h_{tx}'=0.
\end{align}
\\[0.15cm]
\textbf{V(X) case}\\[0.15cm]
\begin{align}
&-2(1-u^2\,V''/V')h_{tx}+u\,h_{tx}'-i\,k\,u\, h_{xy}-\left( k^2\,u+2\,i\,\omega (1-u^2\,V''/V')\right)\,\delta \phi_x+u\,f\, \delta \phi_x''
\nonumber\\&+\left(-2(1-u^2\,V''/V')\,f+u\,(2 i \omega+f')\right)\delta \phi_x'
\,=0\,;\\
&2\,i\,m^2\,u^{2}\omega V'\,\delta \phi_x+u^2\, k\,\omega\, h_{xy}+(6+k^2\,u^2-2 \, m^2 (V-u^2\,V' ) \,\nonumber\\& -6 f+2uf') \, h_{tx}+\left(2\,u\,f-i\,u^2\omega\right)h_{tx}'-u^2\,f\,h_{tx}''\,=\,0 \,\,;\\
&2i\,k\,u\,h_{tx}-iku^2h_{tx}'-2\,i \, k \,m^2 \,u^{2}V'\delta \phi_x+2 h_{xy}\left(3+i\,u \, \omega-3f+uf'- \,m^2(V-u^2 V') \right)
\nonumber\\&-\left(2i\,u^2 \, \omega-2uf+u^2\,f'\right)h_{xy}'-u^2\,f\, h_{xy}''\,=\,0\,;\\
&2\,h_{tx}'-u\,h_{tx}''-2m^2\,u\,V' \, \delta\phi_x'+ik\,u\,h_{xy}'\,=\,0.
\end{align}

\subsection*{Longitudinal sector}
In the longitudinal sector we consider the following set of fluctuations in radial gauge \newline\mbox{$    \{h_{x,\bm{s}}=1/2\,(h_{xx}+h_{yy}),\,h_{x,\bm{a}}=1/2\,(h_{xx}-h_{yy}),\,\delta\phi_y,\, h_{tt},\, h_{ty}\}$}. \\[0.25cm]
\textbf{V(Z) case}\\[0.25cm]
\begin{align}
 &  f'\,\delta\phi_y'\, V'+f\delta\phi_y''\, V'+4 u^3\, f\delta\phi_y'\, V''-\delta\phi_y \left(k^2\, V'+2 u^3\, \left(k^2\, u-2\, i\, \omega \right) V''\right)\nonumber\\
 &-i\, k\, h_{xx}\, \left(2 u^4\, V''+V'\right)+h_{ty}'\, V'+2\, i\, \omega\,  \delta\phi_y'\, V'+4\, u^3\, h_{ty}\, V''=0\\ &
 2 h_{ty} \left(-u f'+3 f-2 m^2 u^4 V'+m^2 V-3\right)+u (u f h_{ty}''+(-2 f+i\, u\, \omega )\, h_{ty}'+i\, k\, u\, h_{tt}'\nonumber \\ &
 -4\, i\, m^2\, u^3 \omega  \delta \phi_y\, V')+k\, u^2\, \omega\,  (h_{x,\bm{s}}+h_{x,\bm{a}})-2\, i\, k\, u\, h_{tt}=0\\
 &6\, h_{tt}+u\, \left(-u\, f'\, h_{x,\bm{s}}'+2\, f\, h_{x,\bm{s}}'+4\, i\, k\, m^2\, u^3\, \delta \phi_y \left(2\, u^4\, V''+V'\right)-i\, k\, u\, h_{ty}'+2\, i\, k\, h_{ty}\right.\nonumber\\
 &\left. -8\, m^2\, u^7\, h_{x,\bm{s}}\, V''+2\, i\, h_{x,\bm{s}} \left(\omega +2\, i\, m^2\, u^3\, V'\right)+u\, h_{tt}''-4\, h_{tt}'-2\, i\, u\, \omega  h_{x,\bm{s}}'\right)=0\\
 &h_{x,\bm{s}}\, \left(2\, u\, f'-6\, f+k^2\, u^2+4\, m^2\, u^4\, V'-2\, m^2\, V+4\, i\, u\, \omega +6\right)-u^2\, f'\, h_{x,\bm{s}}'\nonumber\\ &
 -u^2\, f'\, h_{x,\bm{a}}'+2\, u\, h_{x,\bm{a}}\, f'-u^2\, f\, h_{x,\bm{s}}''-u^2\, f\, h_{x,\bm{a}}''+4\, u\, f\, h_{x,\bm{s}}'+2\, u\, f\, h_{x,\bm{a}}'-6\ f\, h_{x,\bm{a}}\nonumber\\&
 +k^2\, u^2\, h_{x,\bm{a}}-4\, i\, k\, m^2\, u^4\, \delta \phi_y\, V'+2\, i\, k\, u\, h_{ty}-2\, m^2\, h_{x,\bm{a}}\, V-2\, i\, u^2\, \omega\,  h_{x,\bm{s}}'\nonumber \\
 &-2\, i\, u^2\, \omega\,  h_{x,\bm{a}}'-2\, u\, h_{tt}'+6\, h_{tt}+2\, i\, u\, \omega  h_{x,\bm{a}}+6\, h_{x,\bm{a}}=0\\
 &h_{x,\bm{s}} \left(2 u f'-6\, f+k^2\, u^2+4\, m^2\, u^4\, V'-2\, m^2\, V+4\, i\, u\, \omega +6\right)-u^2 f'\, h_{x,\bm{s}}'+u^2\, f'\, h_{x,\bm{a}}'\nonumber \\
 &-2\, u\, h_{x,\bm{a}}\, f'-u^2\, f\, h_{x,\bm{s}}''+u^2\, f h_{x,\bm{a}}''+4\, u\, f\, h_{x,\bm{s}}'-2\, u\, f\, h_{x,\bm{a}}'+6\, f\, h_{x,\bm{a}}+k^2\, u^2\, h_{x,\bm{a}}\nonumber\\
 &-4\,i\, k\, m^2 \,u^4\, \delta \phi_y\, V'-2\, i\, k\, u^2\, h_{ty}'+6\, i\, k\, u\, h_{ty}+2\, m^2\, h_{x,\bm{a}}\, V-2\, i\, u^2\, \omega\,  h_{x,\bm{s}}'\nonumber\\&+2\, i\, u^2\, \omega\,  h_{x,\bm{a}}'-2\, u\, h_{tt}'+6\, h_{tt}-2\, i\, u \,\omega \, h_{x,\bm{a}}-6\, h_{x,\bm{a}}=0\\
 &2\, h_{ty}'-u\, \left(i\, k\, \left(h_{x,\bm{s}}'+h_{x,\bm{a}}'\right)+4\, m^2\, u^2\, \delta \phi_y'\, V'+h_{ty}''\right)=0\\
 &h_{x,\bm{s}}''=0.
\end{align}
\\[0.25cm]
\textbf{V(X) case}\\[0.25cm]
\begin{align}
&u f'\, \delta \phi_y'\, V'\,+2\, u^2\, f\, \delta \phi_y'\, V''+u\, f\, \delta \phi_y''\, V'-2\, f\, \delta \phi_y'\, V'-k^2\, u\, \delta \phi_y\, V'-k^2\, u^3\, \delta \phi_y\, V''\nonumber\\
&+i\, k\, u\, h_{x,\bm{a}}\, V'-i\, k\, u^3\, h_{x,\bm{s}}\, V''+2\, i\, u^2\, \omega\,  \delta \phi_y\, V''+u\, h_{ty}'\, V'+2\, i\, u\, \omega\,  \delta \phi_y'\, V'\nonumber\\
&-2\, i\, \omega\,  \delta \phi_y V'-2\, h_{ty} \left(V'-u^2\, V''\right)=0\\
&u (f \left(u f'\, h_{x,\bm{s}}'-2\, f\, h_{x,\bm{s}}'-2\, i\, k\, m^2\, u^3\, \delta \phi_y\, V''-u\, h_{tt}''+4\, h_{tt}'\right)\nonumber\\
&+k\, h_{ty} \left(i\, u\, f'-2\, i\, f+2\, u\, \omega \right)+h_{x,\bm{s}} \left(2\, m^2\, u^3\, f\, V''+\omega\,  \left(i\, u\, f'-2\, i\, f+2\, u\, \omega \right)\right))\nonumber\\ 
&+h_{tt} \left(u \left(-u f''+4\, f'+2\, m^2\, u\, V'\right)-12\, f+k^2\, u^2-2\, m^2\, V-2\, i\, u\, \omega +6\right)=0\\
&2 h_{ty} \left(u \left(f'+m^2\, u\, V'\right)-3\, f-m^2\, V+3\right)\nonumber\\
&-u \left(u\, f\, h_{ty}''-2\, f\, h_{ty}'+i\, k\, u\, h_{tt}'+k\, u\, \omega\,  h_{x,\bm{s}}+k\, u\, \omega\,  h_{x,\bm{a}}-2\, i\, m^2\, u\, \omega\,  \delta \phi_y\, V'+i\, u\, \omega\,  h_{ty}'\right)\nonumber\\
&+2\, i\, k\, u\, h_{tt}=0\\
&h_{x,\bm{s}} \left(2\, u \left(f'+m^2\, u\, V'\right)-6\, f+k^2\, u^2-2\, m^2\, V+4\, i\, u\, \omega +6\right)-u^2\, f'\, h_{x,\bm{s}}'\nonumber\\
&-u^2\, f'\, h_{x,\bm{a}}'+2\, u\, h_{x,\bm{a}}\, f'-u^2\, f\, h_{x,\bm{s}}''-u^2\, f\, h_{x,\bm{a}}''+4\, u\, f\, h_{x,\bm{s}}'+2\, u\, f\, h_{x,\bm{a}}'-6\, f\, h_{x,\bm{a}}\nonumber\\
&+k^2\, u^2\, h_{x,\bm{a}}+2\, i\, k\, u\, h_{ty}+2\, m^2\, u^2\, h_{x,\bm{a}}\, V'-2\,m^2\, h_{x,\bm{a}}\, V-2\, i\, u^2\, \omega\,  h_{x,\bm{s}}'\nonumber\\
&-2\, i\, u^2\, \omega\,  h_{x,\bm{a}}'
-2\, u\, h_{tt}'+6 h_{tt}(u)+2\, i\, u\, \omega\,  h_{x,\bm{a}}+6 h_{x,\bm{a}}=0\\
&h_{x,\bm{s}} \left(2\, u \left(f'+m^2\, u\, V'\right)-6\, f+k^2\, u^2-2\,m^2\, V+4\, i\, u\, \omega +6\right)-u^2\, f'\, h_{x,\bm{s}}'+u^2\, f'\, h_{x,\bm{a}}'\nonumber\\
&-2\, u\, h_{x,\bm{a}}\, f'-u^2\, f\, h_{x,\bm{s}}''+u^2\, f\, h_{x,\bm{a}}''+4\, u\, f\, h_{x,\bm{s}}'-2\, u\, f h_{x,\bm{a}}'
+6\, f\, h_{x,\bm{a}}+k^2\, u^2\, h_{x,\bm{a}}\nonumber\\
&-4\, i\, k\, m^2\, u^2\, \delta \phi_y\, V'-2\, i\, k\, u^2\, h_{ty}'+6\, i\, k\, u\, h_{ty}-2\, m^2\, u^2\, h_{x,\bm{a}}\, V'+2\, m^2\, h_{x,\bm{a}}\, V\nonumber\\
&-2\, i\, u^2\, \omega\,  h_{x,\bm{s}}'+2\, i\, u^2\, \omega\,  h_{x,\bm{a}}'-2\, u\, h_{tt}'+6\, h_{tt}-2\, i\, u\, \omega\,  h_{x,\bm{a}}-6\, h_{x,\bm{a}}=0\\
&-6\, h_{tt}+u\, (u\, f'\, h_{x,\bm{s}}'-2\, f\, h_{x,\bm{s}}'-2\, i\, k\, m^2\, u^3\, \delta \phi_y\, V''+i\, k\, u\, h_{ty}'-2\, i\, k\, h_{ty}\nonumber\\
&+2\, m^2\, u^3\, h_{x,\bm{s}}\, V''-u\, h_{tt}''+4\, h_{tt}'+2\, i\, u\, \omega\,  h_{x,\bm{s}}'-2\, i\, \omega\,  h_{x,\bm{s}})=0\\
&k\, u \left(h_{x,\bm{s}}'+h_{x,\bm{a}}'\right)-i\, u\, \left(2\,m^2\, \delta \phi_y'\, V'+h_{ty}''\right)+2\, i\, h_{ty}'=0\\
&h_{x,\bm{s}}''=0.
\end{align}

\subsubsection{Time dependent bulk equations}\label{app:HMGtime}
%%%%%%%%%%%%%%%%%%%%%%%%%%%%%%%%%%%%%%%%%%%%%%%%%%
Using the characteristic formulation  the equations of motion are,
\begin{eqnarray}
&&S{}^{''} +\frac{2}{z}S{}^{'}+\frac{{H{}^{'}}^2 }{4}S=0,\label{constraint}\\
&&{d_+S}{}^{'} +\frac{S{}^{'} }{S } {d_+S}=\frac{S}{2 z^2} \left(-3+m^2 2^N  \left(\frac{\cosh (H)}{S^2}\right)^N\right),\label{eq-dpS}\\
&&{d_+H}{}^{'} +\frac{ S^{'} }{S}d_+H=-\frac{H{}^{'} {d_+S} }{S }-m^2 \frac{2^N N  \tanh (H) {\cosh^N (H)}}{z^2 {S^{2N}}},\label{eq-dpH}\\
&&A{}^{''}+\frac{2 }{z}A{}^{'}= \frac{{d_+H}  H{}^{'} }{z^2 }-\frac{4S{}^{'}  {d_+S}  }{z^2 S^2}+m^2 \frac{2^{N+1} N  \cosh^N (H)}{z^4S^{2N}},\label{eq-A}\\
&&4{d_+^2S} + {d_+H}^2 S+2 z^2 A{}^{'}d_+S =0,\label{eq-dppS}
\end{eqnarray}
where $d_+\mathcal{A}:= \dot{\mathcal{A}}-\frac{A}{z^2}\mathcal{A}^{'}$, and prime stands for derivative with respect to radial coordinate $z$ and use dot for derivative with respect to time $t$. Note that $X={2 \cosh (H)}/{S^2}$. 

 We make the following re-definitions of the metric functions
\begin{eqnarray}
&&A=\frac{1+\tilde{A}\ u}{u^2},\quad S=\frac{s_0+\tilde{S}\ u}{u},
\quad H=h_0+\tilde{H}\ u,\quad d_+S=\frac{\widetilde{d_+S}}{u^2},\quad d_+H=\frac{\dot{h_0}}{2}+\widetilde{d_+H}\ u.\nonumber
\end{eqnarray}

We find the following near boundary expansion
\begin{eqnarray}
&&A=\frac{1}{u^2}+\frac{2(s_1-\dot{s_0})}{s_0 \ u}+\left( \frac{s_1^2}{s_0^2}-\frac{2\dot{s_1}}{s_0}-\frac{3\dot{h_0}^2}{4} \right)+a_3 u+\mathcal{O}(u^2),\nonumber\\
&&S=\frac{s_0}{u}+s_1-\frac{s_0\dot{h_0}^2}{8}u+\frac{s_1 \dot{h_0}^2}{8} u^2+\mathcal{O}(u^3),\quad H=h_0+\dot{h_0}u-\frac{s_1 \dot{h_0}}{s_0}u^2+h_3 u^3+\mathcal{O}(z^4).\nonumber
\end{eqnarray}
accompanied by the Ward identity,
\begin{eqnarray*}
 &&\dot{a_3}+\frac{3 a_3 \dot{s_0}}{s_0}-\dot{h_0}{}^2 \left(\frac{\dot{s_0}{}^2-3 s_1{}^2}{2 s_0{}^2}+\frac{\ddot{s_0}}{2 s_0}\right)-\frac{3 \dot{h_0} \ddot{h_0} \dot{s_0}}{2 s_0}+\frac{3}{8} \dot{h_0}{}^4-\frac{3}{2} h_3 \dot{h_0}-\frac{1}{2} \dddot{h_0} \dot{h_0}=0.
\end{eqnarray*}
Using standard holographic renormalization, the boundary stress tensor reads
\begin{eqnarray}\label{stress-tensor}
&&T_{tt} = -a_3,\\
&&T_{xx} = -\frac{{a_3}}{2} - \frac{3  \gamma  {s_1}^2 \dot{\gamma}}{2 \left(1- \gamma ^2\right)^{3/2}}+\frac{3 h_3}{2}-\frac{ \gamma  \left(4 \left( \gamma ^2-1\right)^2 \dddot{\gamma}+3 \left(4  \gamma ^2+1\right) \dot{\gamma}^3-16  \gamma  \left( \gamma ^2-1\right) \dot{\gamma} \ddot{\gamma}\right)}{8 \left( \gamma ^2-1\right)^3}  \gamma,\nonumber \\
&&T_{xy} = -\frac{a_3}{2}  \gamma \!-\! \frac{3 {s_1}^2 \dot{\gamma}}{2 \left(1- \gamma ^2\right)^{3/2}}\! +\! \frac{3 {h_3}}{2}\! -\! \frac{4 \left( \gamma ^2-1\right)^2 \dddot{\gamma}+3 \left(4  \gamma ^2+1\right) \dot{\gamma}^3-16  \gamma  \left( \gamma ^2-1\right) \dot{\gamma} \ddot{\gamma}}{8 \left( \gamma ^2-1\right)^3}.\label{eq:readoffTXY}
\end{eqnarray}
The shear stress to the strain $h_{xy}(t)=\gamma(t)$ is given by the expectation value of $T_{xy}(t)$.
\chapter{Entanglement entropy and $T\bar T$}\label{app:EETT}
In this section, we show the entanglement entropy calculation and the comparison to holography for general dimensions. This appendix is taken from my publication \cite{Grieninger:2019zts}.
\subsection*{$d=2$}
We have
\begin{equation}
        \omega_2=\frac{1+\bm{s}\sqrt{\eta+\frac{c\,\lambda_2}{3\pi\,R^2}}}{4\,\lambda_2}\label{omegad2}
\end{equation}
where we denote the sign of the square root of the $T\bar T$ deformation by $\bm{s}$. With $\omega_2$ at hand, we may compute the partition function of the deformed CFT using eq. \eqref{partf}. As argued in the previous section, we choose the integration constant so that $\log Z_{S^2}(R=0)=0$. This yields
\begin{equation}
    \log Z_{S^2}=-\frac{1}{3\lambda_2}\left(c\bm{s}\,\arcsinhh\!\left(\frac{\sqrt{3\,\pi}\,R}{\sqrt{c\,\lambda_2}}\right)\lambda_2+\eta R\left(3R\pi+\bm{s}\,\sqrt{3\pi}\sqrt{\eta\,3R^2\,\pi+c\,\lambda_2}\right)\right).
\end{equation}
The entanglement entropy for two antipodal points on the sphere follows from the partition function via eq. \eqref{EEreplica} (for the negative sign of the square root)
\begin{equation}
    S_{EE}=\frac{c}{3}\,\arcsinhh\left(\frac{\sqrt{3\,\pi}\,R}{\sqrt{c\,\lambda_2}}\right).\label{EEd222}
\end{equation}
In two dimensions, we may calculate the cutoff independent renormalized entanglement entropy immediately from the knowledge of the derivative of the partition function.
 Plugging \eqref{omegad2} into eq. \eqref{partf} and combining eq. \eqref{EEreplica} and \eqref{runningC}, we find the renormalized entanglement entropy which plays the role of the running $\mathcal C$-function in RG flow as
\begin{equation}
    S_\text{R,EE}=c\left(9\,\eta+\frac{3\,c\,\lambda_2}{R^2\,\pi}\right)^{-1/2}.
\end{equation}
   \subsubsection*{Comparison to the result from holography}
The entanglement entropy from holography is given by eq. \eqref{holoEEd2}. In order to compare to the field theory results, we use the dictionary relating the holography parameters with the field theory ones. This is done by  $4\pi l/\ell_p=c/3$ and $c\,\lambda_2=3\,\pi\,L^2$ 
\begin{equation}
S_\text{EE}=\frac c3\,\arcsinhh\left(\frac{\sqrt{3\,\pi}\,R}{\sqrt{c\,\lambda_2}}\right),
\end{equation}
with the corresponding renormalized entanglement given by the $R$-derivative $\mathcal C=dS/dR$
\begin{equation}
   S_\text{R,EE}= c\left(9\,\eta+\frac{3\,c\,\lambda_2}{R^2\,\pi}\right)^{-1/2}.
\end{equation}
In $d=2$ dimensions, we find the EE from field theory matches the entanglement entropy from holography exactly for $\eta=1,\bm{s}=-1$ (AdS) and $\eta=-1,\bm{s}=-1$ (dS).

\subsection*{$d=3$}
In higher dimensions, the computation is very similar to $d=2$ and we only display the relevant steps; we may read off $\omega_3$ from eq. \eqref{omegaeq}
\begin{equation}
    \omega_3=\frac{R^2+3\,t_3+R\,\bm{s}\,\sqrt{\eta\,R^2+6\,t_3}}{3\,R^2\lambda_3}.
    \end{equation}
It is straightforward to determine the corresponding partition function, given by
    \begin{equation}
    \log Z_{S^3}=-\frac{2\pi^2\,(R^3+9\,R\,t_3+\eta\bm{s}\,(\eta\,R^2+6\,t_3)^{3/2})}{3\,\lambda_3}+\eta\,\bm{s}\frac{4\sqrt{6}\,\pi^2\,t^{3/2}}{3\,\lambda_3}.
\end{equation}
The second term is chosen to ensure $\log Z(R=0)=0$.
    Finally, we find with $\eta^2=1$ and eq. \eqref{EEreplica} and the negative sign of the square root
    \begin{equation}
        S_{EE}=\frac{4\,\pi^2\,t_3}{\lambda_3}\left(- R-\eta\,\sqrt{6t_3}+\eta\,\sqrt{\eta\,R^2+6\,t_3}\right).
    \end{equation}
The scheme independent renormalized entanglement entropy is obtained from the entanglement entropy by using \eqref{runningC} and reads in $d=3$ dimensions
   \begin{equation}
S_{\text{R,EE}}=\frac{4\, \pi^2\,\eta\,t_3^{3/2}}{ \lambda_3 } \left(\sqrt{6}-\frac{6 \sqrt{t_3}}{\sqrt{\eta\, R^2+6\, t_3}}\right).
   \end{equation}
   \subsubsection*{Comparison to the result from holography}
   The entanglement entropy from holography is given by eq. \eqref{holoEEd3} and may be expressed in terms of field theory quantities using eq. \eqref{gravityfieldtheory} (with $6\,\lambda_{3}=\ell_p^{2}\,\sqrt{6\,t_3}, \,L=\sqrt{6\,t_3}$)
   \begin{equation}
      S_{EE}= \frac{4\,\pi^2\,t_3}{\lambda_3}\,\eta\,\left(-\sqrt{6\,t_3}+\sqrt{6\,t_3+\eta\,R^2}\right).
   \end{equation}
   We see that the field theory calculation and the results from holography match up to a scheme dependent area term $\sim -4\,t_3\pi^2\,R/\lambda_3$. We obtain the exact same contribution from the Wald entropy associated with the counterterms given in eq. \eqref{wald1} which yields (in field theory variables) exactly  $\sim -4\,t_3\pi^2\,R/\lambda_3$. The entanglement entropies on both sides match, if the contributions of the counterterms -- which have been added to the field theory side -- are also taken into account in the gravitational theory.
   Similar to the literature, we may compare scheme independent quantities aka the renormalized entanglement entropy.
From the entanglement entropy, we immediately obtain the renormalized entanglement entropy by using eq. \eqref{runningC}
    \begin{equation}
      S_{\text{R,EE}}= \frac{4\, \pi^2\,\eta\,  t_3^{3/2}}{\lambda _3} \left(\sqrt{6}-\frac{6\,\sqrt{t_3}}{\sqrt{\eta  R^2+6\,t_3}}\right).
    \end{equation}
    We see that the results from holography and field theory perfectly match one another for the negative sign of the square root $\eta=1,\bm{s}=-1$ (AdS) and $\eta=-1,\bm{s}=-1$ (dS). 
\subsection*{$d=4$}
In $d=4$ we have using eq. \eqref{omegaeq}
\begin{equation}
\omega_4=\frac{3\,\left(R^2+8\,t_6+R\,\bm{s}\sqrt{\eta\,R^2+16\,t_4}\right)}{8\,R^2\,\lambda_4}.
\end{equation}
    We can compute the sphere partition function by integrating with respect to $R$, where we fix the integration constant by demanding that $\log Z_{S^d}(R=0)=0$
\begin{align}
    \log Z_{S^4}=&-\frac{\pi^2}{\lambda_4}\left(R\left(R^3+16\,R\,t_4+R^2\,\bm{s}\,\sqrt{\eta\,R^2+16\,t_4}+\eta\,8\,\bm{s}\,t_4\sqrt{\eta\,R^2+16\,t_4}\right)\right.\nonumber\\&\left.-128\,\eta\,\bm{s}\,t_4^2\,\arcsinhh\left(\frac{R}{4\,\sqrt{t_4}}\right)\right).
\end{align}
We obtain the entanglement entropy by using the replica trick \eqref{EEreplica}. This gives us
\begin{equation}
S_{4,\text{EE}}=\frac{8\pi^2\,t_4}{\lambda_4}\left(R\left(-R+\eta\,\sqrt{\eta\, R^2+16\,t_4}\right)-16\,\eta\,t_4\,\arcsin(h)\left(\frac{R}{4\,\sqrt{t_4}}\right)\right).\label{EEEE4}
\end{equation}
 In $d=4$ dimensions, the renormalized entanglement entropy follows from eq. \eqref{EEEE4} with eq. \eqref{runningC}
   \begin{equation}
S_{\text{R,EE}}= \frac{128\, \pi ^2\, R^3\,  t_4^2}{\lambda_4  \left(\eta\, R^2+16 t_4\right)^{3/2}}.
   \end{equation}
   \subsubsection*{Comparison to the result from holography}
  In holography, the entanglement entropy in $d=4$ is given by eq. \eqref{holoEEd4} which reads in field theory quantities by relating $8\,\lambda_4=\ell_p^{3}\,\sqrt{16\,t_4}, \,L=\sqrt{16\,t_4}$
  \begin{equation}
       S_\text{EE}=\frac{8\,\pi^2\,t_4}{\lambda_4}\,\eta\,\left(R\sqrt{\eta\,R^2+16\,t_4}-16\,t_4\,\arcsinhh\frac{R}{4\,\sqrt{t_4}}\right).
   \end{equation}
   Again, this matches exactly our field theory computation up to a scheme dependent area term $-8\pi^2 R^2t_4/\lambda_4$ for the negative sign of the the square root. The area term with the negative sign comes from adding counterterms to our action. If we also consider the contributions of the counterterms in the gravitational theory eq. \eqref{wald1}, we see that we observe the exact same term there and thus the results of both sides match.
From the holographic entanglement entropy, we may derive the scheme independent entanglement entropy using eq. \eqref{runningC}
\begin{equation}
  S_{\text{R,EE}}=\frac{128 \pi ^2 R^3 t_4^2}{\lambda _4 \left(\eta\,R^2+16 t_4\right){}^{3/2}}
\end{equation}
We see that the renormalized entanglement entropies from field theory and holography in $d=4$ match perfectly for $\eta=1,\bm{s}=-1$ (AdS) and $\eta=-1,\bm{s}=-1$ (dS).

\subsection*{$d=5$}
In $d=5$, the counterterm proportional to $c_d^{(3)}$ contributes for the first time. We find $\omega_5$ from eq. \eqref{omegaeq}
\begin{equation}
    \omega_5=\frac{30\,R^2\,t_5-225\,t_5^2\,R^3\left(R+\bm{s}\sqrt{\eta R^2+30\,t_5}\right)}{5\,R^4\,\lambda_5}.
\end{equation}
With $\omega_5$, we may compute the partition function by integrating eq. \eqref{partf} with respect to $R$ which results in
\begin{align}
\log Z_{S^5}=&-\frac{\pi^3}{5\,\lambda}\,\left(20\,\eta\,R^2\,\bm{s}\,t_5\sqrt{\eta\,R^2+30\,t_5}+1200\,\bm{s}\,t_5^2\left(\sqrt{30\,t_5}-\sqrt{\eta\,R^2+30\,t_5}\right)\right.\nonumber\\ 
& \left. +\left(2\,R^5+50\,R^3\,t_5-1125\,R\,t_5^2+2\,R^4\,\bm{s}\,\sqrt{\eta\,R^2+30\,t_5}\right)\right),
\end{align}
where we fixed the integration constant so that $\log Z_{S^5}(R=0)=0$. The entanglement entropy follows from the partition function using eq. \eqref{EEreplica}
\begin{align}
  S_{5,\text{EE}}=\frac{4\pi^3\,t_5}{\lambda_5}  \left(-R^3\,+45\,R\,t_5+\eta\,R^2\,\sqrt{\eta\,R^2+30\,t_5}+60\,t_5\left(\sqrt{30\,t_5}-\sqrt{\eta\,R^2+30\,t_5}\right)\right).
\end{align}
 In $d=5$ dimensions, we may compute the renormalized entanglement entropy using \eqref{runningC}
   \begin{equation}
S_{\text{R,EE}}=-\frac{240\, \pi ^3\, t_5^{5/2}\left(900\,t_5^{3/2}\!\!-30t_5\,\sqrt{900\,t_5+30\,\eta\,R^2}+\eta\,R^2\!\left(45\,\sqrt{t_5}-\sqrt{900t_5+\eta\,R^2}\right)\right)}{ \lambda_5  \left(\eta\, R^2+30 t_5\right)^{3/2}}.
   \end{equation}
 
 \subsubsection*{Comparison to the result from holography}
 To compare with the field theory result, we rewrite the result from holography \eqref{holoEEd5} with the dictionary $10\lambda_d=\ell_p^{4}\,\sqrt{30\,t_5}, \,L=\sqrt{30\,t_5}$ in field theory quantities
\begin{equation}
S_\text{EE}=\frac{4\,\pi^3\,t_5}{\lambda_5}\,\left(2\,(30\,t_5)^{3/2}+(\eta\,R^2-60\,t_5)\,\sqrt{30\,t_5+\eta\,R^2}\right).
\end{equation}
We see that the result from holography matches the calculation from field theory up to the scheme dependent terms $\sim 4\pi^3\,t_5/\lambda_5\,(-R^3\,+45\,R\,t_5)$. However, taking the contributions of the counterterms in the gravitational theory into account, we see find the exact same contribution to the entanglement entropy as observed in eq. \eqref{wald4}. The results of both sides hence match. For the sake of completeness, we calculate the renormalized entanglement entropy by using eq. \eqref{runningC}
   \begin{equation}
     S_{\text{R,EE}}=\frac{240 \pi ^3 t_5^2}{\lambda _5 \left(\eta\,  R^2+30\, t_5\right){}^{3/2}} \left( \sqrt{30\,t_5}\,\sqrt{\eta\,  R^2+30\, t_5}\left(30\, t_5 + \eta\,  R^2 \right)-45 \eta  R^2 t_5-900 t_5^2\right),
   \end{equation}
we see that the scheme dependent terms vanish and the results from field theory and holography agree perfectly for $\eta=1,\bm{s}=-1$ (AdS) and $\eta=-1,\bm{s}=-1$ (dS). 

\subsection*{$d=6$}
In $d=6$, $\omega_6$ is given by eq. \eqref{omegaeq}
\begin{equation}
    \omega_6=\frac{5\left(R^4+24\,R^2\,t_6-288\,t_6^2+R^3\,\bm{s}\,\sqrt{\eta\,R^2+48\,t_6}\right)}{12\,R^4\,\lambda_6}.\label{o6}
\end{equation}
The partition function follows by inserting eq. \eqref{o6} into eq. \eqref{partf} and integrating with respect to $R$  
\begin{align*}
\log Z_{S^6}=&-\frac{4 \pi ^3}{9 \lambda_6 } \left(R \left(R^5+36\,R^3\,t_6+R^4\,\bm{s}\sqrt{\eta\,R^2+48\,t_6}-864\,\bm{s}\,t_6^2\,\sqrt{-\eta\,R^2+48\,t_6}\right)\right. \nonumber\\
&\left.-864\,R\,t_6^2+\eta\,12\,R^2\,\bm{s}\,t_6\,\sqrt{\eta\,R^2+48\,t_6}+41472\,\bm{s}\,t_6^3\, \arcsinhh \left(\frac{R}{4 \sqrt{3\,t_6} }\right)\right),
\end{align*}
where we chose the integration constant so that $\log Z_{S^6}(R=0)=0$. The entanglement follows from the partition function by eq. \eqref{EEreplica}
\begin{align}
    S_{EE}=&-\frac{16\,\pi^3\,t_6}{3\lambda_6}\left(R\left(R^3-48\,R\,t_6-\eta\,R^2\sqrt{\eta\,R^2+48\,t_6}+72\,t_6\,\sqrt{\eta\,R^2+48\,t_6}\right)\right.\nonumber\\
    &\left.-3456\,t_6^2\,\arcsinhh\left(\frac{R}{4\,\sqrt{3\,t_6}}\right)\right).
\end{align}
In $d=6$ dimensions, the renormalized entanglement entropy reads (using eq. \eqref{runningC})
   \begin{equation}
 S_{\text{R,EE}}=\frac{18432\, \pi^3\, R^5\, t_6^3}{\lambda  \left(\eta\,R^2+48 t_6\right)^{5/2}}.
   \end{equation}
   
  \subsubsection*{Comparison to the result from holography}
  The entanglement entropy from holography \eqref{holoEEd6} reads in $d=6$ in field theory quantities $12\,\lambda_6=\ell_p^{5}\,\sqrt{48\,t_6}$ and $L=\sqrt{48\,t_6}$
  \begin{equation}
      S_\text{EE}=\frac{8\,\pi^3\,t_6}{3\,\lambda_6}\left(R\,\sqrt{48\,t_6+\eta\,R^2}\,(2\,\eta\,R^2-144\,t_6)+6912 t_6^2\,\arcsinhh\left(\frac {R}{4\, \sqrt{3\,t_6}}\right)\right)\!.
  \end{equation}
       The entanglement entropy from field theory matches the result from holography up to the usual area term $\sim16\,\pi^3\,t_6\,R^4/(3\lambda_6)$ and a scheme dependent term $\sim 256\, \pi^3\, R^2 t_6^2/\lambda_6$.
  The exact same terms arise in the gravitational theory too if we also take the counterterms into account there. The contributions are calculated in eq. \eqref{wald4} and are an exact match to the missing terms. We thus conclude that the entanglement entropies of both sides match. For comparison with similar results in the literature, we are looking at the renormalized entanglement entropy in $d=6$. We find a perfect match between field theory and the result from holography given by
  \begin{equation}
  S_{\text{R,EE}}=   \frac{18432\, \pi^3\, R^5\, t_6^3}{\lambda _6\, \left(\eta\,R^2+48 \,t_6\right){}^{5/2}},
  \end{equation}
 for $\eta=1,\bm{s}=-1$ (AdS) and $\eta=-1,\bm{s}=-1$ (dS).

\bibliographystyle{fullsort}
\bibliography{references}

\chapter*{Acknowledgments}\vspace{-0.5cm}
In the first place, I would like to express my gratitude to my advisor Martin Ammon for his continuous support and over many years (starting with my Bachelor thesis). I am deeply grateful for the countless hours of (patient) explanations, his willingness to discuss problems at any time and his enthusiasm about physics.\vspace{0.35cm}
\\ 
Secondly, I would like to thank Andreas Karch for his warm hospitality during my (not so short) research stay at the University of Washington in Seattle. I greatly appreciated his constant encouragements, enthusiasm while talking about physics \& inspiring insights.\\ \\
I am particularly grateful to...\vspace{0.35cm} \\
.. Matteo for being so impatient...\vspace{0.35cm} \\ 
.. Markus until he sadly left us\vspace{0.35cm} \\ 
.. Martin, S\'ean and Markus (uh here he is again) for proofreading ..\vspace{0.35cm} \\ 
.. Marc for IT support and Se\'an for teaching me how to make coffee \vspace{0.35cm} \\
.. Laura, Julia, Maria, Melanie, Jule, Lisa, Maureen, Hannah, Gloria, Robbie, Jeff, Martha, Fernando, Mar\'ia,
 Friedi for various life lessons; Alona the star of blue stars\vspace{0.35cm} \\
.. Mary Ann and Gerd for great inspiration\vspace{0.35cm} \\
..\! my parents Erika and Ekkehard, my brother  Christian and my friends for endless support\vspace{0.35cm} \\
.. Violeta \& Renly, joy; bliss \& unconditional support. \\ \\ 
Finally, I gratefully acknowledge financial by the  \textit{Fulbright Visiting Scholar Program}, sponsored by the US Department of State and the German-American Fulbright Commission;  by the DAAD  for a \textit{Jahresstipendium f\"ur Doktorandinnen und Doktoranden}; by the Deutsche Forschungsgemeinschaft (project GRK1523/2), and by the University of Jena. 
\thispagestyle{empty}
\chapter*{Ehrenwörtliche Erklärung}
\thispagestyle{empty}
Ich erkläre hiermit ehrenwörtlich, dass ich die vorliegende Arbeit selbständig, ohne unzulässige Hilfe Dritter und ohne Benutzung anderer als der angegebenen Hilfsmittel und Literatur angefertigt habe. Die aus anderen Quellen direkt oder indirekt übernommenen Daten und Konzepte sind unter Angabe der Quelle gekennzeichnet.

Die nachstehend aufgeführten Personen haben mich bei der Auswahl und Auswertung unentgeltlich unterstützt.
\begin{itemize}
    \item Prof. Dr. Martin Ammon, Dr. Matteo Baggioli, Se\'an Gray, Dr. Akash Jain und Dr. Hesam Soltanpanahi (Kapitel 3 basierend auf~\cite{Ammon:2019apj,Baggioli:2019abx,Baggioli:2019mck,Ammon:2020xyv})
    \item Prof. Dr. Martin Ammon, Juan Hernandez, Prof. Dr. Matthias Kaminski, Roshan Koirala, Dr. Julian Leiber und Dr. Jackson Wu (Kapitel 4 basierend auf~\cite{Ammon:2020rvg})
    \item Hao Geng und Prof. Dr. Andreas Karch (Kapitel 5 basierend auf~\cite{Geng:2019bnn})
\end{itemize}

Weitere Personen waren an der inhaltlich-materiellen Erstellung der vorliegenden Arbeit
nicht beteiligt. Insbesondere habe ich hierfür nicht die entgeltliche Hilfe von Vermittlungs- bzw. Beratungsdiensten (Promotionsberater oder andere Personen) in Anspruch genommen. Niemand hat von mir unmittelbar oder mittelbar geldwerte Leistungen für Arbeiten erhalten, die im Zusammenhang mit dem Inhalt der vorgelegten Dissertation stehen.

Die Arbeit wurde bisher weder im In- noch im Ausland in gleicher oder ähnlicher Form einer anderen Prüfungsbehörde vorgelegt.

Die geltende Promotionsordnung der Physikalisch-Astronomischen Fakultät ist mir bekannt.

Ich versichere ehrenwörtlich, dass ich nach bestem Wissen die reine Wahrheit gesagt und nichts verschwiegen habe.
\vfill
Jena,  \hfill \\ \\ \vspace{2cm} \hfill Sebastian Grieninger
\end{document}